\documentclass{amsart}
\usepackage{amssymb,latexsym}
\newtheorem{definition}{Definition}
\newtheorem{theorem}{Theorem}   

\newtheorem{corollary}{Corollary}  
\newtheorem{proposition}{Proposition}

\begin{document} 

\title[Intertwiners and the vector-valued big $q$-Jacobi transform]{Intertwiners of $U'_q\bigl(\widehat{sl}(2)\bigr)$-representations\\
and the vector-valued big $q$-Jacobi transform}
\author{R.M. Gade}
\address{Prangerlstra\ss e 19, 81247 Munich, Germany}
\email{renate.gade@t-online.de}
\thanks{This Research project is supported by M. Krautg\"artner.}

\begin{abstract}
Linear operators $R$ are introduced on tensor products of evaluation
modules of $U'_q\bigl(\widehat{sl}(2)\bigr)$ obtained from the complementary and
strange series representations. The operators $R$ satisfy the intertwining
condition on finite linear combinations of the canonical basis elements of
the tensor products. Infinite sums associated with the action of $R$ on six
pairs of tensor products are evaluated. For two pairs, the sums are related
to the vector-valued big $q$-Jacobi transform of the matrix elements defining
the operator $R$. In one case, the sums specify the action of $R$
on the irreducible representations present in the decomposition of the
underlying indivisible sum of $U_q\bigl(sl(2)\bigr)$-tensor products. 
In both cases, bilinear summation formulae for the matrix 
elements of $R$ provide a generalization of the unitarity 
property. 
\end{abstract}

\maketitle

\section{Introduction}

The strange series representation of the quantum algebra $U_q\bigl(su(1,1)
\bigr)$ has no classical analog among the irreducible
$\ast\,$-representations of the Lie-algebra $su(1,1)$.
Tensor products of the discrete series representations $\pi^+$ and $\pi^-$ 
of $U_q\bigl(su(1,1)\bigr)$ as well as the principal unitary series $\pi^P$
and the complementary series $\pi^C$ have been analyzed in \cite{KJ}-\cite{groene2} (see also references given therein). 
Clebsch-Gordon and Racah coefficients are specified in \cite{KJ}, 
\cite{groene} and \cite{groene1}. Detailed investigations of the tensor
products give rise to various summation formulae for $q$-polynomials
(\cite{groene} and \cite{groene1}).
$U_q$-representations involving the strange series have been considered
in \cite{groene2}, where the decompositions of 
the "indivisible" sums of tensor products
$(\pi^C\otimes\pi^C)\oplus(\pi^S\otimes\pi^S)$ and $(\pi^P\otimes\pi^P)\oplus
(\pi^P\otimes\pi^P)$ are determined. The "indivisible" representations
can be viewed as quantum analogs of the tensor products of 
the complementary series or the principal unitary series of $su(1,1)$. Among
further indivisible sums, the pairs
$(\pi^-\otimes\pi^+)\oplus(\pi^S\otimes\pi^S)$ and $(\pi^-\otimes\pi^S
)\oplus(\pi^S\otimes\pi^+)$ are studied in context with the corepresentation
theory of the locally compact quantum group analog  
of the normalizer of $SU(1,1)$ in $SL(2,\mathbb C)$.

The present study deals with tensor products of evaluation modules of $U'_q\bigl(
\widehat{sl}(2)\bigr)$ associated with the complementary series representations
$\pi^C$ and the strange series representations $\pi^S$. Their $U'_q\bigl(
\widehat{sl}(2)\bigr)$-structure entails a complex parameter $z^{\frac{1}{2}}$ referred to as spectral parameter. The representations are constructed by pulling
back infinite-dimensional representations of the quantum algebra $U_q\bigl(sl(2)\bigr)$ by means of the evaluation homomorphism established in
\cite{jim}. Various linear operators $R$ are 
introduced in terms of matrix elements determined by the intertwining property
formulated with respect to the canonical basis elements of the tensor products. 
In this sense, the operators $R$ are referred to as intertwiners.
The first set of intertwiners maps  the tensor products corresponding to
$\pi^C\otimes\pi^C$ and $\pi^S\otimes\pi^S$ onto tensor products of the same 
type but with inverse spectral parameter. A second set of intertwiners relates
the tensor products corresponding to $\pi^C\otimes\pi^S$ and $\pi^S\otimes\pi^C$. For both sets, the 
$U'_q\bigl(\widehat{sl}(2)\bigr)$-representations are specified by the 
most general choice of parameters compatible with the intertwining property.
In each case, the intertwining property admits two solutions for $R$ 
specified by the sets $\bigl\{R^{k,k+r^*}_{l,l+r^*}\bigr\}_{k,r,l\in\mathbb Z}$ and $\bigl\{\check R^{k,k+r^*}_{l,l+r^*}\bigr\}_{k,r,l\in\mathbb Z}$ 
of matrix elements. A single element is presented in terms of two nonterminating
$_4\phi_3$-series.

The action of the linear operators $R$ on finite linear combinations of the canonical basis elements with coefficients essentially given by big 
$q$-Jacobi functions \cite{bigJ} is examined.
With a suitable restriction of the spectral parameter $z^{\frac{1}{2}}$, the limit of infinite linear combinations gives rise to the bilateral sums 
$\boldsymbol{\tau}^{(r,k)\pm}$ and $\boldsymbol{\tau}^{(r,k)}$ associated
with the set $\bigl\{R^{k,k+r^*}_{l,l+r^*}\bigr\}$, $\bigl\{\check R^{k,k+r^*}_{
l,l+r^*}\bigr\}$ or certain combinations of both. In general, the sums 
$\boldsymbol{\tau}^{(r,k)\pm}$ and $\boldsymbol{\tau}^{(r,k)}$ obey inhomogeneous recurrence relations with respect to the parameters  
$r$ and $k$. However, particular linear combinations 
${\Xi}^{(r,k)\pm}$ and ${\Xi}^{(r,k)}$
of two sums are governed by homogeneous recurrence relations. Two types
of combinations characterized by a parameter $\beta$ taking the values
$1$ or $-1$ are distinguished. For $\beta=1$, tensor products of the same
type are assembled. If $\beta=-1$, the combinations decompose into
the vector-valued big $q$-Jacobi transforms \cite{vbigJac} of the 
elements $R^{k,k+r^*}_{l,l+r^*}$ and $\check R^{k,k+r^*}_{l,l+r^*}$.
In one case,  the linear combinations ${\Xi}^{(r,k)\pm}$ and 
${\Xi}^{(r,k)}$ with $\beta=-1$ have an interpretation related to the decompositions of "indivisible" 
representations found in \cite{groene2}. Viewing the intertwiners $R$ as  operators acting on $\mathcal T=(\pi^C\otimes\pi^C)\oplus(\pi^S\otimes\pi^S)$,  
the transforms describe their action onto an irreducible component in the 
decomposition of $\mathcal T$. 

A summation formula for the evaluation of $\Xi^{(r,k)\pm}$ is derived.
For a particular linear combination $r^{k,k+r^*}_{l,l+r^*}$  
of the matrix elements $R^{k,k+r^*}_{l,l+r^*}$ and 
$\check R^{k,k+r^*}_{l,l+r^*}$, the evaluation of the corresponding sums
$\Xi^{(r,k)\pm}$ and $\Xi^{(r,k)}$ provides simpler expressions. Moreover,  
the element $r^{k,k+r^*}_{l,l+r^*}$ and a further combination
$\mathring r^{k,k+r^*}_{l,l+r^*}$ are distinguished by their 
properties at a particular value $z_0$ of the squared spectral parameter.
In particular, 
at $z=z_0$ they are well-defined for intertwiners mapping tensor products 
of evaluation modules to tensor products with the same parameters.
Up to a normalization, for the same value of $z$ 
a certain linear combination $\bar r^{k,k+r^*}_{
l,l+r^*}$ of $r^{k,k+r^*}_{l,l+r^*}$ and $\mathring r^{k,k+r^*}_{l,l+r^*}$
reduces to $\delta_{k,l}$.
Referring to the nomenclature used in context with the construction of integrable vertex models (see \cite{fad} or \cite{miji}, for example), 
this property is termed the initial condition. At $z=z_0$, the single components $r^{k,k+r^*}_{l,l+r^*}$ and $\mathring r^{k,k+r^*}_{l,l+r^*}$ with $k\neq l$ 
involve a contribution by  terminating $_4\phi_3$-series.

The second part of the paper addresses the construction of inverse intertwiners.
Two linear combinations ${r_{\pm}}^{k,k+r^*}_{l,l+r^*}$ of the matrix elements
$r^{k,k+r^*}_{l,l+r^*}$ and $\mathring r^{k,k+r^*}_{l,l+r^*}$ are introduced.
In the case $\beta=-1$, suitable adjustments of the parameters characterizing the elements $R^{k,k+r^*}_{l,l+r^*}$ 
allow successive applications of the summation formulae for
the sums $\Xi^{(r,k)\pm}$ and $\Xi^{(r,k)}$. Here the subsequent action 
of a second intertwiner mapping back to the original tensor products is considered for $\beta=-1$.  
Specifying the first intertwiner by the set $\bigl\{{r_+}^{k,k+r^*}_{l,l+r^*}
\bigr\}$ and the second by $\bigl\{{r_-}^{k,k+r^*}_{l,l+r^*}\bigr\}$ or
vice versa, the sums $\boldsymbol{\sigma}^{[\pm](r,m)}$ describing the action of both are evaluated. 
Application of the vector-valued big $q$-Jacobi function
transform pair allows to deduce a quadratic bilateral summation formula
obeyed by the elements ${r_+}^{k,k+r^*}_{l,l+r^*}$ and ${r_-}^{k,k+r^*}_{l,l
+r^*}$. Upon adjustment by a simple normalization
the second intertwiner is found to invert the action of the first.
This result can be regarded as generalization of the unitarity property
required for the construction of integrable vertex models from finite-dimensional $U_q\bigl({sl}(2)\bigr)$-modules. 
Since the matrix elements ${r_{\pm}}^{k,k+r^*}_{l,l+r^*}$ satisfy an
initial condition, the generalized unitarity can be expected to present a special case of Yang-Baxter type relations. An account of the latter will 
be given in a separate publication.

The evaluation of the sums $\Xi^{(r,k)\pm}$ and $\Xi^{(r,k)}$
is achieved by means of various transformation properties and summation formulae
of the (bilateral) basic hypergeometric series \cite{GR} and the homogeneous
recurrence relations for ${\Xi}^{(r,k)\pm}$ and ${\Xi}^{(r,k)}$, provided that a
constraint is imposed on the spectral parameter $z^{\frac{1}{2}}$ and the
spectral value of the big $q$-Jacobi functions. Due to two particular relations
satisfied by the combinations ${\Xi}^{(r,k)}$, the condition can be released
sufficiently with respect to the spectral value to accommodate the complete spectrum contributing to the vector-valued big $q$-Jacobi transform. The first relation provides
a contiguous relation with respect to the spectral value. For its derivation, a variant of a procedure employed in \cite{gade1} proves
efficient.

The main results of this investigation are formulated by Theorem \ref{T:tausum2}
in subsection \ref{S:tauev} and by Proposition \ref{P:unitrel} and
Corollary \ref{C:RRsum} in subsection \ref{S:invint}.

The paper is organized as follows. Section \ref{S:def} summarizes the
definitions of basic hypergeometric series and properties of the Jacobi 
theta function. In section \ref{S:U}, the Chevalley basis
of $U'_q\bigl(\widehat{sl}(2)\bigr)$ and the modules corresponding to the
complementary and strange series are introduced. Section \ref{S:R} provides
the definition of the intertwiners $R$. Explicit expressions for the
matrix elements and some of their properties are given. Section \ref{S:tau} 
proceeds to the definition of the sums 
$\boldsymbol{\tau}^{(r,k)\pm}$ and $\boldsymbol{\tau}^{(r,k)}$ and the evaluation of $\Xi^{(r,k)\pm}$ and $\Xi^{(r,k)}$. 
In subsection \ref{S:coef}, the coefficients underlying the sums 
$\boldsymbol{\tau}^{(r,k)\pm}$ and $\boldsymbol{\tau}^{(r,k)}$
are introduced by expressions particularly suited 
to the latter purpose. Their reformulation in terms of big $q$-Jacobi functions is included. The definition of the sums $\boldsymbol{\tau}^{(r,k)\pm}$ and
$\boldsymbol{\tau}^{(r,k)}$ follows in subsection \eqref{S:taudef}, where the
linear combinations ${\Xi}^{(r,k)\pm}$ and ${\Xi}^{(r,k)}$ as well as their homogeneous recurrence relations 
are specified. Subsection \ref{S:bigJac} collects the rewritings of
the combinations ${\Xi}^{(r,k)\pm}$ and ${\Xi}^{(r,k)}$ with $\beta=-1$ in terms
of vector-valued big $q$-Jacobi transforms. Subsection \ref{S:tauev} presents
the evaluation of the sums $\Xi^{(r,k)\pm}$ associated with
the matrix elements $R^{k,k+r^*}_{l,l+r^*}$ and $\check R^{k,k+r^*}_{l,l+r^*}$. 
Then the result is applied to provide expressions for the sums 
$\Xi^{(r,k)\pm}$ and $\Xi^{(r,k)}$ corresponding to the combinations $r^{k,k+r^*}_{l,l+r^*}$. The special case relevant to the initial condition is addressed in section \ref{S:init}.
Section \ref{S:unit} deals with the successive action of intertwiners.
A short general consideration is found in subsection \ref{S:sucint}.
Evaluations of the sums $\Xi^{(r,k)\pm}$ and $\Xi^{
(r,k)}$ related to the elements $\mathring r^{k,k+r^*}_{l,
l+r^*}$ are obtained in subsection \ref{S:mtau}.   
The sums $\boldsymbol{\sigma}^{[\pm](r,m)}$
revealing the generalized unitarity property in the case $\beta=-1$
are investigated in subsection \ref{S:invint}. 

Appendix \ref{A:R} contains the proof of the intertwining property as well as further selected properties of the sets $\bigl\{R^{k,k+r^*}_{l,l+r^*}
\bigr\}$ and $\bigl\{\check R^{k,k+r^*}_{l,l+r^*}\bigr\}$.
For the linear combinations ${\Xi}^{(r,k)}$, the 
contiguous relations with respect to $z^{\frac{1}{2}}$ and the spectral values
of the big $q$-Jacobi functions are derived in Appendix \ref{A:thetacont}.

\section{Definitions}\label{S:def}

According to \cite{GR}, the basic hypergeometric series $_m\phi_n$ with $m\leq 
n+1$ and the bilateral basic hypergeometric series $_n\psi_n$ with base 
$\tilde q$ are defined by
\begin{equation}\label{E:phidef}
_m\phi_n\biggl(\genfrac{}{}{0pt}{}{a_1,\,a_2,\,\ldots,\,a_m}{b_1,\,b_2,\ldots,
b_n};\tilde q,w\biggr)=\sum_{t=0}^{\infty}\frac{(a_1,a_2,\ldots,a_m;\tilde q
)_tw^t}{(\tilde q,b_1,b_2,\ldots,b_n;\tilde q)_t}\Bigl[(-1)^t\tilde q^{\frac{
t(t-1)}{2}}\Bigr]^{n+1-m}
\end{equation}
and
\begin{equation}
\label{E:psidef}
_n\psi_n\biggl(\genfrac{}{}{0pt}{}{a_1,\,a_2,\,\ldots,\,a_n}{b_1,\,b_2,\ldots,
\,b_n};\tilde q,w\biggr)=\sum_{t=-\infty}^{\infty}\frac{(a_1,a_2,\ldots,a_n;
\tilde q)_tw^t}{(b_1,b_2,\ldots,\,b_n;\tilde q)_t},
\end{equation} 
where the $\tilde q$-shifted factorials are given by
\begin{equation*}
\begin{split}
(a_1,a_2,\ldots,a_n;\tilde q)_t&=(a_1;\tilde q)_t(a_2;\tilde q)_t\ldots(a_n;
\tilde q)_t,\\
(a;\tilde q)_t&=\begin{cases}
\frac{(a;\tilde q)_{\infty}}{(a\tilde q^t;\tilde q)_{\infty}},
\qquad(a;\tilde q)_{\infty}=\prod_{s=0}^{\infty}(1-a\tilde q^s)&\quad\text{for}\;t\geq0,\\
&\\
\frac{1}{\bigl(a\tilde q^{t};\tilde q\bigr)_{-t}}&\quad\text{for}\;t<0.
\end{cases}
\end{split}
\end{equation*}
The very well poised basic hypergeometric series with base 
$\tilde q$ is introduced by
\begin{multline*}
_{n+1}W_n(a_1;a_4,a_5,\ldots,a_{n+1};\tilde q,w)\equiv\\
{}_{n+1}\phi_n\biggl(\genfrac{}{}{0pt}{}{a_1,\,\tilde q\sqrt{a_1},\,-\tilde q 
\sqrt{a_1},\,a_4,a_5,\ldots,a_{n+1}}{\sqrt{a_1},\,-\sqrt{a_1},\,\tilde qa_1 
a_4^{-1},\,\tilde qa_1a_5^{-1},\ldots,\tilde qa_1a_{n+1}^{-1}};\tilde q,w\biggr)
\\
=\sum_{t=0}^{\infty}\frac{1-a_1\tilde q^{2t}}{1-a_1}\frac{(a_1,a_4,a_5,\ldots,
a_{n+1};\tilde q)_tw^t}{(\tilde q,\tilde qa_1a_4^{-1},\tilde qa_1a_5^{-1},
\ldots,\tilde qa_1a_{n+1}^{-1};\tilde q)_t}.
\end{multline*}
Throughout the paper, a fixed value of $\tilde q$ subject to $0<\tilde q<1$ is
assumed. The property 
\begin{equation}\label{E:I8}
\bigl(a\tilde q^{-t};\tilde q\bigr)_t=(-1)^ta^t\tilde q^{-t(t+1)}\bigl(a^{-1}
\tilde q;\tilde q\bigr)_t
\end{equation}
will be used without explicit mentioning.

The renormalized Jacobi theta function $\theta_{\tilde q}(x)=(x,x^{-1}
\tilde q;\tilde q)_{\infty}$ satisfies the property
\begin{equation}\label{E:thetaprop}
\theta_{\tilde q}(x\tilde q^t)=(-1)^tx^{-t}\tilde q^{\,-\frac{1}{2}t(t-1)}
\theta_{\tilde q}(x).
\end{equation}
Among the relations for its products denoted by $\theta_{q^2}(
x_1,x_2,\ldots,x_n)=\theta_{q^2}(x_1)\theta_{q^2}(x_2)\ldots\theta_{q^2}(x_n)$,
the simplest given in [\cite{GR}:ex.2.16(i)] reads
\begin{multline}\label{E:ex.2.16}
\theta_{q^2}\bigl(x\lambda,x\lambda^{-1},\mu\nu,\mu\nu^{-1}\bigr)-\theta_{q^2}
\bigl(x\nu,x\nu^{-1},\mu\lambda,\mu\lambda^{-1}\bigr)=\\
\mu\lambda^{-1}\theta_{q^2}\bigl(x\mu,x\mu^{-1},\lambda\nu,\lambda\nu^{-1}
\bigr).
\end{multline}

\section{The quantum affine algebra $U'_q\bigl(\widehat{sl}(2)\bigr)$}\label{S:U}

$U'_q\bigl(\widehat{sl}(2)\bigr)$ is the $\mathbb C$-algebra generated by 
$\{e_i,f_i,q^{\pm h_i}\}_{i=0,1}$ subject to the defining relations
\cite{Ji}, \cite{Dri}
\begin{equation}\label{E:Udef}
\begin{split}
&q^{\pm h_i}q^{\pm h_j}=q^{\pm h_j}q^{\pm h_i},\qquad q^{\pm h_i}q^{\mp h_j}=
q^{\mp h_j}q^{\pm h_i},\\
&q^{h_i}e_j=q^{a_{ij}}e_jq^{h_i},\qquad q^{h_i}f_j=q^{-a_{ij}}f_jq^{h_i},\\
&[e_i,f_j]=\delta_{i,j}\frac{q^{h_i}-q^{-h_i}}{q-q^{-1}},\\
\\
&e_j^3e_i-(1+q^2+q^{-2})(e^2_je_ie_j-e_je_ie_j^2)-e_ie_j^3=0,\\
&f_j^3f_i-(1+q^2+q^{-2})(f^2_jf_if_j-f_jf_if^2_j)-f_if_j^3=0,
\end{split}
\end{equation}
where $a_{00}=a_{11}=-a_{01}=-a_{10}=2$. Restriction of \eqref{E:Udef} to
$i,j=1$ yields the quantum algebra $U_q\bigl(sl(2)\bigr)$.
With the coproduct specified by
\begin{equation}\label{E:co}
\Delta(e_i)=e_i\otimes 1+q^{h_i}\otimes e_i,\quad\Delta(f_i)=f_i\otimes q^{-
h_i}+1\otimes f_i,\quad \Delta\bigl(q^{\pm h_i}\bigr)=q^{\pm h_i}\otimes q^{
\pm h_i},
\end{equation}
the antipode 
\begin{equation*}
S(e_i)=-q^{-h_i}e_i,\quad S(f_i)=-f_iq^{h_i},\quad S(q^{\pm h_i})=q^{\mp h_i}
\end{equation*}
and the counit given by $\epsilon(e_i)=\epsilon(f_i)=\epsilon(h_i)=0$, $\epsilon(1)=1$,
$U'_q\bigl(\widehat{sl}(2)\bigr)$ is equipped with a Hopf algebra structure.

On the $U'_q\bigl(\widehat{sl}(2)\bigr)$-modules $W^{(\epsilon,q^{2\sigma})}$
and $W^{(\epsilon,q^{2\sigma})*}$ with bases $\{w_l^{(\epsilon,q^{2\sigma})}\}_{
l\in\mathbb Z}$ and $\{w^{(\epsilon,q^{2\sigma})*}_l\}_{l\in\mathbb Z}$,
the representations $\pi_W:\,U'_q\bigl(\widehat{sl}(2)\bigr)\to\text{End}(W)$
are defined by
\begin{equation}\label{E:Wmod}
\begin{split}
q^{\pm h_1}w_l^{(\epsilon,q^{2\sigma})}&=q^{\mp h_0}w_l^{(\epsilon,q^{2\sigma})}
=q^{\mp2(l+\epsilon)}w_l^{(\epsilon,q^{2\sigma})},\\
(1-q^2)e_1w_l^{(\epsilon,q^{2\sigma})}&=(1-q^2)f_0w_l^{(\epsilon,q^{2\sigma})}=
-q^{-2(l+\epsilon)+3}s_{l-1}\bigl(q^{2\epsilon},q^{2\sigma}\bigr)w_{l-1}^{(
\epsilon,q^{2\sigma})},\\
(1-q^2)f_1w_l^{(\epsilon,q^{2\sigma})}&=(1-q^2)e_0w^{(\epsilon,q^{2\sigma})}_l
=s_l\bigl(q^{2\epsilon},q^{2\sigma})w_{l+1}^{(\epsilon,q^{2\sigma})}
\end{split}
\end{equation}
for $W=W^{(\epsilon,q^{2\sigma})}$ and
\begin{equation}\label{E:Wdual}
\begin{split}
q^{\pm h_1}w^{(\epsilon,q^{2\sigma})*}&=q^{\mp h_0}w^{(\epsilon,q^{2\sigma})*}
=q^{\pm 2(l+\epsilon)}w_l^{(\epsilon,q^{2\sigma})*},\\
(1-q^2)e_1w_l^{(\epsilon,q^{2\sigma})*}&=(1-q^2)f_0w_l^{(\epsilon,q^{2\sigma})*}
=q^2s_l\bigl(q^{2\epsilon},q^{2\sigma}\bigr)w_{l+1}^{(\epsilon,q^{2\sigma})*},\\
(1-q^2)f_1w_l^{(\epsilon,q^{2\sigma})*}&=(1-q^2)e_0w^{(\epsilon,q^{2\sigma})*}_l=
-q^{-2(l+\epsilon)+1}s_{l-1}\bigl(q^{2\epsilon},q^{2\sigma}\bigr)w_{l-1}^{
(\epsilon,q^{2\sigma})*}
\end{split}
\end{equation}
for $W=W^{(\epsilon,q^{2\sigma})*}$, where
\begin{equation*}
s_l(A,B)=\sqrt{\bigl(1+q^{2l+1}AB\bigr)\bigl(1+q^{2l+1}AB^{-1}\bigr)}
\end{equation*}
and $q^{2\sigma}=\pm q^{2u}$ with $u\in\mathbb R$.
As a $U_q\bigl(sl(2)\bigr)$-representation, $\pi_{W^{(\epsilon,q^{2t})*}}$
with $t\in\mathbb R$ corresponds to the strange series representation
$\pi^S_{\vert t\vert,\epsilon}$ if $\epsilon\in[0,1)$ and $t\neq0$
(see \cite{BK}) while $\pi_{W^{(\epsilon,-q^{2t})*}}$ corresponds to
the complementary series representation $\pi^C_{t-\frac{1}{2},\epsilon}$
if $\epsilon\in[0,\tfrac{1}{2})$ and $t\in(0,\tfrac{1}{2}-\epsilon)$
or $\epsilon\in(\tfrac{1}{2},1)$ and $t\in(0,\epsilon-\tfrac{1}{2})$.
The defining relations for $W^{(\epsilon+m,q^{2\sigma})}$ and $W^{(\epsilon+m,
q^{2\sigma})*}$ with $m\in\mathbb Z$ such that $\epsilon+m\in[0,1)$ are
obtained from the relations \eqref{E:Wmod} and \eqref{E:Wdual} by means of the
substitutions $w_l^{(\epsilon,q^{2\sigma})}\to w_{l-m}^{(\epsilon+m,q^{2\sigma}
)}$ and $w_l^{(\epsilon,q^{2\sigma})*}\to w_{l-m}^{(\epsilon+m,q^{2\sigma})*}$,
respectively. 
Here the modules $W^{(\epsilon,q^{2\sigma})}$ and $W^{(\epsilon,q^{2\sigma})*}$
will be considered for both $\epsilon$ and $u$ real and non-negative such that
$2n<2(\epsilon\pm u)+1<2(n+1)$ with $n\in\mathbb N_0$. Then the square roots
in \eqref{E:Wmod} and \eqref{E:Wdual} refer to the non-negative roots of real
non-negative numbers.

In the special case $q^{2\sigma}=-q^{2\epsilon-1}$, the modules defined by
\eqref{E:Wmod} and \eqref{E:Wdual} decompose (see Section 3 of \cite{gade1}).
The sets $\bigl\{w_l^{(\epsilon,-q^{2\epsilon-1})}\bigr\}_{l\in\mathbb Z_{\geq0}}$ and
$\bigl\{w^{(\epsilon,-q^{2\epsilon-1})*}_l\bigr\}_{l\in\mathbb Z_{\geq0}}$ provide a basis for a highest and a lowest weight module, respectively.
Moreover, the sets $\bigl\{w_l^{(\epsilon,-q^{2\epsilon-1})}\bigr\}_{l\in\mathbb Z_{<0}}$
and $\bigl\{w^{(\epsilon,-q^{2\epsilon-1})*}_l\bigr\}_{l\in\mathbb Z_{<0}}$ furnish a basis of a lowest and a highest weight module, respectively.

The representations $\pi_W:U'_q\bigl(\widehat{sl}(2)\bigr)
\to\text{End}(W)$ defined by \eqref{E:Wmod} and \eqref{E:Wdual} are
constructed from the corresponding $U_q\bigl(sl(2)\bigr)$-representations  
given by \eqref{E:Wmod} and \eqref{E:Wdual} by means of the Jimbo
homomorphism established in \cite{jim}.

Following [\cite{FR}:sect.4], an automorphism $D_{z^{\frac{1}{2}}}$ of
$U'_q\bigl(\widehat{sl}(2)\bigr)\otimes\mathbb C[z^{\frac{1}{2}},z^{-
\frac{1}{2}}]$ is introduced for the formal variable $z^{\frac{1}{2}}$ by
\begin{equation}\label{E:Dz}
D_{z^{\frac{1}{2}}}(e_i)=z^{\frac{1}{2}\delta_{i,0}}e_i,\qquad D_{z^{
\frac{1}{2}}}(f_i)=z^{-\frac{1}{2}\delta_{i,0}}f_i,\qquad D_{z^{\frac{1
}{2}}}(q^{\pm h_i})=q^{\pm h_i}.
\end{equation}
With $z^{\frac{1}{2}}$ specialized to a nonvanishing complex parameter,
the evaluation representation $\pi_{W(z^{\frac{1}{2}})}:U'_q\bigl(
\widehat{sl}(2)\bigr)\to\text{End}(W)\otimes\mathbb C[z^{\frac{1}{2}},
z^{-\frac{1}{2}}]$ is defined by
\begin{equation}\label{E:evrepr}
\pi_{W(z^{\frac{1}{2}})}(X)=\pi_W\bigl(D_{z^{\frac{1}{2}}}(X)\bigr),\;
X\in U'_q\bigl(\widehat{sl}(2)\bigr).
\end{equation}
The evaluation modules $W(z^{\frac{1}{2}})$ satisfy
$\pi_{W(z^{\frac{1}{2}}y^{\frac{1}{2}})}(X)=\pi_{W(z^{\frac{1}{2}})}\bigl(
D_{y^{\frac{1}{2}}}(X)\bigr)$.
In particular, the evaluation modules $W(1)$ with $W=W^{(\epsilon,q^{2\sigma})}$
and $W=W^{(\epsilon,q^{2\sigma})*}$ are equivalent to the modules introduced
by \eqref{E:Wmod} and \eqref{E:Wdual}.

\section{The intertwiners}\label{S:R}

The present section addresses linear operators $R(\tilde z^{\frac{1}{2}},
z^{\frac{1}{2}})$ relating the two tensor products $W^{(\epsilon_1,q^{2\sigma_1
})}(\tilde z^{\frac{1}{2}})\otimes W^{(\epsilon_2,q^{2\sigma_2})*}(\tilde z^{
-\frac{1}{2}})$ and $W^{(\epsilon_4,q^{2\sigma_4})}(z^{-\frac{1}{2}})\otimes 
W^{(\epsilon_3,q^{2\sigma_3})*}(z^{\frac{1}{2}})$, where $\tilde z^{\frac{1}{2}}
=\pm z^{\frac{1}{2}}$. With suitable restrictions on the parameters, the
matrix elements of $R(\tilde z^{\frac{1}{2}},z^{\frac{1}{2}})$ wrt the canonical
basis elements of the tensor products obey the intertwining condition.
A first set of matrix elements is specified by the explicit expressions 
\eqref{E:Rel1} in terms of
two nonterminating $_4\phi_3$-series. Simple replacements of the parameters
provide further sets characterized by the intertwining property.
Linear relations combine any three different sets. All statements concerning the
matrix elements are shown in Appendix \ref{A:R}.

The convergence properties of the sums considered in the following section
as well as the existence of homogeneous relations for particular linear
combinations of two sums are a consequence of the asymptotic behaviour of the matrix
elements of $R(\tilde z^{\frac{1}{2}},z^{\frac{1}{2}})$. At the end of
this section, the asymptotics are obtained from the explicit expressions 
of the matrix elements. 

\vskip 0.5cm

For the evaluation modules associated with the $U'_q\bigl(\widehat{sl}(2)\bigr)$-modules $W^{(\epsilon_1,q^{2\sigma_1})}$,
$W^{(\epsilon_2,q^{2\sigma_2})*}$, $W^{(\epsilon_3,q^{2\sigma_3})*}$, $W^{(
\epsilon_4,q^{2\sigma_4})}$, 
linear operators $R(\tilde z^{\frac{1}{2}},z^{\frac{1}{2}})$:\,
$W^{(\epsilon_1,q^{2\sigma_1})}(\tilde z^{
\frac{1}{2}})\otimes W^{(\epsilon_2,q^{2\sigma_2})*}(\tilde z^{-\frac{1}{2}})
\longrightarrow W^{(\epsilon_4,q^{2\sigma_4})}(z^{-\frac{1}{2}})
\otimes W^{(\epsilon_3,q^{2\sigma_3})*}(z^{\frac{1}{2}})$ 
are specified in the cases
\begin{equation*}
\tilde z^{\frac{1}{2}}=z^{\frac{1}{2}},\quad q^{2\sigma_3}=q^{2\sigma_1},\quad
q^{2\sigma_4}=q^{2\sigma_2}
\end{equation*}
and 
\begin{equation*}
\tilde z^{\frac{1}{2}}=-z^{\frac{1}{2}},\quad q^{2\sigma_3}=-q^{2\sigma_1},\quad 
q^{2\sigma_4}=-q^{2\sigma_2}.
\end{equation*}
In both cases, the restriction
\begin{equation}\label{E:econd}
\epsilon_2-\epsilon_1=\epsilon_3-\epsilon_4
\end{equation}
is imposed.
For either choice, the operators $R(\tilde z^{\frac{1}{2}},z^{\frac{1}{2}})$   
are defined by
\begin{equation*}
\begin{split}
R(\tilde z^{\frac{1}{2}},z^{\frac{1}{2}})\bigl(w_{l_1}^{(\epsilon_1,q^{2\sigma_1
})}\otimes w_{l_2}^{(\epsilon_2,q^{2\sigma_2})*}\bigr)&\equiv
\sum_{k_1,k_2=-\infty}^{\infty}
\mathsf{R}^{k_2,k_1^*}_{l_1,l_2^*}\,w^{(\epsilon_4,q^{2\sigma_4})}_{k_2}\otimes w^{(\epsilon_3,q^{2\sigma_3})*}_{k_1}
\end{split}
\end{equation*}
with the coefficients $\mathsf{R}^{k_2,k^*_1}_{l_1,l^*_2}$
determined by the intertwining property
\begin{multline}\label{E:int}
\Delta_{z^{-\frac{1}{2}}}(X)R(\tilde z^{\frac{1}{2}},z^{\frac{1}{2}})\bigl(
w_{l_1}^{(\epsilon_1,q^{2\sigma_1})}\otimes w^{(\epsilon_2,q^{2\sigma_2})*}_{
l_2}\bigr)=\\
\shoveright{
R(\tilde z^{\frac{1}{2}},z^{\frac{1}{2}})\Delta_{\tilde z^{\frac{1}{2}}}(X)
\bigl(w_{l_1}^{(\epsilon_1,q^{2\sigma_1})}\otimes w_{l_2}^{(\epsilon_2,
q^{2\sigma_2})*}\bigr),}\\
X\in U'_q\bigl(\widehat{sl}(2)\bigr).
\end{multline}
The equations \eqref{E:int} entail the maps
\begin{equation*}
\Delta_{z^{\frac{1}{2}}}(X)=(D_{z^{\frac{1}{2}}}\otimes D_{z^{-\frac{1}{2}}})
\Delta(X),
\end{equation*}
where the automorphism $D_{z^{\frac{1}{2}}}$ is defined by \eqref{E:Dz}.

With the additional restrictions
\begin{subequations}\label{E:Rcond1}
\begin{equation}\label{E:Rcond1a}
q^{2(\epsilon_2-\epsilon_1+\sigma_3-\sigma_4)}\neq q^{2t_1},\,
q^{2(\epsilon_2-\epsilon_1-\sigma_3-\sigma_4)}\neq q^{2t_2},\quad 
t_1,t_2\in\mathbb Z_{>0},
\end{equation}
\begin{equation}\label{E:Rcond1b}
q^{2(\epsilon_2-\epsilon_1-\sigma_3+\sigma_4)}\neq q^{2t},\quad t\in\mathbb Z,
\end{equation}
\end{subequations}
a set of $R$-elements 
$\{\mathsf{R}^{k_2,k_1*}_{l_1,l_2^*}\}$
with the intertwining property \eqref{E:int} is provided by
\begin{multline}\label{E:Rel1}
\quad\;\mathsf{R}^{k_2,k_1^*}_{l_1,l_2^*}=0\quad\text{if}\;k_1-k_2\neq l_2-l_1,\\
\mathsf{R}^{k,k+r^*}_{l,l+r^*}=R^{k,k+r^*}_{l,l+r^*}\equiv R^{k,k+r^*}_{l,l+r^*}\bigl(z,
\alpha;q^{2\epsilon_1},q^{2\epsilon_2};q^{2\epsilon_3},q^{2
\epsilon_4};q^{2\sigma_3},q^{2\sigma_4}\bigr)=\\
\kappa_r\,z^{-k-r}q^{l-k-r}\frac{S_{-k}\bigl(q^{-2\epsilon_4},q^{2\sigma_4}
\bigr)}{S_{-k-r}\bigl(q^{-2\epsilon_3},q^{2\sigma_3}\bigr)}\frac{S_{l+r}
\bigl(q^{2\epsilon_2},\alpha q^{2\sigma_4}\bigr)}{S_l\bigl(q^{2\epsilon_1},
\alpha q^{2\sigma_3}\bigr)}\cdot\\
\Biggl\{\frac{\bigl(\alpha q^{2(l-k-r+\epsilon_1-\epsilon_3+1)},zq^{2(\sigma_4-
\sigma_3+1)},z^{-1}q^{2(\sigma_4-\sigma_3)};q^2\bigr)_{\infty}}{\bigl(\alpha z^{-1}q^{2(l-k+\epsilon_1-\epsilon_4)},q^{-2(\epsilon_2-\epsilon_1+\sigma_3-
\sigma_4)+2},q^{2(r+\epsilon_2-\epsilon_1-\sigma_3+\sigma_4)};q^2\bigr)_{
\infty}}\cdot\\
\bigl(q^{-2(r+\epsilon_2-\epsilon_1-\sigma_3-\sigma_4)+2};q^2\bigr)_r\cdot\\
{}_4\phi_3\biggl(\genfrac{}{}{0pt}{}{-q^{-2(k+r+\epsilon_3-\sigma_3)+1},\,
-\alpha q^{2(l+\epsilon_1+\sigma_3)+1},\,zq^{-2(r+\epsilon_2-\epsilon_1)+2},\,z^{-1}
q^{-2(r+\epsilon_2-\epsilon_1)}}{\alpha q^{2(l-k-r+\epsilon_1-\epsilon_3+1)},\,
q^{-2(r+\epsilon_2-\epsilon_1-\sigma_3+\sigma_4)+2},\,q^{-2(r+\epsilon_2-
\epsilon_1-\sigma_3-\sigma_4)+2}};q^2,q^2\biggr)\\
+\frac{\bigl(\alpha q^{2(l-k+\epsilon_1-\epsilon_4-\sigma_3+\sigma_4+1)},-
q^{-2(k+r+\epsilon_3-\sigma_3)+1},-\alpha q^{2(l+\epsilon_1+\sigma_3)+1};
q^2\bigr)_{\infty}}{\bigl(\alpha z^{-1}q^{2(l-k+\epsilon_1-\epsilon_4)},
-q^{-2(k+\epsilon_4-\sigma_4)+1},-\alpha q^{2(l+r+\epsilon_2+\sigma_4)+1};q^2\bigr)_{\infty}}
\cdot\\
\frac{\bigl(zq^{-2(r+\epsilon_2-\epsilon_1)+2},z^{-1}q^{-2(r+\epsilon_2-
\epsilon_1)},q^{4\sigma_4+2};q^2\bigr)_{\infty}}{\bigl(q^{-2(\epsilon_2-
\epsilon_1+\sigma_3-\sigma_4)+2},q^{-2(\epsilon_2-\epsilon_1-\sigma_3-
\sigma_4)+2},q^{-2(r+\epsilon_2-\epsilon_1-\sigma_3+\sigma_4)};q^2\bigr)_{
\infty}}\cdot\\
{}_4\phi_3\biggl(\genfrac{}{}{0pt}{}{-q^{-2(k+\epsilon_4-\sigma_4)+1},\,
-\alpha q^{2(l+r+\epsilon_2+\sigma_4)+1},\,zq^{2(\sigma_4-\sigma_3+1)},\,z^{-1}q^{2(
\sigma_4-\sigma_3)}}{\alpha q^{2(l-k+\epsilon_1-\epsilon_4-\sigma_3
+\sigma_4+1)},\,q^{4\sigma_4+2},\,q^{2(r+\epsilon_2-\epsilon_1
-\sigma_3+\sigma_4+1)}};q^2,q^2\biggr)\Biggr\}
\end{multline}
where $k,l,r\in\mathbb Z$,
$\alpha=z^{-\frac{1}{2}}\tilde z^{\frac{1}{2}}=q^{2(\sigma_3-\sigma_1)}
=q^{2(\sigma_4-\sigma_2)}$, 
\begin{equation}\label{E:kappa}
\kappa_r=\frac{z^r\alpha^rq^{-2r\sigma_4+r}}{\bigl(zq^{-2(r+\epsilon_2-\epsilon_1
)+2};q^2\bigr)_r}
\end{equation}
and
\begin{equation*}
S_l(A,B)=\sqrt{\bigl(-ABq^{2l+1},-AB^{-1}q^{2l+1};q^2\bigr)_{\infty}}.
\end{equation*}
The intertwining property is demonstrated in Appendix \ref{A:R}. 
Due to \eqref{E:econd},
the condition \eqref{E:Rcond1b} is not satisfied in the case $q^{2\epsilon_3}=-
q^{2\sigma_3+1}$, $q^{2\epsilon_4}=-q^{2\sigma_4+1}$ studied in \cite{gade1}.
However, the conditions \eqref{E:Rcond1a} and \eqref{E:Rcond1b} allow the choice
$q^{2\epsilon_1}=-q^{2\sigma_1+1}=-\alpha q^{2\sigma_3+1}$, $q^{2\epsilon_2}=-
q^{2\sigma_2+1}=-\alpha q^{2\sigma_4+1}$. This special case will be discussed
separately in context with the big $q$-Jacobi function transform of the
$R$-elements.

If $\vert q^{
-2(k+\epsilon_4-\sigma_4)+1}\vert<1$, the rhs of \eqref{E:Rel1} can be obtained from the expression given by \eqref{E:RW2} which is manifestly invariant 
under the exchange $\sigma_3\to-\sigma_3$ and requires the property
\eqref{E:Rcond1a} only. Thus, if both \eqref{E:Rcond1} and the condition \eqref{E:Rcond1b}
with $\sigma_3$ replaced by $-\sigma_3$ are fulfilled, the coefficients
\eqref{E:Rel1} satisfy
\begin{multline}\label{E:Rsig1+-}
R^{k,k+r^*}_{l,l+r^*}\bigl(z,\alpha;q^{2\epsilon_1},q^{2\epsilon_2};q^{2
\epsilon_3},q^{2\epsilon_4};q^{2\sigma_3},q^{2\sigma_4}\bigr)=\\
R^{k,k+r^*}_{l,l+r^*}\bigl(z,\alpha;q^{2\epsilon_1},q^{2\epsilon_2};q^{2
\epsilon_3},q^{2\epsilon_4};q^{-2\sigma_3},q^{2\sigma_4}\bigr)
\end{multline}
if $\vert q^{-2(k+\epsilon_4-\sigma_4)+1}\vert<1$. For other values of $k$,
equation \eqref{E:BC} ensures the property \eqref{E:Rsig1+-}.

If $\sigma_4\neq0$, a different set of solutions is given by
\begin{equation}\label{E:chRel}
\mathsf{R}^{k,k+r^*}_{l,l+r^*}=\check R^{k,k+r^*}_{l,l+r^*}\equiv 
R^{k,k+r^*}_{l,l+r^*}\bigl(z,\alpha;q^{2\epsilon_1},q^{2\epsilon_2};q^{2\epsilon_3},
q^{2\epsilon_4};q^{2\sigma_3},q^{-2\sigma_4}\bigr).
\end{equation}
For any solution $\{\mathsf{R}^{k,k+r}_{l,l+r^*}\}$, 
replacing $\mathsf{R}^{k,k+r^*}_{l,l+r^*}\bigl(z,\alpha;q^{2\epsilon_1},q^{2
\epsilon_2};q^{2\epsilon_3},q^{2\epsilon_4};q^{2\sigma_3},q^{2\sigma_4}\bigr)$
by $\mathsf{R}^{k+r,k^*}_{l+r,l^*}\bigl(z,\alpha;q^{2\epsilon_2},q^{2
\epsilon_1};q^{2\epsilon_4},q^{2\epsilon_3};q^{2\sigma_4},q^{2\sigma_3}\bigr)$
gives another set of coefficients satisfying the conditions \eqref{E:int}. 
This is easily seen writing out the equations \eqref{E:int} explicitly 
according to the definition of the coproduct and the $U'$-modules by \eqref{E:co} and \eqref{E:Wmod}, \eqref{E:Wdual}. Besides \eqref{E:Rel1} and \eqref{E:chRel}, the following investigations involve the
solutions
\begin{multline}\label{E:mrRel}
\mathsf R^{k,k+r^*}_{l,l+r^*}=\mathring R^{k,k+r^*}_{l,l+r^*}\equiv 
\mathring R^{k,k+r^*}_{l,l+r^*}\bigl(z,\alpha;q^{2\epsilon_1},q^{2\epsilon_2};
q^{2\epsilon_3},q^{2\epsilon_4};q^{2\sigma_3},q^{2\sigma_4}\bigr)=\\
R^{k+r,k^*}_{l+r,l^*}\bigl(z,\alpha;q^{2\epsilon_2},q^{2\epsilon_1};q^{2\epsilon_4},
q^{2\epsilon_3};q^{2\sigma_4},q^{2\sigma_3}\bigr)
\end{multline}
and
\begin{multline}\label{E:mrchRel}
\mathsf R^{k,k+r^*}_{l,l+r^*}=\check{\mathring R}^{k,k+r^*}_{l,l+r^*}\equiv
\check{\mathring R}^{k,k+r^*}_{l,l+r^*}\bigl(z,\alpha;q^{2\epsilon_1},q^{2
\epsilon_2};q^{2\epsilon_3},q^{2\epsilon_4};q^{2\sigma_3},q^{2\sigma_4}\bigr)=\\
R^{k+r,k^*}_{l+r,l^*}\bigl(z,\alpha;q^{2\epsilon_2},q^{2\epsilon_1};q^{2
\epsilon_4},q^{2\epsilon_3};q^{2\sigma_4},q^{-2\sigma_3}\bigr).
\end{multline}
As shown in Appendix \ref{A:R}, the coefficients $R^{k,k+r^*}_{l,l+r^*}$,
$\check R^{k,k+r^*}_{l,l+r^*}$ and ${\mathring R}^{k,k+r^*}_{l,l+r^*}$ 
are related by
\begin{multline}\label{E:RmR}
\bigl(zq^{2(\sigma_3-\sigma_4+1)},z^{-1}q^{2(\sigma_3-\sigma_4)},q^{-2(
\epsilon_2-\epsilon_1+\sigma_3-\sigma_4)+2},q^{-2(\epsilon_2-\epsilon_1-\sigma_3
-\sigma_4)+2};q^2\bigr)_{\infty}\cdot\\
\shoveright{q^{2\sigma_4}\theta_{q^2}\bigl(q^{2(\epsilon_2-\epsilon_1+\sigma_3+
\sigma_4+1)}\bigr)R^{k,k+r^*}_{l,l+r^*}}\\
-\bigl(zq^{2(\sigma_3+\sigma_4+1)},z^{-1}q^{2(\sigma_3+\sigma_4)},q^{-2(\epsilon_2
-\epsilon_1+\sigma_3+\sigma_4)+2},q^{-2(\epsilon_2-\epsilon_1-\sigma_3+\sigma_4)
+2};q^2\bigr)_{\infty}\cdot\\
\shoveright{q^{-2\sigma_4}\theta_{q^2}\bigl(q^{2(\epsilon_2-\epsilon_1+\sigma_3-
\sigma_4+1)}\bigr)\check R^{k,k+r^*}_{l,l+r^*}}\\
-\bigl(zq^{2(\epsilon_1-\epsilon_2+1)}
,z^{-1}q^{2(\epsilon_1-\epsilon_2)},q^{2(\epsilon_2-\epsilon_1+\sigma_3+\sigma_4+1
)},q^{2(\epsilon_2-\epsilon_1+\sigma_3-\sigma_4+1)};q^2\bigr)_{\infty}\cdot\\
q^{2\sigma_4}
\theta_{q^2}\bigl(q^{4\sigma_4+2}\bigr)\mathring R^{k,k+r^*}_{l,l+r^*}=0.
\end{multline}

For the coefficients $R^{k,k+r^*}_{l,l+r^*}$, the asymptotic behaviour in the limit $l\to\infty$ is easily inferred from
equation \eqref{E:Rel1}:
\begin{multline}\label{E:Rllim+}
\lim_{l\to\infty}\Bigl(q^{-l}R^{k,k+r^*}_{l,l+r^*}\Bigr)=\kappa_rz^{-k-r}q^{-k-r}
\sqrt{\frac{\bigl(-q^{-2(k+\epsilon_4+\sigma_4)+1};q^2\bigr)_{\infty}}{
\bigl(-q^{-2(k+r+\epsilon_3+\sigma_3)+1};q^2\bigr)_{\infty}}}\cdot\\
\Biggl\{\sqrt{\frac{\bigl(-q^{-2(k+\epsilon_4-\sigma_4)+1};q^2\bigr)_{
\infty}}{\bigl(-q^{-2(k+r+\epsilon_3-\sigma_3)+1};q^2\bigr)_{\infty}}}
\frac{\bigl(zq^{2(\sigma_4-\sigma_3+1)},z^{-1}q^{2(\sigma_4-\sigma_3)};q^2
\bigr)_{\infty}}{\bigl(q^{2(r+\epsilon_2-\epsilon_1-\sigma_3+\sigma_4)},
q^{-2(\epsilon_2-\epsilon_1+\sigma_3-\sigma_4)+2};q^2\bigr)_{\infty}}\cdot\\
\bigl(q^{-2(r+\epsilon_2-\epsilon_1-\sigma_3-\sigma_4)+2};q^2\bigr)_r\cdot\\
{}_3\phi_2\biggl(\genfrac{}{}{0pt}{}{-q^{-2(k+r+\epsilon_3-\sigma_3)+1},\,
zq^{-2(r+\epsilon_2-\epsilon_1)+2},\,z^{-1}q^{-2(r+\epsilon_2-\epsilon_1)}}{
q^{-2(r+\epsilon_2-\epsilon_1-\sigma_3+\sigma_4)+2},\,q^{-2(r+\epsilon_2-
\epsilon_1-\sigma_3-\sigma_4)+2}};q^2,q^2\biggr)\\
\shoveleft{
+\sqrt{\frac{\bigl(-q^{-2(k+r+\epsilon_3-\sigma_3)+1};q^2\bigr)_{\infty}}{\bigl(
-q^{-2(k+\epsilon_4-\sigma_4)+1};q^2\bigr)_{\infty}}}\cdot}\\
\frac{\bigl(zq^{-2(
r+\epsilon_2-\epsilon_1)+2},\,z^{-1}q^{-2(r+\epsilon_2-\epsilon_1)},q^{4\sigma_4
+2};q^2\bigr)_{\infty}}{\bigl(q^{-2(r+\epsilon_2-\epsilon_1-\sigma_3+\sigma_4)},
q^{-2(\epsilon_2-\epsilon_1+\sigma_3-\sigma_4)+2},q^{-2(\epsilon_2-\epsilon_1
-\sigma_3-\sigma_4)+2};q^2\bigr)_{\infty}}\cdot\\
{}_3\phi_2\biggl(\genfrac{}{}{0pt}{}{-q^{-2(k+\epsilon_4-\sigma_4)+1},\,
zq^{2(\sigma_4-\sigma_3+1)},\,z^{-1}q^{2(\sigma_4-\sigma_3)}}{q^{2(r+\epsilon_2
-\epsilon_1-\sigma_3+\sigma_4+1)},\,q^{4\sigma_4+2}};q^2,q^2\biggr)\Biggr\}.
\end{multline}
The rhs depends on the parameters $\epsilon_1$ and $\epsilon_2$ only through
their difference.
If $q^{2(k+r+\epsilon_3+\sigma_3)+1}<1$, equation \eqref{E:RW3}
gives
\begin{multline}\label{E:Rllim-i}
z^{-l}q^{-l} R^{\,k,k+r^*}_{-l,-l+r^*}\sim
\kappa_r\sqrt{\frac{\theta_{q^2}\bigl(-q^{2(r-\epsilon_4-\sigma_4)+1},
-\alpha q^{2(r+\epsilon_2-\sigma_4)+1}\bigr)}{\theta_{q^2}\bigl(- q^{-2(\epsilon_3+\sigma_3)+1},-\alpha q^{2(\epsilon_1-\sigma_3)+1}\bigr)}}\cdot\\
\frac{1}{\theta_{q^2}\bigl(\alpha z^{-1}q^{2(r+\epsilon_1-\epsilon_4)}\bigr)
\bigl(q^{-2(\epsilon_2-\epsilon_1+\sigma_3-\sigma_4)+2};q^2\bigr)_{\infty}}
\cdot\\
\shoveleft{
\Biggl\{\theta_{q^2}\bigl(\alpha q^{2(\epsilon_1-\epsilon_3+1)}\bigr)
\sqrt{\frac{\theta_{q^2}\bigl(-q^{2(r-\epsilon_4+\sigma_4)+1},-\alpha
q^{2(r+\epsilon_2+\sigma_4)+1}\bigr)}{\theta_{q^2}\bigl(-q^{-2(\epsilon_3
-\sigma_3)+1},-\alpha q^{2(\epsilon_1+\sigma_3)+1}\bigr)}}\cdot}\\
\frac{\bigl(zq^{2(\sigma_4-\sigma_3+1)},z^{-1}q^{2(\sigma_4-
\sigma_3)};q^2\bigr)_{\infty}\bigl(q^{-2(r+\epsilon_2-\epsilon_1-\sigma_3-\sigma_4
)+2};q^2\bigr)_r}{\bigl(q^{2(r+\epsilon_2-\epsilon_1-\sigma_3+\sigma_4)}
;q^2\bigr)_{\infty}}\cdot\\
{}_3\phi_2\biggl(\genfrac{}{}{0pt}{}{-q^{-2(k+r+\epsilon_3-\sigma_3)+1},\,
zq^{-2(r+\epsilon_2-\epsilon_1)+2},\,z^{-1}q^{-2(r+\epsilon_2-\epsilon_1)}}{
q^{-2(r+\epsilon_2-\epsilon_1-\sigma_3+\sigma_4)+2},\,q^{-2(r+\epsilon_2-
\epsilon_1-\sigma_3-\sigma_4)+2}},q^2,-q^{2(k+r+\epsilon_3+\sigma_3)+1}
\biggr)\\
+\theta_{q^2}\bigl(\alpha q^{2(r+\epsilon_2+\epsilon_3-\sigma_3+\sigma_4+1)}
\bigr)
\sqrt{\frac{\theta_{q^2}\bigl(-q^{-2(\epsilon_3-\sigma_3)+1},-\alpha q^{2(
\epsilon_1+\sigma_3)+1}\bigr)}{\theta_{q^2}\bigl(-q^{2(r-\epsilon_4+
\sigma_4)+1},-\alpha q^{2(r+\epsilon_2+\sigma_4)+1}\bigr)}}\cdot\\
\frac{\bigl(-q^{2(k+r+\epsilon_3+\sigma_3)+1},
zq^{-2(r+\epsilon_2-\epsilon_1)+2},z^{-1}q^{-2(r+\epsilon_2-\epsilon_1
)},q^{4\sigma_4+2};q^2\bigr)_{\infty}}{\bigl(-q^{2(k+\epsilon_4+
\sigma_4)+1},q^{-2(r+\epsilon_2-\epsilon_1-
\sigma_3+\sigma_4)},q^{-2(\epsilon_2-\epsilon_1-\sigma_3-\sigma_4)+2}
;q^2\bigr)_{\infty}}\cdot\\
{}_3\phi_2\biggl(\genfrac{}{}{0pt}{}{-q^{-2(k+\epsilon_4-\sigma_4)+1},\,
zq^{2(\sigma_4-\sigma_3+1) },\,z^{-1}q^{2(\sigma_4-\sigma_3)}}{q^{2(r+
\epsilon_2-\epsilon_1-\sigma_3+\sigma_4+1)},\,q^{4\sigma_4+2}};q^2,
-q^{2(k+r+\epsilon_3+\sigma_3)+1}\biggr)\Biggr\}\\
+O(q^{2l})
\end{multline}
in the limit $l\to\infty$. A general result can be formulated by
\begin{equation}\label{E:Rllim-}
z^{-l}q^{-l}R^{k,k+r^*}_{-l,-l+r^*}\sim R^{(r)}_k
\end{equation}
as $l\to\infty$, where $R^{(r)}_k$ is given by the rhs of
\eqref{E:Rllim-i} if $q^{2(k+r+\epsilon_3+\sigma_3)+1}<1$.
The element $R^{k,k+r^*}_{-l,-l+r^*}$ can be expressed in terms of
$R^{k,k+r+1^*}_{-l,-l+r+1^*}$ and $R^{k+1,k+r+1^*}_{-l,-l+r^*}$ due to
the relation $B_{k,r,-l}=0$ with $B_{k,r,-l}$ specified by \eqref{E:B}.
Its derivation is given in Appendix \ref{A:R}. The relation
implies
\begin{multline}\label{E:Rllim-rec}
z^{-l}q^{-l}R^{k,k+r^*}_{-l,-l+r^*}\sim \\
\frac{1-zq^{-2(r+\epsilon_2-
\epsilon_1)}}{s_{-k-r-1}(q^{-2\epsilon_3},q^{2\sigma_3})}\,R^{(r+1)}_k
+q^{-2(r+\epsilon_2-\epsilon_1)-1}\frac{s_{-k-1}(q^{-2\epsilon_4},q^{2
\sigma_4})}{s_{-k-r-1}(q^{-2\epsilon_3},q^{2\sigma_3})}\,R^{(r)}_{k+1},
\end{multline}
provided that both the limit \eqref{E:Rllim-} with $r\to r+1$ and \eqref{E:Rllim-}
with $k\to k+1$ exist. Thus the relation establishes the result
\eqref{E:Rllim-} for all values of $k+r$ with the function $R^{(r)}_k$
obtained recursively from the rhs of \eqref{E:Rllim-i} by means of \eqref{E:Rllim-rec} if $q^{2(k+r+\epsilon_3+\sigma_3)+1}>1$.

\section{The sums $\tau^{(r,k)\pm}$ and $\tau^{(r,k)}$}\label{S:tau}

The search for an instructive description of the action of the linear operator $R(\tilde z^{\frac{1}{2}},z^{\frac{1}{2}})$ motivates the consideration 
of pairs of tensor products.

 All subsequent analysis rests upon linear combinations $\rho^{(r;N,M)\pm}$ 
of the canonical basis elements
$w^{(\epsilon_1,q^{2\sigma_1})}_l\otimes w^{(\epsilon_2,q^{2\sigma_2})*}_{
l+r}$ with $-M\leq l\leq N$ and the coefficient of $w^{(\epsilon_1,q^{2
\sigma_1})}_l\otimes w^{(\epsilon_2,q^{2\sigma_2})*}_{l+r}$ given by 
$a^{\pm}_r\rho^{(r)\pm}_l$. The function $\rho^{(r)\pm}_l$ depends on a
variable $e^{i\theta}$ referred to as spectral value. Both functions
$a^{\pm}_r$ and $\rho^{(r)\pm}_l$ are chosen such that the action of
$\Delta(e_1)$ on $\rho^{(r;N,M)\pm}$ yields
a contribution essentially given by $\rho^{(r+1;N-1,M)\pm}$
and two contributions from the border elements $w^{(\epsilon_1,
q^{2\sigma_1})}_N\otimes w^{(\epsilon_2,q^{2\sigma_2})*}_{N+r+1}$ and
$w^{(\epsilon_1,q^{2\sigma_1})}_{-M-1}\otimes w^{(\epsilon_2,q^{2\sigma_2})*}_{
-M+r}$. An analogous statement applies to the application of $\Delta(f_1)$.
This particular result specified by the equations \eqref{E:rhoe1} and
\eqref{E:rhof1} is a consequence of the two contiguous relations
\eqref{E:rhorl+} and \eqref{E:rhorl-} satisfied by $\rho^{(r)\pm}_l$.
Parameter replacements give rise to an alternative choice \eqref{E:ardef2} for the coefficients in the linear combination $\rho^{(r;N,M)\pm}$.

The function $\rho^{(r)\pm}_l$ is introduced in subsection \ref{S:coef} in
a form suited to the
procedure of evaluation in subsection \ref{S:tauev}. Various transformations
of the explicit expression for $\rho^{(r)\pm}_l$ reveal their relation to
big $q$-Jacobi functions and provide a reformulation employed in subsection
\ref{S:tauev}. In addition, subsection \ref{S:coef} includes the analysis of the asymptotic behaviour of $\rho^{(r)\pm}_l$ as $l\to\pm\infty$. The
latter is required in the discussion
of the convergence properties of the sums defined in subsection \ref{S:taudef}
and the homogeneous relations among certain linear combinations thereof.
A particular linear combination $\varsigma^{(r)}_l$ of $\rho^{(r)+}_l$
and $\rho^{(r)-}_l$ is distinguished by a simplification of the asymptotics
for $l\to-\infty$. 

By Definiton \ref{D:taudef1}
in subsection \ref{S:taudef}, the sums $\boldsymbol{\tau}^{(r,k)\pm}$ are introduced as bilateral summations of the matrix elements $\mathsf R^{k,k+r^*
}_{l,l+r^*}$ weighted by the factor $a^{\pm}_r\rho^{(r)\pm}_l$.
Relying on the asymptotics of $\varsigma^{(r)}_l$ as $l\to-\infty$, for two discrete sets of particular spectral 
values $e^{i\theta}$ the sum $\boldsymbol{\tau}^{(r,k)}$ is defined as the
bilateral summation of $\mathsf R^{k,k+r^*}_{l,l+r^*}$ weighted by $a_r\varsigma^{(r)}_l$ . Each set entails an appropriate choice of the 
function $a_r$.

Suitable pairs of parameter sets characterizing two sums $\boldsymbol{\tau}^{
(r,k)\pm}$ and $\boldsymbol{\tau}'^{(r,k)\pm}$ or $\boldsymbol{\tau}^{(r,k)}$
and $\boldsymbol{\tau}'^{(r,k)}$ are considered. Two types of pairs with a
parameter $\beta$ taking the values $1$ or $-1$ are distinguished. Simple 
linear combinations $\Xi^{(r,k)\pm}$ of the sums $\boldsymbol{\tau}^{(r,k)\pm}$ 
and $\boldsymbol{\tau}'^{(r,k)\pm}$ obey the homogeneous linear relations 
\eqref{E:taue1int} and \eqref{E:tauf1int} combining different
values of $r$ and $k$. Analogous relations apply to the combinations $\Xi^{(r,k)}$ of $\boldsymbol{\tau}^{(r,k)}$ and $\boldsymbol{\tau}'^{(r,k)}$. 
Moreover, for $\Xi^{(r,k)}$ a contiguous relation wrt the spectral value needed 
in subsection \ref{S:tauev} is given.

Subsection \ref{S:bigJac} relates the combinations
$\Xi^{(r,k)\pm}$ and $\Xi^{(r,k)}$ with $\beta=-1$ to the vector-valued
big $q$-Jacobi transform of the $R$-elements \eqref{E:Rel1}.
The relations are obtained separately for the spectral value chosen
on the unit circle $\mathbb T$ and from each of the discrete subsets involved in the transform \cite{vbigJac}. They allow to specify the inverse transform
required for the proof of Corollary \ref{C:RRsum} in section \ref{S:unit}.

As a main result of this article, the evaluation of the combination
$\Xi^{(r,k)\pm}$ for the choice of matrix elements $\mathsf R^{k,k+r^*}_{l,l
+r^*}=R^{k,k+r^*}_{l,l+r^*}$ given by the expressions \eqref{E:Rel1}
is presented in subsection \ref{S:tauev} (Theorem \ref{T:tausum2}). The
corresponding result for the choice $\mathsf R^{k,k+r^*}_{l,l+r^*}=
\check R^{k,k+r^*}_{l,l+r^*}$ specified by
\eqref{E:chRel} is readily obtained by a parameter replacement. 
For a particular linear combination $r^{k,k+r^*}_{l,l+r^*}$ of the
matrix elements $R^{k,k+r^*}_{l,l+r^*}$ and $\check R^{k,k+r^*}_{l,l+r^*}$,
the evaluation of $\Xi^{(r,k)\pm}$ for $\mathsf R^{k,k+r^*}_{l,l+r^*}=
r^{k,k+r^*}_{l,l+r^*}$ by means of Theorem \ref{T:tausum2} yields a
remarkably simple expression given by Corollary \ref{C:tausum4}. Within
a partial range of the spectral value $e^{i\theta}$ and the spectral parameter $z^{\frac{1}{2}}$, 
Corollary \ref{C:tausum4} allows an evaluation of $\Xi^{(r,k)}$ for $\mathsf R^{k,k+r^*}_{l,l+r^*}=r^{k,k+r^*}_{l,l+r^*}$. By means of the
two contiguous relations \eqref{E:Xitheta} and \eqref{E:Xiz}
satisfied by $\Xi{(r,k})$, the results given by the equations \eqref{E:tausum5}
and \eqref{E:tausum6} in Corollary \ref{C:tausum5} 
and \ref{C:tausum6} can be extended to the entire range of $z^{\frac{1}{2}}$ 
and $e^{i\theta}$ admitted by the requirement of absolute convergence.
Both relations are derived in Appendix \ref{A:thetacont}.

\subsection{The coefficients}\label{S:coef}

The description of the action of the intertwiner $R(\tilde z^{\frac{1}{2}},
z^{\frac{1}{2}})$ in terms of the sums $\boldsymbol{\tau}^{(r,k)\pm}$ and 
$\boldsymbol{\tau}^{(r,k)}$ involves 
the function $\rho^{(r)\pm}_l$ given by
\begin{multline}\label{E:rho1b}
\rho^{(r)\pm}_l\equiv\rho^{(r)\pm}_l\bigl(q^{2\epsilon_1},q^{2\sigma_1};
q^{2\epsilon_2},q^{2\sigma_2};\cos\theta\bigr)=
q^l\frac{S_l\bigl(q^{2\epsilon_1},q^{2\sigma_1}\bigr)}{S_{l+r}\bigl(
q^{2\epsilon_2},q^{2\sigma_2}\bigr)}\cdot\\
\frac{1}{\bigl(-q^{2(l+
\epsilon_1\pm\sigma_1)+1},q^{2(\epsilon_2-\epsilon_1\pm\sigma_1+\sigma_2+1)},
q^{2(\epsilon_2-\epsilon_1\pm\sigma_1-\sigma_2+1)};q^2\bigr)_{\infty}}\cdot\\
\Biggl\{\frac{\bigl(-q^{2(l+r+\epsilon_2\pm\sigma_1+1)}e^{-i\theta},q^{2(r+
\epsilon_2-\epsilon_1)+1}e^{i\theta},q^{\pm2\sigma_1+2\sigma_2+1}e^{i\theta},
q^{\pm2\sigma_1-2\sigma_2+1}e^{i\theta};q^2\bigr)_{\infty}}{\bigl(e^{2i
\theta};q^2\bigr)_{\infty}}\cdot\\
{}_3\phi_2\biggl(\genfrac{}{}{0pt}{}{q^{2(r+\epsilon_2-\epsilon_1)+1}e^{-i
\theta},\,
q^{\pm2\sigma_1+2\sigma_2+1}e^{-i\theta},\,q^{\pm2\sigma_1-2\sigma_2+1}e^{-i\theta}
}{-q^{2(l+r+\epsilon_2\pm\sigma_1+1)}e^{-i\theta},\,q^2e^{-2i\theta}};q^2,q^2
\biggr)\\
+\text{idem}(\theta,-\theta)\Biggr\}
\end{multline} 
with $r,l\in\mathbb Z$,
if $q^{2\epsilon_i}\neq-q^{2\sigma_i+1}$, $i=1,2$
and $l\in\mathbb Z$ for $r\in \mathbb Z_{\geq0}$,
$q^{2\epsilon_i}=-q^{2\sigma_i+1}$. 
In the case $r\in\mathbb Z_{<0}$, $q^{2\epsilon_i}=-q^{2\sigma_i+1}$, the values of $l$ are restricted to
$l\in\mathbb Z_{<0}$ and $l-\vert r\vert\in\mathbb Z_{\geq0}$. Moreover,
for $\rho^{(r)-}_l$, the cases $l+\epsilon_1-\sigma_1+\tfrac{1}{2}\in\mathbb Z_{
\leq0}$ are excluded.
Throughout the following, the condition $e^{i\theta}\neq q^t$ with 
$t\in\mathbb Z$ is imposed. The contiguous relation
\begin{multline}\label{E:rhorl+}
qs_{l+r}\bigl(q^{2\epsilon_2},q^{2\sigma_2}\bigr)\rho^{(r)\pm}_l-s_l\bigl(
q^{2\epsilon_1},q^{2\sigma_1}\bigr)\rho^{(r)\pm}_{l+1}\\
+q^{2(l+\epsilon_1\pm\sigma_1+1)}\bigl(1-q^{2(r+\epsilon_2-\epsilon_1)+1}e^{i
\theta}\bigr)\bigl(1-q^{2(r+\epsilon_2-\epsilon_1)+1}e^{-i\theta}\bigr)\rho^{
(r+1)\pm}_l=0
\end{multline}
is readily obtained making use of the definition of the $_3\phi_2$-series by
\eqref{E:phidef}.
In the case $\vert q^{2(r+\epsilon_2-\epsilon_1)+1}e^{i\theta}\vert<1$, the
three-term transformation [\cite{GR}:III.34] with $a\to-q^{2(l+\epsilon_1
\pm\sigma_1)+1}$, $b\to q^{\mp2\sigma_1+2\sigma_2+1}e^{-i\theta}$, $c\to q^{
\pm2\sigma_1-2\sigma_2+1}e^{-i\theta}$, $d\to-q^{2(l+r+\epsilon_2\pm
\sigma_1+1)}e^{-i\theta}$, $e\to q^{\pm4\sigma_1+2}$ can be applied to the
expression \eqref{E:rho1b} for $\rho^{(r)\pm}_l$. Then use of [\cite{GR}:III.10]
with $a\to q^{\pm2\sigma_1+2\sigma_2+1}e^{-i\theta}$, $b\to-q^{2(l+
\epsilon_1\pm\sigma_1)+1}$, $c\to q^{\pm2\sigma_1-2\sigma_2+1}e^{-i\theta}$,
$d\to-q^{2(l+r+\epsilon_2\pm\sigma_1+1)}e^{-i\theta}$, $e\to q^{\pm4\sigma_1+2}$
leads to
\begin{multline}\label{E:rho3}
\rho^{(r)\pm}_l=q^l\frac{S_l\bigl(q^{2\epsilon_1},q^{2\sigma_1}\bigr)}{
S_{l+r}\bigl(q^{2\epsilon_2},q^{2\sigma_2}\bigr)}\cdot\\
\frac{\bigl(-q^{2(l+r+\epsilon_2\pm\sigma_1+1)}e^{-i\theta},q^{2(r+\epsilon_2-
\epsilon_1)+1}e^{i\theta},q^{\pm4\sigma_1+2};q^2\bigr)_{\infty}}{\bigl(-q^{2
(l+\epsilon_1\pm\sigma_1)+1},q^{2(\epsilon_2-\epsilon_1\pm\sigma_1+\sigma_2
+1)},q^{2(\epsilon_2-\epsilon_1\pm\sigma_1-\sigma_2+1)};q^2\bigr)_{\infty}}
\cdot\\
{}_3\phi_2\biggl(\genfrac{}{}{0pt}{}{-q^{2(l+\epsilon_1\pm\sigma_1)+1},\,q^{
\pm2\sigma_1+2\sigma_2+1}e^{-i\theta},\,q^{\pm2\sigma_1-2\sigma_2+1}e^{-i
\theta}}{-q^{2(l+r+\epsilon_2\pm\sigma_1+1)}e^{-i\theta},\,q^{\pm4\sigma_1+2}}
;q^2,q^{2(r+\epsilon_2-\epsilon_1)+1}e^{i\theta}\biggr)\\
=q^l\frac{S_l\bigl(q^{2\epsilon_1},q^{2\sigma_1}\bigr)}{S_{l+r}\bigl(
q^{2\epsilon_2},q^{2\sigma_2}\bigr)}\frac{1}{\bigl(q^{2(\epsilon_2-\epsilon_1
\pm\sigma_1+\sigma_2+1)},q^{2(\epsilon_2-\epsilon_1\pm\sigma_1-\sigma_2+1)};
q^2\bigr)_r}\cdot\\
{}_3\phi_2\biggl(\genfrac{}{}{0pt}{}{-q^{-2(l+\epsilon_1\mp\sigma_1)+1},\,q^{2(
r+\epsilon_2-\epsilon_1)+1}e^{i\theta},\,q^{2(r+\epsilon_2-\epsilon_1)+1}
e^{-i\theta}}{q^{2(r+\epsilon_2-\epsilon_1\pm\sigma_1+\sigma_2+1)},\,q^{2(r+
\epsilon_2-\epsilon_1\pm\sigma_1-\sigma_2+1)}};q^2,-q^{2(l+\epsilon_1\pm\sigma_1
)+1}\biggr),
\end{multline}
provided that $q^{2(l+\epsilon_1\pm\sigma_1)+1}<1$. Application of the relation
[\cite{isra}:2.4] with $a\to-q^{-2(l+\epsilon_1\mp\sigma_1)+1}$, $b\to q^{2(
r+\epsilon_2-\epsilon_1)-1}e^{i\theta}$, $c\to q^{2(r+\epsilon_2-\epsilon_1)-1}
e^{-i\theta}$, $d\to q^{2(r+\epsilon_2-\epsilon_1\pm\sigma_1+\sigma_2)}$,
$e\to q^{2(r+\epsilon_2-\epsilon_1\pm\sigma_1-\sigma_2)}$ shows that the
last expression satisfies the relation \eqref{E:rhorl-} shown below.
This ensures its validity for all values of $r$.

The second $_3\phi_2$-series on the rhs of \eqref{E:rho3} is a big $q$-Jacobi function.
Following [\cite{vbigJac}:3.2], the big $q$-Jacobi functions $\varphi_{\gamma}(x)$
and $\varphi_{\gamma}^{\dagger}(x)$ with base $q^2$ are given by 
\begin{equation}\label{E:Jac1}
\begin{split}
\varphi_{\gamma}(x)&\equiv\varphi_{\gamma}(x;a,b,c,d\vert q^2)=
{}_3\phi_2\Biggl(
\genfrac{}{}{0pt}{}{\frac{q^2}{ax},\,s\gamma,\,\frac{s}{\gamma}}{
\frac{
cq^2}{a},\,\frac{dq^2}{a}};q^2,bx\Biggr),\\
\varphi^{\dagger}_{\gamma}(x)&=\varphi_{\gamma}(x;b,a,c,d\vert q^2),
\end{split}
\end{equation}
where $bx<1$ and $s=\sqrt{abcdq^{-2}}$.
For the specifications 
$\gamma\to e^{i\theta}$, $s\to q^{2(r+\epsilon_2-\epsilon_1)+1}$,
$ax\to-q^{2(l+\epsilon_1+\sigma_1)+1}$, $bx\to 
-q^{2(l+\epsilon_1-\sigma_1)+1}$,
$\tfrac{c}{a}\to q^{2(r+\epsilon_2-
\epsilon_1-\sigma_1+\sigma_2)}$, $\tfrac{d}{a}\to q^{2(r+\epsilon_2-
\epsilon_1-\sigma_1-\sigma_2)}$ subject to $q^{2(l+\epsilon_1\pm\sigma_1)+1}
<1$, the big $q$-Jacobi functions can be identified with
\begin{multline}\label{E:varphirho}
q^{-l}\frac{S_{l+r}\bigl(q^{2\epsilon_2},q^{2\sigma_2}\bigr)}{S_l\bigl(
q^{2\epsilon_1},q^{2\sigma_1}\bigr)}
\bigl(q^{2(\epsilon_2-\epsilon_1\pm\sigma_1+\sigma_2+1)},q^{2(\epsilon_2-
\epsilon_1\pm\sigma_1-\sigma_2+1)};q^2\bigr)_r\,\rho^{(r)\pm}_l,
\end{multline}
where the upper and lower sign refer to $\varphi^{\dagger}_{\gamma}(x)$ and $\varphi_{\gamma}(x)$, respectively. 

A further expression for $\rho^{(r)\pm}_l$ proves useful for the
analysis following in subsection \ref{S:tauev}. If $q^{2(l+r+\epsilon_2-
\sigma_2)+1}<1$,
use of [\cite{GR}:III.9] with $a\to q^{\pm2\sigma_1+2\sigma_2+1}
e^{-i\theta}$, $b\to q^{\pm2\sigma_1-2\sigma_2+1}e^{-i\theta}$, $c\to-
q^{2(l+\epsilon_1\pm\sigma_1)+1}$, $d\to q^{\pm4\sigma_1+2}$, $e\to-
q^{2(l+r+\epsilon_2\pm\sigma_1+1)}e^{-i\theta}$ to the first $_3\phi_2$-series
on the rhs of \eqref{E:rho3} leads to
\begin{multline}\label{E:rho2}
\rho^{(r)\pm}_l=q^l\sqrt{\frac{\bigl(-q^{2(l+\epsilon_1\mp\sigma_1)+1},-
q^{2(l+r+\epsilon_2-\sigma_2)+1};q^2\bigr)_{\infty}}{\bigl(-q^{2(l+\epsilon_1
\pm\sigma_1)+1},-q^{2(l+r+\epsilon_2+\sigma_2)+1};q^2\bigr)_{\infty}}}\cdot\\
\frac{\bigl(q^{\pm4\sigma_1+2};q^2\bigr)_{\infty}}{\bigl(q^{2(\epsilon_2-
\epsilon_1\pm\sigma_1-\sigma_2+1)};q^2\bigr)_{\infty}\bigl(q^{2(\epsilon_2-
\epsilon_1\pm\sigma_1+\sigma_2+1)};q^2\bigr)_r}\cdot\\
{}_3\phi_2\biggl(\genfrac{}{}{0pt}{}{-q^{-2(l+\epsilon_1\mp\sigma_1)+1},\,
q^{\pm2\sigma_1+2\sigma_2+1}e^{i\theta},\,q^{\pm2\sigma_1+2\sigma_2+1}e^{
-i\theta}}{q^{2(r+\epsilon_2-\epsilon_1\pm\sigma_1+\sigma_2+1)},\,q^{\pm4
\sigma_1+2}};q^2,-q^{2(l+r+\epsilon_2-\sigma_2)+1}\biggr).
\end{multline}
Writing out the 
$_3\phi_2$-series according to \eqref{E:phidef}, it is readily seen that the expression \eqref{E:rho2} obeys the relation \eqref{E:rhorl+}. This extends the
validity of \eqref{E:rho2} to all values of $r$, provided that $q^{2(l+r+\epsilon_2-\sigma_2)+1}<1$.

The asymptotic behaviour in the limit $l\to\infty$ is easily read from the expression \eqref{E:rho1b}:
\begin{multline}\label{E:rholim+}
\lim_{l\to\infty}\bigl(q^{-l}\rho^{(r)\pm}_l\bigr)=\bigl(q^{2(\epsilon_2-
\epsilon_1\pm\sigma_1+\sigma_2+1)},q^{2(\epsilon_2-\epsilon_1\pm\sigma_1-
\sigma_2+1)};q^2\bigr)_{\infty}^{-1}\cdot\\
\shoveleft{
\Biggl\{\frac{\bigl(q^{2(r+\epsilon_2-\epsilon_1)+1}e^{i\theta},q^{\pm2\sigma_1
+2\sigma_2+1}e^{i\theta},q^{\pm2\sigma_1-2\sigma_2+1}e^{i\theta};q^2\bigr)_{
\infty}}{\bigl(e^{2i\theta},q^2\bigr)_{\infty}}\cdot}\\
{}_3\phi_2\biggl(\genfrac{}{}{0pt}{}{q^{2(r+\epsilon_2-\epsilon_1)+1}e^{-i
\theta},\,q^{\pm2\sigma_1+2\sigma_2)+1}e^{-i\theta},\,q^{\pm2\sigma_1-2\sigma_2
+1}e^{-i\theta}}{q^2e^{-2i\theta},\,0};q^2,q^2\biggr)\\
+\text{idem}(\theta,-\theta)\Biggr\}.
\end{multline}
To evaluate the asymptotic behaviour as $l\to-\infty$,
the three-term transformation [\cite{GR}:III.34] with $a\to q^{-2(r+\epsilon_2
-\epsilon_1)+1}e^{-i\theta}$, $b\to q^{\pm2\sigma_1+2\sigma_2+1}e^{-i\theta}$,
$c\to q^{\pm2\sigma_1-2\sigma_2+1}e^{-i\theta}$, $d\to q^2e^{-2i\theta}$,
$e\to-q^{-2(l+r+\epsilon_2\mp\sigma_1)+2}e^{-i\theta}$ is applied to the first $_3\phi_2$-series on the rhs of \eqref{E:rho1b}. Adding the same
expression with $\theta$ replaced by $-\theta$, a cancellation due
to the property $e^{i\theta}\theta_{q^2}(e^{2i\theta})=-e^{-i\theta}\theta_{q^2}
(e^{-2i\theta})$ leaves
\begin{multline}\label{E:phiPhi1}
\rho^{(r)\pm}_l=\bigl(q^{2(\epsilon_2-\epsilon_1\pm\sigma_1+\sigma_2+1)},
q^{2(\epsilon_2-\epsilon_1\pm\sigma_1-\sigma_2+1)};q^2\bigr)^{-1}_{\infty}\cdot\\
q^l\sqrt{\frac{\theta_{q^2}\bigl(-q^{2(l+\epsilon_1+\sigma_1)+1},-q^{2(l+
\epsilon_1-\sigma_1)+1}\bigr)}{\theta_{q^2}\bigl(-q^{2(l+r+\epsilon_2+
\sigma_2)+1},-q^{2(l+r+\epsilon_2-\sigma_2)+1}\bigr)}}
\frac{\theta_{q^2}
\bigl(-q^{2(l+r+\epsilon_2\pm\sigma_1+1)}e^{-i\theta}\bigr)}{\theta_{q^2}
\bigl(-q^{2(l+\epsilon_1\pm\sigma_1)+1}\bigr)}\cdot\\
\frac{\sqrt{\bigl(-q^{-2(l+\epsilon_1\pm\sigma_1)+1};q^2\bigr)_{\infty}}\,
\bigl(-q^{-2(l+r+\epsilon_2\mp\sigma_1)+2}e^{-i\theta};q^2\bigr)_{\infty}}{
\sqrt{\bigl(-q^{-2(l+r+\epsilon_2+\sigma_2)+1},-q^{-2(l+r+\epsilon_2-
\sigma_2)+1},-q^{-2(l+\epsilon_1\mp\sigma_1)+1};q^2\bigr)_{\infty}}}\cdot\\
\frac{\bigl(q^{2(r+\epsilon_2-\epsilon_1)+1}e^{i\theta},q^{\pm2\sigma_1+2
\sigma_2+1}e^{i\theta},q^{\pm2\sigma_1-\sigma_2+1}e^{i\theta};q^2\bigr)_{
\infty}}{\bigl(e^{2i\theta};q^2\bigr)_{\infty}}\cdot\\
{}_3\phi_2\biggl(\genfrac{}{}{0pt}{}{q^{-2(r+\epsilon_2-\epsilon_1)+1}e^{-i
\theta},\,q^{\pm2\sigma_1+2\sigma_2+1}e^{-i\theta},\,q^{\pm2\sigma_1-2\sigma_2
+1}e^{-i\theta}}{-q^{-2(l+r+\epsilon_2\mp\sigma_1)+2}e^{-i\theta},\,q^2e^{-2i
\theta}};q^2,-q^{-2(l+\epsilon_1\pm\sigma_1)+1}\biggr)\\
+\text{idem}(\theta,-\theta).
\end{multline}
In view of the property \eqref{E:thetaprop}, this yields
\begin{multline}\label{E:rholim-}
\rho^{(r)\pm}_{-l}\sim q^{r^2+2r\epsilon_2}
\bigl(q^{2(\epsilon_2-\epsilon_1\pm\sigma_1+\sigma_2+1)},
q^{2(\epsilon_2-\epsilon_1\pm\sigma_1-\sigma_2+1)};q^2\bigr)^{-1}_{\infty}\cdot\\
\frac{\sqrt{\theta_{q^2}\bigl(-q^{2(\epsilon_1\mp\sigma_1)+1}\bigr)}}{\sqrt{
\theta_{q^2}\bigl(-q^{2(\epsilon_2+\sigma_2)+1},-q^{2(\epsilon_2-\sigma_2)+1},
-q^{2(\epsilon_1\pm\sigma_1)+1}\bigr)}}\cdot\\
\shoveleft{
\Biggl\{e^{-il\theta}\,\theta_{q^2}\bigl(-q^{2(r+\epsilon_2\pm\sigma_1+1)}e^{
-i\theta}\bigr)\cdot}\\
\frac{\bigl(q^{2(r+\epsilon_2-\epsilon_1)+1}e^{i\theta},q^{
\pm2\sigma_1+2\sigma_2+1}e^{i\theta},q^{\pm2\sigma_1-2\sigma_2+1}e^{i\theta};
q^2\bigr)_{\infty}}{\bigl(e^{2i\theta};q^2\bigr)_{\infty}}
+\text{idem}(\theta,-\theta)\Biggr\}.
\end{multline}
A simpler asymptotic behaviour follows for the linear combination
\begin{multline}\label{E:varsigma}
\varsigma^{(r)}_{l}\equiv\varsigma^{(r)}_{l}\bigl(q^{2\epsilon_1},q^{2
\sigma_1};q^{2\epsilon_2},q^{2\sigma_2};e^{i\theta}\bigr)=\\
\theta_{q^2}\bigl(-q^{2(r+\epsilon_2-\sigma_1+1)}e^{-i\theta},-q^{2(
\epsilon_1+\sigma_1)+1}\bigr)
\bigl(q^{2(\sigma_2-\sigma_1)+1}e^{i\theta},q^{-2(\sigma_1+\sigma_2)+1}e^{i
\theta};q^2\bigr)_{\infty}\cdot\\
\shoveright{\bigl(q^{2(\epsilon_2-\epsilon_1+\sigma_1+\sigma_2+1)},q^{2(
\epsilon_2-\epsilon_1+\sigma_1-\sigma_2+1)};q^2\bigr)_{\infty}\,\rho^{(r)+}_{
l}}\\
-\theta_{q^2}\bigl(-q^{2(r+\epsilon_2+\sigma_1+1)}e^{-i\theta},-q^{2(\epsilon_1
-\sigma_1)+1}\bigr)\bigl(q^{2(\sigma_1+\sigma_2)+1}e^{i\theta},q^{2(\sigma_1-
\sigma_2)+1}e^{i\theta};q^2\bigr)_{\infty}\cdot\\
\bigl(q^{2(\epsilon_2-\epsilon_1-\sigma_1+\sigma_2+1)},q^{2(\epsilon_2-\epsilon_1
-\sigma_1-\sigma_2+1)};q^2\bigr)_{\infty}\,\rho^{(r)-}_{l}.
\end{multline}
The transformation [\cite{GR}:III.9] with $a\to q^{-2(r+\epsilon_2-\epsilon_1)
+1}e^{-i\theta}$, $b\to q^{2(\sigma_2-\sigma_1)+1}e^{-i\theta}$, $c\to q^{-2(
\sigma_1+\sigma_2)+1}e^{-i\theta}$, $d\to q^2e^{-2i\theta}$, $e\to-q^{-2(l+r+
\epsilon_2+\sigma_1)+2}e^{-i\theta}$ relates the $_3\phi_2$-series for upper
and lower sign written out on the rhs of \eqref{E:phiPhi1}. Their contributions
to the rhs of \eqref{E:varsigma} cancel.
Application of the corresponding transformation for the remaining two
$_3\phi_2$-series followed by the use of \eqref{E:ex.2.16} with the
specifications 
$x\to-q^{2(r+\epsilon_2+1)}$, $\lambda\to q^{2\sigma_1}e^{i\theta}$, $\mu\to q^{2\sigma_2+1}$, $\nu\to q^{-2\sigma_1}e^{i\theta}$
yields
\begin{multline}\label{E:varsig2}
\varsigma^{(r)}_l=\\-q^{2l^2+2l(r+\epsilon_1+\epsilon_2)
-2(r+\epsilon_2+\sigma_1)}e^{-i(l-1)\theta}\theta_{q^2}
\bigl(q^{4\sigma_1}\bigr)
S_l\bigl(q^{2\epsilon_1},q^{2\sigma_1}\bigr)
S_{l+r}\bigl(q^{2\epsilon_2},q^{2\sigma_2}\bigr)\cdot\\
\bigl(-q^{-2(l+r+\epsilon_2+\sigma_1)+2}e^{i\theta},-q^{-2(l+\epsilon_1-\sigma_1
)+1},q^{2(r+\epsilon_2-\epsilon_1)+1}e^{-i\theta},q^2e^{2i\theta};q^2\bigr)_{
\infty}\cdot\\
{}_3\phi_2\biggl(\genfrac{}{}{0pt}{}{q^{-2(r+\epsilon_2-\epsilon_1)+1}e^{i
\theta},\,q^{2(\sigma_2-\sigma_1)+1}e^{i\theta},\,q^{-2(\sigma_1+\sigma_2)+1}
e^{i\theta}}{-q^{-2(l+r+\epsilon_2+\sigma_1)+2}e^{i\theta},\,q^2e^{2i\theta}}
;q^2,-q^{-2(l+\epsilon_1-\sigma_1)+1}\biggr).
\end{multline}
With the property \eqref{E:thetaprop}, this implies the asymptotic behaviour
\begin{multline}\label{E:rholim-comb}
e^{-il\theta}\varsigma^{(r)}_{-l}\sim -q^{-r^2-2r(\epsilon_2+1)-2(\epsilon_2+
\sigma_1)}e^{i\theta}\theta_{q^2}\bigl(q^{4\sigma_1}\bigr)\cdot\\
\sqrt{\theta_{q^2}\bigl(-q^{2(\epsilon_1+\sigma_1)+1},-q^{2(\epsilon_1-\sigma_1
)+1},-q^{2(\epsilon_2+\sigma_2)+1},-q^{2(\epsilon_2-\sigma_2)+1}\bigr)}\cdot\\
\bigl(q^2e^{2i\theta},q^{2(r+\epsilon_2-\epsilon_1)+1}e^{-i\theta};q^2\bigr)_{
\infty}+O(q^{2l})
\end{multline}
in the limit $l\to\infty$.

For the specifications  
\begin{equation}\label{E:Jacpar}
\begin{split}
&\gamma\to e^{i\theta},\quad s\to q^{2(r+\epsilon_2-\epsilon_1)+1},\quad az_+\to-q^{2(\epsilon_1+
\sigma_1)+1},\quad bz_+\to-q^{2(\epsilon_1-\sigma_1)+1},\\
&\tfrac{c}{a}\to q^{2(r+\epsilon_2-\epsilon_1-\sigma_1+\sigma_2)},\quad
\tfrac{d}{a}\to q^{2(r+\epsilon_2-\epsilon_1-\sigma_1-\sigma_2)},
\end{split}
\end{equation}
and $x\to z_+q^{2l}$ restricted by $q^{2(l+\epsilon_1-\sigma_1)-1}>1$,
the functions $\Phi^+_{\gamma}(x)$ considered in Section 3.5 in \cite{vbigJac} can be expressed in terms of $\varsigma^{(r)}_l$ by
\begin{multline}\label{E:Phi1}
\Phi_{e^{i\theta}}(z_+q^{2l})=
-q^{2(r+\epsilon_2+\sigma_1)}e^{-i\theta}\cdot q^{-l}\,\frac{S_{l+r}\bigl(
q^{2\epsilon_2},q^{2\sigma_2}\bigr)}{S_l\bigl(q^{2\epsilon_1},q^{2\sigma_1}
\bigr)}\cdot\\
\frac{\varsigma^{(r)}_l\bigl(q^{2\epsilon_1},q^{2\sigma_1};q^{2\epsilon_2},
q^{2\sigma_2};e^{i\theta}\bigr)
}{\theta_{q^2}\bigl(-q^{2(r+\epsilon_2+\sigma_2)+1},-q^{2(r+\epsilon_2
-\sigma_2)+1},q^{4\sigma_1}\bigr)\bigl(q^{2(r+\epsilon_2-\epsilon_1)+1}e^{
-i\theta},q^2e^{2i\theta};q^2\bigr)_{\infty}}.
\end{multline}
This follows from Proposition 3.10 in \cite{vbigJac} and the definition
of $\varsigma^{(r)}_l$ given by \eqref{E:rholim-comb}. 

The expression \eqref{E:phiPhi1} for $\rho^{(r)\pm}_l$
corresponds to the formula for the big $q$-Jacobi 
functions with $bx>q^2$ provided by Proposition 3.4 in \cite{vbigJac}. 

The description of the pairs of tensor products
presented below involves the finite linear combinations of basis elements 
$w^{(\epsilon_1,q^{2\sigma_1})}_l\otimes w^{(\epsilon_2,q^{2\sigma_2})*}_{
l+r}$ 
introduced by
\begin{multline}\label{E:rhosumdef1}
\rho^{(r;N,M)\pm}\equiv\rho^{(r;N,M)\pm}\bigl(q^{2\epsilon_1},q^{2\sigma_1};
q^{2\epsilon_2},q^{2\sigma_2};\cos\theta\bigr)=\\
\mathsf a^{\pm}_r\sum_{l=-M}^N\varrho_l^{(r)\pm}\,
w_l^{(\epsilon_1,q^{2\sigma_1})}\otimes w_{l+r}^{(\epsilon_2,q^{2\sigma_2})*},
\end{multline}
where $-M\leq N\in\mathbb Z$ and 
\begin{equation}\label{E:rtable}
\begin{split}
r&=s-n\;\text{with}\;s\in\mathbb Z,n\in\mathbb Z_{\geq0}\quad\text{if}\;e^{
i\theta}=q^{-2(s+\epsilon_2-\epsilon_1)-1},\\
r&=s'+n\;\text{with}\;s'\in\mathbb Z\quad\text{if}\;e^{i\theta}=q^{2(s'+
\epsilon_2-\epsilon_1)-1},\\
r&\in\mathbb Z\quad\text{otherwise}.
\end{split}
\end{equation}

The functions $\varrho^{(r)\pm}_l$ and the prefactors $a_r^{\pm}$ are specified by
\begin{equation}\label{E:ardef}
\begin{split}
\varrho_l^{(r)\pm}&=\rho^{(r)\pm}_l=\rho^{(r)\pm}_l\bigl(q^{2\epsilon_1},q^{2
\sigma_1};q^{2\epsilon_2},q^{2\sigma_2};\cos\theta\bigr),\\
\mathsf a_r^{\pm}&=
a_r^{\pm}\equiv a^{\pm}_r\bigl(q^{2\epsilon_1},q^{2\epsilon_2};q^{2\sigma_1}
;\cos\theta\bigr)=\\
&\begin{cases}
&\frac{(-1)^rq^{\pm 2r\sigma_1+r-s}}{\sqrt{(q^{2(r-s)},q^{4(\epsilon_2-\epsilon_1)
+2(r+s+1)};q^2)_{s-r}}}\;\text{if}\;e^{i\theta}=q^{-2(s+\epsilon_2-\epsilon_1
)-1},\\
\\
&(-1)^rq^{\pm 2r\sigma_1+r-s'}\sqrt{(q^2,q^{4(\epsilon_2-\epsilon_1+s')};q^2)_{r-s'}}\\
&\qquad\qquad\qquad\qquad\qquad\qquad
\text{if}\;e^{i\theta}=q^{2(s'+\epsilon_2-\epsilon_1)-1},\\
&(-q)^rq^{\pm2r\sigma_1}\sqrt{\bigl(q^{2(\epsilon_2-\epsilon_1)+1}e^{i\theta}
,q^{2(\epsilon_2-\epsilon_1)+1}e^{-i\theta};q^2\bigr)_r}\\
&\qquad\qquad\qquad\qquad\qquad\qquad\qquad\qquad\qquad\text{otherwise,}
\end{cases}
\end{split}
\end{equation}
or by
\begin{equation}\label{E:ardef2}
\begin{split}
\varrho^{(r)\pm}_l&=\mathring{\rho}^{(r)\pm}_l=\rho^{(-r)\pm}_{l+r}\bigl(
q^{2\epsilon_2},q^{2\sigma_2};q^{2\epsilon_1},q^{2\sigma_1};\cos\theta\bigr),\\
\mathsf a^{\pm}_r&=\mathring a^{\pm}_r\equiv\mathring a^{\pm}_r\bigl(q^{2
\epsilon_2},q^{2\epsilon_1};q^{2\sigma_2};\cos\theta\bigr)=
(-1)^ra^{\pm}_{-r}\bigl(q^{2\epsilon_2},q^{2\epsilon_1};q^{2\sigma_2};\cos\theta 
\bigr).
\end{split}
\end{equation}

For both choices, the sums $\rho^{(r;N,M)\pm}$ are characterized by the property
\begin{multline}\label{E:rhoe1}
(1-q^2)\Delta(e_1)\rho^{(r;N,M)\pm}=\\
q^2\sqrt{\bigl(1-q^{2(r+\epsilon_2-\epsilon_1)+1}e^{i\theta}\bigr)\bigl(1-
q^{2(r+\epsilon_2-\epsilon_1)+1}e^{-i\theta}\bigr)}\rho^{(r+1;N-1,M)\pm}\\
+\mathsf a_r^{\pm}q^{-2(N+\epsilon_1)+2}s_{N+r}\bigl(q^{2\epsilon_2},
q^{2\sigma_2}\bigr)\,\varrho^{(r)\pm}_Nw_N^{(\epsilon_1,q^{2\sigma_1})}\otimes 
w_{N+r+1}^{(\epsilon_2,q^{2\sigma_2})*}\\
-\mathsf a_r^{\pm}q^{2(M-\epsilon_1)+3}s_{-M-1}\bigl(q^{2\epsilon_1},
q^{2\sigma_1}\bigr)\,\varrho^{(r)\pm}_{-M}w_{-M-1}^{(\epsilon_1,q^{2\sigma_1})}
\otimes w_{-M+r}^{(\epsilon_2,q^{2\sigma_2})*}
\end{multline} 
and
\begin{multline}\label{E:rhof1}
(1-q^2)\Delta(f_1)\rho^{(r;N,M)\pm}=\\
-q^{-2(r+\epsilon_2-\epsilon_1)+1}\sqrt{\bigl(1-q^{2(r+\epsilon_2-\epsilon_1)-1}
e^{i\theta}\bigr)\bigl(1-q^{2(r+\epsilon_2-\epsilon_1)-1}e^{-i\theta}
\bigr)}\rho^{(r-1;N+1,M)\pm}\\
+\mathsf a_r^{\pm}q^{-2(N+r+\epsilon_2)-1}
s_{N+r}\bigl(q^{2\epsilon_2},q^{2\sigma_2}\bigr)\,\varrho^{(r)\pm}_{N+1}
w^{(\epsilon_1,q^{2\sigma_1})}_{N+1}\otimes w_{N+r}^{(\epsilon_2,q^{2\sigma_2})
*}\\
-\mathsf a_r^{\pm}q^{2(M-r-\epsilon_2+1)}s_{-M-1}\bigl(q^{2\epsilon_1},q^{2\sigma_1}
\bigr)\,\varrho^{(r)\pm}_{-M-1}w_{-M}^{(\epsilon_1,q^{2\sigma_1})}\otimes w_{-M
-1+r}^{(\epsilon_2,q^{2\sigma_2})*}
\end{multline}
with $r$ chosen according to \eqref{E:rtable}.
Equation \eqref{E:rhoe1} follows from the relation \eqref{E:rhorl+}. The derivation of \eqref{E:rhof1} entails the contiguous relation 
\begin{equation}\label{E:rhorl-}
s_{l+r-1}\bigl(q^{2\epsilon_2},q^{2\sigma_2}\bigr)\rho^{(r)\pm}_l-qs_{l-1}\bigl(
q^{2\epsilon_1},q^{2\sigma_1}\bigr)\rho^{(r)\pm}_{l-1}+q^{2(l+\epsilon_1\mp\sigma_1
)-1}\rho^{(r-1)\pm}_l=0
\end{equation}
for $\rho^{(r)\pm}_l$ given by \eqref{E:rho1b}. If $\vert q^{-2(l+\epsilon_1\mp 
\sigma_1)+1}\vert<1$, the relation \eqref{E:rhorl-} is easily obtained
referring to the expression \eqref{E:phiPhi1} for $\rho^{(r)\pm}_l$ and
making use of the explicit writing for the $_3\phi_2$-series
on the rhs of \eqref{E:phiPhi1} according to \eqref{E:phidef}. Relation
\eqref{E:rhorl-} with the replacement $l\to l+1$ follows from the equations
\eqref{E:rhorl+} and \eqref{E:rhorl-}. Application of \eqref{E:rhorl+}
and \eqref{E:rhorl+} with $r\to r-1$ yields
\begin{multline}\label{E:rhorl-l+1}
s_l\bigl(q^{2\epsilon_1},q^{2\sigma_1}\bigr)\Bigl\{q^{-1}s_{l+r}\bigl(q^{2\epsilon_2},
q^{2\sigma_2}\bigr)\rho^{(r)\pm}_{l+1}-s_l\bigl(q^{2\epsilon_1},q^{2\sigma_1}\bigr)
\rho^{(r)\pm}_l+q^{2(l+\epsilon_1\mp\sigma_1)}\rho^{(r-1)\pm}_{l+1}\Bigr\}\\
=q^{2(l+\epsilon_1\pm\sigma_1)+1}s_{l+r}\bigl(q^{2\epsilon_2},q^{2\sigma_2}\bigr)
\bigl(1-q^{2(r+\epsilon_2-\epsilon_1)+1}e^{i\theta}\bigr)\bigl(1-q^{2(r+\epsilon_2
-\epsilon_1)+1}e^{-i\theta}\bigr)\rho^{(r+1)\pm}_l\\
+q^{2(l+\epsilon_1\mp\sigma_1)+1}s_{l+r-1}\bigl(q^{2\epsilon_2},q^{2\sigma_2}\bigr)
\rho^{(r-1)\pm}_l+\Bigl\{s_{l+r}\bigl(q^{2\epsilon_2},q^{2\sigma_2}\bigr)^2-
s_l\bigl(q^{2\epsilon_1},q^{2\sigma_1}\bigr)^2+\\
q^{4(l+\epsilon_1)+2}\bigl(1-
q^{2(r+\epsilon_2-\epsilon_1)-1}e^{i\theta}\bigr)\bigl(1-q^{2(r+\epsilon_2-
\epsilon_1)-1}e^{-i\theta}\bigr)\Bigr\}\rho^{(r)\pm}_l.
\end{multline}
Use of \eqref{E:rhorl-} with $r\to r+1$ followed by equation \eqref{E:rhorl+} with $l\to l-1$ allows to express the first contribution to the
rhs of \eqref{E:rhorl-l+1} by
\begin{multline*}
q^{2(l+\epsilon_1\pm\sigma_1)+1}\bigl(1-q^{2(r+\epsilon_2-\epsilon_1)+1}
e^{i\theta}\bigr)\bigl(1-q^{2(r+\epsilon_2-\epsilon_1)+1}e^{-i\theta}\bigr)\cdot\\
\shoveright{
\Bigl\{qs_{l-1}\bigl(q^{2\epsilon_1},q^{2\sigma_1}\bigr)\rho^{(r+1)\pm}_{l-1}-
q^{2(l+\epsilon_1\mp\sigma_1)-1}\rho^{(r)\pm}_l\Bigr\}=}\\
-q^3s_{l-1}\bigl(q^{2\epsilon_1},q^{2\sigma_1}\bigr)
s_{l+r-1}\bigl(q^{2\epsilon_2},q^{2\sigma_2}\bigr)\rho^{(r)\pm}_{l-1}\\
+\Bigl\{q^2s_{l-1}\bigl(q^{2\epsilon_1},q^{2\sigma_1}\bigr)^2-q^{4(l+\epsilon_1)}
\bigl(1-q^{2(r+\epsilon_2-\epsilon_1)+1}e^{i\theta}\bigr)\bigl(1-q^{2(r+
\epsilon_2-\epsilon_1)+1}e^{-i\theta}\bigr)\Bigr\}\rho^{(r)\pm}_l.
\end{multline*}
Finally, eliminating $\rho^{(r)\pm}_{l-1}$ by means of \eqref{E:rhorl-}, the
rhs of \eqref{E:rhorl-l+1} is found to vanish. This confirms equation
\eqref{E:rhorl-} for all values of $l$. 

The relations \eqref{E:rhorl+} and
\eqref{E:rhorl-} can be combined to obtain
\begin{multline}\label{E:rholll}
q^{-2}s_l(q^{2\epsilon_1},q^{2\sigma_1}\bigr)s_{l+r}\bigl(q^{2\epsilon_2},q^{2
\sigma_2}\bigr)\rho^{(r)\pm}_{l+1}+q^2s_{l-1}\bigl(q^{2\epsilon_1},q^{2\sigma_1}
\bigr)s_{l+r-1}\bigl(q^{2\epsilon_2},q^{2\sigma_2}\bigr)\rho^{(r)\pm}_{l-1}\\
-\Bigl\{q^{-1}s_l\bigl(q^{2\epsilon_1},q^{2\sigma_1}\bigr)^2+qs_{l+r-1}
\bigl(q^{2\epsilon_2},q^{2\sigma_2}\bigr)^2-
q^{4(l+\epsilon_1)+1}\bigl(1-q^{2(r+\epsilon_2-\epsilon_1)-1}e^{i\theta}\bigr)
\cdot\\
\bigl(1-q^{2(r+\epsilon_2-\epsilon_1)-1}e^{-i\theta}\bigr)\Bigr\}\rho^{(r)\pm}_l=0.
\end{multline}

The function $\mathring \rho^{(r)+}_l$ specifying the second option for the
linear combinations $\rho^{(r;N,M)+}$ is related to $\rho^{(r)\pm}_l$ by
\begin{multline}\label{E:rhoreli}
\bigl(q^{2(\sigma_2-\sigma_1)+1}e^{i\theta},q^{2(\sigma_2-\sigma_1)+1}e^{-i\theta},
q^{2(\epsilon_2-\epsilon_1+\sigma_1+\sigma_2+1)};q^2\bigr)_{\infty}\cdot\\
\shoveright{
q^{r+2r\sigma_1}\rho^{(r)+}_l\bigl(q^{2\epsilon_1},q^{2\sigma_1};q^{2\epsilon_2},
q^{2\sigma_2};\cos\theta\bigr)}\\
-\frac{\bigl(q^{2(\sigma_1+\sigma_2)+1}e^{i\theta},q^{2(\sigma_1+\sigma_2)+1}
e^{-i\theta},q^{-2(\epsilon_2-\epsilon_1+\sigma_1-\sigma_2)},q^{2(\epsilon_2-
\epsilon_1-\sigma_1+\sigma_2+1)};q^2\bigr)_{\infty}}{\bigl(q^{-2(\epsilon_2-
\epsilon_1-\sigma_1-\sigma_2)};q^2\bigr)_{\infty}}\cdot\\
\shoveright{q^{r-2r\sigma_1}\rho^{(r)-}_l\bigl(q^{2\epsilon_1},q^{2\sigma_1};
q^{2\epsilon_2},q^{2\sigma_2};\cos\theta\bigr)}\\
=\frac{\bigl(q^{2(r+\epsilon_2-\epsilon_1)+1}e^{i\theta},q^{2(r+\epsilon_2-
\epsilon_1)+1}e^{-i\theta},q^{-2(\epsilon_2-\epsilon_1+\sigma_1-\sigma_2)+2}
;q^2\bigr)_{\infty}\theta_{q^2}\bigl(q^{4\sigma_1+2}\bigr)}{\bigl(q^{2(\epsilon_2
-\epsilon_1+\sigma_1-\sigma_2+1)},q^{2(\epsilon_2-\epsilon_1-\sigma_1-\sigma_2)}
;q^2\bigr)_{\infty}}\cdot\\
(-1)^rq^{r(r-1)+2r(\epsilon_2-\epsilon_1-\sigma_2)}\rho^{(-r)+}_{l+r}\bigl(
q^{2\epsilon_2},q^{2\sigma_2};q^{2\epsilon_1},q^{2\sigma_1};\cos\theta\bigr).
\end{multline}
Taking into account the $r$-dependence of $a^{\pm}_r$ specified by \eqref{E:ardef}, a reformulation of relation \eqref{E:rhoreli} 
particularly suited for later use reads
\begin{multline}\label{E:rhorel}
\bigl(q^{2(\sigma_2-\sigma_1)+1}e^{i\theta},q^{2(\sigma_2-\sigma_1)+1}e^{-i
\theta},q^{2(\epsilon_2-\epsilon_1+\sigma_1+\sigma_2+1)};q^2\bigr)_{\infty}
\cdot\\
a^+_r\bigl(q^{2\epsilon_1},q^{2\epsilon_2};q^{2\sigma_1};\cos\theta\bigr)
\rho^{(r)+}_l\bigl(q^{2\epsilon_1},q^{2\sigma_1};q^{2\epsilon_2},q^{2
\sigma_2};\cos\theta\bigr)\\
-\frac{\bigl(q^{2(\sigma_1+\sigma_2)+1}e^{i\theta},q^{2(\sigma_1+\sigma_2)+1}e^{-i
\theta},q^{-2(\epsilon_2-\epsilon_1+\sigma_1-
\sigma_2)},q^{2(\epsilon_2-\epsilon_1-\sigma_1+\sigma_2+1)};q^2\bigr)_{\infty}
}{\bigl(q^{-2(\epsilon_2-\epsilon_1-\sigma_1-\sigma_2)};q^2\bigr)_{\infty}}
\cdot\\
a^-_r\bigl(q^{2\epsilon_1},q^{2\epsilon_2};q^{2\sigma_1};\cos\theta\bigr)
\rho^{(r)-}_l\bigl(q^{2\epsilon_1},q^{2\sigma_1};q^{2\epsilon_2},q^{2
\sigma_2};\cos\theta\bigr)=\\
\shoveleft{\frac{\theta_{q^2}\bigl(q^{4\sigma_1+2}\bigr)\bigl(
q^{-2(\epsilon_2-\epsilon_1+\sigma_1-\sigma_2)+2};q^2\bigr)_{\infty}}{\bigl(
q^{2(\epsilon_2-\epsilon_1+\sigma_1-\sigma_2+1)},q^{2(\epsilon_2-\epsilon_1-
\sigma_1-\sigma_2)};q^2\bigr)_{\infty}}u\bigl(q^{2\epsilon_2},q^{2\epsilon_1};e^{i\theta}
\bigr)
\cdot}\\
(-1)^ra^+_{-r}\bigl(q^{2\epsilon_2},q^{2\epsilon_1};q^{2\sigma_2};\cos\theta\bigr)
\rho^{(-r)+}_{l+r}\bigl(q^{2\epsilon_2},q^{2\sigma_2};
q^{2\epsilon_1},q^{2\sigma_1};\cos\theta\bigr),
\end{multline} 
where
\begin{equation*}
u\bigl(q^{2\epsilon_2},q^{2\epsilon_1};e^{i\theta}\bigr)=\begin{cases}
&0\quad\text{for}\;e^{i\theta}=q^{-2(s+\epsilon_2-\epsilon_1)-1},\\
&
q^{s'^2+2s'(\epsilon_2-\epsilon_1-1)}
\bigl(q^2,q^{4(\epsilon_2-\epsilon_1+s'
)};q^2\bigr)_{\infty}\\
&\qquad\qquad\qquad\qquad\qquad
\text{for}\;e^{i\theta}=q^{2(s'+\epsilon_2-\epsilon_1)-1},\\
&\bigl(q^{2(\epsilon_2-\epsilon_1)+1}e^{i\theta},q^{2(\epsilon_2-\epsilon_1)+1}
e^{-i\theta};q^2\bigr)_{\infty}\;\text{otherwise.}
\end{cases}
\end{equation*}
In the case $\max\bigl(\vert q^{2(r+\epsilon_2-\epsilon_1)+1}e^{i\theta}\vert,\vert q^{2(l+\epsilon_1\mp\sigma_1)+1}\vert\bigr)<1$, equation \eqref{E:rhoreli}
is obtained relating the two sign options of the first $_3\phi_2$-series on the
rhs of \eqref{E:rho3} by the three-term transformation [\cite{GR}:III.33]
with $a\to q^{\pm2\sigma_1-2\sigma_2+1}e^{-i\theta}$, $b\to-q^{2(l+\epsilon_1
\pm\sigma_1)+1}$, $c\to q^{\pm2\sigma_1+2\sigma_2+1}e^{-i\theta}$, $d\to q^{\pm4
\sigma_1+2}$, $e\to-q^{2(l+r+\epsilon_2\pm\sigma_1+1)}e^{-i\theta}$ and
making use of \eqref{E:rho2} with the replacements
$r\to-r$, $l\to l+r$, $\epsilon_1\leftrightarrow\epsilon_2$, $q^{2\sigma_1}
\leftrightarrow q^{2\sigma_2}$. The relation \eqref{E:rhoreli} with $l\to l-1$
follows from equation \eqref{E:rhoreli} and the same equation with $l\to l+1$
due to the contiguous relation \eqref{E:rholll} and \eqref{E:rholll} with
the replacements $r\to-r$, $l\to l+r$, $\epsilon_1\leftrightarrow\epsilon_2$,
$q^{2\sigma_1}\leftrightarrow q^{2\sigma_2}$. 
Thus \eqref{E:rhoreli} is valid for all $l$
provided that $\vert q^{2(r+\epsilon_2-\epsilon_1)+1}e^{i\theta}\vert<1$.
Relation \eqref{E:rhoreli} with the replacement $r\to r-1$ is obtained from
\eqref{E:rhoreli} and the same relation with $l\to l-1$ by means of
the contiguous relation \eqref{E:rhorl-} and by \eqref{E:rhorl+} with
$r\to-r$, $l\to l+r-1$, $\epsilon_1\leftrightarrow\epsilon_2$, $q^{2\sigma_1}
\leftrightarrow q^{2\sigma_2}$.
This shows equation \eqref{E:rhoreli} for all values of $r$ and $l$.

The corresponding relation for $\varsigma^{(r)}_l$ reads
\begin{multline}\label{E:varsig+-r}
\varsigma^{(-r)}_{l+r}\bigl(q^{2\epsilon_2},q^{2\sigma_2};q^{2\epsilon_1},
q^{2\sigma_1};\cos\theta\bigr)=(-1)^rq^{-r^2+4r(\epsilon_1+1)+2(\epsilon_2-
\epsilon_1-\sigma_1+\sigma_2)}e^{-2ir\theta}\cdot\\
\bigl(q^{2(\epsilon_2-\epsilon_1)
+1}e^{i\theta},q^{2(\epsilon_2-\epsilon_1)+1}e^{-i\theta};q^2\bigr)_r
\frac{\theta_{q^2}\bigl(q^{4\sigma_2+2}\bigr)}{\theta_{q^2}\bigl(q^{4\sigma_1+2}
\bigr)}
\frac{\bigl(q^{-2(\epsilon_2-\epsilon_1)+1}e^{-i\theta};q^2\bigr)_{\infty}}{
\bigl(q^{2(\epsilon_2-\epsilon_1)+1}e^{-i\theta};q^2\bigr)_{\infty}}\cdot\\
\varsigma^{(r)}_l\bigl(q^{2\epsilon_1},q^{2\sigma_1};q^{2\epsilon_2},q^{2\sigma_2}
;\cos\theta\bigr).
\end{multline}
If $e^{i\theta}\neq q^{-2(s+\epsilon_2-\epsilon_1)-1}$, the equation is shown
writing out the lhs by means of
\eqref{E:varsigma} with $r\to-r$, $l\to l+r$, $\epsilon_1\leftrightarrow
\epsilon_2$, $q^{2\sigma_1}\leftrightarrow q^{2\sigma_2}$ and making use of
the expansion \eqref{E:rhoreli} to obtain an expression in terms of
$\rho^{(r)\pm}_l(q^{2\epsilon_1},q^{2\sigma_1};q^{2\epsilon_2},q^{2\sigma_2};
\cos\theta)$. Simplifying the prefactors by means of \eqref{E:ex.2.16}
with $x\to iq^{\epsilon_1\pm\sigma_1-2\sigma_2+\frac{1}{2}}$, $\lambda\to -i
q^{-\epsilon_1\pm\sigma_1+\frac{1}{2}}e^{i\theta}$, $\mu\to iq^{2\epsilon_2-
\epsilon_1\mp\sigma_1+\frac{1}{2}}$, $\nu\to-i q^{-\epsilon_1\mp\sigma_1-2
\sigma_2-\frac{1}{2}}$ and recalling equation \eqref{E:varsigma} for $\varsigma^{(r)}_l$ yields the relation \eqref{E:varsig+-r}.
If $e^{i\theta}\neq q^{2(\tilde s+\epsilon_2-\epsilon_1)-1}$. the rhs of
\eqref{E:varsig+-r} can be decomposed according to equation \eqref{E:varsigma}.
Use of the expansion \eqref{E:rhoreli} with $r\to-r$, $l\to l+r$, $\epsilon_1
\leftrightarrow\epsilon_2$, $q^{2\sigma_1}\to q^{2\sigma_2}$, $q^{2\sigma_2}\to 
q^{\pm2\sigma_1}$ followed by the relation \eqref{E:ex.2.16} and 
\eqref{E:varsigma} with $r\to-r$, $l\to l+r$, $\epsilon_1\leftrightarrow 
\epsilon_2$, $q^{2\sigma_1}\leftrightarrow q^{2\sigma_2}$ leads to the lhs of
\eqref{E:varsig+-r}.

\subsection{The definition of the sums}\label{S:taudef}

Throughout the remainder, the case $q^{2\epsilon_1}=-q^{2\sigma_i+1}=
-q^{2u_i+1}$ with $u_i\in\mathbb R$ is excluded.
The following definition entails the function
\begin{equation*}
w_r\equiv w_r\bigl(x;q^{2\epsilon_1},q^{2\epsilon_2};e^{i\theta}\bigr)
\end{equation*}
with the properties
\begin{equation}\label{E:wprop1}
\begin{split}
&w_r=x^{-r}q^{-r(r+2\epsilon_2+1)}e^{ir\theta}\,w_0,\\
&w_r\bigl(x;q^{2\epsilon_1},q^{2\epsilon_2};q^2e^{i\theta}\bigr)=
x q^{2(r+\epsilon_2)}e^{-i\theta}w_r\bigl(x;q^{2\epsilon_1},q^{2
\epsilon_2};e^{i\theta}\bigr)
\end{split}
\end{equation}
and 
\begin{equation}\label{E:wprop2}
x^{r-s}q^{2(r-s)\epsilon_1}w_r\bigl(x;q^{2\epsilon_1},q^{2\epsilon_2};e^{i
\theta}\bigr)=x'^{r-s}q^{2(r-s)\epsilon'_1}
w_r\bigl(x';q^{2\epsilon'_1},q^{2\epsilon'_2};e^{i\theta}\bigr)
\end{equation}
for $e^{i\theta}=\alpha\alpha'q^{2(s+\epsilon'_1+\epsilon_2)+1}$, $x'=\alpha 
\alpha'x$ and $\epsilon'_2-\epsilon'_1=\epsilon_2-\epsilon_1$.
Examples are provided by  $w_r(x;q^{2\epsilon_1},q^{2\epsilon_2};e^{i\theta})=\theta_{q^2}(-x q^{2(r+\epsilon_2+1)}e^{-i
\theta},-xq^{2\epsilon_1+1})$ with $x=\alpha$ or with $x=1$.

\begin{definition}\label{D:taudef1}

The infinite sums $\tau^{(r,k)\pm}$ are introduced by
\begin{multline}\label{E:taudef1}
\boldsymbol{\tau}^{(r,k)\pm}\equiv\boldsymbol{\tau}^{(r,k)\pm}
\bigl(z,\alpha;q^{2\epsilon_1},
q^{2\epsilon_2};q^{2\epsilon_3},q^{2\epsilon_4};q^{2\sigma_3},
q^{2\sigma_4};\cos\theta\bigr)\\
=a_r^{\pm}\bigl(q^{2\epsilon_1},q^{2\epsilon_2};\alpha q^{2\sigma_3};\cos\theta\bigr)
\sum_{l=-\infty}^{\infty}\rho_l^{(r)\pm}\bigl(q^{2\epsilon_1},\alpha q^{2\sigma_3};q^{2\epsilon_2},\alpha q^{2\sigma_4};\cos\theta\bigr)
\,\mathsf{R}^{k,k+r^*}_{l,l+r^*}.
\end{multline}
The infinite sums $\boldsymbol{\tau}^{(r,k)}$ are defined by
\begin{multline}\label{E:taudef2}
\boldsymbol{\tau}^{(r,k)}\equiv \boldsymbol{\tau}^{(r,k)}\bigl(z,x,\alpha;
q^{2\epsilon_1},q^{2\epsilon_2};q^{2\epsilon_3},q^{2\epsilon_4};q^{2\sigma_3},
q^{2\sigma_4};e^{i\theta}\bigr)\\
=a_r\,\sum_{l=-\infty}^{\infty}
\varsigma^{(r)}_l\bigl(q^{2\epsilon_1},\alpha q^{2\sigma_3};
q^{2\epsilon_2},\alpha q^{2\sigma_4};e^{i\theta}\bigr)\,
\mathsf R^{k,k+r^*}_{l,l+r^*},
\end{multline}
where 
\begin{multline}\label{E:ardef3}
a_r\equiv a_r\bigl(x;q^{2\epsilon_1},q^{2\epsilon_2};e^{i\theta}\bigr)=\\
\frac{(-q)^rx^{-r}\sqrt{\bigl(q^{2(\epsilon_2-\epsilon_1)
+1}e^{i\theta},q^{2(\epsilon_2-\epsilon_1)+1}e^{-i\theta};q^2\bigr)_r}}{
w_r\bigl(x;q^{2\epsilon_1},q^{2\epsilon_2};e^{i\theta}\bigr)}
\end{multline}
with $e^{i\theta}=\alpha\alpha'q^{2(s+\epsilon'_1+\epsilon_2)+1}$ or
\begin{equation}\label{E:ardef4}
\begin{split}
x&=1,\\
a_r&\equiv a_r\bigl(\alpha;q^{2\epsilon_1},q^{2\epsilon_2};q^{2\sigma_1};
e^{i\theta}\bigr)=\\
&\quad(-1)^r\alpha^{s'-1}q^{r-s'+(r-s'+1)(r-s'+2\epsilon_1+1)}\frac{
\sqrt{(q^2,q^{4(\epsilon_2-\epsilon_1+s'
)};q^2\bigr)_{r-s'}}}{\theta_{q^2}\bigl(-q^{2(\epsilon_1+\sigma_1)+1},-q^{2(
\epsilon_1-\sigma_1)+1}\bigr)}
\end{split}
\end{equation}
with $e^{i\theta}=q^{2(s'+\epsilon_2-\epsilon_1)-1}$.
For the choice \eqref{E:Rel1} for the coefficients $\mathsf R^{k,k+r^*}_{l,
l+r^*}$,
the sums \eqref{E:taudef1} and \eqref{E:taudef2} will be denoted by
\begin{equation*}
\tau^{(r,k)\pm}\equiv\tau^{(r,k)\pm}\bigl(z,\alpha;q^{2\epsilon_1},
q^{2\epsilon_2};q^{2\epsilon_3},q^{2\epsilon_4};q^{2\sigma_3},q^{2\sigma_4};
\cos\theta\bigr)\\
\end{equation*}
and
\begin{equation*}
\tau^{(r,k)}\equiv\tau^{(r,k)}\bigl(z,x,\alpha;q^{2\epsilon_1},
q^{2\epsilon_2};q^{2\epsilon_3},q^{2\epsilon_4};q^{2\sigma_3},q^{2\sigma_4};
e^{i\theta}\bigr),
\end{equation*}
respectively.
\end{definition}
The sums $\tau^{(r,k)\pm}$ are absolutely convergent for general $\theta$
provided that $\vert zq e^{\pm i\theta}\vert<1$.
This condition can be replaced by $\vert zqe^{i\theta}\vert<1$ if $e^{i\theta}$
is chosen within the discrete set $\Gamma^{\pm}$ given by
\begin{equation}\label{E:Gammapm}
\begin{split}
\Gamma^{\pm}&=\Gamma_{q^{2(r+\epsilon_2-\epsilon_1)+1}}\cup\Gamma_{q^{\pm2
\sigma_1+2\sigma_2+1}}\cup\Gamma_{q^{\pm2\sigma_1-2\sigma_2+1}},\\
\Gamma_a&=\bigl\{a^{-1}q^{-2t}\vert t\in\mathbb Z_{\geq0}\bigr\}.
\end{split}
\end{equation}
Here the upper sign and lower sign refer to $\tau^{(r,k)+}$ and $\tau^{(r,k
)-}$, respectively.
Besides, absolute convergence is ensured by the condition 
$\vert zq^{2(t+r+\epsilon_2+\sigma_1)+1}\vert<1$ for the choice $e^{i\theta}=
-q^{2(t+r+\epsilon_2+\sigma_1)}
\neq q^{-t'}$, where $t\in\mathbb Z$, $t'\in\mathbb Z_{\geq0}$. 
These statements are readily inferred from the asymptotic behaviour of $\rho^{(r)\pm}_l$ and $R^{k,k+r^*}_{l,l+r^*}$ as $l\to\pm\infty$ given
by the equations \eqref{E:rholim+}, \eqref{E:rholim-}, \eqref{E:Rllim+} and
\eqref{E:Rllim-}. In the following, the choice $e^{i\theta}=-q^{2(t+r+\epsilon_2+\sigma_1)}$
won't be considered.

The sums $\tau^{(r,k)}$ are absolutely convergent if $\vert zqe^{i\theta}
\vert<1$ due to \eqref{E:rholim+}, \eqref{E:rholim-comb},
\eqref{E:Rllim+} and \eqref{E:Rllim-}.

Given two sets of parameters 
\begin{equation}\label{E:twosets}
\begin{split}
&\quad(\alpha,x;q^{2\epsilon_1},q^{2\epsilon_2};q^{2\epsilon_3},q^{2\epsilon_4};
q^{2\sigma_3},q^{2\sigma_4}),\notag\\ &(\alpha',x\alpha\alpha';q^{2\epsilon'_1},q^{2\epsilon'_2};q^{2\epsilon_3},
q^{2\epsilon_4};q^{2\sigma_3},q^{2\sigma_4}),\notag
\end{split}
\end{equation}
with
$\epsilon_2-\epsilon_1=\epsilon'_2-\epsilon'_1=\epsilon_3-\epsilon_4$,
the sums attributed to the second set by Definition \ref{D:taudef1} will be abbreviated by
\begin{equation*}
\begin{split}
\boldsymbol{\tau}'^{(r,k)\pm}&=\boldsymbol{\tau}^{(r,k)\pm}\bigl(z,\alpha';
q^{2\epsilon'_1},q^{2\epsilon'_2};q^{2\epsilon_3},q^{2\epsilon_4};q^{2\sigma_3},
q^{2\sigma_4};\cos\theta\bigr),\\
\boldsymbol{\tau}'^{(r,k)}&=\boldsymbol{\tau}^{(r,k)}\bigl(z,x',\alpha';
q^{2\epsilon'_1},q^{2\epsilon'_2};q^{2\epsilon_3},q^{2\epsilon_4};q^{2\sigma_3},
q^{2\sigma_4};e^{i\theta}\bigr),
\end{split}
\end{equation*}
with $x'=x\alpha\alpha'$ for $e^{i\theta}=\alpha\alpha'q^{2(s+\epsilon'_1+
\epsilon_2)+1}$ and $x'=1$ for $e^{i\theta}=q^{2(s'+\epsilon_2-\epsilon_1)-1}$.
The present section addresses the evaluation of the linear combinations
\begin{equation}
\boldsymbol{\Xi}^{(r,k)}=\left(\begin{matrix}
\Xi^{(r,k)+}\\ \Xi^{(r,k)-}\\ \Xi^{(r,k)}\end{matrix}\right)
\end{equation}
given by
\begin{equation}\label{E:Xidef}
\begin{split}
\Xi^{(r,k)\pm}&\equiv \boldsymbol{\tau}^{(r,k)\pm}-\alpha\alpha'q^{2(\epsilon'_1-
\epsilon_1)}\boldsymbol{\tau}'^{(r,k)\pm},\\
\\
\Xi^{(r,k)}&\equiv \boldsymbol{\tau}^{(r,k)}-\alpha\alpha'
q^{2(\epsilon'_1-\epsilon_1)}\boldsymbol{\tau}'^{(r,k)}
\;\text{for}\,e^{i\theta}\in\tilde{\Gamma}\cup\Gamma_{q^{-2(r+\epsilon_2-\epsilon_1
)+1}},
\end{split}
\end{equation}
where 
\begin{equation}\label{E:tilGam}
\tilde{\Gamma}=\Bigl\{\alpha\alpha'q^{2(s+\epsilon_2+\epsilon'_1)+1}\big\vert 
s\in\mathbb Z,\;\vert zq^{2(s+\epsilon_2+\epsilon'_1+1)}\vert<1\Bigr\}.
\end{equation}
For different values of $r$ and $k$, the combinations $\boldsymbol{\Xi}^{(r,k)}$ are related by
\begin{multline}\label{E:taue1int}
\sqrt{\bigl(1-q^{2(r+\epsilon_2-\epsilon_1)+1}e^{i\theta}\bigr)\bigl(1-q^{2(r+
\epsilon_2-\epsilon_1)+1}e^{-i\theta}\bigr)}\,\boldsymbol{\Xi}^{(r+1,k)}=\\
-s_{-k-1}\bigl(q^{-2\epsilon_4},q^{2\sigma_4}\bigr)\,\boldsymbol{\Xi}^{(r,k+1)}
+q^{2(r+\epsilon_2-\epsilon_1)+1}
s_{-k-r-1}\bigl(q^{-2\epsilon_3},q^{2\sigma_3}\bigr)\,\boldsymbol{\Xi}^{(r,k)}
\end{multline}
and
\begin{multline}\label{E:tauf1int}
\sqrt{\bigl(1-q^{2(r+\epsilon_2-\epsilon_1)-1}e^{i\theta}\bigr)\bigl(1-q^{2(r+
\epsilon_2-\epsilon_1)-1}e^{-i\theta}\bigr)}\,\boldsymbol{\Xi}^{(r-1,k)}=\\
-s_{-k}\bigl(q^{-2\epsilon_4},q^{2\sigma_4}\bigr)\,\boldsymbol{\Xi}^{(r,k-1)}
+q^{2(r+\epsilon_2-\epsilon_1)-1}
s_{-k-r}\bigl(q^{-2\epsilon_3},q^{2\sigma_3}\bigr)\,\boldsymbol{\Xi}^{(r,k)}.
\end{multline}
These relations are derived employing the procedure in the
proof of Proposition 3 in \cite{gade1}. The application of the operator 
$R(\tilde z^{\frac{1}{2}},z^{\frac{1}{2}})$ on both sides of
\eqref{E:rhoe1} and \eqref{E:rhof1} is followed by the use of the intertwining
property \eqref{E:int} for $X=e_1$ and $X=f_1$, respectively. For 
the choice $\mathsf R^{k,k+r^*}_{l,l+r^*}=R^{k,k+r^*}_{l,l+r^*}$, an evaluation
of the limits $N,M\to\infty$ by means of the asymptotic behaviour of
the $R$-elements and $\rho^{(r)\pm}_l$ specified by \eqref{E:Rllim+},
\eqref{E:Rllim-} and \eqref{E:rholim+}, \eqref{E:rholim-} provides 
expressions relating $\tau^{(r,k)\pm}$, $\tau^{(r+1,k)\pm}$, $\tau^{(r,k+1)\pm}$
for $X=e_1$ and $\tau^{(r,k)\pm}$, $\tau^{(r-1,k)\pm}$, $\tau^{(r,k-1)
\pm}$ for $X=f_1$. Cancellation 
of the contributions due to the third line of \eqref{E:rhoe1} or \eqref{E:rhof1}
within the particular combination $\Xi^{(r,k)\pm}=\tau^{(r,k)\pm}-\alpha\alpha'
q^{2(\epsilon'_1-\epsilon_1)}\tau'^{(r,k)\pm}$ directly leads to
the relations \eqref{E:taue1int} and \eqref{E:tauf1int}. For the choices
\eqref{E:chRel}, \eqref{E:mrRel} and \eqref{E:mrchRel}, the statements follow 
by means of 
suitable permutations of the parameters and the replacements $\sigma_2\to-
\sigma_2$ or $\sigma_1\to-\sigma_1$. The analogous consideration for 
$\Xi^{(r,k)}$ involves the asymptotic behaviour \eqref{E:rholim-comb}.

Moreover, the linear combinations 
${\Xi}^{(r,k)}\equiv{\Xi}^{(r,k)}(e^{i\theta})$ satisfy the contiguous relations
\begin{multline}\label{E:Xitheta}
\alpha z\lambda_{\pm}e^{\pm3i\theta}\bigl(1-z^{-1}qe^{\pm i\theta}\bigr)
\bigl(1-q^{2(r+\epsilon_2-\epsilon_1)+1}e^{\pm i\theta}\bigr)\frac{a_r(e^{
i\theta})}{a_r(q^{\pm2}e^{i\theta})}{\Xi}^{(r,k)}\bigl(q^{\pm2}
e^{i\theta}\bigr)\\
=zq^{-2(k+\epsilon_4)-1}e^{\pm i\theta}s_k\bigl(q^{2\epsilon_4},q^{2\sigma_4}
\bigr)s_{k+r}\bigl(q^{2\epsilon_3},q^{2\sigma_3}\bigr){\Xi}^{(r,k+1)}(e^{i
\theta})\\
+q^{-2(k+\epsilon_4)+2}s_{k-1}\bigl(q^{2\epsilon_4},q^{2\sigma_4}\bigr)s_{
k+r-1}\bigl(q^{2\epsilon_3},q^{2\sigma_3}\bigr){\Xi}^{(r,k-1)}
(e^{i\theta})\\
-\biggl\{q^{-2(k+\epsilon_4)+1}+zq^{-2(k+\epsilon_4)}e^{\pm i\theta}+q^{2(k+
r+\epsilon_3)}e^{\pm i\theta}+zq^{2(k+r+\epsilon_3)+1}\\
-\frac{q^2e^{\pm2i\theta}\bigl(1-q^{2(r+\epsilon_2-\epsilon_1)-1}e^{\mp i
\theta}\bigr)}{1-q^2e^{\pm2i\theta}}\Bigl[q^{2\sigma_4}+q^{-2\sigma_4}+z\bigl(q^{2
\sigma_3}+q^{-2\sigma_3}\bigr)\Bigr]\\
+\frac{qe^{\pm i\theta}\bigl(1-q^{2(r+\epsilon_2-\epsilon_1)+1}e^{\pm i\theta}\bigr)
}{1-q^2e^{\pm2i\theta}}\Bigl[q^{2\sigma_3}+q^{-2\sigma_3}+z\bigl(q^{2\sigma_4}
+q^{-2\sigma_4}\bigr)\Bigr]\biggr\}{\Xi}^{(r,k)}(e^{i\theta}),
\end{multline}
where $\vert zq^{\pm1}e^{i\theta}\vert<1$,
\begin{equation*}
\begin{split}
&\quad\lambda_-=q^{2(r+\epsilon_2+2)},\\
&\quad\lambda_+=q^{-2(r+\epsilon_2)}e^{-2i\theta}\cdot\\
&\bigl(1-q^{2(\sigma_1+\sigma_2)+1}e^{i\theta}\bigr)\bigl(1-q^{2(\sigma_1-\sigma_2
)+1}e^{i\theta}\bigr)\bigl(1-q^{2(\sigma_2-\sigma_1)+1}e^{i\theta}\bigr)\bigl(1-
q^{-2(\sigma_1+\sigma_2)+1}e^{i\theta}\bigr)
\end{split}
\end{equation*}
and $a_r(e^{i\theta})=a_r$ is given by \eqref{E:ardef3} or \eqref{E:ardef4}. 
The relations \eqref{E:Xitheta} are derived in Appendix \ref{A:thetacont}.

Viewing $(\pi_{W_1}\otimes\pi_{W_2^*})\oplus(\pi_{W'_1}\otimes\pi_{{W'_2}^*})$
specified by the sets \eqref{E:twosets} as sums of $U_q\bigl(sl(2)\bigr)$-representations, the cases $\alpha=-\alpha'$ correspond
to  
\begin{equation}\label{E:piCS}
\bigl(\pi^C\otimes\pi^C\bigr)\oplus\bigl(\pi^S\otimes\pi^S\bigr)\quad\text{and}
\quad \bigl(\pi^C\otimes\pi^S\bigr)\oplus\bigl(\pi^S\otimes\pi^C\bigr),
\end{equation}
provided that $q^2<q^{4\sigma_i}<1$ and the parameters $q^{2\epsilon_i}$,
$q^{2\epsilon'_i}$ are restricted appropriately.

The first case in \eqref{E:piCS} is found among the "indivisible" sums
considered in \cite{groene2}. Its decomposition into irreducible
representations entails Clebsch-Gordon coefficients essentially given by
the big $q$-Jacobi functions \eqref{E:Jac1} and by the functions
$\Phi^+_{\gamma}(x)$ \cite{groene2}.
The combinations $\boldsymbol{\Xi}^{(r,k)}$ describe the action of the
intertwiner $R(\tilde z^{\frac{1}{2}},z^{\frac{1}{2}})$ onto the irreducible components. For $\alpha=\alpha'$, the corresponding $U_q\bigl(sl(2)\bigr)$-representations are $(\pi^C\otimes\pi^C)\oplus(\pi^C 
\otimes\pi^C)$, $(\pi^S\otimes\pi^S)\oplus(\pi^S\otimes\pi^S)$, $(\pi^C
\otimes\pi^S)\oplus(\pi^C\otimes\pi^S)$ and $(\pi^S\otimes\pi^C)\oplus(\pi^S
\otimes\pi^C)$.

\subsection{Relation to the vector-valued big $q$-Jacobi function transform}
\label{S:bigJac}

If $\beta=\alpha\alpha'=-1$ and $\max(q^{4\sigma_1},q^{4\sigma_2})<1$, 
the linear combination $\boldsymbol{\Xi}^{(r,k)}$
can be viewed as a vector-valued
big $q$-Jacobi transform. Following \cite{vbigJac},
for $z_+>0$ and $z_-<0$ and parameters $a,b,c,d$, $s=\sqrt{abcdq^{-2}}$,
the transform involves the discrete set 
\begin{equation}\label{E:bigJset}
\begin{split}
\Gamma&=\Gamma^{\text{fin}}_s\cup\Gamma^{\text{fin}}_{s^{-1}q^2}\cup\Gamma^{
\text{fin}}_{{
dq^2}/{as}}\cup\Gamma^{\text{inf}},\\
\Gamma^{\text{fin}}_e&=\Bigl\{e^{-1}q^{-2t}\big\vert t\in\mathbb Z_{\geq0},\;\;
eq^{2t}>1\Bigr\},\\
\Gamma^{\text{inf}}&=\Bigl\{z_+z_-sq^{2t}\big\vert t\in\mathbb Z,\;-
z_+z_-sq^{2t}<1\Bigr\}.
\end{split}
\end{equation}
The vector-valued big $q$-Jacobi function transform 
$\mathcal F$ is introduced by
\begin{multline}\label{E:Jacdef}
\bigl(\mathcal Ff\bigr)(\gamma)=(1-q^2)\biggl\{z_+\sum_{l=-\infty}^{\infty}
f(z_+q^{2l})\Psi(z_+q^{2l},\gamma)\,w(z_+q^{2l})q^{2l}\\
-z_-\sum_{l=-\infty}^{\infty}
f(z_-q^{2l})\Psi(z_-q^{2l},\gamma)w(z_-q^{2l})q^{2l}\biggr\},
\end{multline}
where
\begin{equation*}
\Psi(x,\gamma)=\begin{cases}\left(\begin{matrix}\varphi_{\gamma}(x)\\\varphi^{
\dagger}_{\gamma}(x)\end{matrix}\right),\quad&\gamma\in \mathbb T
=\bigl\{y\in\mathbb C\big\vert\,\vert y\vert=1\bigr\}\\
\\
\;\Phi^+_{\gamma}(x),\quad&\gamma\in\Gamma,\end{cases}
\end{equation*}
for $x\in\mathbb R^-_q\cup\mathbb R^+_q$ with $R^{\pm}_q=\{z_{\pm}q^{2n}\vert 
n\in\mathbb Z\}$, 
and 
\begin{equation*}
w(x)=\frac{(ax,bx;q^2)_{\infty}}{(cx,dx;q^2)_{\infty}}.
\end{equation*}
The specifications \eqref{E:Jacpar} are supplemented by
\begin{equation}\label{E:Jacpar2}
az_-\to q^{2(\epsilon'_1+\sigma_1)+1},\quad bz_-\to q^{2(\epsilon'_1-\sigma_1)
+1}.
\end{equation}
With the specifications \eqref{E:Jacpar}, the components $\Phi_{e^{i\theta}}(z_+q^{2l})$ and $\varphi_{e^{i\theta}}(z_+q^{2l})$, $\varphi^{\dagger}_{e^{i\theta}}
(z_+q^{2l})$ are given by the rhs of \eqref{E:Phi1} and by the expression \eqref{E:varphirho} with lower and upper sign, 
respectively. According to \eqref{E:Jacpar2}, with the replacements
 $\epsilon_1\to\epsilon'_1$,
$\epsilon_2\to\epsilon'_2$, $q^{2\sigma_1}\to-q^{2\sigma_1}$, $q^{2\sigma_2}\to 
-q^{2\sigma_2}$, the rhs of \eqref{E:Phi1} and \eqref{E:varphirho} provide the expressions for $\Phi_{e^{i\theta}}(z_-q^{2l})$ and $\varphi_{e^{i\theta}}
(z_-q^{2l})$ and $\varphi^{\dagger}_{e^{i\theta}}(z_-q^{2l})$ for lower
and upper sign, respectively.
The specifications \eqref{E:Jacpar} and \eqref{E:Jacpar2}
imply $z_+=-\alpha q^{2\epsilon_1}$, $z_-=-\alpha'q^{2\epsilon'_1}$,
$\tfrac{dq^2}{as}=q^{-2(\sigma_1+\sigma_2)+1}$ and $z_+z_-s=-q^{2(r+\epsilon_2+
\epsilon'_1)+1}$. In particular, due to the restriction $\max(q^{4\sigma_1},q^{4\sigma_2})<1$, the specifications for $\alpha=-\alpha'=-1$ ensure $(a,b,c,d)\in P$ with $P$ defined
in Section 2.1 in \cite{vbigJac}.
Setting 
\begin{equation}\label{E:fzpm}
\begin{split}
f(z_+q^{2l})&=q^{-l}\frac{S_{l+r}\bigl(q^{2\epsilon_2},\alpha q^{2\sigma_4}
\bigr)}{S_l\bigl(q^{2\epsilon_1},\alpha q^{2\sigma_3}\bigr)}
\,R^{k,k+r^*}_{l,l+r^*}\bigl(z,\alpha;q^{2\epsilon_1},q^{2\epsilon_2};
q^{2\epsilon_3},q^{2\epsilon_4};q^{2\sigma_3},q^{2\sigma_4}\bigr),\\
f(z_-q^{2l})&=q^{-l}\frac{S_{l+r}\bigl(q^{2\epsilon'_2},-\alpha q^{2\sigma_4}
\bigr)}{S_l\bigl(q^{2\epsilon'_1},-\alpha q^{2\sigma_3}\bigr)}
\,R^{k,k+r^*}_{l,l+r^*}\bigl(z,-\alpha;q^{2
\epsilon'_1},q^{2\epsilon'_2};q^{2\epsilon_3},q^{2\epsilon_4};q^{2\sigma_3},
q^{2\sigma_4}\bigr)
\end{split}
\end{equation}
and $\alpha=-\alpha'$, Definition \ref{D:taudef1} and the equation 
\eqref{E:Jacdef} imply
\begin{multline}\label{E:Ff1}
\bigl(\mathcal Ff\bigr)(e^{i\theta})=-\alpha q^{2\epsilon_1}(1-q^2)\left(
\begin{matrix}b^-_r&0\\0&b^+_r\end{matrix}\right)\,
\left(\begin{matrix}\tau^{(r,k)-}+q^{2(\epsilon'_1-\epsilon_1)}\tau'^{(r,k)-}\\
\tau^{(r,k)+}+q^{2(\epsilon'_1-\epsilon_1)}\tau'^{(r,k)+}\end{matrix}\right)\;
\text{for}\;e^{i\theta}\in\mathbb T
\end{multline}
with
\begin{equation*}
b^{\pm}_r=(a_r^{\pm})^{-1}
\bigl(q^{2(\epsilon_2-\epsilon_1\pm\sigma_1+\sigma_2+1)},
q^{2(\epsilon_2-\epsilon_1\pm\sigma_1-\sigma_2+1)};q^2\bigr)_r
\end{equation*}
and
\begin{multline}\label{E:Ff2}
\bigl(\mathcal Ff\bigr)(e^{i\theta})=
\bigl(\tau^{(r,k)}+q^{2(\epsilon'_1-\epsilon_1)}\tau'^{(r,k)}\bigr)\cdot\\
\frac{q^{2(r+\epsilon_1+\epsilon_2+\sigma_3)}(1-q^2)e^{-i
\theta}}{a_r\theta_{q^2}\bigl(-q^{2(r+\epsilon_2+\sigma_2)+1},-q^{2(r+\epsilon_2-
\sigma_2)+1},q^{4\sigma_3}\bigr)\bigl(q^2e^{2i\theta},q^{2(r+\epsilon_2-\epsilon_1
)+1}e^{-i\theta};q^2\bigr)_{\infty}}\\
\text{for}\;e^{i\theta}\in \tilde{\Gamma}\cap\Gamma^{\text{inf}} \text{and}\;e^{i\theta}\in\Gamma_{q^{-2(r+\epsilon_2-\epsilon_1)+1}}\cap\Gamma^{
\text{fin}}_{
q^{-2(r+\epsilon_2-\epsilon_1)+1}}.
\end{multline}

The remainder of this subsection specializes to $\alpha=-1$.
If $e^{i\theta}=q^{2(\sigma_1+\sigma_2)-2t-1}$ with $t\in\mathbb Z_{\geq0}$, 
the first contribution on the rhs of \eqref{E:varsigma} vanishes.
Then the equations \eqref{E:Phi1} and \eqref{E:varsigma} lead to
\begin{multline*}
\Phi^+_{\gamma}(z_+q^{2l})\Big\vert_{\gamma=q^{2(\sigma_1+\sigma_2)-2t-1}}=
q^{-l}\frac{S_{l+r}\bigl(q^{2\epsilon_2},q^{2\sigma_2}\bigr)}{S_l\bigl(
q^{2\epsilon_1},q^{2\sigma_1}\bigr)}\cdot\\
(-1)^tq^{-2t^2-2t(r+\epsilon_2-\sigma_2)+4t\sigma_1}
\frac{\bigl(q^{-2t+4(\sigma_1+\sigma_2)};q^2\bigr)_{
\infty}}{\bigl(q^{2(t+1-2\sigma_1)};q^2\bigr)_{\infty}}\cdot\\
\frac{\theta_{q^2}\bigl(-q^{2(\epsilon_1-\sigma_1)+1}\bigr)\bigl(q^{2(\epsilon_2-
\epsilon_1-\sigma_1+\sigma_2+1)},q^{2(\epsilon_2-\epsilon_1-\sigma_1-\sigma_2
+1)};q^2\bigr)_{\infty}}{\theta_{q^2}\bigl(-q^{2(r+\epsilon_2+\sigma_2)+1}\bigr)}
\,\rho^{(r)-}_l.
\end{multline*}
For the specifications \eqref{E:fzpm}, this yields
\begin{multline*}
\bigl(\mathcal Ff\bigr)\bigl(q^{2(\sigma_1+\sigma_2)-2t-1}\bigr)=(-1)^tq^{-2t^2
-2t(r+\epsilon_2-\sigma_2)+4t\sigma_1+2\epsilon_1}\frac{\bigl(q^{-2t+4(\sigma_1+
\sigma_2)};q^2\bigr)_{\infty}}{\bigl(q^{2(t+1-2\sigma_1)};q^2\bigr)_{\infty}}
\cdot\\
\frac{(1-q^2)\theta_{q^2}\bigl(-q^{2(\epsilon_1-\sigma_1)+1}\bigr)\bigl(q^{2(
\epsilon_2-\epsilon_1-\sigma_1+\sigma_2+1)},q^{2(\epsilon_2-\epsilon_1-\sigma_1
-\sigma_2+1)};q^2\bigr)_{\infty}}{a^-_r\theta_{q^2}\bigl(-q^{2(r+\epsilon_2+\sigma_2
)+1}\bigr)}\cdot\\
\bigl(\tau^{(r,k)-}+q^{2(\epsilon'_1-\epsilon_1)}\tau'^{(r,k)-}\bigr)
\Big\vert_{e^{i\theta}=q^{2(\sigma_1+\sigma_2)-2t-1}}.
\end{multline*}
According to the discussion following Definition \ref{D:taudef1}, the
condition $\vert zq^{2(\sigma_1+\sigma_2-t)}\vert<1$ is sufficient to ensure
absolute convergence of the sums on the rhs since $e^{i\theta}\in\Gamma^-$.

For $e^{i\theta}=q^{-2(s+\epsilon_2-\epsilon_1)-1}$ with
$s-r\in\mathbb Z_{\geq0}$,  equation \eqref{E:rho1b} reduces to
\begin{multline*}
\rho^{(r)\pm}_l\Big\vert_{e^{i\theta}=q^{-2(s+\epsilon_2-\epsilon_1)-1}}=
q^l\frac{S_l\bigl(q^{2\epsilon_1},q^{2\sigma_1}\bigr)}{S_{l+r}\bigl(
q^{2\epsilon_2},q^{2\sigma_2}\bigr)}
\cdot\\
\frac{\bigl(q^{4(\epsilon_2-\epsilon_1)+2(s+r+1)},-q^{2(l+r-s+\epsilon_1\pm\sigma_1
)+1};q^2\bigr)_{s-r}}{\bigl(q^{2(\epsilon_2-\epsilon_1\pm\sigma_1+\sigma_2+1)},
q^{2(\epsilon_2-\epsilon_1\pm\sigma_1-\sigma_2+1)};q^2\bigr)_{s}}\cdot\\
{}_3\phi_2\biggl(\genfrac{}{}{0pt}{}{q^{2(r-s)},\,q^{\pm2\sigma_1+2\sigma_2-2(s+
\epsilon_2-\epsilon_1)},\,q^{\pm2\sigma_1-2\sigma_2-2(s+\epsilon_2-\epsilon_1
)}}{-q^{2(l+r-s+\epsilon_1\pm\sigma_1)+1},\,q^{-4(s+\epsilon_2-\epsilon_1)}};q^2,
q^2\biggr).
\end{multline*}
Supplemented by  [\cite{GR}:III.11] with $b\to q^{-2(\sigma_1-\sigma_2+s+\epsilon_2
-\epsilon_1)}$, $c\to q^{-2(\sigma_2+\sigma_2+s+\epsilon_2-\epsilon_1)}$,
$d\to q^{-4(s+\epsilon_2-\epsilon_1)}$, $e\to-q^{2(l+r-s+\epsilon_1-\sigma_1)
+1}$, this yields the relation
\begin{multline}\label{E:rho+-sp}
\rho^{(r)-}_l\Big\vert_{e^{i\theta}=q^{-2(s+\epsilon_2-\epsilon_1)-1}}=
q^{4(r-s)\sigma_1}\frac{\bigl(q^{2(\epsilon_2-\epsilon_1+\sigma_1+\sigma_2+1)},
q^{2(\epsilon_2-\epsilon_1+\sigma_1-\sigma_2+1)};q^2\bigr)_{s}}{\bigl(
q^{2(\epsilon_2-\epsilon_1-\sigma_1+\sigma_2+1)},q^{2(\epsilon_2-\epsilon_1-\sigma_1
-\sigma_2+1)};q^2\bigr)_{s}}\cdot\\
\rho^{(r)+}_l\Big\vert_{e^{i\theta}=q^{-2(s+\epsilon_2-\epsilon_1)-1}}.
\end{multline}
Use of relation \eqref{E:rho+-sp} in \eqref{E:varsigma} and \eqref{E:Phi1} and application of \eqref{E:ex.2.16} with
$x\to q^{2\sigma_2+1}$, $\lambda\to q^{2(r+\epsilon_2-\epsilon_1-\sigma_1)+1}$,
$\mu\to-q^{2(r+\epsilon_2)+3}$, $\nu\to q^{2(r+\epsilon_2-\epsilon_1+\sigma_1
)+1}$ leads to
\begin{multline*}
\Phi^+_{\gamma}(z_+q^{2l})\Big\vert_{\gamma=q^{-2(s+\epsilon_2-\epsilon_1)-1}}=
q^{-(s-r)^2+2(s-r)(\epsilon_1+\sigma_1)-4(s-r)(r+\epsilon_2)}\cdot\\
q^{-l}\frac{S_{l+r}\bigl(q^{2\epsilon_2},q^{2\sigma_2}\bigr)}{S_l\bigl(
q^{2\epsilon_1},q^{2\sigma_1}\bigr)}\frac{\bigl(q^{2(\epsilon_2-
\epsilon_1+\sigma_1+\sigma_2+1)},q^{2(\epsilon_2-\epsilon_1+\sigma_1-\sigma_2
+1)};q^2\bigr)_{s}}{\bigl(q^{4(\epsilon_2-\epsilon_1)+2(s+r+1)};q^2\bigr)_{s-r}}
\cdot\\
\rho^{(r)+}_l\Big\vert_{e^{i\theta}=q^{-2(s+\epsilon_2-\epsilon_1)-1}}.
\end{multline*}
Thus with the specifications \eqref{E:fzpm}, this implies
\begin{multline*}
\bigl(\mathcal Ff\bigr)\bigl(q^{-2(s+\epsilon_2-\epsilon_1)-1}\bigr)=
q^{-(s-r)^2+2(s-r)(\epsilon_1+\sigma_1)-4(s-r)(r+\epsilon_2)+2\epsilon_1}\cdot\\
\frac{(1-q^2)\bigl(q^{2(\epsilon_2-\epsilon_1+\sigma_1+\sigma_2+1)},q^{2(\epsilon_2-
\epsilon_1+\sigma_1-\sigma_2+1)};q^2\bigr)_{s}}{a^+_r\bigl(q^{
4(r+\epsilon_2-\epsilon_1)+2(s+r+1)};q^2\bigr)_{s-r}}\cdot\\
\bigl(\tau^{(r,k)+}+q^{2(\epsilon'_1-\epsilon_1)}\tau'^{(r,k)+}\bigr)
\Big\vert_{e^{i\theta}=q^{-2(s+\epsilon_2-\epsilon_1)-1}}.
\end{multline*}
As $U_q\bigl(sl(2)\bigr)$-modules, $W^{(\epsilon,q^{2\sigma})}$ and $W^{(
\epsilon,q^{2\sigma})*}$ with $q^{2\sigma}<0$ can be identified with
a complementary series representation provided that $q^2<q^{4\sigma}<1$
and $\epsilon$ is restricted according to the value of $q^{4\sigma}$.
If $q^2<q^{4\sigma_i}<1$ for $i=1,2$, the set $\Gamma^{\text{fin}}_{\frac{dq^2}{as}}=
\Gamma^{\text{fin}}_{q^{-2(\sigma_1+\sigma_2)+1}}$ is empty in the case 
$q^2<q^{4(\sigma_1+\sigma_2)}<1$ and contains only one element in the
case $q^4<q^{4(\sigma_1+\sigma_2)}<q^2$.

The inner product $\langle\,,\rangle_{\mathcal H}$ introduced in Section 4 of
\cite{groene2} or Section 5 of \cite{vbigJac} involves a matrix valued
function $\mathbf v$ on $\mathbb T$ and a positive weight function $N(\gamma)$ on the
discrete set $\Gamma$. Denoting the vector space of functions that are 
$\mathbb C^2$-valued on $\mathbb T$ and $\mathbb C$-valued in $\Gamma$ by
$F(\mathbb T\cup\Gamma)$, the Hilbert space $\mathcal H$ consists of
all functions $g(\gamma)$ in $F(\mathbb T\cup\Gamma)$ with the property $g(\gamma)=g(\gamma^{-1})$ almost everywhere on $\mathbb T$ which have 
finite norm with respect to the inner product $\langle\,,\rangle_{\mathcal H}$.
The set $\{\Psi(x,\cdot)\vert x\in\mathbb R^-_q\cup\mathbb R^+_q\}$ forms an orthogonal basis for $\mathcal H$ with the orthogonality relations
\begin{equation}\label{E:Hbas}
\langle\Psi(x',\cdot),\Psi(x,\cdot)\rangle_{\mathcal H}=\delta_{x,x'}x^{-1}
w^{-1}(x).
\end{equation}
The integral transform
$\mathcal G$ defined by
\begin{equation}\label{E:G}
\bigl(\mathcal Gg\bigr)(x)=\langle g,\Psi(x,\cdot)\rangle_{\mathcal H},\qquad
g\in\mathcal H,\quad x\in\mathbb R^-_q\cup\mathbb R^+_q
\end{equation}
is the inverse of $\mathcal F$.

\subsection{Evaluation of the sums}\label{S:tauev}

In context with the sums $\tau^{(r,k)+}$ and $\tau'^{(r,k)+}$, the conditions 
\begin{subequations}\label{E:tausum1cond}
\begin{equation}\label{E:tausum1acond}
\begin{split}
q^{2(\epsilon_2-\epsilon_1-\sigma_3+\sigma_4)}\neq q^{2t_1},\;t_1\in\mathbb Z,
\quad&q^{2(\epsilon_2
-\epsilon_1+\sigma_3-\sigma_4)}\neq q^{2t_2},\;t_2\in\mathbb Z_{\backslash0},\\
q^{2(\epsilon_2-\epsilon_1-\sigma_3-\sigma_4)}\neq q^{2t_3},\;t_3\in
\mathbb Z_{>0},\quad&q^{2(\epsilon_2-
\epsilon_1+\sigma_3+\sigma_4)}\neq q^{2t_4},\;t_4\in\mathbb Z_{<0},
\end{split}
\end{equation}
\begin{equation}\label{E:tausum1bcond}
\begin{split}
\vert zqe^{i\theta}\vert<&1,\\
\vert zqe^{-i\theta}\vert<&1\quad\text{if}\;e^{i\theta}\notin\Gamma^+
\end{split}
\end{equation}
\end{subequations}
are imposed to ensure well-defined quantities $R^{k,k+r^*}_{l,l+r^*}$ and $\rho^{(r)+}_l$ and absolute convergence of the sums. The replacements
$q^{2\sigma_3}\to q^{-2\sigma_3}$ in \eqref{E:tausum1acond} and $\Gamma^+
\to\Gamma^-$ in \eqref{E:tausum1bcond} yield the corresponding conditions
for $\tau^{(r,k)-}$ and $\tau'^{(r,k)-}$.

\begin{theorem}\label{T:tausum2}
The sums $\tau^{(r,k)+}$ and $\tau'^{(r,k)+}$ with the property 
\eqref{E:tausum1cond} 
are related by
\begin{multline}\label{E:tausum2}
\tau^{(r,k)+}-\alpha\alpha'q^{2(\epsilon'_1-\epsilon_1)}\tau'^{(r,k)+}=\\
\frac{\theta_{q^2}\bigl(\alpha\alpha'
q^{2(\epsilon_1-\epsilon'_1+1)}\bigr)}{\theta_{q^2}\bigl(\alpha z^{-1}q^{2(
\epsilon_1-\epsilon_4)},\alpha'z^{-1}q^{2(\epsilon'_1
-\epsilon_4)}\bigr)}\sqrt{\frac{\theta_{q^2}\bigl(- q^{2(\epsilon_3+\sigma_3)+1},-q^{2(\epsilon_4+\sigma_4)+1}\bigr)}{
\theta_{q^2}\bigl(-q^{2(\epsilon_3-\sigma_3)+1},- q^{2(\epsilon_4-\sigma_4)+1}\bigr)}}\cdot\\
\frac{\bigl(zq^{2(\epsilon_1-\epsilon_2+1)},zq^{2(\sigma_4-\sigma_3+1)},
zq^{2(\sigma_3+\sigma_4+1)},q^2;q^2\bigr)_{\infty}}{\bigl(
q^{2(\epsilon_2-\epsilon_1+\sigma_3-\sigma_4+1)},q^{-2(\epsilon_2-\epsilon_1+
\sigma_3-\sigma_4)+2};q^2\bigr)_{\infty}}\cdot\\
\shoveleft{
\Biggl\{\frac{\theta_{q^2}\bigl(zq^{2(\sigma_3-\sigma_4+1)},-\alpha\alpha' z^{-1}q^{2(
\epsilon_1+\epsilon'_1-\epsilon_4+\sigma_3)+1},-q^{2(\epsilon_4-\sigma_4)+1}
\bigr)}{\theta_{q^2}\bigl(q^{2(\epsilon_2-\epsilon_1-\sigma_3+\sigma_4)},-\alpha
q^{2(\epsilon_1+\sigma_3)+1},-\alpha'q^{2(\epsilon'_1+\sigma_3)+1}\bigr)}\cdot}\\
\frac{\bigl(q^{-2(\epsilon_2-\epsilon_1-\sigma_3+\sigma_4)+2};q^2\bigr)_{
\infty}}{\bigl(q^{2(\epsilon_2-\epsilon_1+\sigma_3+\sigma_4+1)};q^2
\bigr)_{\infty}}
\frac{u\bigl(q^{2\epsilon_2},q^{2\epsilon_1};e^{i\theta}\bigr)
}{\bigl(zqe^{i\theta},zqe^{-i\theta};q^2
\bigr)_{\infty}}\cdot\\
(-1)^ra^{+}_{-r}\bigl(q^{2\epsilon_3},q^{2\epsilon_4};q^{2\sigma_3};\cos\theta\bigr)\,
\rho^{(-r)+}_{k+r}\bigl(q^{2\epsilon_3},q^{2\sigma_3};q^{2
\epsilon_4},q^{2\sigma_4};\cos\theta\bigr)\\
+\frac{\theta_{q^2}\bigl(zq^{2(\epsilon_2-\epsilon_1+1)},-\alpha\alpha' z^{-1}q^{2(\epsilon_1+\epsilon'_2-\epsilon_4+\sigma_4)+1},- q^{2(\epsilon_3-
\sigma_3)+1}\bigr)}{\theta_{q^2}\bigl(q^{-2(\epsilon_2-\epsilon_1-\sigma_3
+\sigma_4)}, -\alpha q^{2(\epsilon_2+\sigma_4)+1},-\alpha'q^{2(\epsilon'_2+\sigma_4)+1}\bigr)}\cdot\\
\frac{\bigl(q^{2(\epsilon_2-\epsilon_1-\sigma_3+\sigma_4+1)};q^2\bigr)_{\infty}}{
\bigl(q^{-2(\epsilon_2-\epsilon_1-\sigma_3-\sigma_4)+2};q^2\bigr)_{\infty}}
\frac{\bigl(q^{2(\sigma_3-\sigma_4)+1}e^{i\theta},q^{2(\sigma_3-\sigma_4)+1}
e^{-i\theta};q^2\bigr)_{\infty}}{\bigl(zqe^{i\theta},zqe^{-i\theta};q^2\bigr)_{
\infty}}\cdot\\
a^{+}_r\bigl(q^{2\epsilon_4},q^{2\epsilon_3};q^{2\sigma_4};\cos\theta\bigr)
\rho^{(r)+}_{k}\bigl(q^{2\epsilon_4},q^{2\sigma_4};q^{2
\epsilon_3},q^{2\sigma_3};\cos\theta\bigr)\Biggr\}.
\end{multline}

\end{theorem}

Here $u(q^{2\epsilon_2},q^{2\epsilon_1};e^{i\theta})$ denotes the function
introduced for equation \eqref{E:rhorel}.  
\vskip 0.5cm

\emph{Proof}: 
The assertion is first proven imposing the additional constraint
\begin{equation}\label{E:kcond}
\max\bigl(\vert zq^{2(k+r+\epsilon_3+\sigma_3+1)}e^{\pm i\theta}\vert, \vert 
q^{2(k+\epsilon_4-\sigma_4)+1}\vert,\vert q^{2(k+r+\epsilon_3-\sigma_3)+1}\vert\bigr)<1.
\end{equation}
Insertion of the expressions \eqref{E:Rel1} and \eqref{E:rho1b} for the 
R-elements and $\rho^{(r)+}_l$ into the Definition \ref{D:taudef1} leads to
\begin{multline}\label{E:tauev1}
\tau^{(r,k)+}=\kappa_ra^+_rz^{-k-r}q^{-k-r}\frac{S_{-k}\bigl(q^{-2
\epsilon_4},q^{2\sigma_4}\bigr)}{S_{-k-r}\bigl(q^{-2\epsilon_3},q^{2\sigma_3}
\bigr)}\cdot\\
\frac{\bigl(q^{2(r+\epsilon_2-\epsilon_1)+1}e^{i\theta},q^{2(\sigma_3-\sigma_4)+1}
e^{i\theta},q^{2(\sigma_3+\sigma_4)+1}e^{i\theta};q^2\bigr)_{\infty}}{\bigl(
e^{2i\theta},q^{-2(\epsilon_2-\epsilon_1+\sigma_3-\sigma_4)+2},q^{2(\epsilon_2
-\epsilon_1+\sigma_3+\sigma_4+1)},q^{2(\epsilon_2-\epsilon_1+\sigma_3-\sigma_4+1)}
;q^2\bigr)_{\infty}}\cdot\\
\sum_{l=-\infty}^{\infty}\sum_{m,n=0}^{\infty}\Biggl\{\frac{\bigl(\alpha q^{2(n+l-k-r+\epsilon_1-\epsilon_3+1)},-\alpha q^{2(m+l+r+\epsilon_2+\sigma_3+1)}
e^{-i\theta};q^2
\bigr)_{\infty}q^{2l}}{\bigl(-\alpha q^{2(n+l+\epsilon_1+\sigma_3)+1},\alpha z^{-1}q^{2(l-k+\epsilon_1-\epsilon_4)};q^2\bigr)_{\infty}}\cdot\\
\frac{\bigl(-q^{-2(k+r+\epsilon_3-\sigma_3)+1},zq^{-2(r+\epsilon_2-\epsilon_1)+2},
z^{-1}q^{-2(r+
\epsilon_2-\epsilon_1)};q^2\bigr)_nq^{2n}}{\bigl(q^{-2(r+\epsilon_2-\epsilon_1-
\sigma_3+\sigma_4)+2},q^{-2(r+\epsilon_2-\epsilon_1-\sigma_3-\sigma_4)+2},q^2;q^2
\bigr)_n}\cdot\\
\frac{\bigl(zq^{2(\sigma_4-\sigma_3+1)},z^{-1}q^{2(\sigma_4-\sigma_3)};q^2\bigr)_{
\infty}\bigl(q^{-2(r+\epsilon_2-\epsilon_1-\sigma_3-\sigma_4)+2};q^2\bigr)_r
}{\bigl(q^{2(r+\epsilon_2-\epsilon_1-\sigma_3+\sigma_4)};q^2\bigr)_{\infty}}
\\
+\frac{\bigl(\alpha q^{2(n+l-k+\epsilon_1-\epsilon_4-\sigma_3+
\sigma_4+1)},-\alpha q^{2(m+l+r+\epsilon_2+\sigma_3+1)}e^{-i\theta};q^2
\bigr)_{\infty}q^{2l}}{\bigl(-\alpha q^{2(n+
l+r+\epsilon_2+\sigma_4)+1},\alpha z^{-1}q^{2(l-k+\epsilon_1-\epsilon_4)};
q^2\bigr)_{\infty}}\cdot\\
\frac{\bigl(-q^{-2(k+\epsilon_4-\sigma_4)+1},zq^{2(\sigma_4-\sigma_3+1)},
z^{-1}q^{2(\sigma_4-\sigma_3)};q^2\bigr)_nq^{2n}}{\bigl(q^{2(r+\epsilon_2-\epsilon_1
-\sigma_3+\sigma_4+1)},q^{4\sigma_4+2},q^2;q^2\bigr)_n}\cdot\\
\frac{\bigl(-q^{-2(k+r+\epsilon_3-\sigma_3)+1},zq^{-2(r+\epsilon_2-\epsilon_1)
+2},z^{-1}q^{-2(r+\epsilon_2-\epsilon_1)},q^{4\sigma_4+2};q^2\bigr)_{\infty}}{
\bigl(-q^{-2(k+\epsilon_4-\sigma_4)+1},q^{-2(\epsilon_2-\epsilon_1-
\sigma_3-\sigma_4)+2},q^{-2(r+\epsilon_2-\epsilon_1-\sigma_3+\sigma_4)};q^2
\bigr)_{\infty}}\Biggr\}\cdot\\
\frac{\bigl(q^{2(r+\epsilon_2-\epsilon_1)+1}e^{-i\theta},q^{2(\sigma_3-\sigma_4)
+1}e^{-i\theta},q^{2(\sigma_3+\sigma_4)+1}e^{-i\theta};q^2\bigr)_mq^{2m}}{\bigl(
q^2e^{-2i\theta},q^2;q^2\bigr)_m}\\
+\text{idem}\,(\theta;-\theta).
\end{multline} 
The multiple sum is absolutely convergent since $\vert zqe^{\pm i\theta}\vert<1
$. Interchanging summations, the expression in the third line gives rise to a
$_2\psi_2$-series. Application of Slater's transformation formula 
[\cite{GR}:5.4.3] for the 
$_r\psi_r$-series with $r\to2$, $a_1\to-\alpha q^{2(n+\epsilon_1+\sigma_3)+1}$,
$a_2\to\alpha z^{-1}q^{-2(k-\epsilon_1+\epsilon_4)}$, $b_1\to \alpha q^{2(n-k-r+
\epsilon_1-\epsilon_3+1)}$, $b_2\to-\alpha q^{2(m+r+\epsilon_2+\sigma_3+1)}e^{-i
\theta}$, $z\to q^2$, $c_1\to \alpha\alpha'q^{2(\epsilon_1-\epsilon'_1
+1)}=\alpha\alpha'q^{2(\epsilon_2-\epsilon'_2+1)}$ 
and $c_2\to\alpha z^{-1}q^{-2(k-\epsilon_1+\epsilon_4)+2}$
yields
\begin{multline}\label{E:psitraf1}
\frac{\bigl(\alpha q^{2(n-k-r+\epsilon_1-\epsilon_3+1)},-\alpha q^{2(m+r+\epsilon_2+\sigma_3+1)}e^{-i\theta};q^2\bigr)_{
\infty}}{\bigl(-\alpha q^{2(n+\epsilon_1+\sigma_3)+1},\alpha 
z^{-1}q^{-2(k-\epsilon_1+\epsilon_4)};q^2\bigr)_{\infty}}\cdot\\
{}_2\psi_2\biggl(\genfrac{}{}{0pt}{}{-\alpha q^{2(n+\epsilon_1+\sigma_3
)+1},\,\alpha z^{-1}q^{-2(k-\epsilon_1+\epsilon_4)}}{\alpha q^{2(n-k-r+\epsilon_1-\epsilon_3+1)},\,
-\alpha q^{2(m+r+\epsilon_2+\sigma_3+1)}e^{-i\theta}};q^2,q^2\biggr)=\\
\shoveleft{\alpha\alpha'q^{2(\epsilon'_1-\epsilon_1)}
\frac{\bigl(\alpha' q^{2(n-k-r+\epsilon'_1-\epsilon_3
+1)},-\alpha'q^{2(m+r+\epsilon'_2+\sigma_3+1)}e^{-i\theta};q^2\bigr)_{\infty}}{
\bigl(-\alpha'q^{2(n+\epsilon'_1+\sigma_3)+1},\alpha' z^{-1}q^{-2(k-\epsilon'_1+\epsilon_4)};q^2\bigr)_{\infty}}\cdot}\\
\shoveright{
{}_2\psi_2\biggl(\genfrac{}{}{0pt}{}{-\alpha'q^{2(n+\epsilon'_1+\sigma_3)+1},
\,\alpha' z^{
-1}q^{-2(k-\epsilon'_1+\epsilon_4)}}{
\alpha' q^{2(n-k-r+\epsilon'_1-\epsilon_3+1
)},\,-\alpha'q^{2(m+r+\epsilon'_2+\sigma_3+1)}e^{-i\theta}};q^2,q^2\biggr)}\\
+\frac{\theta_{q^2}\bigl(-\alpha\alpha' z^{-1}q^{2(n-k+\epsilon_1+\epsilon'_1-\epsilon_4
+\sigma_3)+1},\alpha\alpha'q^{2(\epsilon_1-\epsilon'_1+1)}\bigr)}{
\theta_{q^2}\bigl(\alpha z^{-1}q^{-2(k-\epsilon_1+\epsilon_4)},
\alpha' z^{-1}q^{-2(k-\epsilon'_1+\epsilon_4)},-\alpha q^{2(n+\epsilon_1+\sigma_3)+1},-\alpha'q^{2(n+\epsilon'_1+\sigma_3)+1}\bigr)}\cdot\\
\bigl(-z^{-1}q^{-2(n+k+\epsilon_4+\sigma_3)+1},zq^{2n-2(r+\epsilon_2-
\epsilon_1)+2},-zq^{2(m+k+r+\epsilon_3+\sigma_3+1)}e^{-i\theta},q^2;q^2
\bigr)_{\infty}\\
\cdot
{}_2\psi_2\biggl(\genfrac{}{}{0pt}{}{-zq^{2(n+k+\epsilon_4+
\sigma_3)+1},\,1}{zq^{2n-2(r+\epsilon_2-\epsilon_1)+2},\,-
zq^{2(m+k+r+\epsilon_3+\sigma_3+1)}e^{-i\theta}};q^2,q^2\biggr),
\end{multline}
where the property \eqref{E:thetaprop} of the $\theta$-function has been taken 
into account. According to the definitions by \eqref{E:psidef} and \eqref{E:phidef}, the $_2\psi_2$-series in the last line can be rewritten by
\begin{multline}\label{E:psiev1}
{}_2\phi_1\biggl(\genfrac{}{}{0pt}{}{z^{-1}q^{-2n+2(r+\epsilon_2-\epsilon_1)},\,
-z^{-1}q^{-2(m+k+r+\epsilon_3+\sigma_3)}e^{i\theta}}{- z^{-1}q^{-2(n+k+\epsilon_4+\sigma_3)+1}};q^2,zq^{2m+1}e^{-i
\theta}\biggr)=\\
\frac{\bigl(q^{-2n+2(m+r+\epsilon_2-\epsilon_1)+1}e^{-i\theta},
-q^{-2(k+r+\epsilon_3+\sigma_3)+1};q^2\bigr)_{\infty}}{\bigl(-z^{-1}q^{-2(n+
k+\epsilon_4+\sigma_3)+1},zq^{2m+1}e^{-i\theta};q^2\bigr)_{\infty}},
\end{multline}
where the $q$-Gauss sum [\cite{GR}:II.8] has been employed. 

Due to \eqref{E:psiev1} and the property \eqref{E:thetaprop}, the second
contribution to the rhs of \eqref{E:psitraf1} is given by
\begin{multline}\label{E:tauev2}
d_k\frac{\bigl(zq^{2n-2(r+\epsilon_2-\epsilon_1)+2},-zq^{2(m+k+r+
\epsilon_3+\sigma_3+1)}e^{-i\theta},q^{2(m+r+\epsilon_2-\epsilon_1)+1}e^{
-i\theta};q^2\bigr)_{\infty}}{\bigl(zq^{2m+1}e^{-i\theta};q^2\bigr)_{\infty}}\cdot\\
(-1)^nz^nq^{2mn+2n(k+r+\epsilon_3+\sigma_3)}e^{-in\theta}\bigl(q^{-2(m+r+
\epsilon_2-\epsilon_1)+1}e^{i\theta};q^2\bigr)_n\cdot\\
\bigl(-q^{-2(k+r+\epsilon_3+\sigma_3)+1},q^2;q^2\bigr)_{\infty}
\end{multline}
with
\begin{multline*}
d_k=\\
\frac{\theta_{q^2}\bigl(-\alpha\alpha' z^{-1}q^{-2(k-\epsilon_1-\epsilon'_1+
\epsilon_4-\sigma_3)+1},\alpha\alpha'q^{2(\epsilon_1-\epsilon'_1
+1)}\bigr)}{\theta_{q^2}
\bigl(\alpha z^{-1}q^{-2(k-\epsilon_1+\epsilon_4)},\alpha'z^{-1}
q^{-2(k-\epsilon'_1+\epsilon_4)},-\alpha q^{2(\epsilon_1+\sigma_3)+1},
-\alpha'q^{2(\epsilon'_1+\sigma_3)+1}\bigr)}.
\end{multline*}
Analogous steps for the remaining parts on the rhs of \eqref{E:tauev1} show that
the difference $\tau^{(r,k)+}-\alpha\alpha'q^{2(\epsilon'_1-\epsilon_1)}\tau'^{(r,k)+}$ can
be formulated in terms of the contribution due to the second part on the rhs
of \eqref{E:psitraf1} and the corresponding contribution from the remaining
evaluations of the $l$-summations in \eqref{E:tauev1}. Denoting the contribution
due to the second part on the rhs of \eqref{E:psitraf1} by $\Omega(e^{i\theta})$, the evaluation \eqref{E:tauev2} and the requirement
$\vert zq^{2(k+r+\epsilon_3+\sigma_3+1)}e^{-i\theta}
\vert <$ allow to write the sum over $n$ as a $_3\phi_2$-series:
\begin{multline}\label{E:tauev2a}
\Omega(e^{i\theta})=\kappa_ra^+_rd_k\,z^{-k-r}q^{-k-r}\frac{S_{-k}\bigl(
q^{-2\epsilon_4},q^{2\sigma_4}\bigr)}{S_{-k-r}\bigl(q^{-2\epsilon_3},
q^{2\sigma_3}\bigr)}\cdot\\
\frac{\bigl(-q^{-2(k+r+\epsilon_3+\sigma_3)+1},
zq^{-2(r+\epsilon_2-\epsilon_1)+2},zq^{2(\sigma_4-\sigma_3+1)},
z^{-1}q^{2(\sigma_4-\sigma_3)},q^2;q^2\bigr)_{\infty}}{\bigl(q^{-2(\epsilon_2
-\epsilon_1+\sigma_3-\sigma_4)+2},q^{2(\epsilon_2-\epsilon_1+\sigma_3+
\sigma_4+1)},q^{2(\epsilon_2-\epsilon_1+\sigma_3-\sigma_4+1)}
;q^2\bigr)_{\infty}}\cdot\\
\frac{
\bigl(q^{2(r+\epsilon_2-\epsilon_1)+1}e^{i\theta},q^{2(r+\epsilon_2-\epsilon_1)
+1}e^{-i\theta};q^2\bigr)_{\infty}\bigl(q^{-2(r+\epsilon_2-\epsilon_1-\sigma_3
-\sigma_4)+2};q^2\bigr)_r}{\bigl(q^{2(r+\epsilon_2-\epsilon_1-\sigma_3
+\sigma_4)};q^2\bigr)_{\infty}}\cdot\\
\frac{\bigl(-zq^{2(k+r+\epsilon_3+\sigma_3+1)}e^{-i\theta},q^{2(\sigma_3-
\sigma_4)+1}e^{i\theta},q^{2(\sigma_3+\sigma_4)+1}e^{i\theta};q^2\bigr)_{\infty}
}{\bigl(zqe^{-i\theta},e^{2i\theta};q^2\bigr)_{\infty}}\cdot\\
\sum_{m=0}^{\infty}\frac{\bigl(zqe^{-i\theta},q^{2(\sigma_3-\sigma_4)+1}e^{
-i\theta},q^{2(\sigma_3+\sigma_4)+1}e^{-i\theta};q^2\bigr)_mq^{2m}}{\bigl(
-zq^{2(k+r+\epsilon_3+\sigma_3+1)}e^{-i\theta},q^2e^{-2i\theta},q^2;q^2
\bigr)_m}\cdot\\
{}_3\phi_2\biggl(\genfrac{}{}{0pt}{}{-q^{-2(k+r+\epsilon_3-\sigma_3)+1},\,
q^{-2(m+r+\epsilon_2-\epsilon_1)+1}e^{i\theta},\,z^{-1}q^{-2(r+\epsilon_2
-\epsilon_1)}}{q^{-2(r+\epsilon_2-\epsilon_1-\sigma_3+\sigma_4)+2},\,q^{-2(
r+\epsilon_2-\epsilon_1-\sigma_3-\sigma_4)+2}};q^2,-zq^{2(m+k+r+\epsilon_3
+\sigma_3+1)}e^{-i\theta}\biggr).
\end{multline}
Applying the transformation [\cite{GR}:III.9] with $a\to-q^{-2(k+r+\epsilon_3
-\sigma_3)+1}$, $b\to q^{-2(m+r+\epsilon_2-\epsilon_1)+1}e^{i\theta}$, $c\to 
z^{-1}q^{-2(r+\epsilon_2-\epsilon_1)}$, $d\to q^{-2(r+\epsilon_2-\epsilon_1-
\sigma_3-\sigma_4)+2}$ $e\to q^{-2(r+\epsilon_2-\epsilon_1-\sigma_3+\sigma_4
)+2}$, the last two lines are reformulated by
\begin{multline}\label{E:tauev3}
\frac{\bigl(-q^{2(k+\epsilon_4-\sigma_4)+1},zq^{4\sigma_3+3}e^{-i\theta};
q^2\bigr)_{\infty}}{\bigl(-zq^{2(k+r+\epsilon_3+\sigma_3+1)}e^{-i\theta},q^{
-2(r+\epsilon_2-\epsilon_1-\sigma_3+\sigma_4)+2};q^2\bigr)_{\infty}}\cdot\\
\sum_{m=0}^{\infty}\frac{\bigl(zqe^{-i\theta},q^{2(\sigma_3-\sigma_4)+1}e^{
-i\theta},q^{2(\sigma_3+\sigma_4)+1}e^{-i\theta};q^2\bigr)_mq^{2m}}{\bigl(z
q^{4\sigma_3+3}e^{-i\theta},q^2e^{-2i\theta},q^2;q^2\bigr)_m}\cdot\\
{}_3\phi_2\biggl(\genfrac{}{}{0pt}{}{-q^{-2(k+r+\epsilon_3-\sigma_3)+1},\,
q^{2(m+\sigma_3+\sigma_4)+1}e^{-i\theta},\,zq^{2(\sigma_3+\sigma_4+1)}}{zq^{2m
+4\sigma_3+3}e^{-i\theta},\,q^{-2(r+\epsilon_2-\epsilon_1-\sigma_3-\sigma_4
)+2}},q^2,-q^{2(k+\epsilon_4-\sigma_4)+1}\biggr)\\
=\frac{\bigl(-q^{2(k+\epsilon_4-\sigma_4)+1},zq^{4\sigma_3+3}e^{-i\theta}
;q^2\bigr)_{\infty}}{\bigl(-zq^{2(k+r+\epsilon_3+\sigma_3+1)}e^{-i\theta},
q^{-2(r+\epsilon_2-\epsilon_1-\sigma_3+\sigma_4)+2};q^2\bigr)_{\infty}}\cdot\\
\sum_{j=0}^{\infty}\frac{\bigl(-q^{-2(k+r+\epsilon_3-\sigma_3)+1},zq^{2(\sigma_3
+\sigma_4+1)},q^{2(\sigma_3+\sigma_4)+1}e^{-i\theta};q^2\bigr)_j}{\bigl(zq^{4
\sigma_3+3}e^{-i\theta},q^{-2(r+\epsilon_2-\epsilon_1-\sigma_3-\sigma_4)+2},q^2
;q^2\bigr)_j}(-1)^jq^{2j(k+\epsilon_4-\sigma_4)+j}\cdot\\
{}_3\phi_2\biggl(\genfrac{}{}{0pt}{}{q^{2(j+\sigma_3+\sigma_4)+1}e^{-i\theta},
\,zqe^{-i\theta},\,q^{2(\sigma_3-\sigma_4)+1}e^{-i\theta}}{zq^{2j+4\sigma_3
+3}e^{-i\theta},\,q^2e^{-2i\theta}};q^2,q^2\biggr).
\end{multline}
Here the second expression is obtained writing out the first $_3\phi_2$-series
according to \eqref{E:phidef} and interchanging summations. The balanced
$_3\phi_2$-series in the last line and the same series with $\theta$ replaced by $-\theta$ are related by the nonterminating $q$-Saalsch\"utz formula
[\cite{GR}:II.24]. Taking into account the factors on the rhs of \eqref{E:tauev3} and in the third and fourth line of \eqref{E:tauev2a}, the
$q$-Saalsch\"utz formula with $a\to q^{2(j+\sigma_3+\sigma_4)+1}e^{-i\theta}$,
$b\to zqe^{-i\theta}$, $c\to q^{2(\sigma_3-\sigma_4)+1}e^{-i\theta}$, $d\to zq^{2j+4\sigma_3+3}e^{-i\theta}$, $e\to q^2e^{-2i\theta}$ and the
property \eqref{E:thetaprop} lead to
\begin{multline}\label{E:Omeg+-}
\Omega(e^{i\theta})+\Omega(e^{-i\theta})=\kappa_ra^+_r\frac{\theta_{q^2}\bigl(
zq^{2(\sigma_3-\sigma_4+1)}\bigr)}{\theta_{q^2}\bigl(q^{2(r+\epsilon_2-
\epsilon_1-\sigma_3+\sigma_4)}\bigr)}\cdot\\
\frac{\theta_{q^2}\bigl(-\alpha\alpha'z^{-1}q^{2(r+\epsilon_1+\epsilon'_1-
\epsilon_4+\sigma_3)+1},\alpha\alpha'q^{2(\epsilon_1-\epsilon'_1+1)},
-q^{2(r-\epsilon_4+\sigma_4)+1}\bigr)}{\theta_{q^2}
\bigl(\alpha z^{-1}q^{2(r+\epsilon_1-\epsilon_4)},\alpha' z^{-1}q^{2(r+\epsilon'_1-\epsilon_4)},-\alpha q^{2(\epsilon_1+\sigma_3)+1},
-\alpha'q^{2(\epsilon'_1+\sigma_3)+1}\bigr)}\cdot\\
\frac{\bigl(zq^{2(\sigma_4-\sigma_3+1)},zq^{2(\sigma_3+\sigma_4+1)},zq^{-2(r+
\epsilon_2-\epsilon_1)+2},q^{4\sigma_3+2},q^2;q^2\bigr)_{\infty}}{\bigl(
q^{-2(\epsilon_2-\epsilon_1+\sigma_3-\sigma_4)+2},
q^{2(\epsilon_2-\epsilon_1
+\sigma_3+\sigma_4+1)},q^{2(\epsilon_2-\epsilon_1+\sigma_3-\sigma_4+1)};
q^2\bigr)_{\infty}}\cdot\\
\frac{\bigl(q^{2(r+
\epsilon_2-\epsilon_1)+1}e^{i\theta},q^{2(r+\epsilon_2-\epsilon_1)+1}e^{-i
\theta};q^2\bigr)_{\infty}\bigl(q^{-2(r+\epsilon_2-\epsilon_1-\sigma_3-\sigma_4
)+2};q^2\bigr)_r}{\bigl(zqe^{i\theta},zqe^{-i\theta};q^2\bigr)_{\infty}}\cdot\\
q^{2(k+r)(\sigma_3+\sigma_4)+k+r}\sqrt{\frac{\bigl(-q^{-2(k+r+\epsilon_3+\sigma_3)+1},
-q^{-2(k+\epsilon_4+\sigma_4)+1};q^2\bigr)_{\infty}}{\bigl(-q^{-2(k+r+\epsilon_3
-\sigma_3)+1},-q^{-2(k+\epsilon_4-\sigma_4)+1};q^2\bigr)_{\infty}}}\cdot\\
{}_3\phi_2\biggl(\genfrac{}{}{0pt}{}{-q^{-2(k+r+\epsilon_3-\sigma_3)+1},\,
q^{2(\sigma_3+\sigma_4)+1}e^{i\theta},\,q^{2(\sigma_3+\sigma_4)+1}e^{-i\theta}
}{q^{4\sigma_3+2},\,q^{-2(r+\epsilon_2-\epsilon_1-\sigma_3-\sigma_4)+2}};q^2,
-q^{2(k+\epsilon_4-\sigma_4)+1}\biggr).
\end{multline}
Employing a similar procedure for the remaining contributions on the rhs of
\eqref{E:tauev1} gives  
\begin{multline}\label{E:tausum1}
\tau^{(r,k)+}-\alpha\alpha'q^{2(\epsilon'_1-\epsilon_1)}\tau'^{(r,k)+}=\\
\frac{\theta_{q^2}\bigl(\alpha\alpha'q^{2(\epsilon'_1-\epsilon_1)}\bigr)}{
\theta_{q^2}\bigl(\alpha z^{-1}q^{2(r+\epsilon_1-\epsilon_4)},\alpha'z^{-1}q^{2
(r+\epsilon'_1-\epsilon_4)}\bigr)}\sqrt{\frac{\theta_{q^2}\bigl(-q^{2(\epsilon_3
+\sigma_3)+1},-q^{2(r-\epsilon_4-\sigma_4)+1}\bigr)}{\theta_{q^2}\bigl(-q^{2(
\epsilon_3-\sigma_3)+1},-q^{2(r-\epsilon_4+\sigma_4)+1}\bigr)}}\cdot\\
a^+_r\kappa_r\frac{\bigl(zq^{-2(r+\epsilon_2-\epsilon_1)+2},zq^{2(\sigma_4-
\sigma_3+1)},zq^{2(\sigma_3+\sigma_4+1)},q^2;q^2\bigr)_{\infty}}{\bigl(
q^{2(\epsilon_2-\epsilon_1+\sigma_3-\sigma_4+1)},q^{-2(\epsilon_2-\epsilon_1+
\sigma_3-\sigma_4)+2};q^2\bigr)_{\infty}}\cdot\\
q^{k+r}\sqrt{\frac{\bigl(-q^{2(k+r+\epsilon_3-\sigma_3)+1},-q^{2(k+\epsilon_4
-\sigma_4)+1};q^2\bigr)_{\infty}}{\bigl(-q^{2(k+r+\epsilon_3+\sigma_3)+1},-q^{2(
k+\epsilon_4+\sigma_4)+1};q^2\bigr)_{\infty}}}\cdot\\
\Biggl\{\frac{\theta_{q^2}\bigl(zq^{2(\sigma_3-\sigma_4+1)},-\alpha\alpha' 
z^{-1}q^{2(r+\epsilon_1+\epsilon'_1-\epsilon_4+\sigma_3)+1},-q^{2(r-\epsilon_4
+\sigma_4)+1}\bigr)\bigl(q^{4\sigma_3+2};q^2\bigr)_{\infty}}{\theta_{q^2}\bigl(
q^{2(r+\epsilon_2-\epsilon_1-\sigma_3+\sigma_4)},-\alpha q^{2(\epsilon_1+
\sigma_3)+1},-\alpha'q^{2(\epsilon'_1+\sigma_3)+1}\bigr)\bigl(q^{2(\epsilon_2
-\epsilon_1+\sigma_3+\sigma_4+1)};q^2\bigr)_{\infty}}\cdot\\
\frac{\bigl(q^{-2(r+\epsilon_2-\epsilon_1-\sigma_3-\sigma_4)+2};q^2\bigr)_r
\bigl(q^{2(r+\epsilon_2-\epsilon_1)+1}e^{i\theta},q^{2(r+\epsilon_2-
\epsilon_1)+1}e^{-i\theta};q^2\bigr)_{\infty}}{\bigl(zqe^{i\theta},zqe^{-i
\theta};q^2\bigr)_{\infty}}\cdot\\
{}_3\phi_2\biggl(\genfrac{}{}{0pt}{}{-q^{-2(k+r+\epsilon_3-\sigma_3)+1},\,
q^{2(\sigma_3+\sigma_4)+1}e^{i\theta},\,q^{2(\sigma_3+\sigma_4)+1}e^{-i\theta}
}{q^{4\sigma_3+2},\,q^{-2(r+\epsilon_2-\epsilon_1-\sigma_3-\sigma_4)+2}};q^2,
-q^{2(k+\epsilon_4-\sigma_4)+1}\biggr)\\
+\frac{\theta_{q^2}\bigl(zq^{2(r+\epsilon_2-\epsilon_1+1)},-\alpha\alpha'
z^{-1}q^{2(2r+\epsilon_2+\epsilon'_2-\epsilon_3+\sigma_4)+1},-q^{2(\epsilon_3
-\sigma_3)+1}\bigr)}{\theta_{q^2}\bigl(q^{-2(r+\epsilon_2-\epsilon_1-\sigma_3
+\sigma_4)},-\alpha q^{2(r+\epsilon_2+\sigma_4)+1},-\alpha'q^{2(r+\epsilon'_2
+\sigma_4)+1}\bigr)}\cdot\\
\frac{\bigl(q^{4\sigma_4+2};q^2\bigr)_{\infty}}{\bigl(q^{-2(\epsilon_2-
\epsilon_1-\sigma_3-\sigma_4)+2};q^2\bigr)_{\infty}\bigl(q^{2(\epsilon_2-
\epsilon_1+\sigma_3+\sigma_4+1)};q^2\bigr)_r}\cdot\\
\frac{\bigl(q^{2(\sigma_3
-\sigma_4)+1}e^{i\theta},q^{2(\sigma_3-\sigma_4)+1}e^{-i\theta};q^2\bigr)_{
\infty}}{\bigl(zqe^{i\theta},zqe^{-i\theta};q^2\bigr)_{\infty}}\cdot\\
{}_3\phi_2\biggl(\genfrac{}{}{0pt}{}{-q^{-2(k+\epsilon_4-\sigma_4)+1},\,q^{2(
\sigma_3+\sigma_4)+1}e^{i\theta},\,q^{2(\sigma_3+\sigma_4)+1}e^{-i\theta}
}{q^{4\sigma_4+2},\,q^{2(r+\epsilon_2-\epsilon_1+\sigma_3+\sigma_4+1)}};q^2,
-q^{2(k+r+\epsilon_3-\sigma_3)+1}\biggr)\Biggr\}.
\end{multline}
Applying equation \eqref{E:rho2} with $r\to-r$, $l\to k+r$, $\epsilon_1\to 
\epsilon_3$, $\epsilon_2\to\epsilon_4$, $q^{2\sigma_1}\to q^{2\sigma_3}$,
$q^{2\sigma_2}\to q^{2\sigma_4}$ and with $l\to k$, $\epsilon_1\to\epsilon_4$,
$\epsilon_2\to\epsilon_3$, $q^{2\sigma_1}\to q^{2\sigma_4}$, $q^{2\sigma_2}\to 
q^{2\sigma_3}$ allows to express the $_3\phi_2$-series in terms of
the functions $\rho^{(-r)+}_{k+r}(q^{2\epsilon_3},q^{2\sigma_3};q^{2\epsilon_4},
q^{2\sigma_4};\cos\theta)$ and $\rho^{(r)+}_k\bigl(q^{2\epsilon_4},
q^{2\sigma_4};q^{2\epsilon_3},q^{2\sigma_3};\cos\theta)$.
Taking into account the equations \eqref{E:kappa}, \eqref{E:ardef} for
$\kappa_r$ and $a^+_r$, the statement \eqref{E:tausum2} is obtained in the case \eqref{E:kcond}.

Combining \eqref{E:tauf1int} with $r\to r+1$ and \eqref{E:taue1int} leads to
\begin{multline}\label{E:taukrec}
s_{-k}\bigl(q^{-2\epsilon_4},q^{2\sigma_4}\bigr)s_{-k-r}\bigl(q^{-2\epsilon_3},
q^{2\sigma_3}\bigr)\,\Xi^{(r,k-1)\pm}\\
+s_{-k-1}\bigl(q^{-2\epsilon_4},q^{2\sigma_4}\bigr)s_{-k-r-1}\bigl(q^{-2\epsilon_3},
q^{2\sigma_3}\bigr)\,\Xi^{(r,k+1)\pm}\\
\shoveleft{
=\Bigl\{q^{-2(r+\epsilon_2-\epsilon_1)-1}\bigl(1+q^{-2(k+\epsilon_4+
\sigma_4)+1}\bigr)\bigl(1+q^{-2(k+\epsilon_4-\sigma_4)+1}\bigr)}\\
+q^{2(r+\epsilon_2-\epsilon_1)+1}\bigl(1+q^{-2(k+r+\epsilon_3+\sigma_3)-1}
\bigr)\bigl(1+q^{-2(k+r+\epsilon_3-\sigma_3)-1}\bigr)\\
-q^{-2(r+\epsilon_2-\epsilon_1)-1}\bigl(1-q^{2(r+\epsilon_2-\epsilon_1)+1}
e^{i\theta}\bigr)\bigl(1-q^{2(r+\epsilon_2-\epsilon_1)+1}e^{-i\theta}\bigr)
\Bigr\}\,\Xi^{(r,k)\pm}.
\end{multline}
Equation \eqref{E:taukrec} relating $\tau^{(r,k)+}-\alpha\alpha'q^{2(\epsilon'_1-\epsilon_1)} \tau'^{(r,k)+}$ for different values of $k$ is easily confirmed for the expression 
on the rhs of \eqref{E:tausum2} making use of the contiguous relation \eqref{E:rholll}
with the replacements $r\to-r$, $l\to k+r$, $\epsilon_1\to\epsilon_3$, $\epsilon_2\to\epsilon_4$, $q^{2\sigma_1}\to q^{2\sigma_3}$, $q^{2\sigma_2}\to 
q^{2\sigma_4}$ and $l\to k$, $\epsilon_1\to\epsilon_4$, $\epsilon_2\to
\epsilon_3$, $q^{2\sigma_1}\to q^{2\sigma_4}$, $q^{2\sigma_2}\to q^{2\sigma_3}$.
Thus the condition \eqref{E:kcond} can be relaxed.  
This shows the assertion in the case $\vert zqe^{\pm i\theta}\vert<1$.

For $e^{i\theta}\in\Gamma^+$, the contribution written out explicitely on the rhs of \eqref{E:rho1b} vanishes. Therefore the part written out on the rhs of
\eqref{E:tauev1} is not present. The multiple sum in the second part
obtained by the replacement $\theta\to-\theta$ is absolutely convergent provided that $\vert zqe^{i\theta}\vert<1$. 
Similarly, the multiple sum leading to the nonvanishing 
contribution to the second part on the rhs of \eqref{E:tausum1} is
absolutely convergent if $\vert zqe^{i\theta}\vert<1$. Thus in the case
$e^{i\theta}\in\Gamma^+$, the statement remains valid with the second of the conditions \eqref{E:tausum1bcond} suspended.

\vskip 0.5cm

\emph{Remark}: The replacement $\sigma_3\to-\sigma_3$ on the rhs of \eqref{E:tausum2} gives an expression for $\tau^{(r,k)-}-\alpha\alpha' 
q^{2(\epsilon'_1-\epsilon_1)}\tau'^{(r,k)-}$.

The $R$-elements
$R^{k,k+r^*}_{l,l+r^*}$ can be written as the integral transform $\mathcal G$ of a function $g$ given by the rhs of \eqref{E:tausum2} normalized by a simple
factor independent of $k,l,r$. Here the
specifications \eqref{E:Jacpar} and \eqref{E:Jacpar2} are employed in
the explicit expressions in \cite{vbigJac} underlying equation \eqref{E:G}.
This representation of $R^{k,k+r^*}_{l,l+r^*}$ generalizes Proposition 2
in \cite{gade1}.
\vskip 0.5cm

Throughout the remainder, the cases $q^{4\sigma_4}=q^{2n}$ with $n\in 
\mathbb Z_{\geq0}$ are excluded.
For a fixed value of $\epsilon_2+\epsilon'_2\equiv\zeta$ and of $\beta=\alpha\alpha'$, a particular choice of $\mathsf{R}^{k,k+r^*}_{l,
l+r^*}$ in Definition \ref{D:taudef1} gives rise to a simple expression
for the corresponding sum $\Xi^{(r,k)\pm}$.  
Introducing
\begin{multline}\label{E:hatRel}
\mathsf{R}^{k,k+r^*}_{l,l+r^*}=r^{k,k+r^*}_{l,l+r^*}\equiv r^{k,k+r^*}_{
l,l+r^*}\bigl(z,\alpha;\beta,q^{2\zeta};q^{2\epsilon_1},q^{2\epsilon_2};q^{2
\epsilon_3},q^{2\epsilon_4};q^{2\sigma_3},q^{2\sigma_4}\bigr)=\\
\iota(q^{2\sigma_3})\sqrt{\theta_{q^2}\bigl(-q^{2(\epsilon_3
+\sigma_3)+1},
-q^{2(\epsilon_3-\sigma_3)+1}\bigr)}\cdot\\
\Biggl\{c(q^{2\sigma_4})
\theta_{q^2}\bigl(-\alpha q^{2(\epsilon_2+\sigma_4)+1},-\beta\alpha^{-1}q^{2(
\zeta-\epsilon_2+\sigma_4)+1},-\beta z^{-1}q^{2(\zeta-\epsilon_3-\sigma_4)+1}\bigr)R^{k,k+r^*}_{l,l+r^*}\\
-c(q^{-2\sigma_4})\theta_{q^2}\bigl(-\alpha q^{2(\epsilon_2-\sigma_4)+1},-\beta 
\alpha^{-1}q^{2(\zeta-\epsilon_2-\sigma_4)+1},-\beta z^{-1}q^{2(\zeta-
\epsilon_3+\sigma_4)+1}\bigr)\check R^{k,k+r^*}_{l,l+r^*}\Bigr\}
\end{multline}
with $\iota(q^{2\sigma_3})=\vert q^{2\sigma_3}\vert^{-1}q^{2
\sigma_3}$ and
\begin{multline*}
c(q^{2\sigma_4})\equiv c\bigl(z;q^{2\epsilon_3},q^{2
\epsilon_4};q^{2\sigma_3},q^{2\sigma_4}\bigr)=\\
\sqrt{\frac{\theta_{q^2}\bigl(-q^{2(
\epsilon_4-\sigma_4)+1}\bigr)}{\theta_{q^2}\bigl(-q^{2(\epsilon_4+\sigma_4)
+1}\bigr)}}\frac{\bigl(q^{-2(\epsilon_2
-\epsilon_1+\sigma_3-\sigma_4)+2},q^{-2(\epsilon_2-\epsilon_1-\sigma_3-\sigma_4)
+2};q^2\bigr)_{\infty}}{\bigl(zq^{2(\sigma_3+\sigma_4+1)},zq^{2(\sigma_4-\sigma_3
+1)};q^2\bigr)_{\infty}},
\end{multline*}
the sums attributed to \eqref{E:hatRel} by Definition \eqref{D:taudef1}
are denoted by 
\begin{equation}\label{E:hatnot1}
\begin{split}
\widehat{\tau}^{(r,k)\pm}(q^{2\zeta},q^{2\epsilon_2})&\equiv \widehat{\tau}^{(r,k)\pm}
\bigl(z,\alpha;\beta,q^{2\zeta};q^{2\epsilon_1},q^{2\epsilon_2};q^{2\epsilon_3},
q^{2\epsilon_4};q^{2\sigma_3},q^{2\sigma_4};\cos\theta\bigr),\\
\widehat{\tau}^{(r,k)}(q^{2\zeta},q^{2\epsilon_2})&\equiv
\widehat{\tau}^{(r,k)}
\bigl(z,x,\alpha;\beta,q^{2\zeta};q^{2\epsilon_1},q^{2\epsilon_2};q^{2\epsilon_3},
q^{2\epsilon_4};q^{2\sigma_3},q^{2\sigma_4};e^{i\theta}\bigr)
\end{split}
\end{equation}
in the following. For convenience, the notation
\begin{equation}\label{E:hatnot2}
\begin{split}
\widehat{\tau}'^{(r,k)\pm}(q^{2\zeta},q^{2\epsilon'_2})&\equiv\widehat{\tau}^{(
r,k)\pm}\bigl(z,\alpha';\beta,q^{2\zeta};q^{2\epsilon'_1},q^{2\epsilon'_2};q^{2
\epsilon_3},q^{2\epsilon_4};q^{2\sigma_3},q^{2\sigma_4};
\cos\theta\bigr),\\
\widehat{\tau}'^{(r,k)}(q^{2\zeta},q^{2\epsilon'_2})&\equiv\widehat{\tau}^{(
r,k)}\bigl(z,x',\alpha';\beta,q^{2\zeta};q^{2\epsilon'_1},q^{2\epsilon'_2};q^{2
\epsilon_3},q^{2\epsilon_4};q^{2\sigma_3},q^{2\sigma_4};
e^{i\theta}\bigr)
\end{split}
\end{equation}
is adopted.
The sums $\widehat{\tau}^{(r,k)+}(q^{2\zeta},q^{2\epsilon_2})$ and
$\widehat{\tau}'^{(r,k)+}(q^{2\zeta},q^{2\epsilon'_2})$
are well-defined and absolutely convergent provided that
\begin{equation}\label{E:tausum2cond}
\begin{split}
&q^{2(\epsilon_2-\epsilon_1+\sigma_3\pm\sigma_4)}\neq q^{2t_1},\;t_1\in 
\mathbb Z_{\backslash0},\quad q^{2(\epsilon_2-\epsilon_1-\sigma_3\pm\sigma_4)}\neq q^{2t_2},\;t_2\in \mathbb Z\\
&\vert ze^{i\theta}\vert<1,\\
&\vert zqe^{-i\theta}\vert<1\quad\text{if}\;e^{i\theta}\notin\Gamma^+,
\end{split}
\end{equation}
where all four sign choices are independent. Replacing $q^{2\sigma_3}$ by
$q^{-2\sigma_3}$ and $\Gamma^+$ by $\Gamma^-$ yields the corresponding conditions for $\widehat{\tau}^{(r,k)-}(q^{2\zeta},q^{2\epsilon_2})$ and
$\widehat{\tau}'^{(r,k)-}(q^{2\zeta},q^{2\epsilon'_2})$.
The sums $\widehat{\tau}^{(r,k)}
(q^{2\zeta},q^{2\epsilon_2})$ and $\widehat{\tau}'^{(r,k)}(q^{2\zeta},q^{2
\epsilon'_2})$ are well-defined and absolutely convergent if
\begin{equation}\label{E:tausum22cond}
\begin{split}
&q^{2(\epsilon_2-\epsilon_1+\sigma_3\pm\sigma_4)}\neq q^{2t_1},\quad
q^{2(\epsilon_2-
\epsilon_1-\sigma_3\pm\sigma_4)}\neq q^{2t_2},\;t_1,t_2\in\mathbb Z,\\
&\vert zqe^{i\theta}\vert<1.
\end{split}
\end{equation}

\begin{corollary}\label{C:tausum4}
The sums $\widehat{\tau}^{(r,k)+}(q^{2\zeta},q^{2\epsilon_2})$ and
$\widehat{\tau}'^{(r,k)+}(q^{2\zeta},q^{2
\epsilon'_2})$ with the property \eqref{E:tausum2cond} 
satisfy
\begin{multline}\label{E:tausum4}
\widehat{\tau}^{(r,k)+}(q^{2\zeta},q^{2\epsilon_2})-\alpha\alpha'q^{2(
\epsilon'_2-\epsilon_2)}\widehat{\tau}'^{(r,k)+}(q^{2\zeta},q^{2\epsilon'_2})=\\
q^{2(\epsilon_4-\sigma_4)+1}\iota(q^{2\sigma_3})\frac{
\theta_{q^2}\bigl(\alpha\alpha'q^{2(\epsilon'_1-\epsilon_1)},\alpha\alpha' 
z^{-1}q^{2(\epsilon_1+\epsilon'_2+1)},q^{4\sigma_4}
\bigr)}{\theta_{q^2}\bigl(-\alpha q^{2(\epsilon_1+\sigma_3)+1},-\alpha'
q^{2(\epsilon'_1+\sigma_3)+1}\bigr)}\cdot\\
\theta_{q^2}\bigl(-q^{2(\epsilon_3+\sigma_3)+1},-\alpha\alpha'q^{2(
\epsilon_2+\epsilon'_2-\epsilon_3+\sigma_3)+1}\bigr)\cdot\\
\frac{\bigl(zq^{2(\epsilon_1-\epsilon_2+1)},q^{-2(\epsilon_2-\epsilon_1-\sigma_3+
\sigma_4)+2},q^{-2(\epsilon_2-\epsilon_1-\sigma_3-\sigma_4)+2},q^2;q^2\bigr)_{
\infty}}{\bigl(q^{2(\epsilon_2-\epsilon_1+\sigma_3+\sigma_4+1)},q^{2(\epsilon_2-
\epsilon_1+\sigma_3-\sigma_4+1)};q^2\bigr)_{\infty}}\cdot\\
\frac{u\bigl(q^{2\epsilon_2},q^{2\epsilon_1};e^{i\theta}\bigr)
}{\bigl(zqe^{i\theta},zqe^{-i\theta};q^2\bigr)_{\infty}}\cdot\\
(-1)^ra^{+}_{-r}\bigl(q^{2\epsilon_3},q^{2\epsilon_4};q^{2\sigma_3};\cos\theta\bigr)
\rho^{(-r)+}_{k+r}\bigl(q^{2\epsilon_3},q^{2\sigma_3};q^{2\epsilon_4},
q^{2\sigma_4};\cos\theta\bigr).
\end{multline}
\end{corollary}

\emph{Proof}:
Rewriting the relation \eqref{E:rhorel} with the replacements $l\to k$, $\epsilon_1\to\epsilon_4$, $\epsilon_2\to\epsilon_3$, $q^{2\sigma_1} 
\to q^{2\sigma_4}$, $q^{2\sigma_2}\to q^{2\sigma_3}$, the function
$a^+_r\bigl(q^{2\epsilon_4},q^{2\epsilon_3};q^{2\sigma_4};\cos\theta
\bigr)\rho^{(r)+}_k
\bigl(q^{2\epsilon_4},q^{2\sigma_4};q^{2\epsilon_3},q^{2\sigma_3};\cos\theta 
\bigr)$ can be eliminated using Theorem \ref{T:tausum2}. 
Eliminating the function \newline
$a^-_r\bigl(q^{2\epsilon_4},q^{2\epsilon_3};q^{2\sigma_4};\cos\theta\bigr)
\rho^{(r)-}_k\bigl(q^{2\epsilon_4},q^{\sigma_4};q^{2\epsilon_3},q^{2\sigma_3};
\cos\theta\bigr)$ by means of Theorem \ref{T:tausum2}
with $q^{2\sigma_4}\to q^{-2\sigma_4}$,
the third function $(-1)^ra^+_{-r}\bigl(q^{2\epsilon_3},q^{2\epsilon_4};
q^{2\sigma_3};\cos\theta
\bigr)\rho^{(-r)+}_{k+r}\bigl(q^{2\epsilon_3},q^{2\sigma_3};
q^{2\epsilon_4},q^{2\sigma_4};\cos\theta\bigr)$ is expressed as a
linear combination of $\tau^{(r,k)+}-\alpha\alpha'q^{2(\epsilon'_1-\epsilon_1)}
\tau'^{(r,k)+}$ and $\check{\tau}^{(r,k)+}-\alpha\alpha'q^{2(\epsilon'_1-
\epsilon_1)}\check{\tau}'^{(r,k)+}$. Here $\check{\tau}^{(r,k)+}$ and
$\check{\tau}'^{(r,k)+}$ denote the sums $\tau^{(r,k)+}$ and $\tau'^{(r,k)+}$
with $q^{2\sigma_4}$ replaced by $q^{-2\sigma_4}$.
Application of [\cite{GR}:ex.5.22]
with $c\to\alpha z^{-1}q^{2(\epsilon_1-\epsilon_4)}$, $d\to\alpha'z^{-1}
q^{2(\epsilon'_1-\epsilon_4)}$, $e\to-\alpha\alpha'q^{2(\epsilon_1+\epsilon'_2
-\epsilon_4+\sigma_3)+1}$, $f\to\alpha\alpha'z^{-1}q^{2(\epsilon_1+\epsilon'_2)
}$, $b^{-1}g\to q^{2(\epsilon_2-\epsilon_1-\sigma_3+\sigma_4)}$, $h^{-1}b\to 
q^{-4\sigma_4}$, $b^{-1}a\to-\alpha\alpha'z^{-1}q^{2(\epsilon_1+\epsilon'_2-
\epsilon_4+\sigma_4)-1}$
allows to simplify the prefactor of 
$\mathring a^+_r\bigl(q^{2\epsilon_3},q^{2\epsilon_4};q^{2\sigma_3};\cos\theta
\bigr)\rho^{(-r)+}_{k+r}\bigl(q^{2\epsilon_3},
q^{2\sigma_3};q^{2\epsilon_4},q^{2\sigma_4};\cos\theta\bigr)$.
Taking into account the definition of $\widehat{\tau}^{(r,k)+}(q^{2\zeta},q^{2
\epsilon_2})$ by \eqref{E:hatRel} immediately leads to the assertion.

\vskip 0.5cm

Replacing $\sigma_3$ by $-\sigma_3$ on the rhs of \eqref{E:tausum4} and in
the conditions \eqref{E:tausum2cond} 
gives rise to the corresponding expression for 
$\widehat{\tau}^{(r,k)-}(q^{2\zeta},q^{2\epsilon_2})-\alpha\alpha'q^{2(\epsilon'_2
-\epsilon_2)}\widehat{\tau}'^{(r,k)-}(q^{2\zeta},q^{2\epsilon'_2})$.

\begin{corollary}\label{C:tausum5}
For $e^{i\theta}=\alpha\alpha'q^{2(s+\epsilon'_1+\epsilon_2)+1}$,
the sums $\widehat{\tau}^{(r,k)}(q^{2\zeta},q^{2\epsilon_2})$ and 
$\widehat{\tau}'^{(r,k)}(q^{2\zeta},q^{2\epsilon'_2})$ with the property 
\eqref{E:tausum22cond} 
are related by
\begin{multline}\label{E:tausum5}
\widehat{\tau}^{(r,k)}(q^{2\zeta},q^{2\epsilon_2})-\alpha\alpha'q^{2(\epsilon'_2
-\epsilon_2)}\widehat{\tau}'^{(r,k)}(q^{2\zeta},q^{2\epsilon'_2})\Big\vert_{
e^{i\theta}=\alpha\alpha'q^{2(s+\epsilon'_1+\epsilon_2)+1}}=\\
q^{2(\epsilon_4-\sigma_4)+1}\iota(q^{2\sigma_3})
\theta_{q^2}\bigl(\alpha\alpha'q^{2(\epsilon'_2-
\epsilon_2)},\alpha\alpha'z^{-1}q^{2(\epsilon'_1+\epsilon_2+1)},q^{4\sigma_4}
\bigr)\frac{w_0\bigl(\bar x;q^{2\epsilon_3},q^{2\epsilon_4};e^{i\theta_0}
\bigr)}{w_0\bigl(x;q^{2\epsilon_1},q^{2\epsilon_2};e^{i\theta_0}
\bigr)}\cdot\\
\frac{\bigl(zq^{2(\epsilon_1-\epsilon_2+1)},q^2,q^{2(\epsilon_2-\epsilon_1)+1}e^{
i\theta},q^{2(\epsilon_2-\epsilon_1)+1}e^{-i\theta};q^2\bigr)_{\infty}}{
\bigl(zqe^{i\theta},zqe^{-i\theta};q^2\bigr)_{\infty}}\cdot\\
(-1)^ra_{-r}\bigl(\bar x;q^{2\epsilon_3},q^{2\epsilon_4};e^{i\theta}\bigr)
\varsigma^{(-r)}_{k+r}\bigl(q^{2\epsilon_3},q^{2\sigma_3};q^{2\epsilon_4},q^{2
\sigma_4};e^{i\theta}\bigr)\Big\vert_{e^{i\theta}=\alpha\alpha'q^{2(s+
\epsilon'_3+\epsilon_4)+1}},
\end{multline}
where 
\begin{equation}\label{E:epsicond}
\bar x=x\alpha,\quad
\epsilon'_3=\epsilon_2+\epsilon'_1-\epsilon_4.
\end{equation}
and $e^{i\theta_0}=\alpha\alpha'q^{2(\epsilon'_1+\epsilon_2)+1}=\alpha\alpha' 
q^{2(\epsilon'_3+\epsilon_4)+1}$.
\end{corollary}

\emph{Proof}: 
The statement is first proven for a choice of $z$ with the property
\begin{equation}\label{E:C2cond}
\max\bigl(\vert zq^{2(\tilde s+\epsilon'_1+\epsilon_2+1)}\vert,\vert zq^{-2(
\tilde s+\epsilon'_1+\epsilon_2)}\vert\bigr)<1
\end{equation}
for some $\tilde s\in\mathbb Z$.
Specialzing to $e^{i\theta}=\alpha\alpha'q^{2(\tilde s+\epsilon'_1+\epsilon_2)
+1}$, the sum 
$\widehat{\tau}^{(r,k)}(q^{2\zeta},q^{2\epsilon_2})$ can be expressed in terms of $\hat{\tau}^{(r,k)\pm}(q^{2\zeta},q^{2\epsilon_2})$. Definition \ref{D:taudef1}, the property \eqref{E:wprop1} and the definitions of $\varsigma^{(r)}_l$ and the coefficients 
$a^{\pm}_r$, $a_r$ by \eqref{E:varsigma}, \eqref{E:ardef}, \eqref{E:ardef3}
imply
\begin{multline}\label{E:hattaudec}
\boldsymbol{\tau}^{(r,k)}\Big\vert_{e^{i\theta}=
\alpha\alpha'q^{2(\tilde s+\epsilon'_1+\epsilon_2)+1}}=\frac{\theta_{q^2}
\bigl(-\alpha q^{2(\epsilon_2-\sigma_3+1)}e^{-i\theta},-\alpha q^{2(
\epsilon_1+\sigma_3)+1}\bigr)}{w_0\bigl(x;q^{2\epsilon_1},q^{2\epsilon_2};
e^{i\theta}\bigr)}\cdot\\
\bigl(q^{2(\sigma_4-\sigma_3)+1}e^{i\theta},q^{-2(\sigma_3+\sigma_4)+1}e^{i
\theta},q^{2(\epsilon_2-\epsilon_1+\sigma_3+\sigma_4+1)},q^{2(\epsilon_2-
\epsilon_1+\sigma_3-\sigma_4+1)};q^2\bigr)_{\infty}\cdot\\
\boldsymbol{\tau}^{(r,k)+}\Big\vert_{e^{i\theta}=\alpha 
\alpha'q^{2(\tilde s+\epsilon'_1+\epsilon_2)+1}}
-\frac{\theta_{q^2}\bigl(-\alpha q^{2(\epsilon_2+\sigma_3+1)}e^{-i\theta},-
\alpha q^{2(\epsilon_1-\sigma_3)+1}\bigr)}{w_0\bigl(x;q^{2\epsilon_1},q^{2
\epsilon_2};e^{i\theta}\bigr)}\cdot\\
\bigl(q^{2(\sigma_3+\sigma_4)+1}e^{i\theta},q^{2(\sigma_3-\sigma_4)+1}e^{i
\theta},q^{2(\epsilon_2-\epsilon_1-\sigma_3+\sigma_4+1)},q^{2(\epsilon_2-
\epsilon_1-\sigma_3-\sigma_4+1)};q^2\bigr)_{\infty}\cdot\\
\boldsymbol{\tau}^{(r,k)-}\Big\vert_{e^{i\theta}=\alpha\alpha'q^{2(\tilde s+
\epsilon'_1+\epsilon_2)+1}}.
\end{multline}
Taking into account the property \eqref{E:wprop2},
the lhs of \eqref{E:tausum5} can be evaluated by
means of equation \eqref{E:hattaudec} with the specifications $\boldsymbol{\tau}^{(r,k)}\to\widehat{\tau}^{(r,k)}(q^{2\zeta},q^{2\epsilon_2})$,
$\boldsymbol{\tau}^{(r,k)\pm}\to\widehat{\tau}^{(r,k)\pm}(q^{2\zeta},q^{2
\epsilon_2})$ or $\boldsymbol{\tau}^{(r,k)}\to\widehat{\tau}'^{(r,k)}
(q^{2\zeta},q^{2\epsilon'_2})$, $\boldsymbol{\tau}^{(r,k)\pm}
\to\widehat{\tau}'^{(r,k)\pm}(q^{2\zeta},q^{2\epsilon'_2})$ and
Corollary \ref{C:tausum4}. The properties \eqref{E:thetaprop} and
\eqref{E:wprop1} yield
\begin{multline*}
\frac{\theta_{q^2}\bigl(-\alpha q^{2(\epsilon_2-\sigma_3+1)}e^{-i\theta}\bigr)}{
w_0\bigl(x;q^{2\epsilon_1},q^{2\epsilon_2};e^{i\theta}\bigr)}\Big\vert_{e^{
i\theta}=\alpha\alpha'q^{2(\tilde s+\epsilon'_1+\epsilon_2)+1}}=
\frac{\theta_{q^2}\bigl(-\alpha'q^{2(\epsilon'_1+\sigma_3)+1}\bigr)}{w_0\bigl(
x;q^{2\epsilon_1},q^{2\epsilon_2};e^{i\theta_0}\bigr)}\,
(x\alpha q^{2\sigma_3})^{-\tilde s}.
\end{multline*}
Writing
\begin{multline*}
\theta_{q^2}\bigl(-\alpha\alpha'q^{2(\epsilon_2+\epsilon'_2-\epsilon_3\pm\sigma_3
)+1}\bigr)=\theta_{q^2}\bigl(-\alpha\alpha'q^{2(\epsilon'_3\pm\sigma_3)+1}\bigr)\\
=q^{\pm2(\tilde s+r)\sigma_3}\bar x^{s+r}w_0\bigl(\bar x;q^{2\epsilon_3},q^{2
\epsilon_4};e^{i\theta_0}\bigr)\frac{\theta_{q^2}\bigl(q^{-2r+2(\epsilon_4\mp 
\sigma_3+1)}e^{-i\theta}\bigr)}{w_{-r}(\bar x;q^{2\epsilon_3},q^{2\epsilon_4}
;e^{i\theta}\bigr)}\Big\vert_{e^{i\theta}=\alpha\alpha'q^{2(\tilde s+\epsilon'_3
+\epsilon_4)+1}}
\end{multline*}
due to \eqref{E:wprop1}, 
and taking into account the definitions by \eqref{E:varsigma}, \eqref{E:ardef}
and \eqref{E:ardef3} yields the assertion in the case $s=\tilde s$.

Making use of the relations \eqref{E:varsco1a}, \eqref{E:varsco1b} and \eqref{E:varsco2}
with $r\to-r$, $l\to k+r$, $\epsilon_1\to\epsilon_4$, $\epsilon_2\to\epsilon_3$,
$q^{2\sigma_1}\to q^{2\sigma_4}$, $q^{2\sigma_2}\to q^{2\sigma_3}$, it is
straightforward to verify that the expression on the rhs of
\eqref{E:tausum5} satisfies the contiguous relations \eqref{E:Xitheta}
for 
\begin{equation*}
\Xi^{(r,k)}(e^{i\theta})=\widehat{\tau}^{(r,k)}(q^{2\zeta},q^{2\epsilon_2})
-\alpha\alpha'q^{2(\epsilon'_2-\epsilon_2)}\widehat{\tau}'^{(r,k)}(q^{2\zeta},
q^{2\epsilon'_2})\Big\vert_{e^{i\theta}=\alpha\alpha'q^{2(s+\epsilon'_1+\epsilon_2
)+1}}
\end{equation*}
wrt to the argument $e^{i\theta}$. This shows the assertion for all
values of $s$ subject to the conditions \eqref{E:tausum22cond} and $z$ chosen according to \eqref{E:C2cond}.

To indicate the dependence on $z$ and $e^{i\theta}$ in a compact way, the last
sums on the rhs of \eqref{E:hatnot1} and \eqref{E:hatnot2} are denoted by $\widehat{\tau}^{(
r,k)}(q^{2\zeta},q^{2\epsilon_2};z,e^{i\theta})$ and $\widehat{\tau}'^{(r,k)}
(q^{2\zeta},q^{2\epsilon'_2};z,e^{i\theta})$, respectively. 
Writing
$\widehat{\Xi}^{(r,k)}(z,e^{i\theta})\equiv \widehat{\tau}^{(r,k)}(q^{2\zeta},
q^{2\epsilon_2};z,e^{i\theta})-\alpha\alpha'q^{2(\epsilon'_2-\epsilon_2)}
\widehat{\tau}'^{(r,k)}(q^{2\zeta},q^{2\epsilon'_2};z,e^{i\theta})$, the
contiguous relation \eqref{E:Xiz} derived in appendix \ref{A:thetacont} allows to express $\widehat{\Xi}^{(r,k)}(zq^{-2},e^{i\theta})$ in terms of
$\widehat{\Xi}^{(r,k+\bar t)}(z,q^{2t}e^{i\theta})$ with $t\in (-1,0,1)$ and
$\bar t\in (0,1)$.
Application of Corollary \ref{C:tausum5} to $\widehat{\Xi}^{(r,k+\bar t)}(z,
q^{2t}e^{i\theta})$ gives $\widehat{\Xi}^{(r,k)}(zq^{-2},e^{i\theta})$ 
in terms of $a_{-r}(\alpha x;q^{2\epsilon_3},q^{2\epsilon_4};q^{2t}e^{i\theta})$
\newline$\cdot\varsigma^{(-r)}_{k+r+\bar t}\bigl(q^{2\epsilon_3},q^{2\sigma_3};q^{2\epsilon_4},q^{2\sigma_4};q^{2t}e^{i\theta}
)$. For $t=\pm1$, the latter can be expressed in terms
of $\varsigma^{(-r)}_{k+r+\tilde t}(q^{2\epsilon_3},q^{2\sigma_3};
q^{2\epsilon_4},q^{2\sigma_4};e^{i\theta})$ with $\tilde t\in (-1,0,1)$
by means of the relations \eqref{E:varsco1a} and \eqref{E:varsco1b}
with $r\to-r$, $l\to k+r+\bar t$, $q^{2\epsilon_1}\to q^{2\epsilon_3}$, $q^{2
\epsilon_2}\to q^{2\epsilon_4}$, $q^{2\sigma_1}\to q^{2\sigma_3}$, $q^{2\sigma_2}\to q^{2\sigma_4}$. The contribution by $\tilde t=-1$ can be
eliminated using the relation \eqref{E:varsco2}
with $r\to-r$, $l\to k+r$, $q^{2\epsilon_1}\to q^{2\epsilon_3}$, $q^{2\epsilon_2}\to 
q^{2\epsilon_4}$, $q^{2\sigma_1}\to q^{2\sigma_3}$, $q^{2\sigma_2}\to q^{2\sigma_4}$. Then the resulting contributions by $\tilde t=1$ cancel leaving
\begin{multline*}
\widehat{\Xi}^{(r,k)}\bigl(zq^{-2},\alpha\alpha'q^{2(s+\epsilon'_1+\epsilon_2)+1}
\bigr)=\\
-\frac{\alpha\alpha'zq^{-2(\epsilon'_1+\epsilon_2+1)}\bigl(1-zq^{2(\epsilon_1-\epsilon_
2)}\bigr)
}{\bigl(1-zq^{2(s+\epsilon'_1+\epsilon_2)}\bigr)\bigl(1-zq^{-2(s+\epsilon'_1
+\epsilon_2+1)}\bigr)}\,\widehat{\Xi}^{(r,k)}\bigl(z,\alpha\alpha'q^{2(s+\epsilon'_1
+\epsilon_2)+1}\bigr).
\end{multline*}
This confirms the statement \eqref{E:tausum5} for all values of $z$ and $e^{i\theta}$ subject to the conditions \eqref{E:tausum22cond}.

\begin{corollary}\label{C:tausum6}
For $e^{i\theta}=q^{2(s'+\epsilon_2-\epsilon_1)-1}$, the sums $\widehat{\tau}^{(r,k)}(q^{2\zeta},q^{2\epsilon_2})$ and $\widehat{\tau}'^{(r,k)}(q^{2\zeta},q^{2\epsilon'_2})$ with
the property \eqref{E:tausum22cond} satisfy
\begin{multline}\label{E:tausum6}
\widehat{\tau}^{(r,k)}(q^{2\zeta},q^{2\epsilon_2})-\alpha\alpha'q^{2(\epsilon'_2
-\epsilon_2)}\widehat{\tau}'^{(r,k)}(q^{2\zeta},q^{2\epsilon'_2})\Big\vert_{
e^{i\theta}=q^{2(s'+\epsilon_2-\epsilon_1)-1}}=\\
-q^{2\epsilon_4+1}z^{s'}q^{3r^2-2r(r-s'+2\epsilon_4-\epsilon_3
+1)}q^{(s'-1)(s'-2\epsilon_4-1)}\iota(q^{2\sigma_3})\cdot\\
\frac{\theta_{q^2}\bigl(\alpha\alpha'q^{2(\epsilon'_2-\epsilon_2)},\alpha\alpha'z^{-1} 
q^{2(\epsilon'_1+\epsilon_2+1)},-\alpha\alpha'q^{2(\epsilon'_4+\sigma_4)+1},-
\alpha\alpha'q^{2(\epsilon'_4-\sigma_4)+1}\bigr)}{\theta_{q^2}\bigl(-\alpha q^{2(\epsilon_1+\sigma_3)+1},-\alpha q^{2(\epsilon_1-\sigma_3)+1},-\alpha'q^{2(
\epsilon'_1+\sigma_3)+1},-\alpha'q^{2(\epsilon'_1-\sigma_3)+1}\bigr)}\cdot\\
\breve B\,
\frac{u\bigl(q^{2\epsilon_2},q^{2\epsilon_1};e^{i\theta}\bigr)(q^2;q^2)_{\infty}}{
\theta_{q^2}\bigl(q^{4(\epsilon_1
-\epsilon_2)-2(s'-1)}\bigr)\bigl(zq^{2(\epsilon_2-\epsilon_1)};q^2\bigr)_{\infty}}
\frac{\bigl(zq^{2(\epsilon_1-\epsilon_2+1)};q^2\bigr)_{-s'}}{\bigl(z^{-1}q^{2(
\epsilon_1-\epsilon_2+1)};q^2\bigr)_{-s'}}\cdot\\ 
\frac{q^{r(r-1)-s'(s'-1)+2(r-s')(\epsilon_2-\epsilon_1)}}{\sqrt{(q^2;q^{4(s'
+\epsilon_2- \epsilon_1)};q^2)_{r-s'}}}\,
\varsigma^{(-r)}_{k+r}\bigl(q^{2\epsilon_3},q^{2\sigma_3};q^{2\epsilon_4},
q^{2\sigma_4};q^{2(s'+\epsilon_2-\epsilon_1)-1}\bigr),
\end{multline}
where 
\begin{multline}\label{E:breveB}
\breve B=q^{2\sigma_3}\frac{\theta_{q^2}\bigl(-\alpha q^{2(\epsilon_2+\sigma_4)
+1},-\alpha'q^{2(\epsilon'_2+\sigma_4)+1},q^{2(\epsilon_2-\epsilon_1-\sigma_3
-\sigma_4)},q^{2(\epsilon_2-\epsilon_1+\sigma_3-\sigma_4+1)}\bigr)}{\theta_{q^2}
\bigl(-q^{2(\epsilon_4+\sigma_4)+1},-\alpha\alpha'q^{2(\epsilon'_4+\sigma_4)+1}
\bigr)}\\
-\text{idem}(\sigma_4,-\sigma_4).
\end{multline}

\end{corollary}
 
\emph{Proof}: The case $e^{i\theta}\in\Gamma_{q^{-2(r+\epsilon_2-\epsilon_1)
+1}}$ is distinguished by the property
\begin{multline}\label{E:rhopmsp2}
\rho^{(-r)+}_{
k+r}\bigl(q^{2\epsilon_3},q^{2\sigma_3};q^{2\epsilon_4},q^{2\sigma_4};\cos\theta 
\bigr)\Big\vert_{e^{i\theta}=q^{2(s'+\epsilon_2-\epsilon_1)-1}}\\
=q^{4(r-s')\sigma_3}\frac{\bigl(q^{-2(\epsilon_2-\epsilon_1+\sigma_3+\sigma_4)+2},
q^{-2(\epsilon_2-\epsilon_1+\sigma_3-\sigma_4)+2};q^2\bigr)_{-s'}}{\bigl(
q^{-2(\epsilon_2-\epsilon_1-\sigma_3+\sigma_4)+2},q^{-2(\epsilon_2-\epsilon_1-
\sigma_3-\sigma_4)+2};q^2\bigr)_{-s'}}\cdot\\
\rho^{(-r)-}_{
k+r}\bigl(q^{2\epsilon_3},q^{2\sigma_3};q^{2\epsilon_4},q^{2\sigma_4};\cos\theta 
\bigr)\Big\vert_{e^{i\theta}=q^{2(s'+\epsilon_2-\epsilon_1)-1}}
\end{multline}
obtained from equation
\eqref{E:rho+-sp} with the substitutions $r\to-r$, $s\to-s'$, $l\to k+r$, 
$\epsilon_1\to\epsilon_3$, $\epsilon_2\to\epsilon_4$, $q^{2\sigma_1}\to q^{2
\sigma_3}$ and $q^{2\sigma_2}\to q^{2\sigma_4}$. 
Applying the replacements $r\to-r$, $l\to k+r$, $\epsilon_1\to\epsilon_3$, 
$\epsilon_2\to\epsilon_4$, $q^{2\sigma_1}\to q^{2\sigma_3}$, $q^{2\sigma_2}\to
q^{2\sigma_4}$ to \eqref{E:varsigma} 
and making use of \eqref{E:rhopmsp2} and
\eqref{E:ex.2.16} with $x\to iq^{\epsilon_4-2\sigma_3-\sigma_4
+\frac{3}{2}}$, $\lambda\to iq^{-2(s'+\epsilon_2-\epsilon_1)+\epsilon_4+\sigma_4
+\frac{3}{2}}$, $\mu\to i q^{2\epsilon_3-\epsilon_4+\sigma_4+\frac{3}{2}}$,
$\nu\to iq^{\epsilon_4+2\sigma_3-\sigma_4-\frac{1}{2}}$ leads to
\begin{multline}\label{E:varsigrho2}
\varsigma^{(-r)}_{k+r}\bigl(q^{2\epsilon_3},
q^{2\sigma_3};q^{2\epsilon_4},q^{2\sigma_4};q^{2(s'+\epsilon_2-\epsilon_1)-1}
\bigr)=\\
(-1)^{1-s'}q^{-3r^2+2r(r-s'+2\epsilon_4-\epsilon_3+1)}q^{-2s'(s'-1)+2(s'-1)
(2\epsilon_4-\epsilon_3)+s'-1}q^{-2(r-s'+1)\sigma_3}\cdot\\
\bigl(q^{-2(\epsilon_2-\epsilon_1-\sigma_3+\sigma_4)+2},q^{-2(\epsilon_2-
\epsilon_1-\sigma_3-\sigma_4)+2};q^2\bigr)_{-s'}\cdot\\
\theta_{q^2}\bigl(-q^{2(\epsilon_4+\sigma_4)+1},-q^{2(\epsilon_4-\sigma_4)+1},
q^{4(\epsilon_2-\epsilon_1)+2s'},q^{4\sigma_3}\bigr)\cdot\\
\rho^{(-r)+}_{
k+r}\bigl(q^{2\epsilon_3},q^{2\sigma_3};q^{2\epsilon_4},q^{2\sigma_4};\cos\theta 
\bigr)\Big\vert_{e^{i\theta}=q^{2(s'+\epsilon_2-\epsilon_1)-1}}.
\end{multline}
The assertion is first proven for a choice of $z$ such that
\begin{equation}\label{E:C3cond}
\max\bigl(\vert zq^{2(s_0+\epsilon_2-\epsilon_1)}\vert,\vert zq^{-2(s_0+
\epsilon_2-\epsilon_1)+2}\vert\bigr)<1
\end{equation}
for some $s_0\in\mathbb Z$.
Evaluation of the lhs of \eqref{E:tausum6} for $e^{i\theta}=q^{2(s_0+\epsilon_2-\epsilon_1)-1}$ by means of Corollary 
\ref{C:tausum4} taking into account relation \eqref{E:rhopmsp2} yields
\begin{multline}\label{E:tausum6st1}
q^{-2(s_0-1)\sigma_3}
\bigl(q^{2(s_0+\epsilon_2-\epsilon_1-\sigma_3+\sigma_4)},
q^{2(s_0+\epsilon_2-\epsilon_1-\sigma_3-\sigma_4)},q^{2(\epsilon_2-\epsilon_1
+\sigma_3+\sigma_4+1)};q^2\bigr)_{\infty}\cdot\\
\bigl(q^{2(\epsilon_2-\epsilon_1+\sigma_3-\sigma_4+1)};q^2
\bigr)_{\infty}\cdot\\
\Bigl[\widehat{\tau}^{(r,k)+}(q^{2\zeta},q^{2\epsilon_2})-\alpha\alpha' q^{2(
\epsilon'_2-\epsilon_2)}\widehat{\tau}'^{(r,k)+}(q^{2\zeta},q^{2\epsilon'_2})
\Bigr]\Big\vert_{e^{i\theta}=q^{2(s_0+\epsilon_2-\epsilon_1)-1}}\\
\shoveright{-\text{idem}(\sigma_3,-\sigma_3)}\\
=\iota(q^{2\sigma_3})\frac{\theta_{q^2}\bigl(\alpha\alpha' 
q^{2(\epsilon'_2-\epsilon_2)},\alpha\alpha'z^{-1}q^{2(\epsilon'_1+\epsilon_2+1)},
q^{4\sigma_4}\bigr)(q^2;q^2)_{\infty}
}{\theta_{q^2}\bigl(-\alpha q^{2(\epsilon_1+\sigma_3)+1},-
\alpha q^{2(\epsilon_1-\sigma_3)+1},-\alpha'q^{2(\epsilon'_1+\sigma_3)+1},-
\alpha'q^{2(\epsilon'_1-\sigma_3)+1}\bigr)}\cdot\\
\frac{u\bigl(q^{2\epsilon_2},q^{2\epsilon_1};e^{i\theta}\bigr)}{
\bigl(zq^{2(\epsilon_2-\epsilon_1)};q^2
\bigr)_{\infty}}\frac{\bigl(zq^{2(\epsilon_1-\epsilon_2+1)},q^{-2(\epsilon_2-
\epsilon_1+\sigma_3+\sigma_4)+2},q^{-2(\epsilon_2-\epsilon_1+\sigma_3-\sigma_4)
+2};q^2\bigr)_{-s_0}}{\bigl(z^{-1}q^{2(\epsilon_1-\epsilon_2+1)};q^2\bigr)_{-s_0}}
\cdot\\
(-1)^{s_0+1}z^{s_0}q^{2s_0\sigma_3-s_0(s_0-1)-2(s_0-1)(\epsilon_2-\epsilon_1)+2
\epsilon_4-4\sigma_4+1}\cdot\\
\Bigl\{\theta_{q^2}\bigl(-\alpha q^{2(\epsilon_1-\sigma_3)+1},-\alpha'q^{2(\epsilon'_1-\sigma_3)+1},-q^{2(\epsilon_3
+\sigma_3)+1},-\alpha\alpha'q^{2(\epsilon_2+\epsilon'_2-\epsilon_3+\sigma_3)+1},
\\
\shoveright{
q^{-2(\epsilon_2-\epsilon_1-\sigma_3-\sigma_4)},q^{2(\epsilon_2-\epsilon_1-
\sigma_3+\sigma_4)}\bigr)-\text{idem}(\sigma_3,-\sigma_3)\Bigr\}\cdot}\\
(-1)^ra^+_{-r}\bigl(q^{2\epsilon_3},q^{2\epsilon_4};q^{2\sigma_3};\cos\theta\bigr)
\rho^{
(-r)+}_{k+r}\bigl(q^{2\epsilon_3},q^{2\sigma_3};q^{2\epsilon_4},q^{2\sigma_4};
\cos\theta\bigr)\Big\vert_{e^{i\theta}=q^{2(s_0+\epsilon_2-\epsilon_1)-1}}.
\end{multline}
The transformation 
[\cite{GR}:ex.5.22] with $c\to-\alpha q^{-2(\epsilon_2+\sigma_4)+1}$, 
$d\to-\alpha'q^{-2(\epsilon'_2+\sigma_4)+1}$, $e\to-q^{2(\epsilon_4-
\sigma_4)+1}$, $f\to-\alpha\alpha'q^{2(\epsilon'_4-\sigma_4)+1}$, $b^{-1}g\to 
q^{-2(\epsilon_2-\epsilon_1-\sigma_3-\sigma_4)}$, $h^{-1}b\to q^{-4\sigma_3}$,
$b^{-1}a\to q^{-2(\epsilon_2-\epsilon_1-\sigma_3+\sigma_4)}$ allows to
rewrite the expression in the braces on the rhs of \eqref{E:tausum6st1} by
\begin{multline}
q^{2\sigma_3+4\sigma_4}\theta_{q^2}\bigl(-q^{2(\epsilon_4+\sigma_4)+1},-
q^{2(\epsilon_4-\sigma_4)+1},-\alpha\alpha'q^{2(\epsilon'_4+\sigma_4)+1},-
\alpha\alpha'q^{2(\epsilon'_4-\sigma_4)+1}\bigr)\cdot\\
\frac{\theta_{q^2}\bigl(
q^{-4\sigma_3}\bigr)}{\theta_{q^2}\bigl(q^{4\sigma_4}\bigr)}\,\breve B.
\end{multline}
With the relation \eqref{E:varsigrho2}, the assertion for $s'=s_0$ follows immediately.

The relations \eqref{E:varsco1a}, \eqref{E:varsco1b} and
\eqref{E:varsco2} with $r\to-r$, $l\to k+r$, $\epsilon_1\to\epsilon_4$,
$\epsilon_2\to\epsilon_3$, $q^{2\sigma_1}\to q^{2\sigma_4}$, $q^{2\sigma_2}
\to q^{2\sigma_3}$ allow to demonstrate the contiguous relations \eqref{E:Xitheta} for
\begin{equation*}
\Xi^{(r,k)}(e^{i\theta})=\widehat{\tau}^{(r,k)}(q^{2\zeta},q^{2\epsilon_2})-
\alpha\alpha'q^{2(\epsilon'_2-\epsilon_2)}\widehat{\tau}'^{(r,k)}(q^{2\zeta},
q^{2\epsilon'_2})\Big\vert_{e^{i\theta}=q^{2(s'+\epsilon_2-\epsilon_1)-1}}.
\end{equation*}
This shows the assertion for all values of $s'$ restricted by \eqref{E:tausum22cond} and a choice of $z$ giving rise to the property
\eqref{E:C3cond}.

The condition \eqref{E:C3cond} can be released due to equation \eqref{E:Xiz}.
With the replacement $q^{2s}\to \alpha\alpha'q^{2(s'-\epsilon'_1-\epsilon_1-1)}$
applied to the main steps in the last part of the proof of Corollary \ref{C:tausum5}, use of Corollary \ref{C:tausum6} leads to 
\begin{multline*}
\widehat{\Xi}^{(r,k)}\bigl(zq^{-2},q^{2(s'+\epsilon_2-\epsilon_1)-1}\bigr)=\\
-\frac{\alpha \alpha'
zq^{-2(\epsilon'_1+\epsilon_2+1)}\bigl(1-zq^{2(\epsilon_1-\epsilon_2)}
\bigr)}{ 
\bigl(1-zq^{2(s'+\epsilon_2-\epsilon_1-1)}\bigr)\bigl(1-zq^{-2(s'+\epsilon_2-
\epsilon_1)}\bigr)}\widehat{\Xi}^{(r,k)}\bigl(z,q^{2(s'+\epsilon_2-\epsilon_1)-1}
\bigr),
\end{multline*}
provided that Corollary \ref{C:tausum6} can be applied to $\widehat{\Xi}^{(r,k)}
(z,q^{2(s'+\epsilon_2-\epsilon_1)-1})$. Thus the statement \eqref{E:tausum6} is
established for all values of $z$ and $e^{i\theta}$ satisfying the conditions
\eqref{E:tausum22cond}.
\vskip 0.5cm

\section{The initial condition}\label{S:init}

This section turns to a closer examination of the matrix elements $\mathsf R^{
k,k+r^*}_{l,l+r^*}=r^{k,k+r^*}_{l,l+r^*}$ specified by \eqref{E:hatRel}
in the particular cases $z=\varepsilon$, $q^{2\sigma_4}=\varepsilon q^{2
\sigma_3}$ with $\varepsilon=1$ or $\varepsilon=-1$. For a linear combination
$\bar r^{k,k+r^*}_{l,l+r^*}$ of $r^{k,k+r^*}_{l,l+r^*}$ and a second set
$\mathring r^{k,k+r^*}_{l,l+r^*}$ obtained
from $r^{k,k+r^*}_{l,l+r^*}$ by the simple parameter replacements indicated
in \eqref{E:mathrr}, Proposition \ref{P:barrinit} provides the initial condition
valid in the special cases $z=\alpha=\varepsilon$, $\epsilon_1=\epsilon_4$,
$\epsilon_2=\epsilon_3$, $q^{2\sigma_4}=\varepsilon q^{2\sigma_3}$.
Proposition \ref{P:rsp} gives expressions for $r^{k,k+r^*}_{l,l+r^*}$
in the cases $z=\varepsilon$, $q^{2\sigma_4}=\varepsilon q^{2\sigma_3}$
and $z=\alpha=\varepsilon$, $\epsilon_1=\epsilon_4$, $\epsilon_2=\epsilon_3$,
$q^{2\sigma_4}=\varepsilon q^{2\sigma_3}$.

\vskip 0.5cm

Throughout the subsequent analysis, the cases $q^{4\sigma_i}=q^{2n}$
with $n\in\mathbb Z_{\geq0}$ and $i=3,4$ are excluded.
The expression \eqref{E:Rel1} for $R^{k,k+r^*}_{l,l+r^*}$ or $\check R^{k,k+
r^*}_{l,l+r^*}$ entails simple poles located at $z=\alpha q^{2(m+l-k+
\epsilon_1-\epsilon_4)}$ with $m\in\mathbb Z_{\geq0}$. 
In contrast, the singularities are not present in the linear combination $r^{k,k+r^*}_{l,l+r^*}$ defined by \eqref{E:hatRel}.
This feature is easily demonstrated for a particular combination of 
$r^{k,k+r^*}_{l,l+r^*}$ and
\begin{equation}\label{E:mathrr}
\mathring r^{k,k+r^*}_{l,l+r^*}=r^{k+r,k^*}_{l+r,l^*}(z,\alpha;\beta,q^{2(
\epsilon_1+\epsilon'_1)};q^{2\epsilon_2},q^{2\epsilon_1};
q^{2\epsilon_4},q^{2\epsilon_3};q^{2\sigma_4},q^{2\sigma_3})
\end{equation}
given by 
\begin{multline}
\label{E:rrcomb}
\bar r^{k,k+r^*}_{l,l+r^*}\equiv \bar r^{k,k+r^*}_{l,l+r^*}\bigl(z,
\alpha;\beta,q^{2\zeta};q^{2\epsilon_1},q^{2\epsilon_2}
;q^{2\epsilon_3},q^{2\epsilon_4};q^{2\sigma_3},q^{2\sigma_4}\bigr)=\\
\theta_{q^2}\bigl(-zq^{2(\epsilon_3+\sigma_4)+3},-\alpha\alpha'z^{-1}q^{2(
\epsilon_1+\epsilon'_1-\epsilon_4-\sigma_3)+1},-\alpha q^{2(\epsilon_1+\sigma_3)
+1},-\alpha'q^{2(\epsilon'_1+\sigma_3)+1}\bigr)\cdot\\
\shoveright{
\theta_{q^2}\bigl(-q^{2(\epsilon_4+\sigma_4)+1},-q^{2(\epsilon_4-\sigma_4)+1},
q^{4\sigma_3+2}\bigr)\bigl(z^{-1}q^{2(\epsilon_2-\epsilon_1)};q^2\bigr)_{\infty}
\iota(q^{2\sigma_3})r^{k,k+r^*}_{l,l+r^*}}\\
-\theta_{q^2}\bigl(-zq^{2(\epsilon_4+\sigma_3)+3},-\alpha\alpha'z^{-1}q^{2(
\epsilon_2+\epsilon'_2-\epsilon_3-\sigma_4)+1},-\alpha q^{2(\epsilon_2+\sigma_4)
+1},-\alpha'q^{2(\epsilon'_2+\sigma_4)+1}\bigr)\cdot\\
\theta_{q^2}\bigl(-q^{2(\epsilon_3+\sigma_3)+1},-q^{2(\epsilon_3-\sigma_3)+1},
q^{4\sigma_4+2}\bigr)\bigl(z^{-1}q^{2(\epsilon_1-\epsilon_2)};q^2\bigr)_{\infty}
\iota(q^{2\sigma_4})\mathring r^{k,k+r^*}_{l,l+r^*}.
\end{multline}
According to \eqref{E:mrRel} and \eqref{E:mrchRel}, setting
$\mathsf R^{k,k+r^*}_{l,l+r^*}=\bar r^{k,k+r^*}_{l,l+r^*}$ provides a
solution of the intertwining condition \eqref{E:int}.
In terms of $\check R^{k,k+r^*}_{l,l+r^*}$ and $\mathring{\check R}^{k,k+r^*}_{
l,l+r^*}$ introduced by \eqref{E:chRel} and \eqref{E:mrchRel}, the rhs of \eqref{E:rrcomb}
is reformulated as
\begin{multline}\label{E:rrex}
\bar r^{k,k+r^*}_{l,l+r^*}
=-\theta_{q^2}\bigl(\alpha z^{-1}q^{2(\epsilon_1-\epsilon_4)},\alpha'z^{-1}
q^{2(\epsilon'_1-\epsilon_4)},q^{4\sigma_3+2},q^{4\sigma_4+2}\bigr)\cdot\\
\sqrt{\theta_{q^2}(-q^{2(\epsilon_3+\sigma_3)+1},-q^{2(\epsilon_3-
\sigma_3)+1},-q^{2(\epsilon_4+\sigma_4)+1},-q^{2(\epsilon_4-\sigma_4)+1})}
\cdot\\
\Bigl\{\theta_{q^2}\bigl(\alpha\alpha'q^{2(\epsilon_2+\epsilon'_2+1)},-\alpha 
\alpha'z^{-1}q^{2(\epsilon_1+\epsilon'_1-\epsilon_4-\sigma_3)+1},-\alpha q^{2(
\epsilon_1+\sigma_3)+1},-\alpha'q^{2(\epsilon'_1+\sigma_3)+1}\bigr)\cdot\\
\sqrt{\theta_{q^2}(-q^{2(\epsilon_4+\sigma_4)+1},-q^{2(\epsilon_4-
\sigma_4)+1})}\bigl(z^{-1}q^{2(\epsilon_2-\epsilon_1)};q^2\bigr)_{\infty}\cdot
c\bigl(q^{-2\sigma_4}\bigr)\check R^{k,k+r^*}_{l,l+r^*}\\
-\theta_{q^2}\bigl(\alpha\alpha'q^{2(\epsilon_1+\epsilon'_1+1)},-\alpha\alpha'
z^{-1}q^{2(\epsilon_2+\epsilon'_2-\epsilon_3-\sigma_4)+1},-\alpha q^{2(\epsilon_2
+\sigma_4)+1},-\alpha'q^{2(\epsilon'_2+\sigma_4)+1}\bigr)\cdot\\
\sqrt{\theta_{q^2}(-q^{2(\epsilon_3+\sigma_3)+1},-q^{2(\epsilon_3-\sigma_3)+1}
)}\bigl(z^{-1}q^{2(\epsilon_1-\epsilon_2)};q^2\bigr)_{\infty}\cdot
\mathring c\bigl(q^{-2\sigma_3}\bigr)\check{\mathring R}^{k,k+r^*}_{l,l+r^*}
\Bigr\},
\end{multline}
where $\mathring c(q^{-2\sigma_3})\equiv c(z;q^{2\epsilon_4},q^{2\epsilon_3};
q^{2\sigma_4},q^{-2\sigma_3})$. The first $\theta$-function on the
rhs implies the absence of the singularities at $z=\alpha q^{2(m+l-k+\epsilon_1-
\epsilon_4)}$.

Writing out the rhs of \eqref{E:rrcomb} by means of \eqref{E:hatRel}
and the same equation with $r\to-r$, $l\to l+r$, $k\to k+r$, $\epsilon_1
\leftrightarrow\epsilon_2$, $\epsilon_3\leftrightarrow\epsilon_4$,
$q^{2\sigma_3}\leftrightarrow q^{2\sigma_4}$,
the relations \eqref{E:RmR} with $q^{2\sigma_3}\to q^{-2\sigma_3}$
and \eqref{E:RmR} with $r\to-r$, $l\to l+r$, $k\to k+r$,
$\epsilon_1\leftrightarrow\epsilon_2$, $q^{2\sigma_3}
\to q^{2\sigma_4}$, $q^{2\sigma_4}\to q^{-2\sigma_3}$ are employed to obtain an expression in terms of $\check R^{k,k+r^*}_{l,l+r^*}$ and $\check{
\mathring R}^{k,k+r^*}_{l,l+r^*}$. Then application of the relations  \eqref{E:ex.2.16} with
$x\to iq^{\epsilon_3-\sigma_3-2\sigma_4+\frac{3}{2}}$, $\lambda\to-iq^{
\epsilon_3-2\epsilon_4-\sigma_3+\frac{1}{2}}$, $\mu\to izq^{\epsilon_3+\sigma_3
+\frac{3}{2}}$, $\nu\to iq^{\epsilon_3-\sigma_3+2\sigma_4-\frac{1}{2}}$ 
followed by \eqref{E:ex.2.16} with $x\to-\sqrt{\alpha\alpha'}q^{
\epsilon_2+\epsilon'_2+2\sigma_4+1}$, $\lambda\to \alpha q^{\epsilon_2-\epsilon'_2}/
\sqrt{\alpha\alpha'}$, $\mu\to-\sqrt{\alpha\alpha'}q^{\epsilon_2+
\epsilon'_2-2\sigma_4+1}$, $\nu\to\sqrt{\alpha\alpha'}z^{-1}q^{\epsilon_2
-\epsilon'_2-2\epsilon_3}$
to simplify the prefactor of $\check R^{k,k+r^*}_{l,l+r^*}$ gives rise to
the first contribution on the rhs of \eqref{E:rrex}.
Similarly, use of \eqref{E:ex.2.16} with $x\to izq^{\epsilon_4+\sigma_4
+\frac{3}{2}}$, $\lambda\to iq^{\epsilon_3+\epsilon_2-\epsilon_1+\sigma_4-\frac{
1}{2}}$, $\mu\to iq^{\epsilon_4-\sigma_4+2\sigma_3+\frac{3}{2}}$, $\nu\to i
q^{\epsilon_4-2\sigma_3-\sigma_4-\frac{1}{2}}$ and \eqref{E:ex.2.16}
with $x\to-\sqrt{\alpha\alpha'}q^{\epsilon_1+\epsilon'_1-2\sigma_3+1}$, $
\lambda\to\sqrt{\alpha\alpha'}z^{-1}q^{\epsilon_1+\epsilon'_1-2\epsilon_4}$,
$\mu\to-\sqrt{\alpha\alpha'}q^{\epsilon_1+\epsilon'_1+2\sigma_3+1}$, $\nu\to 
\alpha q^{\epsilon_1-\epsilon'_1}/\sqrt{\alpha\alpha'}$ in the prefactor of 
$\check{\mathring R}^{k,k+r^*}_{l,l+r^*}$ leads to the second contribution
on the rhs of \eqref{E:rrex}.

\begin{proposition}\label{P:barrinit}
The solution $\bar r^{k,k+r^*}_{l,l+r^*}$ satisfies
\begin{multline}\label{E:rinit}
\bar r^{k,k+r^*}_{l,l+r^*}\bigl(1,1;\alpha',q^{2\zeta};q^{2\epsilon_1},q^{2
\epsilon_2};q^{2\epsilon_2},q^{2\epsilon_1};q^{2\sigma_3},q^{2\sigma_3}\bigr)
=-\delta_{k,l}(q^2;q^2)_{\infty}\cdot\\
\theta_{q^2}\bigl(\alpha'q^{2(\epsilon'_1+\epsilon_2+1)},\alpha'q^{2(\epsilon'_1
-\epsilon_1)},q^{2(\epsilon_2-\epsilon_1)},-q^{2(\epsilon_1+\sigma_3)+1},-
q^{2(\epsilon_2+\sigma_3)+1},q^{4\sigma_3+2}\bigr)\cdot\\
\Bigl\{\theta_{q^2}\bigl(q^{2(\epsilon_2-\epsilon_1+2\sigma_3+1)},-q^{2(
\epsilon_1+\sigma_3)+1},-\alpha'q^{2(\epsilon'_1+\sigma_3)+1},-q^{2(
\epsilon_2-\sigma_3)+1},-\alpha'q^{2(\epsilon'_2-\sigma_3)+1}\bigr)\\
\shoveleft{-q^{-4\sigma_3}\cdot}\\
\theta_{q^2}\bigl(q^{2(\epsilon_2-\epsilon_1-2\sigma_3+1)},-q^{2(\epsilon_1-
\sigma_3)+1},-\alpha'q^{2(\epsilon'_1-\sigma_3)+1},-q^{2(\epsilon_2+\sigma_3
)+1},-\alpha'q^{2(\epsilon'_2+\sigma_3)+1}\bigr)\Bigr\}
\end{multline}
and
\begin{multline}\label{E:rinit-}
\bar r^{k,k+r^*}_{l,l+r^*}\bigl(-1,-1;-\alpha',q^{2\zeta};q^{2\epsilon_1},
q^{2\epsilon_2};q^{2\epsilon_2},q^{2\epsilon_1};q^{2\sigma_3},-q^{2\sigma_3}
\bigr)=\\
(-1)^k\bar r^{k,k+r^*}_{l,l+r^*}\bigl(1,1;\alpha';-q^{2\zeta};-q^{2\epsilon_1},
q^{2\epsilon_2};q^{2\epsilon_2},-q^{2\epsilon_1};q^{2\sigma_3},q^{2\sigma_3}
\bigr).
\end{multline}
\end{proposition}

\emph{Proof}:
For $z=\varepsilon$ and $q^{2\sigma_4}=\varepsilon q^{2\sigma_3}$ with
$\varepsilon=1$ or $\varepsilon=-1$, use of [\cite{GR}:II.24]
with $a\to-q^{-2(k+r+\epsilon_3-\sigma_3)+1}$, $b\to-\alpha q^{2(l+
\epsilon_1+\sigma_3)+1}$, $c\to\varepsilon q^{-2(r+\epsilon_2-\epsilon_1)}$, $e\to \varepsilon
q^{-2(r+\epsilon_2-\epsilon_1-2\sigma_3)+2}$, $f\to\alpha q^{2(l-k-r+\epsilon_1
-\epsilon_3+1)}$ in the expression \eqref{E:Rel1} with $q^{2\sigma_4}\to 
\varepsilon q^{-2\sigma_3}$ yields
\begin{multline}\label{E:checkRsp}
\theta_{q^2}\bigl(\varepsilon\alpha q^{2(l-k+\epsilon_1-\epsilon_4)}\bigr)
\check R^{k,k+r^*}_{l,l+r^*}
\bigl(\varepsilon,\alpha;q^{2\epsilon_1},q^{2\epsilon_2};q^{2\epsilon_3},
q^{2\epsilon_4};q^{2\sigma_3},\varepsilon q^{2\sigma_3}\bigr)=\\
\varepsilon^{k+r}
\alpha^rq^{l-k+2r\sigma_3}\sqrt{\frac{\bigl(-q^{-2(k+r+\epsilon_3+\sigma_3)
+1},-\varepsilon q^{-2(k+\epsilon_4-\sigma_3)+1};q^2\bigr)_{\infty}}{\bigl(-q^{-2(k+r+
\epsilon_3-\sigma_3)+1},-\varepsilon q^{-2(k+\epsilon_4+\sigma_3)+1};q^2\bigr)_{
\infty}}}\cdot\\
\sqrt{\frac{\bigl(-\alpha q^{2(l+\epsilon_1-\sigma_3)+1},-\varepsilon\alpha q^{2(l+r+\epsilon_2+\sigma_3)+1};q^2\bigr)_{\infty}}{\bigl(-\alpha q^{2(l
+\epsilon_1+\sigma_3)+1},-\varepsilon \alpha q^{2(l+r+\epsilon_2-\sigma_3)+1};q^2\bigr)_{\infty}}}
\cdot\\
\frac{\bigl(\varepsilon\alpha q^{2(l-k+\epsilon_1-\epsilon_4+1)},
\varepsilon\alpha q^{2(k-l-\epsilon_1+\epsilon_4+1)},q^{-4\sigma_3+2};q^2\bigr)_{\infty}}{\bigl(
\varepsilon q^{-2(\epsilon_2-\epsilon_1+2\sigma_3)+2};q^2\bigr)_{\infty}}.
\end{multline}
Insertion of \eqref{E:checkRsp} and the same equation with $r\to-r$, $k\to k+r$,
$l\to l+r$, $\epsilon_1\leftrightarrow\epsilon_2$, $\epsilon_3\leftrightarrow 
\epsilon_4$ to the rhs of \eqref{E:rrex} and specialization to $\alpha=z$,
$\epsilon_1=\epsilon_4$, $\epsilon_2=\epsilon_3$ leads to equation \eqref{E:rinit-} and
\begin{multline*}
\bar r^{k,k+r^*}_{l,l+r^*}\bigl(1,1;\alpha',q^{2\zeta};q^{2\epsilon_1},
q^{2\epsilon_2};q^{2\epsilon_2},q^{2\epsilon_1};q^{2\sigma_3},q^{2\sigma_3}
\bigr)=-\delta_{l,k}\,(q^2;q^2)_{\infty}\cdot\\
\theta_{q^2}\bigl(\alpha' q^{2(\epsilon'_1-\epsilon_1)},q^{2(
\epsilon_2-\epsilon_1)},-q^{2(\epsilon_1+\sigma_3)+1},-q^{2(\epsilon_2+\sigma_3)+1},
q^{4\sigma_3+2},q^{4\sigma_3+2}\bigr)\cdot\\
\Bigl\{\theta_{q^2}\bigl(\alpha'q^{2(\epsilon_2+\epsilon'_2+1)},-q^{2(\epsilon_1
+\sigma_3)+1},-\alpha'q^{2(\epsilon'_1+\sigma_3)+1},-q^{2(\epsilon_1-\sigma_3)+1},
-\alpha'q^{2(\epsilon'_1-\sigma_3)+1}\bigr)\\
\shoveleft{+q^{2(\epsilon_1-\epsilon_2)}\cdot}\\
\theta_{q^2}\bigl(\alpha'q^{2(\epsilon_1+\epsilon'_1+1)},
-q^{2(\epsilon_2+\sigma_3)+1},-\alpha'q^{2(\epsilon'_2+\sigma_3)+1},-q^{2(
\epsilon_2-\sigma_3)+1},-\alpha'q^{2(\epsilon'_2-\sigma_3)+1}\bigr)\Bigr\}.\\
\end{multline*}
According to relation [\cite{GR}:ex.5.23] with $n\to4$, $\tfrac{a_1}{b_1}\to-\varepsilon q^{2(\epsilon_1+\sigma_3)+1}$, $\tfrac{a_1}{b_2}\to-\varepsilon\alpha'q^{-2(
\epsilon'_2-\sigma_3)+1}$, $\tfrac{a_1}{b_3}\to\varepsilon q^{2(\epsilon_2-\epsilon_1+2\sigma_3)}$, $\tfrac{a_1}{b_4}\to 
\varepsilon q^{2(\epsilon_2-\epsilon_1)}$, $\tfrac{a_1}{
a_2}\to q^{4\sigma_3}$, $\tfrac{a_1}{a_3}\to-q^{2(\epsilon_2+\sigma_3)+1}$,
$\tfrac{a_1}{a_4}\to-\alpha'q^{-2(\epsilon'_1-\sigma_3)+1}$, the last expression
equals the rhs of equation \eqref{E:rinit}.

\vskip 0.5cm

The property specified by \eqref{E:rinit} and \eqref{E:rinit-} is referred to as initial condition.

Suitable transformations applied to the $_8W_7$-series in \eqref{E:RW1} and 
\eqref{E:RW3}-\eqref{E:breveR} given for $R^{k,k+r^*}_{l,l+r^*}$ in Appendix \ref{A:R} allow to derive expressions for $r^{k,k+r^*}_{l,l+r^*}$ indicating 
directly
the analytic behaviour described above. For the present purpose however it is
sufficient to restrict the remainder of the section to the case $\alpha=
\varepsilon$, $q^{2\sigma_4}=\varepsilon q^{2\sigma_3}$ with $\varepsilon=1$
or $\varepsilon=-1$. With the additional restrictions $\alpha=\varepsilon$,
$\epsilon_1=\epsilon_4$, $\epsilon_2=\epsilon_3$, the solution $r^{k,k+r^*}_{
l,l+r^*}$ reduces to a terminating $_4\phi_3$-series if $k\neq l$.
Convenient abbreviations suited to these choices are introduced by
\begin{multline}\label{E:skrl}
v_{k,r,l}\equiv
v_{k,r,l}\bigl(\varepsilon,\alpha;q^{2\epsilon_1},q^{2\epsilon_2};q^{2\epsilon_3},
q^{2\epsilon_4};q^{2\sigma_3}\bigr)=\\
\shoveleft{\frac{\bigl(1+\varepsilon
q^{2(k+\epsilon_4+\sigma_3)+1}\bigr)\bigl(1+q^{2(k+r+\epsilon_3
-\sigma_3)+1}\bigr)}{\bigl(1+\varepsilon q^{2(k+\epsilon_4-\sigma_3)+3
}\bigr)\bigl(1+q^{2(k+r+\epsilon_3+\sigma_3)-1}\bigr)}\cdot}\\
{}_4\phi_3\biggl(\genfrac{}{}{0pt}{}{\varepsilon
\alpha q^{-2(k-l+\epsilon_4-\epsilon_1-1)},\,\varepsilon
q^{-2(r+\epsilon_2-\epsilon_1)+2},\,q^{-4\sigma_3+2},\,q^2}{-q^{-2(k+r+\epsilon_3
+\sigma_3)+3},\,-\alpha q^{2(l+\epsilon_1-\sigma_3)+3},\,q^4};q^2,q^2\biggr)\\
+\frac{\bigl(\varepsilon
\alpha q^{-2(k-l+\epsilon_4-\epsilon_1-1)},\alpha q^{2(k+l+r+
\epsilon_1+\epsilon_3+1)},-q^{2(k+r+\epsilon_3+\sigma_3)+3};q^2\bigr)_{\infty}}{
\bigl(-\varepsilon
q^{2(k+\epsilon_4+\sigma_3)+3},-q^{2(k+r+\epsilon_3-\sigma_3)+3},-q^{-2(k+r+
\epsilon_3+\sigma_3)+1};q^2\bigr)_{\infty}}\cdot\\
\frac{\bigl(\varepsilon q^{-2(r+\epsilon_2-\epsilon_1)+2},q^{-4\sigma_3+2};
q^2\bigr)_{\infty}}{\bigl(-\alpha q^{2(l+\epsilon_1-\sigma_3)+1},-\alpha q^{2(l+r+\epsilon_3+\sigma_3)+1};q^2\bigr)_{\infty}}\frac{1-q^2}{1+
\varepsilon q^{2(k+\epsilon_4-\sigma_3)+3}}\cdot\\
{}_3\phi_2\biggl(\genfrac{}{}{0pt}{}{-\varepsilon q^{2(k+\epsilon_4+\sigma_3)
+1},\,-q^{2(k+r+\epsilon_3-\sigma_3)+1},\,-\varepsilon \alpha q^{2(l+r+
\epsilon_2+\sigma_3)+1}}{\alpha q^{2(k+l+r+\epsilon_1+\epsilon_3+1)},
\,-q^{2(k+r+\epsilon_3+\sigma_3)+3}};q^2,q^2\biggr)
\end{multline}
and
\begin{equation*}
V_{k,r,l}=\begin{cases}
t_{k,r,l}\bigl(\varepsilon;q^{2\epsilon_1},q^{2\epsilon_2};q^{2\sigma_3}\bigr)&
\quad\text{if}\;k>l,\\
t_{l+r,-r,k+r}\bigl(\varepsilon;q^{2\epsilon_2},q^{2\epsilon_1};q^{2\sigma_3}
\bigr)&\quad 
\text{if}\;l>k,\\
v_{k,r,l}\bigl(\varepsilon,\varepsilon;q^{2\epsilon_1},q^{2\epsilon_2};q^{2
\epsilon_2},q^{2\epsilon_1};q^{2\sigma_3}\bigr)&\quad\text{if}\;l=k,
\end{cases}
\end{equation*}
where
\begin{multline*}
t_{k,r,l}\bigl(\varepsilon;q^{2\epsilon_1},q^{2\epsilon_2};q^{2\sigma_3}\bigr)
=\frac{1}{\bigl(1+q^{2(k+r+\epsilon_2+\sigma_3)-1}\bigr)
\bigl(1+q^{2(l+\epsilon_1-\sigma_3)+1}\bigr)}\cdot\\
{}_4\phi_3\biggl(\genfrac{}{}{0pt}{}{q^{-2(k-l-1)},\,\varepsilon q^{-2(r+\epsilon_2-\epsilon_1
)+2},\,q^{-4\sigma_3+2},\,q^2}{-q^{-2(k+r+\epsilon_2+\sigma_3)+3},\,-
\varepsilon q^{2(l+\epsilon_1-\sigma_3)+3},\,q^4};q^2,q^2\biggr).
\end{multline*}

\begin{proposition}\label{P:rsp}
In the cases $z=\varepsilon$, $q^{2\sigma_4}=\varepsilon q^{2\sigma_3}$ with
$\varepsilon=1$ or $\varepsilon=-1$, the solution $r^{k,k+r^*}_{l,l+r^*}$ is given by
\begin{multline}\label{E:rz1}
r^{k,k+r^*}_{l,l+r^*}\bigl(1,\alpha;\beta,q^{2\zeta};q^{2\epsilon_1},q^{2\epsilon_2};
q^{2\epsilon_3},q^{2\epsilon_4};q^{2\sigma_3},q^{2\sigma_3}\bigr)=\\
q^{2(\epsilon_3-\sigma_3)+1}\iota(q^{2\sigma_3})\frac{\bigl(q^{2(\epsilon_1-\epsilon_2+
1)};q^2\bigr)_{\infty}}{(q^2;q^2)_{\infty}}\cdot\\
\sqrt{\frac{\bigl(-q^{2(k+\epsilon_4+\sigma_3)+1},-q^{2(k+r+\epsilon_3-\sigma_3
)+1},-\alpha q^{2(l+\epsilon_1-\sigma_3)+1},-\alpha q^{2(l+r+\epsilon_2+\sigma_3
)+1};q^2\bigr)_{\infty}}{\bigl(-q^{2(k+\epsilon_4-\sigma_3)+1},-q^{2(k+r+
\epsilon_3+\sigma_3)+1},-\alpha q^{2(l+\epsilon_1+\sigma_3)+1},-\alpha q^{2(l+r+
\epsilon_2-\sigma_3)+1};q^2\bigr)_{\infty}}}\cdot\\
\shoveleft{\biggl\{
(-\alpha)^{l-k}q^{(l-k)^2+2(l-k)(\epsilon_1-\epsilon_4)}\bigl(\alpha q^{2(l-k+\epsilon_1-\epsilon_4+1)},\alpha q^{2(k-l+\epsilon_4-\epsilon_1+1)};
q^2\bigr)_{\infty}\cdot}\\
\shoveright{\theta_{q^2}\bigl(\alpha\alpha'q^{2(\epsilon_2+\epsilon'_2+1)},\alpha' 
q^{2(\epsilon'_2-\epsilon_3)},q^{4\sigma_3}\bigr)}\\
\shoveleft{+q^{l+k+2(\epsilon_1-\epsilon_2)}
\,v_{k,r,l}\cdot}\\
\theta_{q^2}\bigl(-\alpha q^{2(\epsilon_2+\sigma_3)+1},-\alpha'q^{2(\epsilon'_2
+\sigma_3)+1},-q^{2(\epsilon_3-\sigma_3)+1},-\alpha\alpha'q^{2(\epsilon_2+
\epsilon'_2-\epsilon_3-\sigma_3)+1}\bigr)\cdot\\
\frac{\bigl(1+q^{2(k+\epsilon_4-\sigma_3)+3}\bigr)\bigl(1-q^{2(r+\epsilon_2-
\epsilon_1)}\bigr)\bigl(1-q^{4\sigma_3}\bigr)}{\bigl(1+q^{2(k+\epsilon_4+
\sigma_3)+1}\bigr)\bigl(1+q^{2(k+r+\epsilon_3-\sigma_3)+1}\bigr)\bigl(1+\alpha 
q^{2(l+\epsilon_1-\sigma_3)+1}\bigr)(1-q^2)}\biggr\}
\end{multline}
and
\begin{multline}\label{E:rz1-}
r^{k,k+r^*}_{l,l+r^*}\bigl(-1,\alpha;\beta,q^{2\zeta};q^{2\epsilon_1},q^{2
\epsilon_2};q^{2\epsilon_3},q^{2\epsilon_4};q^{2\sigma_3},-q^{2\sigma_3}\bigr)=\\
(-1)^{k}\,r^{k,k+r^*}_{l,l+r^*}\bigl(1,\alpha;\beta,q^{2\zeta};q^{2\epsilon_1},
-q^{2\epsilon_2};q^{2\epsilon_3},-q^{2\epsilon_4};q^{2\sigma_3},q^{2\sigma_3}
\bigr).
\end{multline}
With the further restriction $\alpha=\varepsilon$ and $\epsilon_1=\epsilon_4$, $\epsilon_2=\epsilon_3$, the expression for $r^{k,k+r^*}_{l,l+r^*}$ reduces to
\begin{multline}\label{E:rz1eps14}
r^{k,k+r^*}_{l,l+r^*}\bigl(\varepsilon,\varepsilon;\varepsilon\alpha',q^{2\zeta};
q^{2\epsilon_1},q^{2\epsilon_2};
q^{2\epsilon_2},q^{2\epsilon_1};q^{2\sigma_3},\varepsilon q^{2\sigma_3}\bigr)=
\delta_{k,l}\,\varepsilon^k\cdot\\
q^{2(\epsilon_2-\sigma_3)+1}\iota(q^{2\sigma_3})
\theta_{q^2}\bigl(\varepsilon\alpha'q^{2(\epsilon_2+\epsilon'_2+1)},
\varepsilon\alpha'q^{2(\epsilon_2-\epsilon'_2+1)},q^{4\sigma_3}\bigr)
\bigl(\varepsilon q^{2(\epsilon_1-\epsilon_2+1)},q^2;q^2\bigr)_{\infty}\\
+V_{k,r,l}\cdot\varepsilon^{k+1}
\theta_{q^2}\bigl(-q^{2(\epsilon_2+\sigma_3)+1},-\varepsilon
\alpha'q^{2(\epsilon'_2+\sigma_3)+1},-q^{2(\epsilon_2-\sigma_3)+1},
-\varepsilon \alpha'q^{2(\epsilon'_2-\sigma_3)+1}\bigr)\cdot\\
  q^{2(\epsilon_1-\sigma_3)+1}\iota(q^{2\sigma_3})\frac{\bigl(\varepsilon q^{2(\epsilon_1
-\epsilon_2+1)};q^2
\bigr)_{\infty}\bigl(1-\varepsilon q^{2(r+\epsilon_2-\epsilon_1)}
\bigr)\bigl(1-q^{4\sigma_3}\bigr)}{(q^2;q^2)_{\infty}(1-q^2)}\cdot\\
q^{k+l}\sqrt{\frac{(-\varepsilon q^{2(\epsilon_1-\sigma_3)+1},-q^{2(r+\epsilon_2
+\sigma_3)+1};q^2)_k(-\varepsilon q^{2(\epsilon_1+\sigma_3)+1},
-q^{2(r+\epsilon_2-\sigma_3)+1};q^2)_l}{(-\varepsilon q^{2(\epsilon_1+\sigma_3)+1},-q^{2(r+\epsilon_2-\sigma_3)+1};q^2
)_k(-\varepsilon q^{2(\epsilon_1-\sigma_3)+1},-q^{2(r+\epsilon_2+\sigma_3)+1};q^2)_l}}.
\end{multline}

\end{proposition}

\emph{Proof}: In the case $z=\varepsilon$, $q^{2\sigma_4}=\varepsilon
q^{2\sigma_3}$, 
the expressions \eqref{E:checkRsp} and \eqref{E:checkRsp} with $q^{2\sigma_3}
\to q^{-2\sigma_3}$ can be employed in the definition of $r^{k,k+r^*}_{l,l+r^*}$
by \eqref{E:hatRel}. Making use of the relation
\begin{multline}\label{E:Wfrac}
\frac{\bigl(1+\varepsilon q^{2(k+\epsilon_4+\sigma_3)+1}\bigr)\bigl(1+q^{2(k+r+
\epsilon_3-\sigma_3)+1}\bigr)\bigl(1+\alpha q^{2(l+\epsilon_1-\sigma_3)+1}
\bigr)\bigl(1-q^2\bigr)}{\bigl(1-\varepsilon\alpha q^{2(l-k+
\epsilon_1-\epsilon_4)}\bigr)\bigl(1+\varepsilon q^{2(k+\epsilon_4
-\sigma_3)+3}\bigr)\bigl(1-\varepsilon q^{2(r+\epsilon_2-\epsilon_1)}\bigr)
\bigl(1-q^{-4\sigma_3}\bigr)}\cdot\\
\varepsilon q^{-2(k+\epsilon_4+\sigma_3)-1}\Biggl\{1-
\frac{\bigl(-\varepsilon q^{2(k+\epsilon_4-\sigma_3)+1},
-q^{2(k+r+\epsilon_3+\sigma_3)+1},-\alpha q^{2(l+\epsilon_1+\sigma_3)+1}
;q^2\bigr)_{\infty}}{\bigl(-\varepsilon q^{2(k+\epsilon_4
+\sigma_3)+1},-q^{2(k+r+\epsilon_3-\sigma_3)+1},-\alpha q^{2(l+\epsilon_1-
\sigma_3)+1};q^2\bigr)_{\infty}}\\
\shoveright{\cdot\frac{\bigl(-\varepsilon \alpha q^{2(l+r+\epsilon_2-\sigma_3)+1};q^2\bigr)_{\infty}}{\bigl(-\varepsilon
\alpha q^{2(l+r+\epsilon_2+\sigma_3)+1};q^2\bigr)_{\infty}}\Biggr\}}\\
=v_{k,r,l}\bigl(\varepsilon,\alpha;q^{2\epsilon_1},q^{2\epsilon_2};
q^{2\epsilon_3},q^{2\epsilon_4};q^{2\sigma_3}\bigr)=\hat v_{k,r,l}
\bigl(\varepsilon,\alpha;q^{2\epsilon_1},
q^{2\epsilon_2};q^{2\epsilon_3},q^{2\epsilon_4};q^{2\sigma_3}\bigr),
\end{multline}
where
\begin{multline*}
\hat v_{k,r,l}\bigl(\varepsilon,
\alpha;q^{2\epsilon_1},q^{2\epsilon_2};q^{2\epsilon_3},
q^{2\epsilon_4};q^{2\sigma_3}\bigr)=\\
v_{l+r,-r,k+r}\bigl(\varepsilon,
\alpha;\alpha q^{2\epsilon_3},\alpha q^{2\epsilon_4};\alpha q^{2
\epsilon_1},\alpha q^{2\epsilon_2};q^{2\sigma_3}\bigr)\cdot\\
\frac{\bigl(1+\varepsilon q^{2(k+\epsilon_4+\sigma_3)+1}\bigr)\bigl(
1+\varepsilon\alpha q^{2(l+r+\epsilon_2-\sigma_3)+3}\bigr)}{\bigl(1+\varepsilon q^{2(k+\epsilon_4-\sigma_3)+3}\bigr)\bigl(1+\varepsilon \alpha q^{2(l+r+\epsilon_2+\sigma_3)+1}\bigr)},
\end{multline*}
and equation \eqref{E:ex.2.16} with $x\to-\sqrt{\alpha\alpha'}q^{\epsilon_2+\epsilon'_2+2\sigma_3+1}$, 
$\lambda\to\varepsilon\alpha q^{\epsilon_2-\epsilon'_2}/ 
\sqrt{\alpha\alpha'}$, $\mu\to-\sqrt{\alpha\alpha'}q^{\epsilon_2+\epsilon'_2
-2\sigma_3+1}$, $\nu\to\sqrt{\alpha\alpha'}q^{\epsilon_2+\epsilon'_2-2
\epsilon_3}$ gives rise to the assertions \eqref{E:rz1} and \eqref{E:rz1-}. 
With the special choices $\alpha=\varepsilon$, $\epsilon_1=\epsilon_4$, $\epsilon_2=\epsilon_3$ in the expression \eqref{E:skrl},
equation \eqref{E:rz1eps14} for $k\geq l$ follows from the statements \eqref{E:rz1} and \eqref{E:rz1-}. In view of
the last line in \eqref{E:Wfrac}, use of $\hat v_{k,r,l}(\varepsilon;
q^{2\epsilon_1},q^{2\epsilon_2};q^{2\epsilon_2},q^{2\epsilon_1};q^{2\sigma_3})$ 
in \eqref{E:rz1} leads to equation \eqref{E:rz1eps14} in the case $l>k$.

To complete the proof of the proposition, equation \eqref{E:Wfrac} remains to
be shown. Since the statement \eqref{E:Wfrac} for
$\varepsilon=-1$ is obtained from the case $\varepsilon=1$ by means of
the replacements $q^{2\epsilon_2}\to-q^{2\epsilon_2}$, $q^{2\epsilon_4}\to 
-q^{2\epsilon_4}$, it is sufficient to derive equation \eqref{E:Wfrac} for the choice $\varepsilon=1$.
If $\vert q^{2(l+r+\epsilon_2+\sigma_3)+1}\vert<1$, the three-term 
transformation formula [\cite{GR}:III.36] with $a\to-q^{2(k+\epsilon_4-
\sigma_3)+3}$, $c\to-q^{2(k+\epsilon_4-\sigma_3)+1}$, $d\to q^2$,
$e\to q^{-4\sigma_3+2}$,  
$b\to\alpha q^{2(k-l+\epsilon_4-\epsilon_1+1)}$, $f\to q^{-2(r+\epsilon_2-
\epsilon_1)+2}$,
or $b\to q^{-2(r+\epsilon_2-\epsilon_1)+2}$, $f\to\alpha q^{2(k-l+\epsilon_4-
\epsilon_1+1)}$ yields 
\begin{multline}
v_{k,r,l}\bigl(\alpha;q^{2\epsilon_1},q^{2\epsilon_2};q^{2\epsilon_3},q^{2
\epsilon_4};q^{2\sigma_3}\bigr)=\hat v_{k,r,l}\bigl(\alpha;q^{2\epsilon_1},
q^{2\epsilon_2};q^{2\epsilon_3},q^{2\epsilon_4};q^{2\sigma_3}\bigr)=\\
{}_8W_7\bigl(-q^{2(k+\epsilon_4-\sigma_3)+3};\alpha q^{2(k-l+\epsilon_4-
\epsilon_1+1)},-q^{2(k+\epsilon_4-\sigma_3)+1},q^{-2(r+\epsilon_2-\epsilon_1)
+2},q^{-4\sigma_3+2},q^2;\\
q^2,-\alpha q^{2(l+r+\epsilon_2+\sigma_3)+1}\bigr).
\end{multline}
An evaluation of the $_8W_7$-series by means of the contiguous relation
[\cite{isra}:2.2] with $a\to-q^{2(k+\epsilon_4-\sigma_3)+3}$, $b\to-q^{2(k+
\epsilon_4-\sigma_3)+3}$, $c\to q^2$, $d\to\alpha q^{2(k-l+\epsilon_4-\epsilon_1
+1)}$, $e\to q^{-2(r+\epsilon_2-\epsilon_1)+2}$, $f\to q^{-4\sigma_3+2}$
and the summation formula [\cite{GR}:II.20] with $a\to-q^{2(k+\epsilon_4-
\sigma_3)+3}$, $b\to\alpha q^{2(k-l+\epsilon_4-\epsilon_1+1)}$, $c\to q^{-2(
r+\epsilon_2-\epsilon_1)+2}$, $f\to q^{-4\sigma_3+2}$ gives rise to the lhs of 
\eqref{E:Wfrac}, provided that $\vert q^{2(l+r+\epsilon_2+\sigma_3)-1}\vert<1$.
The contiguous relation [\cite{isra}:2.2] with $a\to-
q^{2(k+\epsilon_4-\sigma_3)+3}$, $b\to\alpha q^{2(k-l+\epsilon_4-\epsilon_1
+2)}$, $c\to q^{-2(r+\epsilon_2-\epsilon_1)+2}$, $d\to-q^{2(k+\epsilon_4-
\sigma_3)+1}$, $e\to q^{-4\sigma_3+2}$, $f\to q^2$
allows to extend the derivation of \eqref{E:Wfrac} to the case $1<\vert q^{2(l+r+\epsilon_2+\sigma_3)-1}\vert<q^{-2}$.

Combining the formula
[\cite{GR}:III.36] with $a\to-q^{-2(k+\epsilon_4+\sigma_3)+3}$, $b\to-q^{-2(k+
\epsilon_4+\sigma_3)+1}$, $c\to q^{2(r+\epsilon_2-\epsilon_1+1)}$, $d\to\alpha q^{2(l-k+\epsilon_1-\epsilon_4+1)}$, $e\to q^{-4\sigma_3+2}$, $f\to q^2$
and the nonterminating $q$-Saalsch\"utz formula [\cite{GR}:II.24]
with $a\to-q^{-2(k+\epsilon_4
-\sigma_3)+1}$, $b\to-\alpha q^{-2(l+\epsilon_1+\sigma_3)+1}$, $c\to-\alpha q^{-2(l+r+\epsilon_2-\sigma_3)+1}$, $e\to\alpha q^{-2(k+l+r
+\epsilon_1+\epsilon_3)+2}$, $f\to-\alpha q^{-2(l+\epsilon_1-\sigma_3)+3}$
leads to an expression for $v_{k,r,l}(\alpha;q^{2\epsilon_1},q^{2\epsilon_2};
q^{2\epsilon_3},q^{2\epsilon_4};q^{2\sigma_3})$ in the case $\vert q^{-2(l+
r+\epsilon_2+\sigma_3)+1}\vert<1$. 
Alternatively, formula [\cite{GR}:III.36] with $a\to-q^{-2(k+\epsilon_4+\sigma_3)
+3}$, $b\to-q^{-2(k+\epsilon_4+\sigma_3)+1}$, $c\to\alpha q^{2(l-k+\epsilon_4
-\epsilon_1+1)}$, $d\to q^{2(r+\epsilon_2-\epsilon_1+1)}$, $e\to q^{-4\sigma_3
+2}$, $f\to q^2$ supplemented by [\cite{GR}:II.24] with $a\to -q^{-2(k+
\epsilon_4-\sigma_3)+1}$, $b\to-q^{-2(k+r+\epsilon_3+\sigma_3)+1}$, $c\to-\alpha 
q^{-2(l+r+\epsilon_2-\sigma_3)+1}$, $e\to\alpha q^{-2(k+l+r+\epsilon_1+
\epsilon_3)+2}$, $f\to-q^{-2(k+r+\epsilon_3-\sigma_3)+3}$ yields the same
expression for $\hat v_{k,r,l}(\alpha;q^{2\epsilon_1},q^{2\epsilon_2};q^{2
\epsilon_3},q^{2\epsilon_4};q^{2\sigma_3})$. Both manipulations are 
summarized by
\begin{multline}
v_{k,r,l}\bigl(\alpha;q^{2\epsilon_1},q^{2\epsilon_2};q^{2\epsilon_3},q^{2
\epsilon_4};q^{2\sigma_3}\bigr)=
\hat v_{k,r,l}\bigl(\alpha;q^{2\epsilon_1},
q^{2\epsilon_2};q^{2\epsilon_3},q^{2\epsilon_4};q^{2\sigma_3}\bigr)=\\
\frac{\bigl(1+q^{2(k+\epsilon_4+\sigma_3)+1}\bigr)\bigl(1+q^{-2(k+\epsilon_4
+\sigma_3)+3}\bigr)\bigl(1+q^{2(k+r+\epsilon_3-\sigma_3)+1}\bigr)\bigl(1+\alpha
q^{-2(l+\epsilon_1-\sigma_3)-1}\bigr)}{\bigl(
1+q^{2(k+\epsilon_4-\sigma_3)+3}\bigr)\bigl(1+q^{-2(k+\epsilon_4-\sigma_3)+1}
\bigr)\bigl(1+q^{2(k+r+\epsilon_3+\sigma_3)-1}\bigr)\bigl(1+\alpha q^{-2(l+
\epsilon_1+\sigma_3)+1}\bigr)}\\
\quad\cdot{}_8W_7\bigl(-q^{-2(k+\epsilon_4+\sigma_3)+3};\alpha q^{2(l-k+\epsilon_1-
\epsilon_4+1)},-q^{-2(k+\epsilon_4+\sigma_3)+1},q^{2(r+\epsilon_2-\epsilon_1+1)},
q^{-4\sigma_3+2},q^2;\\
\shoveright{q^2,-\alpha q^{-2(l+r+\epsilon_2-\sigma_3)+1}\bigr)}\\
\shoveleft{+\frac{(1-q^2)\theta_{q^2}\bigl(q^{4\sigma_3+2},q^{2(r+\epsilon_2-
\epsilon_1+1)}\bigr)}{\bigl(1-q^{4\sigma_3}\bigr)
\bigl(1-q^{-2(r+\epsilon_2-\epsilon_1)}\bigr)\bigl(1+q^{2(k+\epsilon_4-\sigma_3)+3}
\bigr)}\cdot}\\
\frac{\bigl(\alpha q^{2(l-k+\epsilon_1-\epsilon_4+1)},\alpha q^{2(k-l+\epsilon_4
-\epsilon_1+1)};q^2\bigr)_{\infty}\theta_{q^2}\bigl(\alpha q^{2(k+l+\epsilon_2+
\epsilon_4+1)}\bigr)}{\bigl(-q^{-2(k+\epsilon_4-\sigma_3)+1},-q^{2(k+\epsilon_4+
\sigma_3)+3},-q^{2(k+r+\epsilon_3-\sigma_3)+3},-q^{-2(k+r+\epsilon_3+\sigma_3)+1}
;q^2\bigr)_{\infty}}\cdot\\
\frac{1}{\bigl(-\alpha q^{2(l+\epsilon_1-\sigma_3)+3},-\alpha q^{-2(l+\epsilon_1
+\sigma_3)+1},-\alpha q^{-2(l+r+\epsilon_2-\sigma_3)+1},-\alpha q^{2(l+r+
\epsilon_2+\sigma_3)+1};q^2\bigr)_{\infty}}.
\end{multline}
If $\vert q^{-2(l+r+\epsilon_2-\sigma_3)-1}\vert<1$,
evaluation of the $_8W_7$-series using [\cite{isra}:2.2] with $a\to -q^{-2(
k+\epsilon_4+\sigma_3)+3}$, $b\to-q^{-2(k+\epsilon_4+\sigma_3)+3}$, $c\to q^2$,
$d\to\alpha q^{2(l-k+\epsilon_1-\epsilon_4+1)}$, $e\to q^{2(r+\epsilon_2-
\epsilon_1+1)}$, $f\to q^{-4\sigma_3+2}$ and [\cite{GR}:II.20] with $a\to 
-q^{-2(k+\epsilon_4+\sigma_3)+3}$, $b\to\alpha q^{2(l-k+\epsilon_1-\epsilon_4
+1)}$, $c\to q^{2(r+\epsilon_2-\epsilon_1+1)}$, $d\to q^{-4\sigma_3+2}$
and application of the relation
\eqref{E:ex.2.16} with $x\to-q^{\epsilon_3+\epsilon_4+1}$ $\lambda\to q^{
\epsilon_3-\epsilon_4-2\sigma_3}$, $\mu\to-\alpha q^{\epsilon_1+\epsilon_2+1}$,
$\nu\to q^{\epsilon_1-\epsilon_2-2\sigma_3}$ leads to the lhs of \eqref{E:Wfrac}. The case $1<\vert q^{-2(l+r+\epsilon_2-\sigma_3)-1}\vert<q^{
-2}$ is included using the relation
[\cite{isra}:2.2] with $a\to-q^{-2(k+\epsilon_4+\sigma_3)+3}$,
$b\to\alpha q^{2(l-k+\epsilon_1-\epsilon_4+2)}$, $c\to q^{2(r+\epsilon_2-
\epsilon_1+1)}$, $d\to-q^{-2(k+\epsilon_4+\sigma_3)+1}$, $e\to q^{-4\sigma_3+2}$, $f\to q^2$. This proves equation \eqref{E:Wfrac}.

\section{Generalized unitarity}\label{S:unit}

The present section aims at the construction of operators inverting
the action of the intertwiner $R(\tilde z^{\frac{1}{2}},z^{\frac{1}{2}})$
in the case $\alpha=-\alpha'$.
With respect to a given pair $(q^{2\epsilon_1},q^{2\epsilon_2},\alpha q^{2\sigma_3},\alpha q^{2\sigma_4})$ and $(q^{2\epsilon'_1},q^{2\epsilon'_2},-\alpha q^{2\sigma_3},
-\alpha q^{2\sigma_4})$, a particular 4-tuple $(q^{2\epsilon'_3},q^{2
\epsilon'_4},-q^{2\sigma_3},-q^{2\sigma_4})$ is 
attributed to a given 4-tuple $(q^{2\epsilon_3},q^{2\epsilon_4},q^{2
\sigma_3},q^{2\sigma_4})$ to enable the subsequent application of Theorem 
\ref{T:tausum2} and Corollary \ref{C:tausum4}-\ref{C:tausum6} to the pair 
formed by $(q^{2\epsilon_3},q^{2\epsilon_4},q^{2\sigma_3},q^{2\sigma_4})$ and $(q^{2\epsilon'_3},q^{2\epsilon'_4},-q^{2\sigma_3},-q^{2\sigma_4})$. 
In subsection \ref{S:sucint}, the parameters 
$q^{2\epsilon'_3}$ and $q^{2\epsilon'_4}$ are specified. To complete the 
preparations for the investigation of two particular intertwiners 
in subsection \ref{S:invint}, the results given by Corollary \ref{C:tausum4} and \ref{C:tausum6} for the combinations $\Xi^{(r,k)\pm}$ or $\Xi^{(r,k)}$ 
of the sums $\boldsymbol{\tau}^{(r,k)\pm}$ and $\boldsymbol{\tau}'^{(r,k)\pm}$
or $\boldsymbol{\tau}^{(r,k)}$ and $\boldsymbol{\tau}'^{(r,k)}$ related to 
$r^{k,k+r^*}_{l,l+r^*}$ are supplemented by the corresponding statements 
for the matrix elements $\mathring r^{k,k+r^*}_{l,l+r^*}$ defined by
\eqref{E:mathrr}. Subsection \ref{S:mtau} provides the results 
\eqref{E:hattau} and \eqref{E:mhattau0} replacing Corollary \ref{C:tausum4} 
and \ref{C:tausum6} for this choice.

Two linear combinations ${r_{\pm}}^{k,k+r^*}_{l,l+r^*}$ of the matrix elements
$r^{k,k+r^*}_{l,l+r^*}$ and $\mathring r^{k,k+r^*}_{l,l+r^*}$ are 
introduced in subsection \ref{S:invint} for $\alpha=\alpha'$ or $\alpha=-
\alpha'$. In the special case $z=\alpha=
\varepsilon$, $\epsilon_1=\epsilon_4$, $\epsilon_2=\epsilon_3$, $q^{2\sigma_4}
=\varepsilon q^{2\sigma_3}$ with $\varepsilon=1$ or $\varepsilon=-1$,
the matrix elements ${r_{\pm}}^{k,k+r^*}_{l,l+r^*}$ are shown to obey the
initial condition \eqref{E:Runinit}. An evaluation of the combination 
$\Xi^{(r,k)\pm}$ associated with the matrix elements ${r_{\pm}}^{k,k+r^*}_{
l,l+r^*}$ yields the expression \eqref{E:tauunit}. For $\Xi^{(r,k)}$
with $e^{i\theta}=\alpha\alpha'q^{2(\tilde s+\epsilon'_1+\epsilon_2)+1}$,
the corresponding result is given by equation \eqref{E:tauunit2}. The
derivation of \eqref{E:tauunit} and \eqref{E:tauunit2}
involves Corollary \ref{C:tausum4} and \ref{C:tausum5} as well as 
the result \eqref{E:hattau} in subsection \ref{S:mtau}.

Definition \ref{D:unsigdef} introduces the sums $\sigma^{[\pm](r,m)\pm}$ and
$\sigma^{[\pm](r,m)}$ as the bilateral summations of the combinations $\Xi^{(r,k)\pm}$ and $\Xi^{(r,k)}$ associated with the matrix elements 
${r_{\pm}}^{k,k+r^*}_{l,l+r^*}$ weighted by a factor ${r_{\mp}}^{m,m+r^*}_{
k,k+r^*}$ in the case $\alpha=-\alpha'$. 
Here the factor ${r_{\mp}}^{m,m+r^*}_{k,k+r^*}$ depends 
on a separate set of parameters related to the original set for 
$\Xi^{(r,k)\pm}$ and $\Xi^{(r,k)}$ by inversion of the squared
spectral parameter $z$, permutation of the parameters $q^{2\epsilon_i}$ and
the replacement $q^{2\sigma_3}\to\alpha q^{2\sigma_4}$. 
The sums $\sigma^{[\pm](r,m)\pm}$ and $\sigma^{[\pm](r,m)}$
allow for the search of an inverse action following the application 
of the intertwiner $R(\tilde z^{\frac{1}{2}},z^{\frac{1}{2}})$.
Proposition \ref{P:unitrel} specifying the evaluation of the sums 
$\sigma^{[\pm](r,m)\pm}$ and $\sigma^{[\pm](r,m)}$ formulates the second main 
result of this article. Its proof relies on the formulae \eqref{E:tauunit} 
and \eqref{E:tauunit2} for the combinations $\Xi^{(r,k)\pm}$ and $\Xi^{(r,k)}$ 
and on the equations \eqref{E:tausum7} and \eqref{E:mhattau0} in subsection 
\ref{S:mtau}.

Referring to the vector-valued big $q$-Jacobi function transform
pair associated with the identifications in subsection \ref{S:bigJac}, a pair of
quadratic bilateral summation formulae for the matrix elements ${r_{\pm}}^{
k,k+r^*}_{l,l+r^*}$ is obtained from Proposition \ref{P:unitrel}. 
According to the latter result,
the intertwiner associated with ${r_{\mp}}^{m,m+r^*}_{k,k+r^*}$ inverts
the action of the first intertwiner related to ${r_{\pm}}^{k,k+r^*}_{l,l+r^*}$,
provided one or both intertwiners are equipped with simple normalizations.
The summation formulae specified by Corollary \ref{C:RRsum} 
state the third main result of the publication.

\subsection{Successive action of the intertwiner}\label{S:sucint}

Given a pair $(q^{2\epsilon_1},q^{2\epsilon_2})$ and $(q^{2\epsilon'_1},q^{2
\epsilon'_2})$ with $\epsilon_2-\epsilon_1=\epsilon'_2-\epsilon'_1$, Corollary
\ref{C:tausum5} in subsection \ref{S:tauev}
associates a tuple $(q^{2\epsilon'_3},q^{2\epsilon'_4})$ to any
tuple $(q^{2\epsilon_3},q^{2\epsilon_4})$ restricted by $\epsilon_3-\epsilon_4=\epsilon_2-\epsilon_1$. Its parameters are determined by
\begin{equation}\label{E:epscond3}
\epsilon'_3+\epsilon_4=\epsilon'_4+\epsilon_3=
\epsilon'_1+\epsilon_2=\epsilon'_2+\epsilon_1,
\end{equation}
which implies 
\begin{equation}
\epsilon_3+\epsilon'_3=\epsilon_2+\epsilon'_2=\zeta.
\end{equation}
Referring to both $(q^{2\epsilon_3},q^{2\epsilon_4})$ and $(q^{2
\epsilon'_3},q^{2\epsilon'_4})$ in a symmetric way, equation \eqref{E:tausum4} in Corollary \ref{C:tausum4} is reformulated by
\begin{multline*}
\widehat{\tau}^{(r,k)+}\bigl(z,\alpha\beta';\beta,q^{2\zeta};q^{2
\epsilon_1},q^{2\epsilon_2};q^{2\zeta_3},q^{2\zeta_4};\beta' q^{2\sigma_3},
\beta' q^{2\sigma_4};\cos\theta\bigr)\\
-\alpha\alpha'q^{2(\epsilon'_2-\epsilon_2)}\widehat{\tau}'^{(r,k)+}\bigl(
z,\alpha'\beta';\beta,q^{2\zeta};q^{2\epsilon'_1},q^{2\epsilon'_2};q^{2\zeta_3},
q^{2\zeta_4};\beta'q^{2\sigma_3},\beta'q^{2\sigma_4};\cos\theta\bigr)
\Bigr]\\
=q^{2(\zeta_4-\sigma_4)+1}\iota(q^{2\sigma_3})\theta_{q^2}\bigl(\alpha\alpha'q^{2(
\epsilon'_2-\epsilon_2)
},\alpha\alpha'z^{-1}q^{2(\epsilon_1+\epsilon'_2+1)},q^{4\sigma_4}\bigr)\cdot\\
\frac{\theta_{q^2}\bigl(-q^{2(\epsilon_3+\sigma_3)+1},-\alpha\alpha'q^{2(
\epsilon'_3+\sigma_3)+1}\bigr)}{\theta_{q^2}\bigl(-\alpha q^{2(\epsilon_1+
\sigma_3)+1},-\alpha'q^{2(\epsilon'_1+\sigma_3)+1}\bigr)}\cdot\\
\frac{\bigl(zq^{2(\epsilon_1-\epsilon_2+1)},q^{-2(\epsilon_2-\epsilon_1-
\sigma_3+\sigma_4)+2},q^{-2(\epsilon_2-\epsilon_1-\sigma_3-\sigma_4)+2},q^2;
q^2\bigr)_{\infty}}{\bigl(q^{2(\epsilon_2-\epsilon_1+\sigma_3+\sigma_4+1)},
q^{2(\epsilon_2-\epsilon_1+\sigma_3-\sigma_4+1)};q^2\bigr)_{\infty}}\cdot\\
\frac{u\bigl(q^{2\epsilon_2},q^{2\epsilon_1};e^{i\theta}\bigr)}{\bigl(zqe^{i\theta},
zqe^{-i\theta};q^2\bigr)_{\infty}}\cdot\\
(-1)^ra^+_{-r}\bigl(q^{2\zeta_3},q^{2\zeta_4};{\beta}'q^{2\sigma_3};\cos\theta
\bigr)
\rho^{(-r)+}_{k+r}\bigl(q^{2\zeta_3},{\beta}'q^{2\sigma_3};q^{2\zeta_4},{\beta}'
q^{2\sigma_4};\cos\theta \bigr)
\end{multline*}
for $(\zeta_3,\zeta_4,{\beta}')=(\epsilon_3,\epsilon_4,1)$ or $(\zeta_3,\zeta_4,{\beta}')=(\epsilon'_3,\epsilon'_4,
\alpha\alpha')$. Thus for $\alpha'=-\alpha$, Corollary \ref{C:tausum4} can be applied in context with
the subsequent action of the intertwiners
$R(z^{-\frac{1}{2}},\tilde{\alpha}z^{-\frac{1}{2}}):
W^{(\zeta_4,\beta'q^{2\sigma_4})}_{\beta'z^{-\frac{1}{2}}}\otimes W^{(\zeta_3,\beta'q^{2\sigma_3})*}_{\beta'z^{
\frac{1}{2}}}\to W^{(\epsilon_5,\tilde{\alpha}q^{2\sigma_3})}_{\tilde\alpha z^{\frac{1}{2}}}\otimes W^{(\epsilon_6,
\tilde{\alpha}q^{2\sigma_4})*}_{\tilde{\alpha}z^{-\frac{1}{2}}}$.

The present section addresses the existence of inverse intertwiners.
Therefore $W^{(\epsilon_5,\tilde{\alpha}q^{2\sigma_3})}_{
\tilde{\alpha}z^{\frac{1}{2}}}\otimes W^{(\epsilon_6,\tilde{\alpha}q^{2
\sigma_4})*}_{\tilde{\alpha}z^{-\frac{1}{2}}}$ is specialized to the
initial tensor products of the previous section given by
$W^{(\epsilon_1,q^{2\sigma_1})}_{\tilde z^{\frac{1
}{2}}}\otimes W^{(\epsilon_2,q^{2\sigma_2})*}_{\tilde z^{-\frac{1}{2}}}$.
A description of the successive application of the intertwiners is provided in
terms of the sums introduced by the Definition \ref{D:unsigdef} in subsection
\ref{S:invint}. For the second choice $W^{(\epsilon'_1,\alpha\alpha'q^{2\sigma_1})}_{\alpha\alpha'
\tilde z^{\frac{1}{2}}}\otimes W^{(\epsilon'_2,\alpha\alpha'q^{2\sigma_2})*}_{
\alpha\alpha'\tilde z^{-\frac{1}{2}}}$, all statements are obtained from
the results for the first choice by means of obvious parameter replacements
and adjustment of simple prefactors.

To facilitate notation, the abbreviations
\begin{equation*}
\begin{split}
\gamma(q^{2\sigma_3})&=\frac{\theta_{q^2}\bigl(-q^{2(\epsilon_3+\sigma_3)+1},-
\alpha\alpha'q^{2(\epsilon'_3+\sigma_3)+1}\bigr)}{\theta_{q^2}\bigl(-\alpha 
q^{2(\epsilon_1+\sigma_3)+1},-\alpha'q^{2(\epsilon'_1+\sigma_3)+1}\bigr)},\\
\delta(q^{2\sigma_4})&=\frac{\theta_{q^2}\bigl(-\alpha q^{2(\epsilon_2+\sigma_4
)+1},-\alpha'q^{2(\epsilon'_2+\sigma_4)+1}\bigr)}{\theta_{q^2}\bigl(-q^{2(
\epsilon_4+\sigma_4)+1},-\alpha\alpha'q^{2(\epsilon'_4+\sigma_4)+1}\bigr)},\\
H_{\theta}&=\bigl(\gamma(q^{2\sigma_3})\gamma(q^{-2\sigma_3})\delta(q^{2\sigma_4}
)\delta(q^{-2\sigma_4})\bigr)^{-1}
\end{split}
\end{equation*}
and
\begin{multline*}
c=q^{2(\epsilon_2+\epsilon_4+1)}\cdot\\
\theta_{q^2}\bigl(\alpha\alpha'q^{2(\epsilon'_2-\epsilon_2)},\alpha\alpha'
z^{-1}q^{2(\epsilon'_1+\epsilon_2+1)},\alpha\alpha'q^{2(\epsilon'_3-\epsilon_3)},
\alpha\alpha'zq^{2(\epsilon'_3+\epsilon_4+1)}\bigr)
\end{multline*}
are introduced. 

Throughout this section, the cases $\alpha q^{2(\epsilon_i-
\sigma_{i+2})}=-q^{2t+1}$ or $\alpha'q^{2(\epsilon'_i-\sigma_{i+2})}=-q^{2t'+1}$
with $i=1,2$ and $t,t'\in\mathbb Z$ are excluded.

\subsection{The sums $\mathring{\widehat{\tau}}^{(r,k)\pm}(q^{2\zeta},q^{2\epsilon_2})$ and 
$\mathring{\widehat{\tau}}^{(r,k)}(q^{2\zeta},q^{2\epsilon_2})$}\label{S:mtau}

The sums associated with the choice
\begin{multline*}
\boldsymbol{R}^{k,k+r^*}_{l,l+r^*}=\mathring r^{k,k+r^*}_{l,l+r^*}\equiv 
\mathring r^{k,k+r^*}_{l,l+r^*}\bigl(z,\alpha;\beta,q^{2\zeta};
q^{2\epsilon_1},q^{2\epsilon_2};
q^{2\epsilon_3},q^{2\epsilon_4};q^{2\sigma_3},q^{2\sigma_4}\bigr)=\\
r^{k+r,k^*}_{l+r,l^*}\bigl(z,\alpha;\beta,q^{2(\epsilon_1+\epsilon'_1)};
q^{2\epsilon_2},q^{2\epsilon_1};q^{2
\epsilon_4},q^{2\epsilon_3};q^{2\sigma_4},q^{2\sigma_3}\bigr)
\end{multline*}
in Definition \ref{D:taudef1} (equations \eqref{E:taudef1} and \eqref{E:taudef2} in subsection \ref{S:taudef}) will be denoted by
\begin{equation*}
\begin{split}
\mathring{\widehat{\tau}}^{(r,k)\pm}(q^{2\zeta},q^{2\epsilon_2})&\equiv
\mathring{\widehat{\tau}}^{(r,k)\pm}\bigl(z,\alpha;\beta,q^{2\zeta};q^{2\epsilon_1},
q^{2\epsilon_2};q^{2\epsilon_3},q^{2\epsilon_4};q^{2\sigma_3},q^{2\sigma_4};
\cos\theta\bigr),\\
\mathring{\widehat{\tau}}^{(r,k)}(q^{2\zeta},q^{2\epsilon_2})&\equiv
\mathring{\widehat{\tau}}^{(r,k)}\bigl(z,x,\alpha;\beta,q^{2\zeta};q^{2\epsilon_1},
q^{2\epsilon_2};q^{2\epsilon_3},q^{2\epsilon_4};q^{2\sigma_3},q^{2\sigma_4};
e^{i\theta}\bigr).
\end{split}
\end{equation*}
A second set of parameters gives rise to the sums
\begin{equation*}
\begin{split}
\mathring{\widehat{\tau}}'^{(r,k)\pm}\bigl(q^{2\zeta},q^{2\epsilon'_2})&
\equiv
\mathring{\widehat{\tau}}^{(r,k)\pm}\bigl(z,\alpha';\beta,q^{2\zeta};q^{2
\epsilon'_1},q^{2\epsilon'_2};q^{2\epsilon_3},q^{2\epsilon_4};q^{2\sigma_3},
q^{2\sigma_4};\cos\theta\bigr),\\
\mathring{\widehat{\tau}}'^{(r,k)}\bigl(q^{2\zeta},q^{2\epsilon'_2}\bigr)&
\equiv
\mathring{\widehat{\tau}}^{(r,k)}\bigl(z,x',\alpha';\beta,q^{2\zeta};
q^{2\epsilon'_1},q^{2\epsilon'_2};q^{2\epsilon_3},q^{2\epsilon_4};q^{2
\sigma_3},q^{2\sigma_4};e^{i\theta}\bigr),
\end{split}
\end{equation*}
where $\epsilon'_2-\epsilon'_1=\epsilon_2-\epsilon_1$.
If
\begin{equation}\label{E:tausum5cond}
\begin{split}
&q^{2(\epsilon_2-\epsilon_1\pm\sigma_3-\sigma_4)}\neq q^{2t_1},\;
t_1\in\mathbb Z_{\backslash0},\quad q^{2(\epsilon_2-\epsilon_1\pm\sigma_3+\sigma_4)}\neq  
q^{2t_2},\;t_2\in\mathbb Z,\\
&\vert zqe^{i\theta}\vert<1,\\
&\vert zqe^{-i\theta}\vert<1\quad\text{for}\;e^{i\theta}\notin\Gamma^+
\end{split}
\end{equation}
with $\Gamma^{\pm}$ defined by \eqref{E:Gammapm},
the sums $\mathring{\widehat{\tau}}^{(r,k)+}(q^{2\zeta},q^{2\epsilon_2})$ 
and $\mathring{\widehat{\tau}}'^{(r,k)+}(q^{2\zeta},q^{2\epsilon'_2})$ are
well-defined and absolutely convergent. 
For $\mathring{\widehat{\tau}}^{(r,k)-}(q^{2\zeta},q^{2\epsilon_2})$
and $\mathring{\widehat{\tau}}'^{(r,k)-}(q^{2\zeta},q^{2\epsilon'_2})$,
the corresponding conditions are obtained by the replacements $q^{2\sigma_3}
\to q^{-2\sigma_3}$ and $\Gamma^+\to\Gamma^-$.

The sums $\mathring{\widehat{\tau}}^{(r,k)}(q^{2\zeta},q^{2\epsilon_2})$
are well-defined and absolutely convergent if
\begin{equation}\label{E:tausum5conda}
\begin{split}
&q^{2(\epsilon_2-\epsilon_1+\sigma_3\pm\sigma_4)}\neq q^{2t_1},\quad q^{2(\epsilon_2
-\epsilon_1-\sigma_3\pm\sigma_4)}\neq q^{2t_2},\;t_1,t_2\in\mathbb Z,\\ 
&\vert zqe^{i\theta}\vert<1.
\end{split}
\end{equation}
All sign choices in \eqref{E:tausum5cond} and \eqref{E:tausum5conda} are
independent.

The sums $\mathring{\widehat{\tau}}^{(r,k)+}(q^{2\zeta},q^{2\epsilon_2})$
and $\mathring{\widehat{\tau}}'^{(r,k)+}(q^{2\zeta},q^{2\epsilon'_2})$
are related by
\begin{multline}\label{E:hattau}
\mathring{\widehat{\tau}}^{(r,k)+}(q^{2\zeta},q^{2\epsilon_2})-\alpha\alpha'q^{2(
\epsilon'_2-\epsilon_2)}\mathring{\widehat{\tau}}'^{(r,k)+}(q^{2\zeta},q^{2
\epsilon'_2})=\\
q^{2(\epsilon_3-\sigma_3)+1}\iota(q^{2\sigma_4})\theta_{q^2}\bigl(\alpha\alpha'
q^{2(\epsilon'_1-\epsilon_1)},\alpha\alpha'z^{-1}q^{2(\epsilon_1+\epsilon'_2+1)},
q^{4\sigma_3}\bigr)\frac{\bigl(zq^{2(
\epsilon_2-\epsilon_1+1)},q^2;q^2\bigr)_{\infty}}{\bigl(zqe^{i\theta},zqe^{
-i\theta};q^2\bigr)_{\infty}}\cdot\\
\Biggl\{\frac{\bigl(q^{2(\sigma_3-\sigma_4)+1}e^{i\theta},q^{2(\sigma_3-\sigma_4)
+1}e^{-i\theta},q^{2(\epsilon_2-\epsilon_1-\sigma_3+\sigma_4+1)};q^2\bigr)_{
\infty}\theta_{q^2}\bigl(q^{2(\epsilon_2-\epsilon_1+\sigma_3+\sigma_4+1)}\bigr)}{
\theta_{q^2}\bigl(q^{4\sigma_4+2}\bigr)
\bigl(q^{2(\epsilon_2-\epsilon_1+\sigma_3-\sigma_4+1)};q^2\bigr)_{\infty}
\delta(q^{2\sigma_4})}\cdot\\
\shoveright{
a^+_r\bigl(q^{2\epsilon_4},q^{2\epsilon_3};q^{2\sigma_4};\cos\theta
\bigr)\rho^{(r)+}_k\bigl(q^{2\epsilon_4},q^{2\sigma_4};
q^{2\epsilon_3},q^{2\sigma_3};\cos\theta\bigr)}\\
+\frac{
\bigl(q^{2(\sigma_3+\sigma_4)+1}e^{i\theta},q^{2(\sigma_3+\sigma_4)+1}e^{-i\theta},
q^{2(\epsilon_2-\epsilon_1-
\sigma_3-\sigma_4+1)};q^2\bigr)_{\infty}
\theta_{q^2}\bigl(q^{2(\epsilon_2-\epsilon_1+\sigma_3-\sigma_4+1)}\bigr)}{
\theta(q^{-4\sigma_4+2})\bigl(q^{2(\epsilon_2-\epsilon_1+\sigma_3+\sigma_4+1)}
;q^2\bigr)_{\infty}\delta(q^{-2\sigma_4})}\cdot\\
a^-_r\bigl(q^{2\epsilon_4},q^{2\epsilon_3};q^{2\sigma_4};\cos\theta
\bigr)\rho^{(r)-}_k\bigl(
q^{2\epsilon_4},q^{2\sigma_4};q^{2\epsilon_3},q^{2\sigma_3};\cos\theta\bigr)\Biggr\}.
\end{multline}
Since 
\begin{multline*}
\mathring r^{k,k+r^*}_{l,l+r^*}(z,\alpha;\beta,q^{2\zeta};q^{2\epsilon_1},q^{2\epsilon_2};
q^{2\epsilon_3},q^{2\epsilon_4};q^{-2\sigma_3},q^{2\sigma_4})=\\
-\mathring r^{
k,k+r^*}_{l,l+r^*}(z,\alpha;\beta,q^{2\zeta};q^{2\epsilon_1},q^{2\epsilon_2};
q^{2\epsilon_3},q^{2\epsilon_4};q^{2\sigma_3},q^{2\sigma_4})
\end{multline*}
according to \eqref{E:hatRel},
the replacements $\sigma_1\to-\sigma_1$, $\sigma_3\to-\sigma_3$ on the rhs
of \eqref{E:hattau} and multiplication by $-1$ yield an
expression for $\mathring{\widehat{\tau}}^{(r,k)-}(q^{2\zeta},q^{2\epsilon_2})
-\alpha\alpha'q^{2(\epsilon'_2-\epsilon_2)}\mathring{\widehat{\tau}}'^{(r,k)-}
(q^{2\zeta},q^{2\epsilon'_2})$.

Writing $\mathring r^{k,k+r^*}_{l,l+r^*}$ in terms of $R^{k,k+r^*}_{l,l+r}$
and $\check R^{k,k+r^*}_{l,l+r^*}$ by means of \eqref{E:RmR} and the same
relation with $q^{2\sigma_3}\to q^{-2\sigma_3}$, the lhs of \eqref{E:hattau}
can be evaluated making use of Theorem \ref{T:tausum2} (equation \eqref{E:tausum2}) and the same result
with $q^{2\sigma_4}$ replaced by $q^{-2\sigma_4}$. The prefactor of
$\rho^{(-r)+}_{k+r}(q^{2\epsilon_3},q^{2\sigma_3};q^{2\epsilon_4},q^{2
\sigma_4};\cos\theta)$ can be simplified employing relation \eqref{E:ex.2.16}
with $x\to zq^2$, $\lambda\to q^{2(\sigma_4-\sigma_3)}$, $\mu\to q^{2(
\epsilon_2-\epsilon_1)}$, $\nu\to q^{2(\sigma_3+\sigma_4)}$. Use of
the expansion \eqref{E:rhorel} with $l\to k$, $\epsilon_1\to\epsilon_4$,
$\epsilon_2\to\epsilon_3$, $q^{2\sigma_1}\to q^{2\sigma_4}$, $q^{2\sigma_2}
\to q^{2\sigma_3}$ allows to express the lhs of \eqref{E:hattau}
as a linear combination of $a^+_r(q^{2\epsilon_4},q^{2\epsilon_3};q^{2
\sigma_4};\cos\theta)\rho^{(r)+}_k(q^{2\epsilon_4},q^{2\sigma_4};q^{2\epsilon_3},
q^{2\sigma_3};\cos\theta)$ and $a^-_r(q^{2\epsilon_4},q^{2\epsilon_3};
q^{2\sigma_4};\cos\theta)\rho^{(r)-}_k(q^{2\epsilon_4},q^{2\sigma_4};q^{2
\epsilon_3},q^{2\sigma_3};\cos\theta)$. Then application of [\cite{GR}:ex.5.23]
with $n\to4$, $\tfrac{a_1}{a_2}\to q^{4\sigma_3+2}$, $\tfrac{a_1}{a_3}\to 
q^{-2(\epsilon_2-\epsilon_1-\sigma_3\pm\sigma_4)}$, $\tfrac{a_1}{a_4}\to-\alpha 
\alpha'z^{-1}q^{2(\epsilon'_4+\sigma_3)+1}$, $\tfrac{a_1}{b_1}\to-q^{-2(
\epsilon_3-\sigma_3)+1}$, $\tfrac{a_1}{b_2}\to z^{-1}q^{2(\sigma_3\mp\sigma_4)}$,
$\tfrac{a_1}{b_3}\to-\alpha q^{2(\epsilon_1+\sigma_3)+1}$, $\tfrac{a_1}{b_4}
\to-\alpha'q^{2(\epsilon'_1+\sigma_3)+1}$ to simplify the coefficients yields the rhs of equation \eqref{E:hattau}.

For $e^{i\theta}\in\Gamma_{q^{-2(r+\epsilon_2-\epsilon_1)+1}}$, equation \eqref{E:varsig+-r} yields
\begin{multline}\label{E:varsigrel2}
u\bigl(q^{2\epsilon_2},q^{2\epsilon_1};e^{i\theta}\bigr)
\cdot\\\frac{(-1)^r\theta_{q^2}\bigl(q^{4\sigma_3+2}\bigr)}{\sqrt{\bigl(
q^{2(s'-r)},q^{4(\epsilon_1-\epsilon_2)-2(s'+r-1)};q^2\bigr)_{r-s'}}}\,\varsigma^{(-r)}_{
l+r}\bigl(q^{2\epsilon_2},\alpha q^{2\sigma_4};q^{2\epsilon_1},\alpha q^{2\sigma_3};q^{2(s'+\epsilon_2-\epsilon_1)-1}\bigr)\\
=q^{2(\epsilon_2-\epsilon_1-\sigma_3+\sigma_4)}q^{2r(\epsilon_1+\epsilon_2+1)}
q^{-4r(s'+\epsilon_2-\epsilon_1)+2(2r-s')}\theta_{q^2}\bigl(q^{4\sigma_4+2},
q^{4(\epsilon_2-\epsilon_1)+2s'}\bigr)\cdot\\
\sqrt{\bigl(q^2,q^{4(\epsilon_2-\epsilon_1+s')};q^2\bigr)_{r-s'}}\;\varsigma^{(r)}_l
\bigl(q^{2\epsilon_1},\alpha q^{2\sigma_3};q^{2\epsilon_2},\alpha q^{2\sigma_4};
q^{2(s'+\epsilon_2-\epsilon_1)-1}\bigr).
\end{multline}
Making use
of the relation \eqref{E:varsigrel2} with $l\to k$, $\epsilon_1\to\epsilon_4$,
$\epsilon_2\to\epsilon_3$, $q^{2\sigma_3}\leftrightarrow\alpha q^{2\sigma_4}$,
$e^{i\theta}\to q^{2(s'+\epsilon_2-\epsilon_1)-1}$, equation \eqref{E:tausum6} in Corollary \ref{C:tausum6} is rewritten by
\begin{multline}\label{E:tausum7}
\widehat{\tau}^{(r,k)}(q^{2\zeta},q^{2\epsilon_2})-\alpha\alpha'q^{2(\epsilon'_2
-\epsilon_2)}\widehat{\tau}'^{(r,k)}(q^{2\zeta},q^{2\epsilon'_2})
\Big\vert_{e^{i\theta}=q^{2(s'+\epsilon_2-
\epsilon_1)-1}}=\\
-\alpha q^{2\epsilon_4+1}\cdot q^{2(\epsilon_2-\epsilon_1+\sigma_3-\sigma_4)}
\iota(q^{2\sigma_3})\frac{\theta_{q^2}\bigl(\alpha\alpha'q^{2(\epsilon'_2-\epsilon_2)},
\alpha\alpha' z^{-1}q^{2(\epsilon'_1+\epsilon_2+1)},q^{4\sigma_4+2}\bigr)}{
\theta_{q^2}\bigl(q^{4\sigma_3+2}\bigr)}\cdot\\
\frac{\theta_{q^2}
\bigl(-q^{2(\epsilon_4+\sigma_4)+1},-q^{2(\epsilon_4-\sigma_4)+1},-\alpha 
\alpha'q^{2(\epsilon'_4+\sigma_4)+1},-\alpha\alpha'q^{2(\epsilon'_4-\sigma_4)
+1}\bigr)}{\theta_{q^2}\bigl(-\alpha q^{2(\epsilon_1+\sigma_3)+1},-\alpha q^{2(
\epsilon_1-\sigma_3)+1},-\alpha'q^{2(\epsilon'_1+\sigma_3)+1},-\alpha' 
q^{2(\epsilon'_1-\sigma_3)+1}\bigr)}\cdot\\
(\alpha z)^{s'}\frac{\bigl(zq^{2(\epsilon_1-\epsilon_2
+1)};q^2\bigr)_{-s'}}{\bigl(z^{-1}q^{2(\epsilon_1-\epsilon_2+1)};q^2\bigr)_{
-s'}}\frac{(q^2;q^2)_{\infty}}{\bigl(zq^{2(\epsilon_2-\epsilon_1)};q^2\bigr)_{
\infty}}\,\breve B\cdot\\
a_r\bigl(\alpha;q^{2\epsilon_4},q^{2\epsilon_3};q^{2\sigma_4};
q^{2(s'+\epsilon_3-\epsilon_4)-1}
\bigr)\varsigma^{(r)}_k\bigl(q^{2\epsilon_4},q^{2\sigma_4};q^{2\epsilon_3},
q^{2\sigma_3};q^{2(s'+\epsilon_3-\epsilon_4)-1}\bigr).
\end{multline}
Here the abbreviation $\breve B$ specified by \eqref{E:breveB} has been employed. Use of relation \eqref{E:varsigrel2} followed by
\eqref{E:varsigrho2} with $k\to l$,
$\epsilon_3\to\epsilon_2$, $\epsilon_4\to\epsilon_1$,
$q^{2\sigma_3}\to q^{2\sigma_2}$, $q^{2\sigma_4}\to q^{2\sigma_1}$ 
gives
\begin{multline*}
a_r\bigl(\alpha;q^{2\epsilon_1},q^{2\epsilon_2};q^{2\sigma_1};
q^{2(s'+\epsilon_2-\epsilon_1)-1}\bigr)
\varsigma^{(r)}_l\bigl(q^{2\epsilon_1},q^{2\sigma_1};q^{2\epsilon_2},
q^{2\sigma_2};q^{2(s'+\epsilon_2-\epsilon_1)-1}\bigr)\\
=-(-\alpha)^{s'-1}q^{2s'\sigma_2+2\sigma_1}\theta_{q^2}\bigl(q^{4\sigma_1+2}\bigr)
\bigl(q^2,q^{4(\epsilon_2-\epsilon_1+s')};q^2\bigr)_{\infty}\cdot\\
\bigl(q^{-2(\epsilon_2-\epsilon_1+\sigma_1-\sigma_2)+2},
q^{-2(\epsilon_2-\epsilon_1-\sigma_1-\sigma_2)+2};q^2\bigr)_{-s'}\cdot\\
a^+_{-r}\bigl(q^{2\epsilon_2},q^{2\epsilon_1};q^{2\sigma_2};\cos\theta
\bigr)\rho^{(-r)+}_{l+r}\bigl(q^{2\epsilon_2},q^{2\sigma_2};q^{2\epsilon_1},
q^{2\sigma_1};\cos\theta\bigr)\Big\vert_{e^{i\theta}=q^{-2(-s+\epsilon_1-\epsilon_2
)-1}}.
\end{multline*} 
Thus Corollary
\ref{C:tausum4} for $e^{i\theta}\in\Gamma_{q^{2(r+\epsilon_2-\epsilon_1)+1}}$
with the replacements $r\to-r$, $k\to k+r$, $\epsilon_1\leftrightarrow
\epsilon_2$, $\epsilon'_1\leftrightarrow\epsilon'_2$, $\epsilon_3\leftrightarrow 
\epsilon_4$, $\epsilon'_3\leftrightarrow\epsilon'_4$, $q^{2\sigma_3}\to 
q^{-2\sigma_4}$, $q^{2\sigma_4}\to q^{2\sigma_3}$ accounts for the evaluation of
the linear combination $\mathring{\widehat{\tau}}^{(r,k)}(q^{2\zeta},q^{2\epsilon_2})-\alpha\alpha'q^{2(
\epsilon'_2-\epsilon_2)}\mathring{\widehat{\tau}}'^{(r,k)}(q^{2\zeta},q^{2
\epsilon'_2})$.
With these substitutions, equation \eqref{E:tausum4}
implies
\begin{equation}\label{E:mhattau0}
\mathring{\widehat{\tau}}^{(r,k)}(q^{2\zeta},q^{2\epsilon_2})-\alpha\alpha'
q^{2(\epsilon'_2-\epsilon_2)}\mathring{\widehat{\tau}}'^{(r,k)}(q^{2\zeta},
q^{2\epsilon'_2})\Big\vert_{e^{i\theta}=q^{2(s'+\epsilon_2-\epsilon_1)-1}}=0.
\end{equation}

\subsection{Inverse Intertwiners}\label{S:invint}

For a pair of particular combinations of $r^{k,k+r^*}_{l,l+r^*}$ and $\mathring r^{k,k+r^*}_{l,l+r^*}$, the intertwiner giving rise to
the inverse action in the case $\alpha=-\alpha'$ can be specified in 
terms of simple parameter replacements. 
The linear combinations are given by the choices
\begin{multline}\label{E:Run}
\mathsf R^{k,k+r^*}_{l,l+r^*}={r_{\pm}}^{k,k+r^*}_{l,l+r^*}\equiv
{r_{\pm}}^{k,k+r^*}_{l,l+r^*}\bigl(z,\alpha;\beta,q^{2\zeta};q^{2\epsilon_1},
q^{2\epsilon_2};q^{2\epsilon_3},q^{2\epsilon_4};q^{2\sigma_3},q^{2\sigma_4}\bigr)=\\
\shoveleft{
r^{k,k+r^*}_{l,l+r^*}\bigl(z,\alpha;\beta,q^{2\zeta};q^{2\epsilon_1},q^{2\epsilon_2};
q^{2\epsilon_3},q^{2\epsilon_4};q^{2\sigma_3},q^{2\sigma_4}\bigr)}\\
\shoveleft{+\mathring r^{k,k+r^*}_{l,l+r^*}\bigl(z,\alpha;\beta,
q^{2\zeta};q^{2\epsilon_1},q^{2
\epsilon_2};q^{2\epsilon_3},q^{2\epsilon_4};q^{2\sigma_3},q^{2\sigma_4}\bigr)
\cdot}\\
\frac{q^{2(\epsilon_2-\epsilon_1)}\iota(q^{2\sigma_3})\iota(q^{2\sigma_4})
E_{[\pm]}}{2\theta_{q^2}\bigl(zq^{2(\epsilon_2-
\epsilon_1)},q^{4\sigma_3},q^{4\sigma_3+2}\bigr)\bigl(zq^{2(\epsilon_2-\epsilon_1
+1)},z^{-1}q^{2(\epsilon_2-\epsilon_1)};q^2\bigr)_{\infty}},
\end{multline}
where
\begin{multline*}
E_{[\pm]}=-zF\pm z\sqrt{F^2-4q^{-4(\sigma_3+\sigma_4)}H^{-1}_{\theta}\theta_{q^2}\bigl(zq^{2
(\epsilon_1-\epsilon_2+1)},z^{-1}q^{2(\epsilon_1-\epsilon_2)},
q^{4\sigma_3},q^{4\sigma_4}\bigr)^2}
\end{multline*}
and
\begin{multline}\label{E:F}
F\equiv F(z)
=\theta_{q^2}\bigl(zq^{2(\sigma_3+\sigma_4+1)},z^{-1}q^{2(\sigma_3+\sigma_4)}
\bigr)\cdot\\
\shoveleft{
\bigl[\gamma(q^{2\sigma_3})\delta(q^{-2\sigma_4})\theta_{q^2}\bigl(q^{2(
\epsilon_2-\epsilon_1-\sigma_3+\sigma_4)},q^{2(\epsilon_2-\epsilon_1-\sigma_3+
\sigma_4+1)}\bigr)}\\
\shoveright{+\gamma(q^{-2\sigma_3})\delta(q^{2\sigma_4})\theta_{q^2}\bigl(
q^{2(\epsilon_2-\epsilon_1+\sigma_3-\sigma_4)},q^{2(\epsilon_2-\epsilon_1+
\sigma_3-\sigma_4+1)}\bigr)\bigr]}\\
-\text{idem}(q^{2\sigma_4},q^{-2\sigma_4}).
\end{multline} 
For $z=\alpha=\varepsilon$, $\epsilon_1=\epsilon_4$, $\epsilon_2=\epsilon_3$, $q^{2\sigma_3}=\varepsilon q^{2\sigma_4}$ with $\varepsilon=1$ or 
$\varepsilon=-1$, the solution ${r_{\pm}}^{k,k+r^*}_{l,l+r^*}$ satisfies
an initial condition. At $z=\varepsilon$, $q^{2\sigma_4}=\varepsilon q^{2\sigma_3}$, the expression \eqref{E:Run} reduces to
\begin{equation}\label{E:rpmred}
\begin{split}
&{r_+}^{k,k+r^*}_{l,l+r^*}\bigl(\varepsilon,\alpha;\beta,
q^{2\zeta};q^{2\epsilon_1},q^{2\epsilon_2};
q^{2\epsilon_3},q^{2\epsilon_4};q^{2\sigma_3},\varepsilon q^{2\sigma_3}\bigr)=\\
&\Biggl\{r^{k,k+r^*}_{l,l+r^*}-\varepsilon q^{2(\epsilon_2-
\epsilon_1)}\iota(q^{2\sigma_3})\iota(q^{2\sigma_4})\gamma(q^{2\sigma_3})
\delta(\varepsilon q^{-2\sigma_3})\frac{\bigl(\varepsilon q^{2(\epsilon_1
-\epsilon_2)};q^2\bigr)_{\infty}}{\bigl(\varepsilon q^{2(\epsilon_2-\epsilon_1)};q^2\bigr)_{
\infty}}\cdot\\
&\qquad\qquad\qquad\qquad\qquad\qquad\qquad\qquad\qquad\qquad\qquad
\mathring r^{k,k+r^*}_{l,l+r^*}\biggr\}\bigg\vert_{z=
\varepsilon,q^{2\sigma_4}=\varepsilon q^{2\sigma_3}},\\
&{r_-}^{k,k+r^*}_{l,l+r^*}\bigl(\varepsilon,\alpha;\beta,q^{2\zeta};q^{2\epsilon_1},
q^{2\epsilon_2};q^{2\epsilon_3},q^{2\epsilon_4};q^{2\sigma_3},\varepsilon q^{2\sigma_3}\bigr)=\\
&\biggl\{r^{k,k+r^*}_{l,l+r^*}-\varepsilon q^{2(\epsilon_2-\epsilon_1)}\iota(q^{2\sigma_3})\iota(q^{2\sigma_4})
\gamma(q^{-2\sigma_3})\delta(\varepsilon q^{2\sigma_3})
\frac{\bigl(\varepsilon q^{2(\epsilon_1-\epsilon_2)};q^2\bigr)_{\infty}}{
\bigl(\varepsilon q^{2(\epsilon_2-\epsilon_1)};q^2\bigr)_{\infty}}\,
\cdot\\
&\qquad\qquad\qquad\qquad\qquad\qquad\qquad\qquad\qquad\qquad\qquad
\mathring r^{k,k+r^*}_{l,l+r^*}\biggr\}\bigg\vert_{z=\varepsilon,
q^{2\sigma_4}=\varepsilon q^{2\sigma_3}}.
\end{split}
\end{equation}
Application of \eqref{E:ex.2.16} with $x\to-\varepsilon\sqrt{\alpha\alpha'}q^{
\epsilon_4+\epsilon'_4+2\sigma_3+1}$, $\lambda\to\sqrt{\alpha\alpha'} q^{
\epsilon'_4-\epsilon_4}$, $\mu\to-\varepsilon \sqrt{\alpha\alpha'}q^{\epsilon_1+
\epsilon'_1-2\sigma_3+1}$, $\nu\to\varepsilon\alpha q^{\epsilon_1-\epsilon'_1}/\sqrt{\alpha\alpha'}$
gives
\begin{multline*}
{r_-}^{k,k+r^*}_{l,l+r^*}\bigl(\varepsilon,\alpha;\beta,q^{2\zeta};q^{2\epsilon_1},
q^{2\epsilon_2};q^{2\epsilon_3},q^{2\epsilon_4};q^{2\sigma_3},
\varepsilon q^{2\sigma_3}\bigr)=\\
\frac{q^{2(\epsilon_4+\sigma_3)+1}
}{\theta_{q^2}\bigl(-\alpha q^{2(\epsilon_1-\sigma_3)+1},-\alpha'q^{2(
\epsilon'_1-\sigma_3)+1},-\varepsilon q^{2(\epsilon_4+\sigma_3)+1},-
\varepsilon\alpha\alpha'q^{2(\epsilon'_4+\sigma_3)+1}\bigr)}\cdot\\
\shoveleft{
\biggl\{r^{k,k+r^*}_{l,l+r^*}\cdot\varepsilon\theta_{q^2}\bigl(\varepsilon\alpha q^{2(\epsilon_1-\epsilon_4)},\varepsilon\alpha'q^{2(\epsilon'_1-\epsilon_4)},
\alpha\alpha'q^{2(\epsilon_1+\epsilon'_1+1)},q^{4\sigma_3+2}\bigr)}\\
+\frac{q^{2(\epsilon_2-\epsilon_1)}\iota(q^{2\sigma_3})\,\bar r^{k,k+r^*}_{l,
l+r^*}}{\theta_{q^2}\bigl(-\varepsilon
q^{2(\epsilon_4+\sigma_3)+1},-q^{2(\epsilon_3+\sigma_3)+1},q^{4\sigma_3+2}
\bigr)\bigl(\varepsilon q^{2(\epsilon_2-\epsilon_1)};q^2\bigr)_{\infty}}
\biggr\}\bigg\vert_{z=\varepsilon,q^{2\sigma_4}=\varepsilon q^{2\sigma_3}}
\end{multline*}
with $\bar r^{k,k+r^*}_{l,l+r^*}$ defined by \eqref{E:rrcomb}. Making use of
Proposition \ref{P:barrinit} and \ref{P:rsp} in section \ref{S:init},
specialization to $\alpha=\varepsilon$, $\epsilon_1=\epsilon_4$, $\epsilon_2=\epsilon_3$ or to $\alpha'=\varepsilon$, $\epsilon'_1=\epsilon_4$, $\epsilon'_2=\epsilon_3$ and comparison of the prefactors of $\mathring r^{
k,k+r^*}_{l,l+r^*}$ on both rhs of \eqref{E:rpmred}
yields the initial conditions
\begin{multline}\label{E:Runinit}
{r_{\pm}}^{k,k+r^*}_{l,l+r^*}\bigl(\varepsilon,\varepsilon;\varepsilon
\alpha',q^{2\zeta};q^{2\epsilon_1},q^{2\epsilon_2};q^{2\epsilon_2},q^{2\epsilon_1};
q^{2\sigma_3},\varepsilon q^{2\sigma_3}\bigr)=\\
\delta_{l,k}\cdot\varepsilon^{k}
q^{2\epsilon_2+1}\iota(q^{2\sigma_3})\theta_{q^2}\bigl(\alpha'q^{2(\epsilon'_1+
\epsilon_2+1)},\varepsilon\alpha'q^{2(\epsilon'_1-\epsilon_1)}\bigr)\bigl(
\varepsilon q^{2(\epsilon_1-\epsilon_2+1)},q^2;q^2\bigr)_{\infty}
\cdot\\
\Bigl\{q^{-2\sigma_3}\gamma(q^{2\sigma_3})\theta_{q^2}\bigl(\varepsilon q^{2(
\epsilon_2-\epsilon_1-2\sigma_3+1)}\bigr)-q^{2\sigma_3}\gamma(q^{-2
\sigma_3})\theta_{q^2}\bigl(\varepsilon q^{2(\epsilon_2-
\epsilon_1+2\sigma_3+1)}\bigr)\Bigr\}\Big\vert_{\substack{\alpha=\varepsilon\\
\epsilon_1=\epsilon_4,\epsilon_2=\epsilon_3}},\\
\shoveleft{
{r_{\pm}}^{k,k+r^*}_{l,l+r^*}\bigl(\varepsilon,\alpha;\varepsilon\alpha,q^{2
\zeta};q^{2\epsilon_1},q^{2\epsilon_2};q^{2\epsilon'_2},q^{2\epsilon'_1};q^{2
\sigma_3},\varepsilon q^{2\sigma_3}\bigr)=0.\qquad\qquad\qquad\quad}
\end{multline}
The sums attributed to the choice \eqref{E:Run} by Definition \ref{D:taudef1}
will be denoted by
\begin{equation*}
\begin{split}
\tau^{[\pm](r,k)\pm}\bigl(q^{2\zeta},q^{2\epsilon_2}\bigr)&\equiv\tau^{[\pm](r,k)\pm}
\bigl(z,\alpha;\beta,q^{2\zeta};q^{2\epsilon_1},q^{2\epsilon_2};q^{2\epsilon_3},
q^{2\epsilon_4};q^{2\sigma_3},q^{2\sigma_4};\cos\theta\bigr),\\
\tau^{[\pm](r,k)}\bigl(q^{2\zeta},q^{2\epsilon_2}\bigr)&\equiv\tau^{[\pm](r,k)}
\bigl(z,x,
\alpha;\beta,q^{2\zeta};q^{2\epsilon_1},q^{2\epsilon_2};q^{2\epsilon_3},q^{2\epsilon_4}
;q^{2\sigma_3},q^{2\sigma_4};e^{i\theta}\bigr).
\end{split}
\end{equation*}
A second set of sums $\tau'^{[\pm](r,k)\pm}(q^{2\zeta},q^{2\epsilon_2})$ and
$\tau'^{[\pm](r,k)}(q^{2\zeta},q^{2\epsilon_2})$ is obtained by the
substitutions $\alpha\leftrightarrow\alpha'$, $x\to x'$, $\epsilon_1\to\epsilon'_1$, $\epsilon_2\to\epsilon'_2$.
Rewriting the last line of equation \eqref{E:tausum4} in Corollary \ref{C:tausum4} by means of the relation 
\eqref{E:rhorel} with $l\to k$, $\epsilon_1\to\epsilon_4$, $\epsilon_2\to
\epsilon_3$, $q^{2\sigma_1}\to q^{2\sigma_4}$, $q^{2\sigma_2}\to q^{2\sigma_3}$
and making use of equation \eqref{E:hattau} gives rise to the evaluation
\begin{multline}\label{E:tauunit}
{\tau}^{[\pm](r,k)\pm}(q^{2\zeta},q^{2\epsilon_2})-\alpha\alpha'q^{2(\epsilon'_2-
\epsilon_2)}\tau'^{[\pm](r,k)\pm}(q^{2\zeta},q^{2\epsilon'_2})=\\
-q\iota(q^{2\sigma_3})\theta_{q^2}\bigl(\alpha\alpha'q^{2(\epsilon'_1-\epsilon_1)},
\alpha\alpha'z^{-1}q^{2(\epsilon'_1+\epsilon_2+1)}\bigr)\frac{\bigl(q^2,zq^{2(\epsilon_1
-\epsilon_2+1)};q^2\bigr)_{\infty}}{\bigl(zqe^{i\theta},zqe^{-i\theta}
;q^2\bigr)_{\infty}}\cdot\\
\Biggl\{\frac{\bigl(q^{\pm2\sigma_3-2\sigma_4+1}e^{i\theta},q^{\pm2\sigma_3-2\sigma_4
+1}e^{-i\theta},q^{2(\epsilon_2-\epsilon_1\mp\sigma_3+\sigma_4+1)};q^2\bigr)_{
\infty}}{\bigl(q^{2(\epsilon_2-\epsilon_1\pm\sigma_3-\sigma_4+1)};q^2\bigr)_{
\infty}}\cdot\\
\shoveleft{
\biggl[\gamma(q^{\pm2\sigma_3})q^{2(\epsilon_4+\sigma_4)}\theta_{q^2}\bigl(q^{2(
\epsilon_2-\epsilon_1\mp\sigma_3-\sigma_4)}\bigr)}\mp E_{[\pm]}
q^{2(\epsilon_2-\epsilon_1+\epsilon_3\mp\sigma_3)}\cdot\\
\shoveright{\frac{\theta_{q^2}\bigl(q^{2(\epsilon_2-
\epsilon_1\pm\sigma_3+\sigma_4+1)}\bigr)}{2\delta(q^{2\sigma_4})\theta_{q^2}\bigl(
zq^{2(\epsilon_2-\epsilon_1)},z^{-1}q^{2(\epsilon_2-\epsilon_1)},
q^{\pm4\sigma_3+2},q^{4\sigma_4+2}\bigr)}\biggr]\cdot}\\
\shoveright{a^+_r\bigl(q^{2\epsilon_4},
q^{2\epsilon_3};q^{2\sigma_4};\cos\theta\bigr)\rho^{(r)+}_k\bigl(q^{2\epsilon_4},
q^{2\sigma_4};q^{2\epsilon_3},q^{2\sigma_3};\cos\theta\bigr)}\\
-\frac{\bigl(q^{2(\sigma_4\pm\sigma_3)+1}e^{i\theta},q^{2(\sigma_4\pm\sigma_3)+1}
e^{-i\theta},q^{2(\epsilon_2-\epsilon_1\mp\sigma_3-\sigma_4+1)};q^2\bigr)_{
\infty}}{\bigl(q^{2(\epsilon_2-\epsilon_1\pm\sigma_3+\sigma_4+1)};q^2\bigr)_{
\infty}}\cdot\\
\shoveleft{\biggl[\gamma(q^{\pm2\sigma_3})q^{2(\epsilon_4-\sigma_4)}\theta_{q^2}
\bigl(q^{2(\epsilon_2-\epsilon_1\mp\sigma_3+\sigma_4)}\bigr)
\pm E_{[\pm]}q^{2(\epsilon_2-\epsilon_1+\epsilon_3\mp\sigma_3)}\cdot}\\
\shoveright{\frac{
\theta_{q^2}\bigl(q^{2(\epsilon_2-\epsilon_1\pm\sigma_3-\sigma_4+1)}\bigr)}{2\delta(
q^{-2\sigma_4})\theta_{q^2}\bigl(zq^{2(\epsilon_2-\epsilon_1)},z^{-1}q^{2(
\epsilon_2-\epsilon_1)},q^{\pm4\sigma_3+2},q^{4\sigma_4}\bigr)}
\biggr]\cdot}\\
a^-_r\bigl(q^{2\epsilon_4},q^{2\epsilon_3};q^{2\sigma_4};\cos\theta\bigr)
\rho^{(r)-}_k\bigl(q^{2\epsilon_4},q^{2\sigma_4};q^{2\epsilon_3},q^{2\sigma_3};
\cos\theta\bigr)\Biggr\}.
\end{multline} 
Equation \eqref{E:tauunit} and the decomposition
\eqref{E:hattaudec} with the specifications
$\boldsymbol{\tau}^{(r,k)}\to\tau^{[\pm](r,k)}(q^{2\zeta},q^{2\epsilon_2})$
and $\boldsymbol{\tau}^{(r,k)\pm}\to\tau^{[\pm](r,k)\pm}(q^{2\zeta},q^{2
\epsilon_2})$ lead to the corresponding result for $\tau^{[\pm](r,k)}(
q^{2\zeta},q^{2\epsilon_2})$, provided that the
condition $\max\bigl(\vert zqe^{i\theta}\vert,\vert zqe^{-i\theta}\vert\bigr)<1$ is satisfied.
Making use of the equations \eqref{E:tauunit} in the decomposition
\eqref{E:hattaudec}, the relation
\eqref{E:ex.2.16} with $x\to -i\alpha\alpha'q^{\epsilon'_1+\epsilon_2+
\epsilon'_3+\sigma_4+\frac{3}{2}}$, $\lambda\to iq^{\epsilon'_1+\epsilon_2
-\epsilon'_3-2\sigma_3+\sigma_4+\frac{1}{2}}$, $\mu\to iq^{\epsilon_2-\epsilon_1
+\epsilon_3-\sigma_4+\frac{1}{2}}$, $\nu\to iq^{\epsilon_4+2\sigma_3+
\sigma_4+\frac{1}{2}}$ and with $x\to \sqrt{-\alpha}q^{\epsilon_2
-2\sigma_3+\sigma_4+\frac{3}{2}}$, $\lambda\to q^{-\epsilon_2+\sigma_4-
\frac{1}{2}}/\sqrt{-\alpha}$, $\mu\to\sqrt{-\alpha}q^{\epsilon_2+2\sigma_3+
\sigma_4+\frac{3}{2}}$, $\nu\to q^{\epsilon_2-2\epsilon_1+\sigma_4+\frac{1}{2}}
/\sqrt{-\alpha}$
allows to simplify the prefactor of $\rho^{(r)+}_k\bigl(q^{2\epsilon_4},q^{2
\sigma_4};q^{2\epsilon_3},q^{2\sigma_3};\alpha\alpha'q^{2(\tilde s+\epsilon_3+
\epsilon'_4)+1})$. The replacement $q^{2\sigma_4}\to q^{-2\sigma_4}$ yields the corresponding choices for the prefactors of $\rho^{(r)-}_k
(q^{2\epsilon_4},q^{2\sigma_4};q^{2\epsilon_3},q^{2\sigma_3};\alpha\alpha'q^{2
(\tilde s+\epsilon_3+\epsilon'_4)+1})$. Writing
\begin{multline*}
\theta_{q^2}\bigl(-\alpha\alpha'q^{2(\epsilon'_4\pm\sigma_4)+1}\bigr)=\\
(\alpha\alpha'
)^{\tilde s}q^{\tilde s^2+2\tilde s(\epsilon'_2\pm\sigma_4)}\theta_{q^2}\bigl(
-q^{2(\epsilon_3\mp\sigma_4+1)}e^{-i\theta}\bigr)\Big\vert_{e^{i\theta}=\alpha\alpha'
q^{2(\tilde s+\epsilon_3+\epsilon'_4)+1}}
\end{multline*} 
and using the relation
\eqref{E:varsigma} with $l\to k$, $\epsilon_1\to\epsilon_4$,
$\epsilon_2\to\epsilon_3$, $q^{2\sigma_3}\leftrightarrow \alpha q^{2\sigma_4}$
and the property \eqref{E:wprop1} leads to
\begin{multline}\label{E:tauunit2}
\tau^{[\pm](r,k)}(q^{2\zeta},q^{2\epsilon_2})-\alpha\alpha'q^{2(\epsilon'_2-
\epsilon_2)}\tau'^{[\pm](r,k)}(q^{2\zeta},q^{2\epsilon'_2})\Big\vert_{e^{i\theta}
=\alpha\alpha'q^{2(\tilde s+\epsilon'_1+\epsilon_2)+1}}=\\
-q^{2\epsilon_3+1}\iota(q^{2\sigma_3}) 
\theta_{q^2}\bigl(\alpha\alpha'q^{2(\epsilon'_2-\epsilon_2)},
\alpha\alpha'z^{-1}q^{2(\epsilon'_1+\epsilon_2+1)}\bigr)\frac{w_0\bigl(x\alpha;
q^{2\epsilon_4},q^{2\epsilon_3};e^{i\theta_0}\bigr)}{w_0\bigl(x;q^{2\epsilon_1},
q^{2\epsilon_2};e^{i\theta_0}\bigr)}\cdot\\
\shoveleft{
\frac{\bigl(q^2,zq^{2(\epsilon_1-\epsilon_2+1)};q^2\bigr)_{\infty}}{\bigl(zqe^{
i\theta},zqe^{-i\theta};q^2\bigr)_{\infty}}\biggl\{q^{2\sigma_3+2\tilde s(
\epsilon_2-\epsilon_1)}\theta_{q^2}\bigl(q^{4\sigma_3+2},q^{2(\epsilon_2-
\epsilon_1)+1}e^{i\theta}\bigr)}\\
\shoveright{
-E_{[\pm]}q^{-2\sigma_4-2\tilde s(\epsilon_2-\epsilon_1)}\frac{\theta_{q^2}
\bigl(q^{2(\epsilon_2-\epsilon_1)+1}e^{-i\theta}\bigr)}{2\theta_{q^2}\bigl(q^{4
\sigma_4+2},zq^{2(\epsilon_2-\epsilon_1)}z^{-1}q^{2(\epsilon_2-\epsilon_1)}\bigr)}
\biggr\}}\\
a_r\bigl(x\alpha;q^{2\epsilon_4},q^{2\epsilon_3};e^{i\theta}\bigr)\,
\varsigma^{(r)}_k\bigl(q^{2\epsilon_4},q^{2\sigma_4};q^{2\epsilon_3},
q^{2\sigma_3};e^{i\theta}\bigr)\Big\vert_{e^{i\theta}=\alpha\alpha'q^{2(\tilde s
+\epsilon_3+\epsilon'_4)+1}}
\end{multline}
in the case $\vert zqe^{\pm i\theta}\vert<1$. Alternatively, the expression
\eqref{E:tauunit2} is obtained from Corollary \ref{C:tausum5}. 
Transformation of the last line of \eqref{E:tausum5} by means of \eqref{E:varsig+-r} with $l\to k$, $\epsilon_1\to\epsilon_4$, $\epsilon_2\to 
\epsilon_3$, $q^{2\sigma_1}\to q^{2\sigma_4}$, $q^{2\sigma_2}\to q^{2\sigma_3}$
and evaluation of $\mathring{\widehat{\tau}}^{(r,k)}(q^{2\zeta},q^{2\epsilon_2})
-\alpha\alpha'q^{2(\epsilon'_2-\epsilon_2)}\mathring{\widehat{\tau}}'^{(r,k)}(
q^{2\zeta},q^{2\epsilon'_2})$ employing the relation \eqref{E:varsig+-r} followed by
application of Corollary \ref{C:tausum5} with $r\to-r$, $k\to k+r$, $\epsilon_1\leftrightarrow\epsilon_2$, $\epsilon'_1\leftrightarrow\epsilon'_2$,
$\epsilon_3\leftrightarrow\epsilon_4$, $q^{2\sigma_3}\leftrightarrow q^{2\sigma_4}$
yields \eqref{E:tauunit2} in the case $\vert zqe^{i\theta}\vert<1$.
\vskip 0.5cm

Use of relation \eqref{E:ex.2.16} with $x\to q^{2(\epsilon_2-\epsilon_1)+1}$, $\lambda\to zq$, $\mu\to q^{2(\sigma_3+
\sigma_4)+1}$, $\nu\to q^{2(\sigma_3-\sigma_4)+1}$ in the expression \eqref{E:F}
for $F$ to eliminate the factor $\theta_{q^2}\bigl(zq^{2(\sigma_3+\sigma_4+1)},
z^{-1}q^{2(\sigma_3+\sigma_4)}\bigr)$ leads to
\begin{multline}\label{E:F2}
F=\frac{\theta_{q^2}\bigl(zq^{2(\epsilon_2-\epsilon_1+1)},z^{-1}q^{2(\epsilon_2-
\epsilon_1)},q^{4\sigma_3+2},q^{4\sigma_4}\bigr)}{\theta_{q^2}\bigl(q^{2(
\epsilon_2-\epsilon_1+\sigma_3-\sigma_4+1)},q^{2(\epsilon_2-\epsilon_1-\sigma_3
+\sigma_4)}\bigr)}\cdot\\
\shoveleft{
\Bigl\{\gamma(q^{2\sigma_3})\delta(q^{-2\sigma_4})\theta_{q^2}\bigl(q^{2(
\epsilon_2-\epsilon_1-\sigma_3+\sigma_4)},q^{2(\epsilon_2-\epsilon_1-\sigma_3
+\sigma_4+1)}\bigr)}\\
\shoveright{+\gamma(q^{-2\sigma_3})\delta(q^{2\sigma_4})\theta_{q^2}\bigl(q^{2(
\epsilon_2-\epsilon_1+\sigma_3-\sigma_4)},q^{2(\epsilon_2-\epsilon_1+\sigma_3
-\sigma_4+1)}\bigr)\Bigr\}}\\
+\breve B\breve C\frac{\theta_{q^2}\bigl(zq^{2(\sigma_3-\sigma_4+1)},z^{-1}q^{2(
\sigma_3-\sigma_4)}\bigr)}{\theta_{q^2}\bigl(q^{2(\epsilon_2-\epsilon_1+
\sigma_3-\sigma_4+1)},q^{2(\epsilon_2-\epsilon_1-\sigma_3+\sigma_4+1)}\bigr)}
\end{multline}
with $\breve B$ given by equation \eqref{E:breveB} in subsection \ref{S:tauev} and
\begin{multline}
\label{E:breveC}
\breve C=q^{-2(\epsilon_2-\epsilon_1-\sigma_4)}\cdot\\
\Bigl\{\gamma(q^{2\sigma_3})\theta_{q^2}\bigl(q^{2(\epsilon_2-\epsilon_1-\sigma_3
+\sigma_4+1)},q^{2(\epsilon_2-\epsilon_1-\sigma_3-\sigma_4)}\bigr)-\text{idem}
(\sigma_3,-\sigma_3)\Bigr\}.
\end{multline}

\begin{definition}\label{D:unsigdef}
The vector-valued functions 
\begin{equation}\label{E:unsigdef}
\boldsymbol{\sigma}^{[\pm](r,m)}\equiv\left(\begin{matrix}\sigma^{[\pm](r,m)+}\\
\sigma^{[\pm](r,m)-}\\\sigma^{[\pm](r,m)}\end{matrix}\right)
\end{equation}
are defined by
\begin{multline}\label{E:sigunit}
\boldsymbol{\sigma}^{[\pm](r,m)}\equiv\boldsymbol{\sigma}^{[\pm](r,m)}\bigl(z,
\alpha,\alpha';q^{2\zeta};q^{2\epsilon_1},q^{2\epsilon_2};q^{2\epsilon_3},
q^{2\epsilon_4};q^{2\sigma_3},q^{2\sigma_4};e^{i\theta}\bigr)=\\
\shoveleft{
\sum_{k=-\infty}^{\infty}{r_{\mp}}^{m,m+r^*}_{k,k+r^*}\bigl(z^{-1},\alpha;\beta,
q^{2\zeta};q^{2\epsilon_4},q^{2\epsilon_3};q^{2\epsilon_2},q^{2\epsilon_1};
\alpha q^{2\sigma_4},\alpha q^{2\sigma_3}\bigr)\cdot}\\
\Bigl[\boldsymbol{\tau}^{[\pm](r,k)}\bigl(q^{2\zeta},q^{2\epsilon_2}\bigr)-
\alpha\alpha'q^{2(\epsilon'_2-\epsilon_2)}\boldsymbol{\tau}'^{[\pm](r,k)}
\bigl(q^{2\zeta},q^{2\epsilon'_2}\bigr)\Bigr],
\end{multline}
where
\begin{equation}\label{E:tauunvec}
\boldsymbol{\tau}^{[\pm](r,k)}\bigl(q^{2\zeta},q^{2\epsilon_2}\bigr)=\left(
\begin{matrix}\tau^{[\pm](r,k)+}\bigl(q^{2\zeta},q^{2\epsilon_2}\bigr)\\
\tau^{[\pm](r,k)-}\bigl(q^{2\zeta},q^{2\epsilon_2}\bigr)\\\tau^{[\pm](r,k)}
\bigl(q^{2\zeta},q^{2\epsilon_2}\bigr)\end{matrix}\right)
\end{equation}
and $\boldsymbol{\tau}'^{[\pm](r,k)}(q^{2\zeta},q^{2\epsilon'_2})$ is given 
by the rhs of \eqref{E:tauunvec} with the replacements $\alpha\leftrightarrow\alpha'$, $x\to x'$, ${\epsilon_1}\to 
{\epsilon'_1}$, ${\epsilon_2}\to {\epsilon'_2}$.

The lowest component of $\boldsymbol{\sigma}^{[\pm](r,m)}$ is considered
for $e^{i\theta}\in\tilde{\Gamma}\cup\Gamma_{q^{-2(r+\epsilon_2-\epsilon_1)
+1}}$.
\end{definition}
 
Both sets $\tilde{\Gamma}$ and $\Gamma_a$ are introduced in subsection
\ref{S:tauev} by \eqref{E:Gammapm} and \eqref{E:tilGam}.
Besides $\boldsymbol{\sigma}^{[\pm](r,m)}$, the following result involves
the functions
\begin{equation*}
\boldsymbol{\sigma}'^{[\pm](r,m)}=\boldsymbol{\sigma}^{[\pm](r,m)}\bigl(
z,\alpha',\alpha;q^{2\zeta};q^{2\epsilon_1},q^{2\epsilon_2};q^{2\epsilon'_3},
q^{2\epsilon'_4};\alpha\alpha'q^{2\sigma_3},\alpha\alpha'q^{2\sigma_4};e^{i
\theta}\bigr).
\end{equation*}

The components ${\sigma}^{[\pm](r,m)+}$ and ${\sigma}'^{
[\pm](r,m)+}$ are well-defined and absolutely convergent if
\begin{equation}\label{E:tausum7cond}
\begin{split}
&q^{2(\epsilon_2-\epsilon_1+\sigma_3\pm\sigma_4)}\neq q^{2t_1},\quad q^{2(\epsilon_2-
\epsilon_1-\sigma_3\pm\sigma_4)}\neq q^{2t_2},\;t_1,t_2\in\mathbb Z,\\
&\max\bigl(\vert zqe^{i\theta}\vert,\vert z^{-1}qe^{i\theta}\vert\bigr)<1,\\
&\max\bigl(\vert zqe^{-i\theta}\vert,\vert z^{-1}qe^{-i\theta}\vert\bigr)
<1\quad\text{if}\;e^{i\theta}\notin\Gamma^+
\end{split}
\end{equation}
with $\Gamma^{\pm}$ defined by \eqref{E:Gammapm}. 
For $\sigma^{[\pm](r,m)-}$ and $\sigma'^{[\pm](r,m)-}$, 
the replacement $\Gamma^+\to\Gamma^-$ gives
rise to the corresponding conditions.
The lowest component $\sigma^{[\pm](r,m)}$ or $\sigma'^{[\pm](r,m)}$ is
well-defined and absolutely convergent if
\begin{equation}\label{E:tausum8cond}
\begin{split}
&q^{2(\epsilon_2-\epsilon_1+\sigma_3\pm\sigma_4)}\neq q^{2t_1},\quad q^{2(
\epsilon_2-\epsilon_1-\sigma_3\pm\sigma_4)}\neq q^{2t_2},\;t_1,t_2\in
\mathbb Z,\\
&\max\bigl(\vert zqe^{i\theta}\vert,\vert z^{-1}qe^{i\theta}\vert\bigr)<1.
\end{split}
\end{equation}
All sign choices in \eqref{E:tausum7cond} and \eqref{E:tausum8cond} are 
independent.

\begin{proposition}\label{P:unitrel}

The vector-valued sums $\boldsymbol{\sigma}^{[\pm](r,m)}$ and 
$\boldsymbol{\sigma}'^{[\pm](r,m)}$ with $\alpha=-\alpha'$ and the property \eqref{E:tausum7cond} or \eqref{E:tausum8cond} are related by
\begin{multline}\label{E:unitrel}
\boldsymbol{\sigma}^{[\pm](r,m)}-\boldsymbol{\sigma}'^{[\pm](r,m)}=
c\tilde B\tilde Cq^{4(\epsilon_2-\epsilon_1)}
\frac{(q^2,q^2;q^2)_{\infty}}{\bigl(zq^{2(
\epsilon_2-\epsilon_1)},z^{-1}q^{2(\epsilon_2-\epsilon_1)};q^2\bigr)_{\infty}}
\left(\begin{matrix}a^+_r\rho^{(r)+}_m\\a^-_r\rho^{(r)-}_m\\
a_r\varsigma^{(r)}_m\end{matrix}\right)
\end{multline}
with $\tilde B=\iota(q^{2\sigma_3})\breve B$ and $\tilde C=
\iota(q^{2\sigma_4})\breve B$.

\end{proposition}

\emph{Proof}: For the upper two components,
insertion of the expression \eqref{E:tauunit} for the bracket on the
rhs of \eqref{E:sigunit} allows to perform the $k$-summation employing
the evaluation \eqref{E:tauunit} with the replacements $k\to m$, $z\to z^{-1}$,
$\epsilon_1\leftrightarrow\epsilon_4$, $\epsilon_2\leftrightarrow\epsilon_3$,
$\epsilon'_1\leftrightarrow\epsilon'_4$, $\epsilon'_2\leftrightarrow\epsilon'_3$,
$q^{2\sigma_3}\leftrightarrow\alpha q^{2\sigma_4}$. The contributions to the prefactor of $a^{\mp}_r\rho^{(r)\mp}_m$ cancel leaving
\begin{multline}\label{E:sigunst1}
\sigma^{[\pm](r,m)\pm}-\sigma'^{[\pm](r,m)\pm}
=q^{2(\epsilon_2+\epsilon_4+1)}\iota(q^{2\sigma_3})\iota(q^{2\sigma_4})(q^2,q^2;
q^2)_{\infty}\cdot\\
\frac{\theta_{q^2}\bigl(\alpha\alpha'q^{2(\epsilon'_2-\epsilon_2)},\alpha\alpha' 
z^{-1}q^{2(\epsilon'_1+\epsilon_2+1)},\alpha\alpha'q^{2(\epsilon'_3-\epsilon_3)},
\alpha\alpha'zq^{2(\epsilon'_3+\epsilon_4+1)}\bigr)}{\theta_{q^2}\bigl(zqe^{
i\theta},zqe^{-i\theta}\bigr)}\cdot\\
\shoveleft{\biggl[\frac{zF\,q^{2(\epsilon_2-\epsilon_1)}
}{\theta_{q^2}\bigl(q^{4\sigma_3+2},q^{4\sigma_4+2)}\bigr)\bigl(
zq^{2(\epsilon_2-\epsilon_1)},z^{-1}q^{2(\epsilon_2-\epsilon_1)};q^2\bigr)_{
\infty}}\cdot}\\
\shoveleft{\Bigl\{\theta_{q^2}\bigl(q^{2(\sigma_3-\sigma_4)+1}e^{i\theta},q^{2(
\sigma_3-\sigma_4)+1}e^{-i\theta},q^{2(\epsilon_2-\epsilon_1-\sigma_3-\sigma_4)},
q^{2(\epsilon_2-\epsilon_1+\sigma_3+\sigma_4+1)}\bigr)\qquad\qquad}\\
\shoveright{
-q^{-4\sigma_4}\theta_{q^2}\bigl(q^{2(\sigma_3+\sigma_4)+1}e^{i\theta},q^{2(
\sigma_3+\sigma_4)+1}e^{-i\theta},q^{2(\epsilon_2-\epsilon_1-\sigma_3+\sigma_4)},
q^{2(\epsilon_2-\epsilon_1+\sigma_3-\sigma_4+1)}\bigr)\Bigr\}}\\ 
\shoveright{
-\bigl(zq^{2(\epsilon_1-\epsilon_2+1)},z^{-1}q^{2(\epsilon_1-\epsilon_2+1)};
q^2\bigr)_{\infty}\bigl(A^+-A^-\bigr)\biggr]\cdot}\\
a^{\pm}_r\bigl(q^{2\epsilon_1},q^{2\epsilon_2};q^{2\sigma_1};\cos\theta\bigr)
\rho^{(r)\pm}_m\bigl(q^{2\epsilon_1},q^{2\sigma_1};q^{2\epsilon_2},q^{2\sigma_2};
\cos\theta\bigr),
\end{multline}
where
\begin{multline}\label{E:A+}
A^+\equiv A^+(q^{2\sigma_3})=\frac{\theta_{q^2}\bigl(-q^{2(\epsilon_3+
\sigma_3)+1},-\alpha\alpha'q^{2(\epsilon'_3+\sigma_3)+1}\bigr)}{\theta_{q^2}\bigl(
-\alpha q^{2(\epsilon_1+\sigma_3)+1},-\alpha'q^{2(\epsilon'_1+\sigma_3)+1}
\bigr)}\cdot\\
\Bigl\{\delta(q^{2\sigma_4})
\theta_{q^2}\bigl(q^{2(\sigma_3-\sigma_4)+1}e^{i\theta},q^{2(\sigma_3-\sigma_4
)+1}e^{-i\theta},q^{2(\epsilon_2-\epsilon_1-\sigma_3-\sigma_4)},q^{2(\epsilon_2-
\epsilon_1-\sigma_3-\sigma_4+1)}\bigr)\\
-\text{idem}(\sigma_4,-\sigma_4)\Bigr\}
\end{multline}
and $A^-\equiv A^-(q^{2\sigma_3})=A^+(q^{-2\sigma_3})$. Evaluation of the first braces in \eqref{E:sigunst1} by means of \eqref{E:ex.2.16} with
$x\to q^{2(\sigma_3-\sigma_4)+1}$, $\lambda\to e^{i\theta}$, $\mu\to q^{2(
\epsilon_2-\epsilon_1)+1}$, $\nu\to q^{2(\sigma_3+\sigma_4)+1}$ yields
\begin{equation}\label{E:sigunst2}
\theta_{q^2}\bigl(q^{2(\epsilon_2-\epsilon_1)+1}e^{i\theta},q^{2(\epsilon_2-
\epsilon_1)+1}e^{-i\theta},q^{4\sigma_3+2},q^{4\sigma_4+2}\bigr).
\end{equation}
Relation \eqref{E:ex.2.16} with $x\to q^{2(\sigma_3-\sigma_4)+1}$,
$\lambda\to e^{i\theta}$, $\mu\to q^{2(\epsilon_2-\epsilon_1)+1}$, $\nu\to zq$ reads
\begin{multline}\label{E:ex2}
\theta_{q^2}\bigl(q^{2(\sigma_3-\sigma_4)+1}e^{i\theta},q^{2(\sigma_3-\sigma_4)
+1}e^{-i\theta},zq^{2(\epsilon_2-\epsilon_1+1)},z^{-1}q^{2(\epsilon_2-\epsilon_1)}
\bigr)=\\
\theta_{q^2}\bigl(q^{2(\epsilon_2-\epsilon_1)+1} e^{i\theta},q^{2(\epsilon_2-
\epsilon_1)+1}e^{-i\theta},zq^{2(\sigma_3-\sigma_4+1)},z^{-1}q^{2(\sigma_3-
\sigma_4)}\bigr)\\
-z^{-1}q^{2(\epsilon_2-\epsilon_1)}\theta_{q^2}\bigl(zqe^{i\theta},zqe^{-i\theta},
q^{2(\epsilon_2-\epsilon_1+\sigma_3-\sigma_4+1)},q^{2(\epsilon_2-\epsilon_1-\sigma_3
+\sigma_4+1)}\bigr).
\end{multline}
Equation \eqref{E:ex2} and the same relation with $\sigma_4\to-\sigma_4$
allow to eliminate the factors $\theta_{q^2}\bigl(
q^{2(\sigma_3-\sigma_4)+1}e^{i\theta},q^{2(\sigma_3-\sigma_4)+1}e^{-i\theta}\bigr)$
and $\theta_{q^2}\bigl(q^{2(\sigma_3+\sigma_4)+1}e^{i\theta},q^{2(\sigma_3
+\sigma_4)+1}e^{-i\theta}\bigr)$ from the rhs of \eqref{E:A+} to obtain
\begin{multline*}
A^+=\frac{\theta_{q^2}\bigl(q^{2(\epsilon_2-\epsilon_1)+1}e^{i\theta},q^{2(
\epsilon_2-\epsilon_1)+1}e^{-i\theta}\bigr)}{\theta_{q^2}\bigl(zq^{2(\epsilon_2
-\epsilon_1+1)},z^{-1}q^{2(\epsilon_2-\epsilon_1)}\bigr)}\,\gamma(q^{2\sigma_3})
\cdot\\
\Bigl\{\delta(q^{2\sigma_4})\theta_{q^2}\bigl(zq^{2(\sigma_3-\sigma_4+1)},z^{-1}
q^{2(\sigma_3-\sigma_4)},q^{2(\epsilon_2-\epsilon_1-\sigma_3-\sigma_4)},q^{2(
\epsilon_2-\epsilon_1-\sigma_3-\sigma_4+1)}\bigr)\\
\shoveright{-\text{idem}\,(\sigma_4;-\sigma_4)\Bigr\}}\\
+\frac{\theta_{q^2}\bigl(zqe^{i\theta},zqe^{-i\theta},q^{2(\epsilon_2-\epsilon_1
-\sigma_3+\sigma_4+1)},q^{2(\epsilon_2-\epsilon_1-\sigma_3-\sigma_4+1)}\bigr)}{
\theta_{q^2}\bigl(zq^{2(\epsilon_2-\epsilon_1+1)},z^{-1}q^{2(\epsilon_2-
\epsilon_1+1)}\bigr)}\,\gamma(q^{2\sigma_3})\cdot\\
\Bigl\{\delta(q^{2\sigma_4})\theta_{q^2}\bigl(q^{2(\epsilon_2-\epsilon_1-
\sigma_3-\sigma_4)},q^{2(\epsilon_2-\epsilon_1+\sigma_3-\sigma_4+1)}\bigr)
-\text{idem}\,(\sigma_4;-\sigma_4)\Bigr\}.
\end{multline*}
Replacing $\sigma_3$ by $-\sigma_3$ on the rhs yields the corresponding
expression for $A^-$. Then the difference is given by
\begin{multline}\label{E:A++A-}
A^+-A^-=-F
\frac{\theta_{q^2}\bigl(q^{2(\epsilon_2-\epsilon_1)+1}e^{i\theta},q^{2(\epsilon_2-
\epsilon_1)+1}e^{-i\theta}\bigr)}{\theta_{q^2}\bigl(zq^{2(\epsilon_2-\epsilon_1+1)},
z^{-1}q^{2(\epsilon_2-\epsilon_1)}\bigr)}\\
-\breve B\breve C\,
\frac{\theta_{q^2}\bigl(zqe^{i\theta},zqe^{-i\theta}\bigr)}{\theta_{q^2}\bigl(
zq^{2(\epsilon_2-\epsilon_1+1)},z^{-1}q^{2(\epsilon_2-\epsilon_1+1)}\bigr)}.
\end{multline}
Upon insertion into the rhs of equation \eqref{E:sigunst1}, the first part cancels the contribution by \eqref{E:sigunst2}. The second part gives rise to
the rhs of \eqref{E:unitrel} for both upper components.

For the lowest component in the case $e^{i\theta}=\alpha\alpha'q^{2(\tilde s+\epsilon'_1+\epsilon_2)+1}$, the statement \eqref{E:unitrel} follows from
the decomposition \eqref{E:hattaudec} with $\boldsymbol{\tau}^{(r,k)}\to
\tau^{[\pm](r,k)}(q^{2\zeta},q^{2\epsilon_2})$, $\boldsymbol{\tau}^{(r,k)\pm}
\to\tau^{[\pm](r,k)\pm}(q^{2\zeta},q^{2\epsilon_2})$, equation
\eqref{E:unitrel} for the two upper components and relation \eqref{E:varsigma},
provided that the conditions \eqref{E:tausum7cond} are satisfied.
Alternatively, using the expression \eqref{E:tauunit2} for the bracket on the rhs of \eqref{E:sigunit}, the $k$-summation is evaluated employing \eqref{E:tauunit2}
with the replacements $k\to m$, $z\to z^{-1}$, $\epsilon_1\leftrightarrow
\epsilon_4$, $\epsilon_2\leftrightarrow\epsilon_3$, $\epsilon'_1\leftrightarrow 
\epsilon'_4$, $\epsilon'_2\leftrightarrow\epsilon'_3$, $q^{2\sigma_3}
\leftrightarrow\alpha q^{2\sigma_4}$. This leads to
\begin{multline*}
\sigma^{[\pm](r,m)}-\sigma'^{[\pm](r,m)}\Big\vert_{e^{i\theta}=\alpha\alpha' 
q^{2(\tilde s+\epsilon'_1+\epsilon_2)+1}}=\\
cq^{2(\epsilon_2-\epsilon_1)}\iota(q^{2\sigma_3})\iota(q^{2\sigma_4})
\frac{\bigl(q^2,q^2,zq^{2(\epsilon_1-\epsilon_2+1)},z^{-1}q^{2(\epsilon_1-
\epsilon_2+1)};q^2\bigr)_{\infty}}{\theta_{q^2}\bigl(zqe^{i\theta},z^{-1}q
e^{i\theta}\bigr)}\cdot\\
q^{-4\tilde s(\epsilon_2-\epsilon_1)}\biggl\{q^{2(\sigma_3+\sigma_4)+4\tilde s
(\epsilon_2-\epsilon_1)}\theta_{q^2}\bigl(q^{2(\epsilon_2-\epsilon_1)+1}
e^{i\theta},q^{2(\epsilon_2-\epsilon_1)+1}e^{i\theta},q^{4\sigma_3+2},q^{4
\sigma_4+2}\bigr)\\
\shoveleft{
+q^{2(\sigma_3+\sigma_4)-4(\tilde s+1)(\epsilon_2-\epsilon_1)}\gamma(q^{2
\sigma_3})\gamma(q^{-2\sigma_3})\delta(q^{2\sigma_4})\delta(q^{-2\sigma_4})\cdot}\\
\shoveright{
\theta_{q^2}\bigl(q^{2(\epsilon_2-\epsilon_1)+1}e^{-i\theta},q^{2(\epsilon_2-
\epsilon_1)+1}e^{-i\theta},q^{4\sigma_3+2},q^{4\sigma_4+2}\bigr)}\\
+zF\frac{\theta_{q^2}\bigl(q^{2(\epsilon_2-\epsilon_1)+1}e^{i\theta},qe^{(
\epsilon_2-\epsilon_1)+1}e^{-i\theta}\bigr)}{\theta_{q^2}\bigl(zq^{2(\epsilon_2-
\epsilon_1)},z^{-1}q^{2(\epsilon_2-\epsilon_1)}\bigr)}\biggr\}\cdot\\
a_r\bigl(x;q^{2\epsilon_1},q^{2\epsilon_2};e^{i\theta}\bigr)\,
\varsigma^{(r)}_m\bigl(q^{2\epsilon_1},q^{2\sigma_1};q^{2\epsilon_2},q^{2
\sigma_2};e^{i\theta}\bigr)\Big\vert_{e^{i\theta}=\alpha\alpha'q^{2(\tilde s+
\epsilon'_1+\epsilon_2)+1}}.
\end{multline*}
Use of equation \eqref{E:ex2} to eliminate the factors $\theta_{q^2}\bigl(zq^{2(
\sigma_3+\sigma_4+1)},z^{-1}q^{2(\sigma_3+\sigma_4)}\bigr)$ and $\theta_{q^2}
\bigl(zq^{2(\sigma_3-\sigma_4+1)},z^{-1}q^{2(\sigma_3-\sigma_4)}\bigr)$ in the
expression \eqref{E:F} for $F$ gives
\begin{multline}\label{E:sigunitsp1}
\sigma^{[\pm](r,m)}-\sigma'^{[\pm](r,m)}-\frac{cq^{4(\epsilon_2-
\epsilon_1)}\tilde B\tilde C(q^2,q^2;q^2)_{\infty}}{
\bigl(zq^{2(\epsilon_2-\epsilon_1)},z^{-1}q^{2(\epsilon_2-\epsilon_1)};q^2
\bigr)_{\infty}}\,a_r\varsigma^{(r)}_m\bigg\vert_{e^{i\theta}=\alpha\alpha'
q^{2(\tilde s+\epsilon'_1+\epsilon_2)+1}}\\
=\frac{c\iota(q^{2\sigma_3})\iota(q^{2\sigma_4})(q^2,q^2,zq^{2(\epsilon_1-\epsilon_2+1)},
z^{-1}q^{2(\epsilon_1-\epsilon_2+1)};q^2)_{\infty}}{\theta_{q^2}\bigl(zqe^{
i\theta},zqe^{-i\theta}\bigr)}\cdot\\
\Bigl\{q^{2(\sigma_3+\sigma_4)+2(2\tilde s+1)(\epsilon_2-
\epsilon_1)}\theta_{q^2}\bigl(q^{2(\epsilon_2-\epsilon_1)+1}e^{i\theta},q^{2(
\epsilon_2-\epsilon_1)+1}e^{i\theta},q^{4\sigma_3+2},q^{4\sigma_4+2}\bigr)\\
\shoveleft{
+q^{2(\sigma_3+\sigma_4)-2(2\tilde s+1)(\epsilon_2-\epsilon_1)}\gamma(q^{2
\sigma_3})\gamma(q^{-2\sigma_3})\delta(q^{2\sigma_4})\delta(q^{-2\sigma_4})\cdot}\\
\shoveright{
\theta_{q^2}\bigl(q^{2(\epsilon_2-\epsilon_1)+1}e^{-i\theta},q^{2(\epsilon_2-
\epsilon_1)+1}e^{-i\theta},q^{4\sigma_3+2},q^{4\sigma_4+2}\bigr)}\\
-F(q^{-1}e^{i\theta})\Bigr\}a_r\varsigma^{(r)}_m.
\end{multline}
Insertion of \eqref{E:F2} with $zq\to e^{i\theta}$ for the last line yields
\begin{multline}\label{E:sigunitsp2}
cq^{2(\sigma_3+\sigma_4)}\iota(q^{2\sigma_3})\iota(q^{2\sigma_4})\cdot\\
\frac{\bigl(q^2,q^2,zq^{2(
\epsilon_1-\epsilon_2+1)},z^{-1}q^{2(\epsilon_1-\epsilon_2+1)};q^2\bigr)_{
\infty}\theta_{q^2}\bigl(q^{4\sigma_3+2},q^{4\sigma_4+2}\bigr)}{\theta_{
q^2}\bigl(zqe^{i\theta},zqe^{-i\theta}\bigr)}a_r\varsigma^{(r)}_m\cdot\\
\biggl\{q^{2(\epsilon_2-\epsilon_1)}G(q^{2\sigma_3},q^{2\sigma_4})G(q^{2\sigma_4},
q^{2\sigma_3})-
\frac{\breve B\breve C\,\theta_{q^2}\bigl(q^{2(\sigma_3-\sigma_4)+1}
e^{i\theta},q^{2(\sigma_3-\sigma_4)+1}e^{-i\theta}\bigr)}{\theta_{q^2}\bigl(
q^{2(\epsilon_2-\epsilon_1+\sigma_3-\sigma_4+1)},q^{2(\epsilon_2-\epsilon_1-
\sigma_3+\sigma_4+1)}\bigr)}\biggr\}
\end{multline}
for the rhs of \eqref{E:sigunitsp1}, where
\begin{multline*}
G(q^{2\sigma_3},q^{2\sigma_4})=q^{2\tilde s(\epsilon_2-\epsilon_1)}\theta_{q^2}
\bigl(q^{2(\epsilon_2-\epsilon_1)+1}e^{i\theta}\bigr)\\
+\gamma(q^{2\sigma_3})\delta(q^{-2\sigma_4})\,q^{-2(\tilde s+1)(\epsilon_2-
\epsilon_1)+2(\sigma_3-\sigma_4)}
\frac{\theta_{q^2}\bigl(q^{2(\epsilon_2-\epsilon_1)+1}e^{-i\theta},
q^{2(\epsilon_2-\epsilon_1-\sigma_3+\sigma_4)}\bigr)}{\theta_{q^2}\bigl(
q^{2(\epsilon_2-\epsilon_1+\sigma_3-\sigma_4)}\bigr)}.
\end{multline*}
According to the proof of the statement \eqref{E:unitrel} for the lowest
component in the case $e^{i\theta}=\alpha\alpha'q^{2(\tilde s+\epsilon'_1+
\epsilon_2)+1}$ subject to the restriction \eqref{E:tausum7cond}, the rhs
of \eqref{E:sigunitsp1} vanishes if $\max(\vert zqe^{\pm i\theta}\vert,\vert 
z^{-1}qe^{\pm i\theta}\vert)<1$. Since the braces on the rhs of \eqref{E:sigunitsp1} are independent of $z$, this confirms the assertion in
the more general case specified by the conditions \eqref{E:tausum8cond}.

Taking to account the antisymmetry of the expression \eqref{E:F} for $F$
under $q^{2\sigma_4}\to q^{-2\sigma_4}$, repeating the previous steps
yields the lhs of \eqref{E:sigunitsp1} equated to the expression \eqref{E:sigunitsp2} with the replacement $q^{2\sigma_4}\to q^{-2\sigma_4}$ within the braces. This implies
\begin{multline}\label{E:GGBC}
q^{2(\epsilon_2-\epsilon_1)}G(q^{2\sigma_3},q^{2\sigma_4})G(q^{2\sigma_4},
q^{2\sigma_3})-\frac{\breve B\breve C\,\theta_{q^2}\bigl(q^{2(\sigma_3-\sigma_4)
+1}e^{i\theta},q^{2(\sigma_3-\sigma_4)+1}e^{-i\theta}\bigr)}{\theta_{q^2}\bigl(
q^{2(\epsilon_2-\epsilon_1+\sigma_3-\sigma_4+1)},q^{2(\epsilon_2-\epsilon_1-
\sigma_3+\sigma_4+1)}\bigr)}=\\
q^{2(\epsilon_2-\epsilon_1)}G(q^{2\sigma_3},q^{-2\sigma_4})G(q^{-2\sigma_4},
q^{2\sigma_3})-
\frac{\breve B\breve C\,\theta_{q^2}\bigl(q^{2(\sigma_3+\sigma_4)+1}e^{i\theta},
q^{2(\sigma_3+\sigma_4)+1}e^{-i\theta}\bigr)}{\theta_{q^2}\bigl(q^{2(\epsilon_2
-\epsilon_1+\sigma_3+\sigma_4+1)},q^{2(\epsilon_2-\epsilon_1-\sigma_3-\sigma_4
+1)}\bigr)}\\
=0
\end{multline}
for $e^{i\theta}=\alpha\alpha'q^{2(\tilde s+\epsilon'_1+\epsilon_2)+1}$.

Due to equation \eqref{E:mhattau0}, for $e^{i\theta}=q^{2(s'+\epsilon_2-
\epsilon_1)-1}$ the lowest component of the bracket on the rhs of \eqref{E:sigunit} reduces to
\begin{multline*}
\tau^{[\pm](k,r)}\bigl(q^{2\zeta},q^{2\epsilon_2}\bigr)-\alpha\alpha'q^{2(
\epsilon'_2-\epsilon_2)}\tau'^{[\pm](r,k)}\bigl(q^{2\zeta},q^{2\epsilon'_2}
\bigr)\Big\vert_{e^{i\theta}=q^{2(s'+\epsilon_2-\epsilon_1)-1}}=\\
\widehat{\tau}^{(r,k)}\bigl(q^{2\zeta},q^{2\epsilon_2}\bigr)-\alpha\alpha' q^{2
(\epsilon'_2-\epsilon_2)}\widehat{\tau}'^{(r,k)}\bigl(q^{2\zeta},q^{2\epsilon'_2}
\bigr)\Big\vert_{e^{i\theta}=q^{2(s'+\epsilon_2-\epsilon_1)-1}}.
\end{multline*}
Using the expression \eqref{E:tausum7} for the rhs, the $k$-summation on the
rhs of \eqref{E:sigunit} can be evaluated by means of equation \eqref{E:tausum7}
with the replacements $k\to m$, $z\to z^{-1}$, $\epsilon_1\leftrightarrow
\epsilon_4$, $\epsilon_2\leftrightarrow\epsilon_3$, $\epsilon'_1\leftrightarrow 
\epsilon'_4$, $\epsilon'_2\leftrightarrow\epsilon'_3$, $q^{2\sigma_3}\leftrightarrow \alpha q^{2\sigma_4}$. This step immediately
leads to the rhs of \eqref{E:sigunit} for $e^{i\theta}\in\Gamma_{q^{-2(r+\epsilon_2-\epsilon_1)+1}}$.
\vskip 0.5cm

\emph{Remark 1}: 
Equation [\cite{GR}:ex.5.23] with $n\to4$, $\tfrac{a_1}{b_1}\to-q^{2(\epsilon_3+
\sigma_3)+1}$, $\tfrac{a_1}{b_2}\to-\alpha\alpha'q^{2(\epsilon'_3+\sigma_3)+1}$,
$\tfrac{a_1}{b_3}\to-\alpha q^{-2(\epsilon_1-\sigma_3)+1}$, $\tfrac{a_1}{b_4}\to-
\alpha'q^{2(\epsilon'_1-\sigma_3)+1}$, $\tfrac{a_1}{a_2}\to q^{4\sigma_3}$,
$\tfrac{a_1}{a_3}\to q^{2(\epsilon_2-\epsilon_1+\sigma_3-\sigma_4+1)}$, 
$\tfrac{a_1}{a_4}\to q^{2(\epsilon_2-\epsilon_1+\sigma_3+\sigma_4+1)}$ yields
\begin{multline*}
\theta_{q^2}\bigl(q^{4\sigma_3},-q^{2(\epsilon_4+\sigma_4)+1},-q^{2(\epsilon_4-
\sigma_4)+1},-\alpha\alpha'q^{2(\epsilon'_4+\sigma_4)+1},-\alpha\alpha'q^{2(
\epsilon'_4-\sigma_4)+1}\bigr)\breve B=\\
\shoveleft{q^{2(\epsilon_2-\epsilon_1+\sigma_3-\sigma_4)}\cdot}\\
\theta_{q^2}\bigl(q^{4\sigma_4},-\alpha q^{2(\epsilon_1+\sigma_3)+1},-\alpha 
q^{2(\epsilon_1-\sigma_3)+1},-\alpha'q^{2(\epsilon'_1+\sigma_3)+1},-\alpha'
q^{2(\epsilon'_1-\sigma_3)+1}\bigr)\breve C,
\end{multline*}
where $\breve B$ and $\breve C$ are given by \eqref{E:breveB} and \eqref{E:breveC}.
Moreover, equation [\cite{GR}:ex.5.23] gives rise to a similar relation
for $G(q^{2\sigma_3},q^{2\sigma_4})$ and $G(q^{2\sigma_4},q^{2\sigma_3})$.

\vskip 0.5cm

\emph{Remark 2}:
If $\breve B=0$, the rhs of equation \eqref{E:unitrel} in Proposition \ref{P:unitrel} 
and the last contribution on the rhs of \eqref{E:F2} specifying $F$ vanish.
Choosing $\alpha=1$, $\epsilon_2=\epsilon_4$, $\epsilon'_2=
\epsilon'_4$ or $\alpha'=1$, $\epsilon_2=\epsilon'_4$, $\epsilon'_2=\epsilon_4$
yields $\delta(q^{2\sigma_4})=\delta(q^{-2\sigma_4})=1$ and $q^{-2\sigma_3}\breve B=\theta_{q^2}(q^{2(\epsilon_2-\epsilon_1
-\sigma_3-\sigma_4)},
q^{2(\epsilon_2-\epsilon_1+\sigma_3-\sigma_4+1)})-\theta_{q^2}(q^{2(\epsilon_2-
\epsilon_1-\sigma_3+\sigma_4)},q^{2(\epsilon_2-\epsilon_1+\sigma_3+\sigma_4+1)})$.
The requirement $\epsilon_1=\epsilon_2=\epsilon_4$, $\epsilon'_2=\epsilon'_4$
implies $\epsilon_1=\epsilon_2=\epsilon_3=\epsilon_4$ and $\epsilon'_1=
\epsilon'_2=\epsilon'_3=\epsilon'_4$ due to the condition \eqref{E:epscond3}.
Similarly, the requirement $\epsilon_1=\epsilon_2=\epsilon'_4$, $\epsilon'_2=
\epsilon_4$ implies $\epsilon_1=\epsilon_2=\epsilon'_3=\epsilon'_4$ and $\epsilon'_1=\epsilon'_2=\epsilon_3=\epsilon_4$. Thus $\breve B=0$ is found in
the cases 
\begin{equation*}
\alpha=1,\quad\epsilon_1=\epsilon_2=\epsilon_3=\epsilon_4,\quad \epsilon'_1
=\epsilon'_2=\epsilon'_3=\epsilon'_4
\end{equation*}
and 
\begin{equation*}
\alpha'=1,\quad\epsilon_1=\epsilon_2=\epsilon'_3=\epsilon'_4,\quad
\epsilon'_1=\epsilon'_2=\epsilon_3=\epsilon_4.
\end{equation*}

\vskip 0.5cm
If $\alpha=-\alpha'$, the relation to the vector-valued big $q$-Jacobi function
transform specified in subsection \eqref{S:bigJac}
allows to deduce a pair of quadratic summation formulae for the elements \eqref{E:Run} from Proposition \ref{P:unitrel}.

\begin{corollary}\label{C:RRsum}
In the case $\alpha=-\alpha'=-1$ and $\max(q^{4\sigma_1},q^{4\sigma_2})<1$,
the elements ${r_{\pm}}^{k,k+r^*}_{l,l+r^*}$ fulfill
\begin{multline}\label{E:RRsum1}
\sum_{k=-\infty}^{\infty}{r_{\mp}}^{m,m+r^*}_{k,k+r^*}\bigl(z^{-1},\alpha;-1,
q^{2\zeta};q^{2\epsilon_4},q^{2\epsilon_3};q^{2\epsilon_2},q^{2\epsilon_1};
\alpha q^{2\sigma_4},\alpha q^{2\sigma_3}\bigr)\cdot\\
\shoveright{
{r_{\pm}}^{k,k+r^*}_{l,l+r^*}\bigl(z,\alpha;-1,q^{2\zeta};q^{2\epsilon_1},
q^{2\epsilon_2};q^{2\epsilon_3},q^{2\epsilon_4};q^{2\sigma_3},q^{2\sigma_4}
\bigr)}\\
+\sum_{k=-\infty}^{\infty}{r_{\mp}}^{m,m+r^*}_{k,k+r^*}\bigl(z^{-1},-\alpha;
-1,q^{2\zeta};q^{2\epsilon'_4},q^{2\epsilon'_3};q^{2\epsilon_2},q^{2\epsilon_1};
\alpha q^{2\sigma_4},\alpha q^{2\sigma_3}\bigr)\cdot\\
\shoveright{{r_{\pm}}^{k,k+r^*}_{l,l+r^*}\bigl(z,-\alpha;-1,q^{2\zeta};q^{2
\epsilon_1},q^{2\epsilon_2};q^{2\epsilon'_3},q^{2\epsilon'_4};-q^{2\sigma_3},
-q^{2\sigma_4}\bigr)}\\
=\delta_{l,m}\cdot c\tilde B\tilde Cq^{4(\epsilon_2-\epsilon_1)}
\frac{(1-q^2)(q^2,q^2;q^2)_{\infty}}{\bigl(zq^{2(\epsilon_2-\epsilon_1)},
z^{-1}q^{2(\epsilon_2-\epsilon_1)};q^2\bigr)_{\infty}}
\end{multline}
and
\begin{multline}\label{E:RRsum2}
\sum_{k=-\infty}^{\infty}{r_{\mp}}^{m,m+r^*}_{k,k+r^*}\bigl(z^{-1},\alpha;-1,
q^{2\zeta};q^{2\epsilon_4},q^{2\epsilon_3};q^{2\epsilon_2},q^{2\epsilon_1};
\alpha q^{2\sigma_4},\alpha q^{2\sigma_3}\bigr)\cdot\\
\shoveright{{r_{\pm}}^{k,k+r^*}_{l,l+r^*}\bigl(z,-\alpha;-1,q^{2\zeta};q^{2
\epsilon'_1},q^{2\epsilon'_2};q^{2\epsilon_3},q^{2\epsilon_4};q^{2\sigma_3},
q^{2\sigma_4}\bigr)}\\
+\sum_{k=-\infty}^{\infty}{r_{\mp}}^{m,m+r^*}_{k,k+r^*}\bigl(z^{-1},-\alpha;
-1,q^{2\zeta};q^{2\epsilon'_4},q^{2\epsilon'_3};q^{2\epsilon_2},q^{2\epsilon_1};
\alpha q^{2\sigma_4},\alpha q^{2\sigma_3}\bigr)\cdot\\
{r_{\pm}}^{k,k+r^*}_{l,l+r^*}\bigl(z,\alpha;-1,q^{2\zeta};q^{2\epsilon'_1},
q^{2\epsilon'_2};q^{2\epsilon'_3},q^{2\epsilon'_4};-q^{2\sigma_3},-q^{2\sigma_4}
\bigr)=0.
\end{multline}
\end{corollary}

\emph{Proof}:
By means of Proposition \ref{P:unitrel} and the identifications \eqref{E:varphirho} and \eqref{E:Phi1}, the orthogonality relations \eqref{E:Hbas} for the vector-valued big $q$-Jacobi functions
with $x'=z_+q^{2m}$, $x=z_+q^{2l}$ or $x=z_-q^{2l}$
and the parameters specified by \eqref{E:Jacpar} and \eqref{E:Jacpar2} can be
reformulated as the integral transform \eqref{E:G} with the 
same parameters and the function $g(\gamma)$
given by
\begin{multline*}
q^{-l}\frac{S_{l+r}\bigl(q^{2\epsilon_2},q^{2\sigma_2}\bigr)}{S_l\bigl(
q^{2\epsilon_1},q^{2\sigma_1}\bigr)}\,G(z)\,
\left(\begin{matrix} b^-_r\bigl(\sigma^{[\pm](r,m)-}-\sigma'^{[\pm]
(r,m)-}\bigr)\\
b^+_r\bigl(\sigma^{[\pm](r,m)+}-\sigma'^{[\pm](r,m)+}\bigr)
\end{matrix}\right)
\end{multline*}
for $\gamma\in\mathbb T$, by
\begin{multline*}
q^{-l}\frac{S_{l+r}(q^{2\epsilon_2},q^{2\sigma_2})}{S_l(q^{2\epsilon_1},q^{2
\sigma_1})}q^{-(s-r)^2+2(s-r)(\epsilon_1+\sigma_1)-4(s-r)(r+\epsilon_2)}\cdot\\
\frac{\bigl(q^{2(\epsilon_2-\epsilon_1+\sigma_1+\sigma_2+1)},q^{2(\epsilon_2
-\epsilon_1+\sigma_1-\sigma_2+1)};q^2\bigr)_s}{a^+_r\bigl(q^{4(\epsilon_2-
\epsilon_1)+2(s+r+1)};q^2\bigr)_{s-r}}\,G(z)\,\bigl(\sigma^{[\pm](r,m)+}-\sigma'^{[
\pm](r,m)+}\bigr),
\end{multline*}
for $\gamma=q^{-2(s+\epsilon_2-\epsilon_1)-1}$, by
\begin{multline*}
q^{-l}\frac{S_{l+r}\bigl(q^{2\epsilon_2},q^{2\sigma_2}\bigr)}{S_l\bigl(
q^{2\epsilon_1},q^{2\sigma_1}\bigr)}(-1)^tq^{-2t^2-2t(r+\epsilon_2-\sigma_2)
+4t\sigma_1}\frac{\bigl(q^{-2t+4(\sigma_1+\sigma_2)};q^2\bigr)_{\infty}}{
\bigl(q^{2(t+1-2\sigma_1)};q^2\bigr)_{\infty}}\cdot\\
\frac{\theta_{q^2}\bigl(-q^{2(\epsilon_1-\sigma_1)+1}\bigr)\bigl(q^{2(
\epsilon_2-\epsilon_1-\sigma_1+\sigma_2+1)},q^{2(\epsilon_2-\epsilon_1-\sigma_1
-\sigma_2+1)};q^2\bigr)_{\infty}}{a^-_r\theta_{q^2}\bigl(-q^{2(r+\epsilon_2
+\sigma_2)+1}\bigr)}\cdot\\
G(z)\,\bigl(\sigma^{[\pm](r,m)-}-\sigma'^{[\pm](r,m)-}\bigr)
\end{multline*}
for $\gamma=q^{2(\sigma_1+\sigma_2)-2t-1}$ and by
\begin{multline*}
-q^{-l}\frac{S_{l+r}\bigl(q^{2\epsilon_2},q^{2\sigma_2}\bigr)}{S_l\bigl(
q^{2\epsilon_1},q^{2\sigma_1}\bigr)}
\frac{q^{2(r+\epsilon_2+\sigma_3)}e^{-i\theta}
}{a_r\theta_{q^2}\bigl(-q^{2(r+\epsilon_2+\sigma_2)+1},-q^{2(r+\epsilon_2
-\sigma_2)+1},q^{4\sigma_1}\bigr)}\cdot\\
\frac{G(z)}{\bigl(q^{2(r+\epsilon_2-\epsilon_1)+1}e^{-i\theta},q^2e^{2i
\theta};q^2\bigr)_{\infty}}
\,\bigl(\sigma^{[\pm](r,m)}-\sigma'^{[\pm](r,m)}\bigr)
\end{multline*}
for $\gamma\in\tilde{\Gamma}\cap\Gamma^{\text{inf}}$ or $\gamma\in\Gamma_{q^{-2(
r+\epsilon_2-\epsilon_1)+1}}\cap\Gamma^{\text{fin}}_{q^{-2(r+\epsilon_2-\epsilon_1)
+1}}$. The four sets collecting the values of $\gamma$ in the last case
are defined by \eqref{E:tilGam} and \eqref{E:Gammapm} in subsection \ref{S:taudef} and by \eqref{E:bigJset} in subsection \ref{S:bigJac}.
Each rhs involdes the function
\begin{equation*}
G(z)=\frac{q^{4(\epsilon_1-\epsilon_2)}}{c\tilde B\tilde C}
\frac{\bigl(zq^{2(\epsilon_2-\epsilon_1)},z^{-1}q^{2(\epsilon_2-\epsilon_1)};
q^2\bigr)_{\infty}}{(q^2,q^2;q^2)_{\infty}}.
\end{equation*}
Due to the restrictions \eqref{E:tausum7cond} and
\eqref{E:tausum8cond}, both the $k$-summation in the Definition \ref{D:unsigdef}
and the summation implicit in the brackets on the rhs of \eqref{E:sigunit}
are absolutely convergent. Interchanging the summations and making use of
the identifications \eqref{E:Ff1} and \eqref{E:Ff2}, the function $g(\gamma)$ can be
expressed by $g(\gamma)=\bigl(\mathcal Ff^{[\pm]}\bigr)(\gamma)$ with the parameters of $\mathcal F$ given by \eqref{E:Jacpar} and \eqref{E:Jacpar2}. 
The function $f^{[\pm]}$ can be written out explicitly making use
of the definitions by \eqref{E:sigunit} and \eqref{E:Run} and the relations 
\eqref{E:Ff1}, \eqref{E:Ff2}. Since the integral transform $\mathcal G$ inverts the vector-valued big $q$-Jacobi transform $\mathcal F$, the assertions 
\eqref{E:RRsum1} and \eqref{E:RRsum2} follow immediately for $x=z_+q^{2l}$ and $x=z_-q^{2l}$, respectively.

\vskip 0.5cm

The bilateral summations in Corollary \ref{C:RRsum} are a generalization of the
unitarity property satisfied by R-matrices related to finite-dimensional $U'_q\bigl(\widehat{sl}(2)\bigr)$-modules.

\appendix

\section{The $R$-elements}\label{A:R}

In subsection \ref{A:int}, the set of matrix elements $\{R^{k,k+r^*}_{l,l+r^*}\}$ specified by \eqref{E:Rel1} is shown to satisfy the intertwining condition \eqref{E:int}. The symmetry \eqref{E:Rsig1+-} 
under $q^{2\sigma_3}\to q^{-2\sigma_3}$ and equation \eqref{E:RmR} relating $R^{k,k+r^*}_{l,l+r^*}$, $\check R^{k,k+r^*}_{l,l+r^*}$ and 
$\mathring R^{k,k+r^*}_{l,l+r^*}$ are proven in subsection \ref{A:Rprop}.

\subsection{Proof of the intertwining property}\label{A:int}

For $X=q^{\pm h_i}$, the property \eqref{E:int} is ensured by the restriction
\eqref{E:econd}. The remaining conditions can be written out by
\begin{equation}\label{E:intwrite}
\begin{array}{lllll}
&X=e_0:&\frac{zD-q^{-2(r+\epsilon_2-\epsilon_1)}A}{1-z^{-1}q^{-2(r+\epsilon_2
-\epsilon_1)}}=0,\qquad
&X=e_1:&\frac{C-B}{1-z^{-1}q^{2(r+\epsilon_2-\epsilon_1)}}=0,\\
&&&&\\
&X=f_0:&\frac{Cq^{2(r+\epsilon_2-\epsilon_1)}-zB}{1-z^{-1}q^{2(r+\epsilon_2-
\epsilon_1)}}=0,\qquad
&X=f_1:&\frac{D-A}{1-z^{-1}q^{-2(r+\epsilon_2-\epsilon_1)}}=0,
\end{array}
\end{equation}
where
\begin{multline}\label{E:A}
A\equiv A_{k,r,l}
=-q^{-2(k+\epsilon_4)}\bigl(1-z^{-1}q^{-2(r+\epsilon_2-\epsilon_1)}\bigr)
s_{l+r-1}\bigl(q^{2\epsilon_2},q^{2\sigma_2}\bigr)\,R^{k+1,k+r^*}_{l,l+r-1^*}\\
+\alpha z^{-1}\bigl(1-\alpha zq^{2(l-k+\epsilon_1-\epsilon_4)}\bigr)s_{-k-1}\bigl(
q^{-2\epsilon_4},q^{2\sigma_4}\bigr)\,R^{k,k+r^*}_{l,l+r^*}\\
-\alpha q\bigl(1-\alpha q^{2(l-k+r+\epsilon_2-\epsilon_4-1)}\bigr)s_{-k-r-1}\bigl(
q^{-2\epsilon_3},q^{2\sigma_3}\bigr)\,R^{k+1,k+r+1^*}_{l,l+r^*},
\end{multline}
\begin{multline}\label{E:B}
B\equiv B_{k,r,l}
=-q^{-2(k+r+\epsilon_3)-1}\bigl(1-z^{-1}q^{2(r+\epsilon_2-\epsilon_1)}\bigr)
s_{l+r}\bigl(q^{2\epsilon_2},q^{2\sigma_2}\bigr)\,R^{k,k+r+1^*}_{l,l+r+1^*}
\\
-\alpha q\bigl(1-\alpha z^{-1}q^{2(l-k+\epsilon_1-\epsilon_4-1)}\bigr)s_{-k-1}
\bigl(q^{-2\epsilon_4},q^{2\sigma_4}\bigr)\, R^{k+1,k+r+1^*}_{l,l+r^*}\\
+\alpha z^{-1}\bigl(1-\alpha q^{2(l-k+r+\epsilon_2-\epsilon_4)}\bigr)s_{-k-r-1}\bigl(
q^{-2\epsilon_3},q^{2\sigma_3}\bigr)\, R^{k,k+r^*}_{l,l+r^*},
\end{multline}
\begin{multline}
C\equiv C_{k,r,l}
=-q^{-2(k+r+\epsilon_3)}\bigl(1-z^{-1}q^{2(r+\epsilon_2-\epsilon_1)}\bigr)s_{l-1}
\bigl(q^{2\epsilon_1},q^{2\sigma_1}\bigr)\,R^{k,k+r+1^*}_{l-1,l+r^*}\\
-\alpha q\bigl(1-\alpha q^{2(l-k-r+\epsilon_1-\epsilon_3-1)}\bigr)s_{-k-1}\bigl(q^{-2
\epsilon_4},q^{2\sigma_4}\bigr)\,R^{k+1,k+r+1^*}_{l,l+r^*}\\
+\alpha z^{-1}\bigl(1-\alpha zq^{2(l-k+\epsilon_1-\epsilon_4)}\bigr)s_{-k-r-1}\bigl(
q^{-2\epsilon_3},q^{2\sigma_3}\bigr)\,R^{k,k+r^*}_{l,l+r^*}
\end{multline}
and
\begin{multline}\label{E:D}
D\equiv D_{k,r,l}
=-q^{-2(k+\epsilon_4)-1}\bigl(1-z^{-1}q^{-2(r+\epsilon_2-\epsilon_1}\bigr)
s_l\bigl(q^{2\epsilon_1},q^{2\sigma_1}\bigr)\,R^{k+1,k+r^*}_{l+1,l+r^*}\\
+\alpha z^{-1}\bigl(1-\alpha q^{2(l-k-r+\epsilon_1-\epsilon_3)}\bigr)s_{-k-1}\bigl(
q^{-2\epsilon_4},q^{2\sigma_4}\bigr)\,R^{k,k+r^*}_{l,l+r^*}\\
-\alpha q\bigl(1-\alpha z^{-1}q^{2(l-k+\epsilon_1-\epsilon_4-1)}\bigr)s_{-k-r-1}\bigl(
q^{-2\epsilon_3},q^{2\sigma_3}\bigr)\,R^{k+1,k+r+1^*}_{l,l+r^*}.
\end{multline}
The contiguous relation
\begin{multline*}
(1-b_1)(a_1-a_2){}_4\phi_3\biggl(\genfrac{}{}{0pt}{}{a_1,\,a_2,\,a_3,\,a_4}{b_1,
\,b_2,\,b_3};q^2,q^2\biggr)=(1-a_1)(b_1-a_2)\cdot\\
{}_4\phi_3\biggl(\genfrac{}{}{0pt}{}{a_1q^2,\,a_2,\,a_3,\,a_4}{
b_1q^2,\,b_2,\,b_3};q^2,q^2\biggr)
+(1-a_2)(a_1-b_1){}_4\phi_3\biggl(\genfrac{}{}{0pt}{}{a_1,\,a_2q^2,\,a_3,\,a_4}{
b_1q^2,\,b_2,\,b_3};q^2,q^2\biggr)
\end{multline*}
is an immediate consequence of the definition of the series $_m\phi_n$ by equation \eqref{E:phidef}.
Applying the relation with $a_1\to-q^{-2(k+r+\epsilon_3-\sigma_3)+1}$, $a_2\to-\alpha 
q^{2(l+\epsilon_1+\sigma_3)+1}$, $a_3\to zq^{-2(r+\epsilon_2-\epsilon_1)+2}$, $a_4\to z^{-1}q^{-2(r+\epsilon_2-\epsilon_1)}$, $b_1\to\alpha q^{2(l-k-r+\epsilon_1-
\epsilon_3+1)}$, $b_2\to q^{-2(r+\epsilon_2-\epsilon_1-\sigma_3+\sigma_4)+2}$,
$b_3\to q^{-2(r+\epsilon_2-\epsilon_1-\sigma_3-\sigma_4)+2}$ to the first $_4\phi_3$-series in the expression \eqref{E:Rel1} and with $a_1\to-q^{-2(k+
\epsilon_4-\sigma_4)+1}$, $a_2\to-\alpha q^{2(l+r+\epsilon_2+\sigma_4)+1}$, $a_3\to zq^{2(\sigma_4-\sigma_3+1)}$, $a_4\to z^{-1}q^{2(\sigma_4
-\sigma_3)}$, $b_1\to\alpha 
q^{2(l-k+\epsilon_1-\epsilon_4-\sigma_3+\sigma_4+1)}$, $b_2\to q^{4\sigma_4+2}$,
$b_3\to q^{2(r+\epsilon_2-\epsilon_1-\sigma_3+\sigma_4+1)}$ to the second 
$_4\phi_3$-series yields 
\begin{multline}\label{E:Fkrl}
0=F_{k,r,l}\equiv
\bigl(1-\alpha z^{-1}q^{2(l-k+\epsilon_1-\epsilon_4)}\bigr)\bigl(1-\alpha q^{2
(l+k+r+\epsilon_2+\epsilon_4)}\bigr)R^{k,k+r^*}_{l,l+r^*}\\
+\alpha z^{-1}q^{2(l+k+r+\epsilon_2+\epsilon_4)-1}s_{-k}\bigl(q^{-2\epsilon_4},
q^{2\sigma_4}\bigr)s_{-k-r}\bigl(q^{-2\epsilon_3},q^{2\sigma_3}\bigr)R^{k-1,k+r
-1^*}_{l,l+r^*}\\
-q^{-1}s_l\bigl(q^{2\epsilon_1},q^{2\sigma_1}\bigr)s_{l+r}\bigl(q^{2\epsilon_2},
q^{2\sigma_2}\bigr)R^{k,k+r^*}_{l+1,l+r+1^*}.
\end{multline}
According to \eqref{E:A}-\eqref{E:D}, the rhs of \eqref{E:Fkrl} is found in the linear combination
\begin{multline}\label{E:ABCD0}
s_{l+r}\bigl(q^{2\epsilon_2},q^{2\sigma_2}\bigr)s_{-k-r}\bigl(q^{-2\epsilon_3},
q^{2\sigma_3}\bigr)\cdot\\
\Bigl\{s_l\bigl(q^{2\epsilon_1},q^{2\sigma_1}\bigr)A_{k-1,r,l+1}
-qs_{l+r}\bigl(q^{2\epsilon_2},q^{2\sigma_2}\bigr)D_{k-1,r,l}\Bigr\}+\\
s_l\bigl(q^{2\epsilon_1},q^{2\sigma_1}\bigr)s_{-k}\bigl(q^{-2\epsilon_4},q^{2
\sigma_4}\bigr)\Bigl\{qs_l\bigl(q^{2\epsilon_1},q^{2\sigma_1}\bigr)B_{k-1,r,l}
-s_{l+r}\bigl(q^{2\epsilon_2},q^{2\sigma_2}\bigr)C_{k-1,r,l+1}\Bigr\}\\
=\alpha q^{-2(k+r+\epsilon_3)+3}F_{k,r,l}\,\Bigl\{q^{-2(k+r+\epsilon_3)+1}
\bigl(1-q^{4(r+\epsilon_2-\epsilon_1)}\bigr)\bigl(1+\alpha q^{2(k+l+r+\epsilon_2
+\epsilon_4)}\bigr)\\
+\bigl(1-\alpha q^{2(l-k+r+\epsilon_2-\epsilon_4+1)}\bigr)\bigl(q^{2\sigma_3}+
q^{-2\sigma_3}\bigr)\\
-q^{2(r+\epsilon_2-\epsilon_1)}\bigl(1-\alpha q^{2(l-k-r+\epsilon_1-\epsilon_3+1)}
\bigr)\bigl(q^{2\sigma_4}+q^{-2\sigma_4}\bigr)\Bigr\}=0.
\end{multline}
If $\vert q^{-2(k+r+\epsilon_3-\sigma_1)+1}\vert<1$,
the three-term transformation [\cite{GR}:III.36] with $a\to-\alpha q^{2(l+\epsilon_1-\sigma_3+2\sigma_4)+1}$, $b\to-\alpha q^{2(l+\epsilon_1-
\sigma_3)+1}$, $c\to- q^{2(k+\epsilon_4+\sigma_4)+1}$, $d\to -\alpha q^{2(l+r+\epsilon_2+\sigma_4)+1}$,
$e\to zq^{2(\sigma_4-\sigma_3+1)}$, $e\to z^{-1}q^{2(\sigma_4-\sigma_3)}$
allows to express the R-element given by \eqref{E:Rel1} by
\begin{multline}\label{E:RW1}
R^{k,k+r^*}_{l,l+r^*}\bigl(z,\alpha;q^{2
\epsilon_1},q^{2\epsilon_2};q^{2\epsilon_3},q^{2\epsilon_4};q^{2\sigma_3},
q^{2\sigma_4}\bigr)=
\bigl(q^{-2(r+\epsilon_2-\epsilon_1+\sigma_3-\sigma_4)+2};q^2\bigr)_r
\cdot\\
\kappa_rz^{-k-r}q^{l-k-r}\frac{S_{-k}\bigl(q^{-2\epsilon_4},q^{2\sigma_4}
\bigr)}{S_{-k-r}\bigl(q^{-2\epsilon_3},q^{2\sigma_3}\bigr)}\frac{S_{l+r}
\bigl(q^{2\epsilon_2},\alpha q^{2\sigma_4}\bigr)}{S_l\bigl(q^{2
\epsilon_1},\alpha q^{2\sigma_3}\bigr)}\cdot\\
\shoveleft{
\frac{\bigl(\alpha q^{2(l-k+\epsilon_1-\epsilon_4-\sigma_3+\sigma_4+1)},
-q^{-2(k+r+\epsilon_3-\sigma_3)+1},-\alpha zq^{2(l+\epsilon_1+\sigma_4)+3};
q^2\bigr)_{\infty}}{\bigl(\alpha z^{-1}q^{2(l-k+\epsilon_1-\epsilon_4)},
-q^{-2(k+\epsilon_4-\sigma_4)+1},-\alpha q^{2(l+\epsilon_1-\sigma_3+2\sigma_4
)+3};q^2\bigr)_{\infty}}\cdot}\\
\shoveright{
\frac{\bigl(-\alpha z^{-1}q^{2(l+\epsilon_1+\sigma_4)+1},q^{4\sigma_4+2};q^2\bigr)_{
\infty}}{\bigl(-\alpha q^{2(l+r+\epsilon_2+\sigma_4)+1},
q^{-2(\epsilon_2-\epsilon_1-\sigma_3-\sigma_4)+2};q^2\bigr)_{\infty}}\cdot}\\
{}_8W_7\bigl(-\alpha q^{2(l+\epsilon_1-\sigma_3+2\sigma_4)+1};-\alpha q^{2(l+\epsilon_1-\sigma_3)+1},-q^{2(k+\epsilon_4+\sigma_4)+1},-\alpha 
q^{2(l+r+\epsilon_2+\sigma_4)+1},\\
zq^{2(\sigma_4-\sigma_3+1)},z^{-1}q^{2(\sigma_4-\sigma_3)};q^2,
-q^{-2(k+r+\epsilon_3-\sigma_3)+1}\bigr).
\end{multline}
Various contiguous relations for $_8W_7$-series are found in \cite{isra}.
Taking advantage of \eqref{E:RW1}, the
relation [\cite{isra}:2.2] with $a\to-\alpha q^{2(l+\epsilon_1-\sigma_3+
2\sigma_4)+1}$, $b\to-\alpha q^{2(l+r+\epsilon_2+\sigma_4)+1}$, $c\to- q^{2(k+\epsilon_4
+\sigma_4)+3}$, $d\to-\alpha q^{2(l+\epsilon_1-\sigma_3)+1}$, $e\to zq^{2(\sigma_4-\sigma_3+1)}$, $e\to z^{-1}q^{2(\sigma_4-\sigma_3)}$ yields
\begin{equation}\label{E:A=0}
A=0
\end{equation}
for $\vert q^{-2(k+r+\epsilon_3-\sigma_3)+1}\vert>1$.
Moreover, relation [\cite{isra}:2.3] with 
$a\to-\alpha q^{2(l+\epsilon_1-\sigma_3+2\sigma_4)+1}$, $b\to-\alpha q^{2(l+r+\epsilon_2+
\sigma_4)+3}$, $c\to-q^{2(k+\epsilon_4+\sigma_4)+1}$, $d\to-\alpha q^{2(l+
\epsilon_1-\sigma_3)+1}$, $e\to zq^{2(\sigma_4-\sigma_3+1)}$, $f\to z^{-1}
q^{2(\sigma_4-\sigma_3)}$ implies
\begin{equation}\label{E:B=0}
B=0
\end{equation}
for $\vert q^{-2(k+r+\epsilon_3-\sigma_3)+1}\vert>1$.

In the case $\vert q^{2(k+r+\epsilon_3+\sigma_3)+1}\vert<1$, combining 
the three-term transformations [\cite{GR}:III.36] with
$a\to-\alpha q^{-2(l+2r+2\epsilon_2-\epsilon_1-\sigma_3)+1}$, $b\to-\alpha q^{-2(l+r+
\epsilon_2-\sigma_4)+1}$, $c\to-\alpha q^{-2(l+r+\epsilon_2+\sigma_4)+1}$, $d\to 
-q^{-2(k+r+\epsilon_3-\sigma_3)+1}$, $e\to zq^{-2(r+\epsilon_2-\epsilon_1)+2}$,
$f\to z^{-1}q^{-2(r+\epsilon_2-\epsilon_1)}$ and with $a\to-\alpha q^{-2(l+\epsilon_1
+\sigma_3-2\sigma_4)+1}$, $b\to-\alpha q^{-2(l+r+\epsilon_2-\sigma_4)+1}$, $c\to-\alpha
q^{-2(l+\epsilon_1+\sigma_3)+1}$, $d\to-q^{-2(k+\epsilon_4-\sigma_4)+1}$,
$e\to zq^{2(\sigma_4-\sigma_3+1)}$, $f\to z^{-1}q^{2(\sigma_4-\sigma_3 )}$
yields
\begin{multline}\label{E:RW3}
R^{k,k+r^*}_{l,l+r^*}\bigl(z,\alpha;q^{2
\epsilon_1},q^{2\epsilon_2};q^{2\epsilon_3},q^{2\epsilon_4};q^{2\sigma_3},
q^{2\sigma_4}\bigr)=\bigl(q^{-2(\epsilon_2-\epsilon_1+\sigma_3-\sigma_4)+2}
;q^2\bigr)_{\infty}^{-1}\cdot\\
\Biggl\{h_{\theta}\frac{\theta_{q^2}\bigl(\alpha q^{2(\epsilon_1-\epsilon_3+1)}
\bigr)}{\theta_{q^2}\bigl(\alpha z^{-1}q^{2(\epsilon_1-\epsilon_4)}\bigr)}
\frac{\bigl(zq^{2(\sigma_4-\sigma_3+1)},z^{-1}q^{2(\sigma_4-\sigma_3)};q^2
\bigr)_{\infty}
}{\bigl(q^{2(\epsilon_2-\epsilon_1-\sigma_3+\sigma_4)};q^2\bigr)_{
\infty}}\hat R^{k,k+r^*}_{l,l+r^*}\\
\shoveleft{
+h^{-1}_{\theta}\frac{\theta_{q^2}\bigl(\alpha q^{2(\epsilon_1-\epsilon_4-\sigma_3+
\sigma_4+1)},-q^{2(\epsilon_4+\sigma_4)+1},-\alpha q^{2(
\epsilon_2-\sigma_4)+1}\bigr)}{\theta_{q^2}\bigl(
\alpha z^{-1}q^{2(\epsilon_1-\epsilon_4)},-q^{2(\epsilon_3+\sigma_3
)+1},-\alpha q^{2(\epsilon_1-\sigma_3)+1}\bigr)}\cdot}\\
\frac{\bigl(zq^{-2(\epsilon_2-\epsilon_1)+2},z^{-1}q^{-2(\epsilon_2
-\epsilon_1)},q^{4\sigma_4+2};q^2\bigr)_{\infty}}{\bigl(q^{-2(\epsilon_2-\epsilon_1-
\sigma_3+\sigma_4)},q^{-2(\epsilon_2-\epsilon_1-\sigma_3-\sigma_4)+2};q^2
\bigr)_{\infty}}\breve R^{k,k+r^*}_{l,l+r^*}\Biggr\},
\end{multline} 
where 
\begin{equation*}
h_{\theta}=\sqrt{\frac{\theta_{q^2}\bigl(-q^{2(\epsilon_4+\sigma_4)+1},
-q^{2(\epsilon_4-\sigma_4)+1},-\alpha q^{2(\epsilon_2+\sigma_4)+1},-\alpha q^{2(
\epsilon_2-\sigma_4)+1}\bigr)}{\theta_{q^2}\bigl(-q^{2(\epsilon_3+
\sigma_3)+1},-q^{2(\epsilon_3-\sigma_3)+1},-\alpha q^{2(\epsilon_1+\sigma_3)+1},
-\alpha q^{2(\epsilon_1-\sigma_3)+1}\bigr)}},
\end{equation*}
\begin{multline}\label{E:hatR}
\hat R^{k,k+r^*}_{l,l+r^*}\equiv\hat R^{k,k+r^*}_{l,l+r^*}\bigl(z,\alpha;q^{2
\epsilon_1},q^{2\epsilon_2};q^{2\epsilon_3},q^{2\epsilon_4};q^{2\sigma_3},
q^{2\sigma_4}\bigr)=\\
=\hat{\kappa}_rz^{-l}q^{k+r-l}\frac{S_{k+r}\bigl(q^{2\epsilon_3},q^{2\sigma_3}
\bigr)}{S_k\bigl(q^{2\epsilon_4},q^{2\sigma_4}\bigr)}\frac{S_{-l}\bigl(
q^{-2\epsilon_1},\alpha q^{2\sigma_3}\bigr)}{S_{-l-r}\bigl(q^{-2\epsilon_2},
\alpha q^{2\sigma_4}\bigr)}\cdot\\
\frac{\bigl(\alpha q^{-2(l-k+r+\epsilon_2-\epsilon_4)+2},-\alpha zq^{-2(l+r+\epsilon_2-\sigma_3)+3},-\alpha z^{-1}q^{-2(l+r+\epsilon_2-\sigma_3)+1};q^2\bigr)_{\infty}}{
\bigl(\alpha z^{-1}q^{-2(l-k+\epsilon_1-\epsilon_4)},-\alpha q^{-2(l+\epsilon_1-\sigma_3)+1},-\alpha q^{-2(l+2r+2\epsilon_2-\epsilon_1-\sigma_3)+3};q^2\bigr)_{\infty}}\cdot\\
\shoveleft{
{}_8W_7\bigl(-\alpha q^{-2(l+2r+2\epsilon_2-\epsilon_1-\sigma_3)+1};-
\alpha q^{-2(l+r+
\epsilon_2+\sigma_4)+1},-\alpha q^{-2(l+r+\epsilon_2-\sigma_4)+1},}\\
-q^{-2(k+r+\epsilon_3-\sigma_3)+1},
zq^{-2(r+\epsilon_2-\epsilon_1)+2},z^{-1}q^{-2(r+\epsilon_2-\epsilon_1)};q^2,
-q^{2(k+r+\epsilon_3+\sigma_3)+1}\bigr) 
\end{multline}
and
\begin{multline}\label{E:breveR}
\breve R^{k,k+r^*}_{l,l+r^*}\equiv \breve R^{k,k+r^*}_{l,l+r^*}\bigl(z,\alpha;
q^{2\epsilon_1},q^{2\epsilon_2};q^{2\epsilon_3},q^{2\epsilon_4};q^{2\sigma_3},
q^{2\sigma_4}\bigr)=\\
\breve{\kappa}_rz^{-l}q^{k+r-l}
\sqrt{\frac{\bigl(-q^{2(k+
\epsilon_4-\sigma_4)+1},-q^{2(k+r+\epsilon_3+\sigma_3)+1};q^2\bigr)_{\infty}}{
\bigl(-q^{2(k+\epsilon_4+\sigma_4)+1},-q^{2(k+r+\epsilon_3-\sigma_3)
+1};q^2\bigr)_{\infty}}}\frac{S_{-l-r}\bigl(q^{-2\epsilon_2},\alpha q^{2
\sigma_4}\bigr)}{S_{-l}\bigl(q^{-2\epsilon_1},\alpha q^{2\sigma_3}\bigr)}\cdot\\
\frac{\bigl(\alpha q^{-2(l-k+\epsilon_1-\epsilon_4+\sigma_3-\sigma_4)
+2},-\alpha zq^{-2(l+\epsilon_1-\sigma_4)+3},-\alpha z^{-1}q^{-2(l+\epsilon_1-\sigma_4)+1};q^2\bigr)_{\infty}}{\bigl(\alpha z^{-1}
q^{-2(l-k+\epsilon_1-\epsilon_4)},-\alpha q^{-2(l+r+\epsilon_2-\sigma_4)+1},
-\alpha q^{-2(l+\epsilon_1+\sigma_3-2\sigma_4)+3};q^2\bigr)_{\infty}}\cdot\\
{}_8W_7\bigl(-\alpha q^{-2(l+\epsilon_1+\sigma_3-2\sigma_4)+1};-\alpha q^{-2(l+r+\epsilon_2-\sigma_4)+1},-\alpha q^{-2(l+\epsilon_1+\sigma_3)
+1},-q^{-2(k+\epsilon_4-\sigma_4)+1},\\
zq^{2(\sigma_4-\sigma_3+1)},z^{-1}q^{2(\sigma_4-\sigma_3)};q^2,-q^{2(k+r+
\epsilon_3+\sigma_3)+1}\bigr)
\end{multline}
with
\begin{equation*}
\begin{split}
\hat{\kappa}_r&=z^{-r}(-1)^rq^{-r^2-2r(\epsilon_2-\epsilon_1-\sigma_3)}\,\frac{\bigl(q^{2
(\epsilon_2-\epsilon_1-\sigma_3+\sigma_4)},q^{2(\epsilon_2-\epsilon_1-\sigma_3
-\sigma_4)};q^2\bigr)_r}{\bigl(z^{-1}q^{2(\epsilon_2-\epsilon_1)};q^2\bigr)_r},\\
\breve{\kappa}_r&=z^{-r}q^{2r\sigma_4-r}\frac{\bigl(zq^{2(\epsilon_2-\epsilon_1
+1)};q^2\bigr)_r}{\bigl(q^{2(\epsilon_2-\epsilon_1-\sigma_3+\sigma_4+1)};q^2
\bigr)_r}. 
\end{split}
\end{equation*}
The expressions obtained replacing $R^{k_1,k_2^*}_{l_1,l_2^*}$ by
$\hat R^{k_1,k_2^*}_{l_1,l_2^*}$ or $\breve R^{k_1,k_2^*}_{l_1,l_2^*}$
on the rhs of \eqref{E:A} and \eqref{E:B} will be denoted by $\hat A$, 
$\hat B$ and $\breve A$, $\breve B$.

Application of the contiguous relations [\cite{isra}:2.18] and [\cite{isra}:2.17] with $A^2\to-\alpha q^{-2(l+2r+2\epsilon_2-\epsilon_1-
\sigma_3)+3}$, $A\lambda^{-1}\to-q^{-2(k+r+\epsilon_3-\sigma_3)-1}$, $A\mu^{-1}
\to-\alpha q^{-2(l+r+\epsilon_2+\sigma_4)+1}$, $A\nu^{-1}\to-\alpha q^{-2(
l+r+\epsilon_2-\sigma_4)+1}$, $A\rho^{-1}\to zq^{-2(r+\epsilon_2-\epsilon_1)
+2}$, $A\sigma^{-1}\to z^{-1}q^{-2(r+\epsilon_2-\epsilon_1)}$ yields
\begin{equation*}
\hat A=\hat B=0
\end{equation*}
for $\vert q^{2(k+r+\epsilon_3+\sigma_3)+1}\vert<1$.
Use of relation [\cite{isra}:2.2] with $a\to-\alpha q^{-2(l+
\epsilon_1+\sigma_3-2\sigma_4)+1}$, $b\to-\alpha q^{-2(l+r+\epsilon_2-\sigma_4
)+1}$, $c\to-q^{-2(k+\epsilon_4-\sigma_4)+1}$, $d\to-\alpha q^{-2(l+\epsilon_1
+\sigma_3)+1}$, $e\to zq^{2(\sigma_4-\sigma_3+1)}$, $f\to z^{-1}q^{2(\sigma_4
-\sigma_3)}$ leads to
\begin{equation*}
\breve B=0
\end{equation*}
for $\vert q^{2(k+r+\epsilon_3+\sigma_3)+1}\vert<1$.
The relation [\cite{isra}:2.3] with $a\to-\alpha q^{-2(l+\epsilon_1+\sigma_3 
-2\sigma_4)+1}$, $b\to-\alpha q^{-2(l+r+\epsilon_2-\sigma_4)+3}$, $c\to-q^{-2(
k+\epsilon_4-\sigma_4)-1}$, $d\to-\alpha q^{-2(l+\epsilon_1+\sigma_3)+1}$,
$e\to zq^{2(\sigma_4-\sigma_3+1)}$, $f\to z^{-1}q^{2(\sigma_4-\sigma_3)}$
implies
\begin{equation*}
\breve A=0
\end{equation*}
for $\vert q^{2(k+r+\epsilon_3+\sigma_3)+1}\vert<1$. According to \eqref{E:RW3}, this implies $A=B=0$ for $\vert q^{2(k+r+\epsilon_3+\sigma_3)+1}
\vert<1$. Since $\sigma_3\geq0$, the above steps show 
\begin{equation}\label{E:AB0}
A=B=0.
\end{equation}
The expression \eqref{E:Rel1} for $R^{k,k+r^*}_{l,l+r^*}$ enjoys the
property
\begin{multline}\label{E:Rel1sym}
R^{k,k+r^*}_{l,l+r^*}=R^{k,k+r^*}_{l,l+r^*}\bigl(z,\alpha;q^{2\epsilon_1},
q^{2\epsilon_2};q^{2\epsilon_3},q^{2\epsilon_4};q^{2\sigma_3},q^{2\sigma_4}\bigr)=\\
z^{-(k+l+r)}R^{\,-l-r,-l^*}_{-k-r,-k^*}\bigl(z,\alpha;q^{-2\epsilon_3},q^{-2\epsilon_4};
q^{-2\epsilon_1},q^{-2\epsilon_2};\alpha q^{2\sigma_3},\alpha q^{2\sigma_4}\bigr).
\end{multline}
Therefore, the substitutions 
\begin{equation*}
\begin{split}
\hat R^{k,k+r^*}_{l,l+r^*}&\to\hat R^{\,-l-r,-l^*}_{-k-r,-k^*}\bigl(z,
\alpha;q^{-2\epsilon_3},q^{-2\epsilon_4};q^{-2\epsilon_1},q^{-2\epsilon_2};
\alpha q^{2\sigma_3},\alpha q^{2\sigma_4}\bigr),\\
\breve R^{k,k+r^*}_{l,l+r^*}&\to\breve R^{\,-l-r,-l^*}_{-k-r,
-k^*}\bigl(z,\alpha;q^{-2\epsilon_3},q^{-2\epsilon_4};q^{-2\epsilon_1},q^{-2
\epsilon_2};\alpha q^{2\sigma_3},\alpha q^{-2\sigma_4}\bigr)
\end{split}
\end{equation*}
on the rhs of \eqref{E:RW3} followed by multiplication with 
$z^{-(k+l+r)}$ yield a further expression for $R^{k,k+r^*}_{
l,l+r^*}\bigl(z,\alpha;q^{2\epsilon_1},q^{2\epsilon_2};q^{2\epsilon_3},q^{2
\epsilon_4};q^{2\sigma_3},q^{2\sigma_4}\bigr)$.
For the first replacement, the relations [\cite{isra}2.17] 
and [\cite{isra}:2.18] with $A^2\to-q^{2(k-r-\epsilon_3+2\epsilon_4+\sigma_3)+3}$,
$A\mu^{-1}\to-q^{2(k+\epsilon_4+\sigma_4)+1}$, $A\nu^{-1}\to-q^{2(k+\epsilon_4-
\sigma_4)+1}$, $A\rho^{-1}\to zq^{-2(r+\epsilon_2-\epsilon_1)+2}$, $A\sigma^{-1}
\to z^{-1}q^{-2(r+\epsilon_2-\epsilon_1)}$ and
$A\lambda^{-1}\to-\alpha q^{2(l+\epsilon_1+\sigma_3)+1}$ for [\cite{isra}:2.17],
$A\lambda^{-1}\to-\alpha q^{2(l+\epsilon_1+\sigma_3)-1}$ for [\cite{isra}:2.18]
can be employed. To the second replacement, the relations [\cite{isra}:2.2]
with $b\to-q^{2(k+\epsilon_4+\sigma_4)+1}$, $c\to-\alpha q^{2(l+r+\epsilon_2+
\sigma_4)+3}$ and [\cite{isra}:2.3] with $b\to -q^{2(k+\epsilon_4+\sigma_4)+3}$,
$c\to-\alpha q^{2(l+r+\epsilon_2+\sigma_4)-1}$ and $a\to-q^{2(k+r+\epsilon_3-
\sigma_3+2\sigma_4)+1}$, $d\to-q^{2(k+r+\epsilon_3-\sigma_3)+1}$, $e\to zq^{2(
\sigma_4-\sigma_3+1)}$, $f\to z^{-1}q^{2(\sigma_4-\sigma_3)}$ for both apply.
Making use of \eqref{E:RW3}, this yields 
\begin{multline}\label{E:AD}
-q^{-2(k+\epsilon_4)+2}\Bigl\{\bigl(1-\alpha z^{-1}q^{2(l-k+\epsilon_1-
\epsilon_4)}\bigr)s_{l+r}\bigl(q^{2\epsilon_2},q^{2\sigma_2}\bigr)R^{k,k+r^*}_{
l,l+r^*}\\
-q^{-1}\bigl(1-\alpha q^{2(l-k+r+\epsilon_2-\epsilon_4+1)}\bigr)s_l\bigl(
q^{2\epsilon_1},q^{2\sigma_1}\bigr)R^{k,k+r^*}_{l+1,l+r+1^*}\\
+q^{2(l+\epsilon_1)}\bigl(1-z^{-1}q^{2(r+\epsilon_2-\epsilon_1)}\bigr)
s_{-k}\bigl(q^{-2\epsilon_4},q^{2\sigma_4}\bigr)R^{k-1,k+r^*}_{l,l+r+1^*}\Bigr\}=
\bigl(1-z^{-1}q^{-2(r+\epsilon_2-\epsilon_1+1)}\bigr)^{-1}\\
\cdot\Bigl\{\bigl(1-\alpha z^{-1}q^{2(l-k+\epsilon_1-\epsilon_4)}\bigr)A_{k-1,r+1,l}-\bigl(1-\alpha q^{2(
l-k+r+\epsilon_2-\epsilon_4+1)}\bigr)D_{k-1,r+1,l}\Bigr\}=0
\end{multline}
for $\vert q^{2(l+\epsilon_1-\sigma_3)+1}\vert>1$ and 
\begin{multline}\label{E:BC}
-q^{-2(k+r+\epsilon_3)+1}\Bigl\{\bigl(1-\alpha zq^{2(l-k+\epsilon_1-\epsilon_4)}
\bigr)s_{l+r-1}\bigl(q^{2\epsilon_2},q^{2\sigma_2}\bigr)R^{k,k+r^*}_{l,l+r^*}\\
-q\bigl(1-\alpha q^{2(l-k+r+\epsilon_2-\epsilon_4-1)}\bigr)s_{l-1}\bigl(
q^{2\epsilon_1},q^{2\sigma_1}\bigr)R^{k,k+r^*}_{l-1,l+r-1^*}\\
+q^{2(l+\epsilon_1)}\bigl(1-zq^{2(r+\epsilon_2-\epsilon_1)}\bigr)s_{-k-1}
\bigl(q^{-2\epsilon_4},q^{2\sigma_4}\bigr)R^{k+1,k+r^*}_{l,l+r-1^*}\Bigr\}=
\bigl(1-z^{-1}q^{2(r+\epsilon_2-\epsilon_1-1)}\bigr)^{-1}\\
\cdot
\Bigl\{\bigl(1-\alpha zq^{2(l-k+\epsilon_1-\epsilon_4)}\bigr)B_{k,r-1,l}-\bigl(
1-\alpha q^{2(l-k+r+\epsilon_2-\epsilon_4-1)}\bigr)C_{k,r-1,l}\Bigr\}=0
\end{multline}
for $\vert q^{2(l+\epsilon_1-\sigma_3)-1}\vert>1$. For $\vert q^{2(l+
\epsilon_1+\sigma_3)+1}\vert<1$, the transformation
[\cite{GR}:III.36] with $a\to-q^{-2(k+r+\epsilon_3+\sigma_3-2\sigma_4)+1}$, 
$b\to-\alpha q^{-2(l+r+\epsilon_2-\sigma_4)+1}$, $c\to-q^{-2(k+r+\epsilon_3
+\sigma_3)+1}$, $d\to-q^{-2(k+\epsilon_4-\sigma_4)+1}$, $e\to zq^{2(\sigma_4
-\sigma_3+1)}$, $f\to z^{-1}q^{2(\sigma_4-\sigma_3)}$ yields
\begin{multline*}
R^{k,k+r^*}_{l,l+r^*}\bigl(z,\alpha;q^{2\epsilon_1},q^{2\epsilon_2};
q^{2\epsilon_3},q^{2\epsilon_4};q^{2\sigma_3},q^{2\sigma_4}\bigr)=\\
\bigl(q^{-2(r+\epsilon_2-\epsilon_1+\sigma_3-\sigma_4)+2};q^2\bigr)_r 
\kappa_rz^{-k-r}q^{-k-r}\frac{S_{-k}\bigl(q^{-2\epsilon_4},q^{2\sigma_4}
\bigr)}{S_{-k-r}\bigl(q^{-2\epsilon_3},q^{2\sigma_3}\bigr)}\cdot\\
q^l\sqrt{\frac{\bigl(-\alpha q^{2(l+\epsilon_1+\sigma_3)+1},-\alpha q^{2(l+r+
\epsilon_2-\sigma_4)+1};q^2\bigr)_{\infty}}{\bigl(-\alpha q^{2(l+\epsilon_1-
\sigma_3)+1},-\alpha q^{2(l+r+\epsilon_2+\sigma_4)+1};q^2\bigr)_{\infty}}}
\frac{\bigl(q^{4\sigma_4+2};q^2\bigr)_{\infty}}{\bigl(q^{-2(\epsilon_2-
\epsilon_1-\sigma_3-\sigma_4)+2};q^2\bigr)_{\infty}}\cdot\\
\frac{\bigl(\alpha q^{2(l-k+\epsilon_1-\epsilon_4-\sigma_3+\sigma_4+1)},-z
q^{-2(k+r+\epsilon_3-\sigma_4)+3},-z^{-1}q^{-2(k+r+\epsilon_3-\sigma_4)+1}
;q^2\bigr)_{\infty}}{\bigl(\alpha z^{-1}q^{2(l-k+\epsilon_1-
\epsilon_4)},-q^{-2(k+\epsilon_4-\sigma_4)+1},-q^{-2(k+r+\epsilon_3+\sigma_3
-2\sigma_4)+3};q^2\bigr)_{\infty}}\cdot\\
_8W_7\bigl(-q^{-2(k+r+\epsilon_3+\sigma_3-2\sigma_4)+1};-q^{-2(k+r+\epsilon_3
+\sigma_3)+1},-q^{-2(k+\epsilon_4-\sigma_4)+1},\\
-\alpha q^{-2(l+r+\epsilon_2-\sigma_4)+1},
zq^{2(\sigma_4-\sigma_3+1)},z^{-1}q^{2(\sigma_4-\sigma_3)};q^2,-\alpha q^{2(l+
\epsilon_1+\sigma_3)+1}\bigr).
\end{multline*}
The relations [\cite{isra}:2.2] and [\cite{isra}:2.3] with
$a\to -q^{-2(k+r+\epsilon_3+\sigma_3-2\sigma_4)+1}$, $c\to-q^{-2(k+\epsilon_4-
\sigma_4)+1}$, $d\to-q^{-2(k+r+\epsilon_3+\sigma_3)+1}$, 
$e\to zq^{2(\sigma_4-\sigma_3+1)}$, $f\to z^{-1} q^{2(\sigma_4-\sigma_3)}$ and
$b\to-\alpha q^{-2(l+r+\epsilon_2-\sigma_4)+3}$ for [\cite{isra}:2.2],
$b\to-\alpha q^{-2(l+r+\epsilon_2-\sigma_4)+1}$ for [\cite{isra}:2.3] give rise to \eqref{E:BC} and \eqref{E:AD}, respectively. Here  
$\vert q^{2(l+\epsilon_1+\sigma_3)-1}\vert<1$ is required for \eqref{E:BC}
and $\vert q^{2(l+\epsilon_1+\sigma_3)+1}\vert<1$ for \eqref{E:AD}.
Since $\sigma_3\geq0$, this proves the relations \eqref{E:AD} and 
\eqref{E:BC}. Combined with the results \eqref{E:AB0}, the 
latter imply
\begin{equation}\label{E:CD}
C=D=0
\end{equation}
unless
\begin{equation*}\label{E:eps24}
\alpha=1,\qquad \epsilon_2-\epsilon_4=m,\;m\in\mathbb Z.
\end{equation*}
In this case, the combinations $C_{k,r-1,k-r-m+1}$ and $D_{k-1,r+1,k-r-m-1}$ are left undetermined. 
Taking into account the results \eqref{E:AB0}, \eqref{E:AD} and \eqref{E:BC}, the relation \eqref{E:ABCD0} shows
\begin{equation*}
C_{k,r-1,k-r-m+1}=D_{k-1,r+1,k-r-m-1}=0,
\end{equation*}
which proves the result \eqref{E:CD} in general.
This establishes the equations \eqref{E:intwrite}.

\subsection{Further Properties}\label{A:Rprop}

For $\vert q^{-2(k+\epsilon_4-\sigma_4)+1}\vert<1$, the transformation 
[\cite{GR}:III.36] with $a\to-\alpha q^{2(l+r+2\epsilon_1-\epsilon_2+
\sigma_4)+1}$, $b\to-\alpha q^{2(l+\epsilon_1-\sigma_3)+1}$, $c\to-q^{2(k+
\epsilon_4+\sigma_4)+1}$, $d\to-\alpha q^{2(l+\epsilon_1+\sigma_3)+1}$,
$e\to zq^{-2(r+\epsilon_2-\epsilon_1)+2}$, $f\to z^{-1}q^{-2(r+\epsilon_2-
\epsilon_1)}$ gives rise to
\begin{multline}\label{E:RW2}
R^{k,k+r^*}_{l,l+r^*}\bigl(z,\alpha;q^{2\epsilon_1},q^{2\epsilon_2};q^{2\epsilon_3
},q^{2\epsilon_4};q^{2\sigma_3},q^{2\sigma_4}\bigr)=\\
\kappa_r\bigl(q^{-2(r+\epsilon_2-\epsilon_1+\sigma_3-\sigma_4)+2},q^{-2(r+
\epsilon_2-\epsilon_1-\sigma_3-\sigma_4)+2};q^2\bigr)_r\cdot\\
z^{-k-r}q^{l-k-r}\frac{S_{-k}\bigl(q^{-2\epsilon_4},q^{2\sigma_4}\bigr)}{
S_{-k-r}\bigl(q^{-2\epsilon_3},q^{2\sigma_3}\bigr)}\frac{S_{l+r}\bigl(
q^{2\epsilon_2},\alpha q^{2\sigma_4}\bigr)}{S_l\bigl(q^{2\epsilon_1},
\alpha q^{2\sigma_3}\bigr)}\cdot\\
\frac{\bigl(\alpha q^{2(l-k-r+\epsilon_1-\epsilon_3+1)},-\alpha zq^{2(l+\epsilon_1+\sigma_4)+3},-\alpha z^{-1}q^{2(l+\epsilon_1+\sigma_4)+1}
;q^2\bigr)_{\infty}}{\bigl(\alpha z^{-1}q^{2(l-k+\epsilon_1-\epsilon_4)},-\alpha q^{2(l+r+\epsilon_2+\sigma_4)+1},-\alpha q^{2(l-r+2\epsilon_1-\epsilon_2+
\sigma_4)+3};q^2\bigr)_{\infty}}\cdot\\
{}_8W_7\bigl(-\alpha q^{2(l-r+2\epsilon_1-\epsilon_2+\sigma_4)+1};-\alpha q^{2(l+\epsilon_1+\sigma_3)+1},-\alpha q^{2(l+\epsilon_1-\sigma_3)+1},
-q^{2(k+\epsilon_4+\sigma_4)+1},\\
zq^{-2(r+\epsilon_2-\epsilon_1)+2},z^{-1}q^{-2(r+\epsilon_2-\epsilon_1)};q^2,
-q^{-2(k+\epsilon_4-\sigma_4)+1}\bigr).
\end{multline}
The expression on the rhs is manifestly invariant under 
$\sigma_3\to-\sigma_3$. Since the coefficients of the R-elements on the lhs
of \eqref{E:BC} share this property, the invariance extends to all values
of $k$. This establishes the symmetry \eqref{E:Rsig1+-}.

In the case $\vert q^{-2(k+\epsilon_4-\sigma_4)+1}\vert<1$, equation \eqref{E:RW1} yields
\begin{multline*}
\mathring{\check R}^{k,k+r^*}_{l,l+r^*}=R^{k+r,k^*}_{l+r,l^*}\bigl(z,\alpha;
q^{2\epsilon_2},q^{2\epsilon_1};q^{2\epsilon_4},q^{2\epsilon_3};q^{2\sigma_4},
q^{-2\sigma_3}\bigr)=\\
\frac{\alpha^rz^{-k-r}q^{l-k-2r\alpha_3}}{\bigl(zq^{2(r+\epsilon_2-\epsilon_1+1)};
q^2\bigr)_{-r}}\frac{S_{-k-r}\bigl(q^{-2\epsilon_3},q^{2\sigma_3}\bigr)}{S_{-k}
\bigl(q^{-2\epsilon_4},q^{2\sigma_4}\bigr)}\frac{S_{l}\bigl(q^{2\epsilon_1},
\alpha q^{2\sigma_3}\bigr)}{S_{l+r}\bigl(q^{2\epsilon_2},\alpha q^{2\sigma_4}
\bigr)}\cdot\\
\frac{\bigl(\alpha q^{2(l-k+\epsilon_1-\epsilon_4-\sigma_3-\sigma_4+1)},-q^{-2
(k+\epsilon_4-\sigma_4)+1},-\alpha zq^{2(l+r+\epsilon_2-\sigma_3)+3};q^2\bigr)_{
\infty}}{\bigl(\alpha z^{-1}q^{2(l-k+\epsilon_1-\epsilon_4)},-q^{-2(k+r+
\epsilon_3+\sigma_3)+1},-\alpha q^{2(l+r+\epsilon_2-2\sigma_3-\sigma_4)+3};q^2
\bigr)_{\infty}}\cdot\\
\frac{\bigl(-\alpha z^{-1}q^{2(l+r+\epsilon_2-\sigma_3)+1},q^{-4\sigma_3+2};q^2
\bigr)_{\infty}}{\bigl(-\alpha q^{2(l+\epsilon_1-\sigma_3)+1},q^{2(\epsilon_2
-\epsilon_1-\sigma_3+\sigma_4+1)};q^2\bigr)_{\infty}\bigl(q^{2(\epsilon_2-\epsilon_1
-\sigma_3-\sigma_4+1)};q^2\bigr)_r}\cdot\\
\shoveleft{
_8W_7\bigl(-\alpha q^{2(l+r+\epsilon_2-2\sigma_3-\sigma_4)+1};-\alpha q^{2(l+
\epsilon_1-\sigma_3)+1},-\alpha q^{2(l+r+\epsilon_2-\sigma_4)+1},}\\
-q^{2(k+r+\epsilon_3-\sigma_3)+1},
zq^{-2(\sigma_3+\sigma_4)+2},z^{-1}q^{-2(\sigma_3+\sigma_4)};q^2,q^{-2(k+
\epsilon_4-\sigma_4)+1}\bigr).
\end{multline*}
Application of [\cite{GR}:III.36] with $a\to-\alpha q^{2(l+r+\epsilon_2-2
\sigma_3-\sigma_4)+1}$, $b\to-\alpha q^{2(l+\epsilon_1-\sigma_3)+1}$, $c\to-
q^{2(k+r+\epsilon_3-\sigma_3)+1}$, $d\to-\alpha q^{2(l+r+\epsilon_2-\sigma_4)
+1}$, $e\to zq^{-2(\sigma_3+\sigma_4)+2}$, $f\to z^{-1}q^{-2(\sigma_3+
\sigma_4)}$  leads to
\begin{multline}\label{E:RmchR}
\bigl(zq^{2(\epsilon_1-\epsilon_2+1)},z^{-1}q^{2(\epsilon_1-
\epsilon_2)},q^{2(\epsilon_2-\epsilon_1-\sigma_3+\sigma_4+1)},q^{2(\epsilon_2-
\epsilon_1-\sigma_3-\sigma_4+1)};q^2\bigr)_{\infty}\cdot\\
\shoveright{q^{2\sigma_4}\theta_{q^2}\bigl(q^{4\sigma_4+2}\bigr)\mathring{
\check R}^{k,k+r^*}_{l,l+r^*}}\\
+\bigl(zq^{2(\sigma_4-\sigma_3+1)},z^{-1}q^{2(\sigma_4-\sigma_3)},q^{-2(
\epsilon_2-\epsilon_1+\sigma_3+\sigma_4)+2},q^{-2(\epsilon_2-\epsilon_1-
\sigma_3+\sigma_4)+2};q^2\bigr)_{\infty}\cdot\\
\shoveright{q^{-2\sigma_4}\theta_{q^2}\bigl(q^{2(\epsilon_2-\epsilon_1-\sigma_3
-\sigma_4+1)}\bigr)\check R^{k,k+r^*}_{l,l+r^*}=}\\
\shoveleft{
\bigl(zq^{-2(\sigma_3+\sigma_4)+2},z^{-1}q^{-2(\sigma_3+\sigma_4)};q^2\bigr)_{
\infty}\cdot}\\
\shoveright{
\alpha ^rz^{-k-r}q^{l-k}\frac{S_{-k}\bigl(q^{-2\epsilon_4},q^{2\sigma_4}
\bigr)}{S_{-k-r}\bigl(q^{-2\epsilon_3},q^{2\sigma_3}\bigr)}\frac{S_{l+r}
\bigl(q^{2\epsilon_2},\alpha q^{2\sigma_4}\bigr)}{S_l\bigl(q^{2\epsilon_1},
\alpha q^{2\sigma_3}\bigr)}\cdot}\\
\Biggl\{q^{2(\sigma_4-r\sigma_3)}
\frac{\bigl(\alpha q^{2(l-k+\epsilon_1-\epsilon_4-\sigma_3+\sigma_4+1)},
-q^{-2(k+r+\epsilon_3-\sigma_3)+1},-\alpha q^{2(l+\epsilon_1+\sigma_3)+1};
q^2\bigr)_{\infty}}{\bigl(\alpha z^{-1}q^{2(l-k+\epsilon_1-\epsilon_4)},
-q^{-2(k+\epsilon_4-\sigma_4)+1},-\alpha q^{2(l+r+\epsilon_2+\sigma_4)+1};
q^2\bigr)_{\infty}}\cdot\\
\bigl(zq^{2(\epsilon_1-\epsilon_2+1)},z^{-1}q^{2(\epsilon_1-\epsilon_2)},q^{2(r+
\epsilon_2-\epsilon_1-\sigma_3+\sigma_4+1)},q^{4\sigma_4+2};q^2\bigr)_{\infty}\bigl(
zq^{2(\epsilon_2-\epsilon_1+1)};q^2\bigr)_r\cdot\\
{}_4\phi_3\biggl(\genfrac{}{}{0pt}{}{-q^{-2(k+\epsilon_4-\sigma_4)+1},\,-
\alpha q^{2(l+r+\epsilon_2+\sigma_4)+1},\,zq^{2(\sigma_4-\sigma_3+1)},\,z^{-1}
q^{2(\sigma_4-\sigma_3)}}{\alpha q^{2(l-k+\epsilon_1-\epsilon_4-\sigma_3+
\sigma_4+1)},\,q^{2(r+\epsilon_2-\epsilon_1-\sigma_3+\sigma_4+1)},\,q^{4\sigma_4
+2}};q^2,q^2\biggr)\\
+\bigl(\alpha q^{2(l-k-r+\epsilon_1-\epsilon_3+1)},
zq^{2(\sigma_4-\sigma_3+1)},z^{-1}q^{2(\sigma_4-\sigma_3)},q^{-2(\epsilon_2-
\epsilon_1-\sigma_3+\sigma_4)+2};q^2\bigr)_{\infty}\cdot\\
\frac{\theta_{q^2}\bigl(q^{2(\epsilon_2-\epsilon_1-\sigma_3-\sigma_4+1)}\bigr)
\bigl(q^{-2(r+\epsilon_2-\epsilon_1-\sigma_3+\sigma_4)+2};q^2\bigr)_r}{\bigl(
q^{2(r+\epsilon_2-\epsilon_1-\sigma_3-\sigma_4)};q^2\bigr)_{\infty}\bigl(zq^{
-2(r+\epsilon_2-\epsilon_1)+2};q^2\bigr)_r}\,z^rq^{2(r-1)\sigma_4}\cdot\\
{}_4\phi_3\biggl(\genfrac{}{}{0pt}{}{-q^{-2(k+r+\epsilon_3-\sigma_3)+1},\,
-\alpha q^{2(l+\epsilon_1+\sigma_3)+1},\,zq^{-2(r+\epsilon_2-\epsilon_1)+2},
\,z^{-1}q^{-2(r+\epsilon_2-\epsilon_1)}}{\alpha q^{2(l-k-r+\epsilon_1-\epsilon_3
+1)},\,q^{-2(r+\epsilon_2-\epsilon_1-\sigma_3+\sigma_4)+2},\,q^{-2(r+\epsilon_2
-\epsilon_1-\sigma_3-\sigma_4)+2}};q^2,q^2\biggr)\Biggr\}\\
=\bigl(zq^{-2(\sigma_3+\sigma_4+1)},z^{-1}q^{-2(\sigma_3+\sigma_4)},q^{-2(
\epsilon_2-\epsilon_1+\sigma_3-\sigma_4)+2},q^{-2(\epsilon_2-\epsilon_1-
\sigma_3-\sigma_4)+2};q^2\bigr)_{\infty}\cdot\\
q^{2\sigma_4}\theta_{q^2}\bigl(q^{2(\epsilon_2-\epsilon_1-\sigma_3+\sigma_4)}
\bigr)R^{k,k+r^*}_{l,l+r^*}.
\end{multline}
Here equation \eqref{E:Rel1} with $q^{2\sigma_4}\to-q^{-2\sigma_4}$ has been
employed to express $\check R^{k,k+r^*}_{l,l+r^*}$ in terms of two $_4\phi_3$-series. Making use of \eqref{E:thetaprop} and \eqref{E:I8}, the last
step follows easily from \eqref{E:Rel1}. Replacing $q^{2\sigma_3}\to q^{-2\sigma_3}$ in \eqref{E:RmchR} yields the relation \eqref{E:RmR}
for $q^{-2(k+\epsilon_4-\sigma_4)+1}<1$.

The equations \eqref{E:AB0} and \eqref{E:AD} imply
\begin{multline*}
\frac{1}{1-z^{-1}q^{-2(r+\epsilon_2-\epsilon_1)}}\Bigl\{
\bigl(1-\alpha zq^{2(l-k+\epsilon_1-\epsilon_4)}\bigr)D_{k,r,l}-\bigl(
1-\alpha q^{2(l-k-r+\epsilon_1-\epsilon_3)}A_{k,r,l}\bigr)\Bigr\}=\\
-q^{-2(k+\epsilon_4)-1}\Bigl\{\bigl(1-\alpha zq^{2(l-k+\epsilon_1-\epsilon_4)}
\bigr)s_l\bigl(q^{2\epsilon_1},q^{2\sigma_1}\bigr)R^{k+1,k+r^*}_{l+1,l+r^*}\\
-q\bigl(1-\alpha q^{2(l-k-r+\epsilon_1-\epsilon_3)}\bigr)s_{l+r-1}\bigl(
q^{2\epsilon_2},q^{2\sigma_2}\bigr)R^{k+1,k+r^*}_{l,l+r-1^*}\\
+q^{2(l+r+\epsilon_2)}\bigl(1-zq^{-2(r+\epsilon_2-\epsilon_1)+2}\bigr)s_{-k-r-1}
\bigl(q^{-2\epsilon_3},q^{2\sigma_3}\bigr)R^{k+1,k+r+1^*}_{l,l+r^*}\Bigr\}=0.
\end{multline*}
With the substitutions $k\to k+r-1$, $l\to l+r-1$, $r\to-r+1$, $\epsilon_1
\leftrightarrow\epsilon_2$, $\epsilon_3\leftrightarrow\epsilon_4$,
$q^{2\sigma_3}\leftrightarrow q^{2\sigma_4}$ and the definition of
$\mathring R^{k,k+r^*}_{l,l+r^*}$ by \eqref{E:mrRel}, this yields
\begin{multline}\label{E:mRBC}
\bigl(1-\alpha zq^{2(l-k+\epsilon_1-\epsilon_4)}\bigr)s_{l+r-1}\bigl(q^{2
\epsilon_2},q^{2\sigma_2}\bigr)\mathring R^{k,k+r^*}_{l,l+r^*}\\
-q\bigl(1-\alpha q^{2(l-k+r+\epsilon_2-\epsilon_4-1)}\bigr)s_{l-1}\bigl(
q^{2\epsilon_1},q^{2\sigma_1}\bigr)\mathring R^{k,k+r^*}_{l-1,l+r-1^*}\\
+q^{2(l+\epsilon_1)}\bigl(1-zq^{2(r+\epsilon_2-\epsilon_1)}\bigr)s_{-k-1}
\bigl(q^{-2\epsilon_4},q^{2\sigma_4}\bigr)\mathring R^{k+1,k+r^*}_{l,l+r-1^*}=0.
\end{multline}
The equations \eqref{E:BC}, the same equation with $q^{2\sigma_4}\to q^{-2\sigma_4}$ and \eqref{E:mRBC} can be combined to extend \eqref{E:RmR}
to all values of $k$.

\section{Contiguous relations for $\Xi^{(r,k)}$}\label{A:thetacont}

The proof of Corollary \ref{C:tausum5} and \ref{C:tausum6} in subsection
\ref{S:tauev} involves the
contiguous relations \eqref{E:Xitheta} and \eqref{E:Xiz} for $\Xi^{(r,k)}$
derived in subsection \ref{S:Xitheta} and \ref{S:Xiz}, respectively.

\subsection{The relation \eqref{E:Xitheta}}\label{S:Xitheta}
Throughout this section, the $e^{i\theta}$-dependence of the quantities
introduced by \eqref{E:rho1b}, \eqref{E:ardef}, \eqref{E:varsigma},
\eqref{E:ardef2}, \eqref{E:Xidef} and Definition \ref{D:taudef1} will be indicated writing $\rho^{(r)\pm}_l(e^{i\theta})$, $a^{\pm}_r(
e^{i\theta})$, $\varsigma^{(r)}_l(e^{i\theta})$, $a_r(e^{i\theta})$, ${\Xi}^{(r,k)}(e^{i\theta})$ and $\boldsymbol{\tau}^{(r,k)}(e^{i\theta})$. 
The proof involves the finite sums
\begin{equation}\label{E:taufin}
\boldsymbol{\tau}^{(r,k;N,M)}(e^{i\theta})\equiv a_r(e^{i\theta})\sum_{l=-M}^N\varsigma^{
(r)}_l(e^{i\theta})\mathsf R^{k,k+r^*}_{l,l+r^*}.
\end{equation}

The contiguous relations [\cite{isra}:2.2] and [\cite{isra}:2.3] for $_8W_7$-series with the specifications $b\to Aq^2$, $c\to B$, $d\to C$,
$e\to\tfrac{aq^2}{D}$, $f\to\tfrac{aq^2}{E}$ and 
$b\to Aq^2$, $c\to C$,
$d\to B$, $e\to\tfrac{aq^2}{D}$, $f\to\tfrac{aq^2}{E}$ provide 
relations among $_3\phi_2$-series when specialized to $a=0$. A combination
of these suited for use in this subsection is given by
\begin{multline}\label{E:3phi2cont1}
B\bigl(1-\tfrac{Aq^2}{B}\bigr)\bigl(1-\tfrac{DE}{ABCq^2}\bigr)\cdot\\
\Biggl\{(1-C){}_3\phi_2
\biggl(\genfrac{}{}{0pt}{}{A,\,B,\,Cq^2}{D,\,E};q^2,\frac{DE}{ABCq^2}\biggr)
-(A-C){}_3\phi_2\biggl(\genfrac{}{}{0pt}{}{A,\,B,\,C}{D,\,E};q^2,\frac{DE}{
ABC}\biggr)\Biggr\}\\
=B(1-A)\bigl(1-\tfrac{D}{B}\bigr)\bigl(1-\tfrac{E}{B}\bigr)\,{}_3\phi_2
\biggl(\genfrac{}{}{0pt}{}{Aq^2,\,Bq^{-2},\,C}{D,\,E};q^2,\frac{DE}{ABC}\biggr)\\
-Aq^2(1-A)\bigl(1-\tfrac{D}{Aq^2}\bigr)\bigl(1-\tfrac{E}{Aq^2}\bigr)\,
{}_3\phi_2\biggl(\genfrac{}{}{0pt}{}{A,\,B,\,C}{D,\,E};q^2,\frac{DE}{ABC}
\biggr).
\end{multline}
The contiguous relation
\begin{multline}\label{E:3phi2cont4}
\frac{(1-A)(1-B)(1-C)}{(1-D)(1-E)\bigl(1-Eq^2\bigr)}\frac{DE}{ABC}\,{}_3\phi_2
\biggl(\genfrac{}{}{0pt}{}{Aq^2,\,Bq^2,\,Cq^2}{Dq^2,\,Eq^4};q^2,\frac{DE}{
ABC}\biggr)=\\
{}_3\phi_2\biggl(\genfrac{}{}{0pt}{}{A,\,B,\,C}{D,\,E};q^2,\frac{DE}{ABC}
\biggr)-{}_3\phi_2\biggl(\genfrac{}{}{0pt}{}{A,\,B,\,C}{D,\,Eq^2};q^2,\frac{
DEq^2}{ABC}\biggr)
\end{multline}
is a direct consequence of the definition of the $_m\phi_n$-series by 
\eqref{E:phidef}. A further relation for $_3\phi_2$-series is obtained from
[\cite{isra}:2.2] setting $b\to\tfrac{aq^2}{D}$, $c\to\tfrac{aq^2}{E}$,
$d\to A$, $e\to B$, $f\to C$ and specializing to $a=0$.
Combined with equation \eqref{E:3phi2cont4} it gives rise
to the contiguous relation
\begin{multline}\label{E:3phi2cont2}
\frac{D(1-A)(1-B)(1-C)
\bigl(1-\frac{E}{A}\bigr)\bigl(1-\frac{E}{B}\bigr)\bigl(1-\frac{E}{C}
\bigr)}{(1-D)(1-E)\bigl(1-Eq^2\bigr)}\cdot\\
\shoveright{{}_3\phi_2\biggl(\genfrac{}{}{0pt}{}{
Aq^2,\,Bq^2,\,Cq^2}{Dq^2,\,Eq^4};q^2,\frac{DE}{ABC}\biggr)=}\\
-\frac{\frac{ABC}{D}(1-E)
\bigl(1-\frac{D}{A}\bigr)\bigl(1-\frac{D}{B}\bigr)\bigl(1-\frac{D}{C}
\bigr)}{1-D}{}_3\phi_2\biggl(\genfrac{}{}{0pt}{}{A,\,B,\,C}{Dq^2,\,E};q^2,
\frac{DEq^2}{ABC}\biggr)\\
+\Bigl\{D(1-E)\bigl(1+\tfrac{ABC}{D^2}\bigr)-BC(1-A)\bigl(1+\frac{E}{BC}
\bigr)-A\bigl(1-\tfrac{E}{A}\bigr)(B+C)\Bigr\}\cdot\\
{}_3\phi_2\biggl(\genfrac{}{}{0pt}{}{A,\,B,\,C}{D,\,E};q^2,\frac{DE}{ABC}
\biggr).
\end{multline}
Another relation for $_3\phi_2$-series follows from [\cite{isra}:2.2]
with $b\to A$, $c\to \tfrac{aq^2}{D}$, $d\to B$, $e\to C$, 
$f\to\tfrac{aq^2}{E}$ setting $a\to0$.
Supplemented by the relation [\cite{gupta}:2.4] with $a\to Aq^{-2}$,
$b\to Bq^{-2}$, $c\to Cq^{-2}$, $d\to Dq^{-2}$, $e\to Eq^{-2}$ and
equation \eqref{E:3phi2cont4} with $A\to Aq^{-2}$, $B\to Bq^{-2}$, $C\to C
q^{-2}$, $D\to Dq^{-2}$, $E\to Eq^{-4}$ it leads to
\begin{multline}\label{E:3phi2cont3}
\bigl(1-Dq^{-2}\bigr)\bigl(1-Eq^{-2}\bigr)\bigl(1-Eq^{-4}\bigr){}_3\phi_2
\biggl(\genfrac{}{}{0pt}{}{Aq^{-2},\,Bq^{-2},\,Cq^{-2}}{Dq^{-2},\,Eq^{-4}}
;q^2,\frac{DE}{ABC}\biggr)=\\
-\frac{Eq^{-2}\bigl(1-Eq^{-4}\bigr)\bigl(1-\frac{D}{A}\bigr)\bigl(1-\frac{
D}{B}\bigr)\bigl(1-\frac{D}{C}\bigr)}{1-D}{}_3\phi_2\biggl(\genfrac{}{}{0pt}{}{
A,\,B,\,C}{Dq^2,\,E};q^2,\frac{DEq^2}{ABC}\biggr)\\
-\Bigl\{\bigl(1-Eq^{-4}\bigr)\bigl(1+\tfrac{D^2E}{ABCq^2}\bigr)-Dq^{-2}
\bigl(1-\tfrac{E}{Aq^2}\bigr)\bigl(1+\tfrac{E}{BC}\bigr)-\tfrac{DE}{ABCq^2}
\bigl(1-Aq^{-2}\bigr)(B+C)\Bigr\}\cdot\\
{}_3\phi_2\biggl(\genfrac{}{}{0pt}{}{A,\,B,\,C}{D,\,E};q^2,\frac{DE}{ABC}
\biggr).
\end{multline}
Relation \eqref{E:3phi2cont1} with $A\to q^{2(r+\epsilon_2-\epsilon_1)+1}e^{
i\theta}$, $B\to q^{2(r+\epsilon_2-\epsilon_1)+1}e^{-i\theta}$, $C\to-
q^{-2(l+\epsilon_1\mp\sigma_1)+1}$, $D\to q^{2(r+\epsilon_2-\epsilon_1\pm
\sigma_1+\sigma_2+1)}$, $E\to q^{2(r+\epsilon_2-\epsilon_1\pm\sigma_1-\sigma_2
+1)}$ implies 
\begin{multline}\label{E:rhoco1}
\bigl(1-q^{\pm2\sigma_1+2\sigma_2+1}e^{i\theta}\bigr)\bigl(1-q^{\pm2\sigma_1-2
\sigma_2+1}e^{i\theta}\bigr)
\bigl(1-q^{2(r+\epsilon_2-\epsilon_1)+1}e^{i
\theta}\bigr)\rho^{(r)\pm}_l\bigl(q^{2}e^{i\theta}\bigr)\\
=q^{-2(l+\epsilon_1\mp\sigma_1)+2}\bigl(1-q^2e^{2i\theta}\bigr)s_{l-1}\bigl(
q^{2\epsilon_1},q^{2\sigma_1}\bigr)s_{l+r-1}\bigl(q^{2\epsilon_2},q^{2\sigma_2}
\bigr)\rho^{(r)\pm}_{l-1}\bigl(e^{i\theta}\bigr)\\
-\Bigl\{q^{-2(l+\epsilon_1\mp\sigma_1)+1}\bigl(1-q^2e^{2i\theta}\bigr)\bigl(
1+q^{2(l+\epsilon_1+\sigma_2)}e^{i\theta}\bigr)\bigl(1+q^{2(l+r+\epsilon_2-
\sigma_2)-1}\bigr)\\
+q^{\pm2\sigma_1+2\sigma_2+3}e^{3i\theta}\bigl(1-q^{2(\sigma_1-\sigma_2)-1}
e^{-i\theta}\bigr)\bigl(1-q^{-2(\sigma_1+\sigma_2)-1}e^{-i\theta}\bigr)\bigl(1-
q^{2(r+\epsilon_2-\epsilon_1)-1}e^{-i\theta}\bigr)\Bigr\}\\
\cdot\rho^{(r)\pm}_l\bigl(e^{i\theta}\bigr)
\end{multline}
with $\rho^{(r)\pm}_l(e^{i\theta})$ given by \eqref{E:rho3}. 
Multiplying equation \eqref{E:rhoco1} with
\begin{multline*}
q^{\mp2\sigma_1}\theta_{q^2}\bigl(-q^{2(r+\epsilon_2\mp\sigma_1+1)}e^{-i\theta},
-q^{2(\epsilon_1\pm\sigma_1)+1}\bigr)\cdot\\
\bigl(q^{\mp2\sigma_1+2\sigma_2+1}e^{i\theta},q^{\mp2\sigma_1-2\sigma_2+1}e^{i
\theta},q^{2(\epsilon_2-\epsilon_1+\sigma_1+\sigma_2+1)},q^{2(\epsilon_2-
\epsilon_1+\sigma_1-\sigma_2+1)};q^2\bigr)_{\infty},
\end{multline*}
the difference between the two sign options reads
\begin{multline}\label{E:varsco1a}
\bigl(1-q^{2(\sigma_1+\sigma_2)+1}e^{i\theta}\bigr)\bigl(1-q^{2(\sigma_1-
\sigma_2)+1}e^{i\theta}\bigr)\bigl(1-q^{2(\sigma_2-\sigma_1)+1}e^{i\theta}
\bigr)\bigl(1-q^{-2(\sigma_1+\sigma_2)+1}e^{i\theta}\bigr)\cdot\\
q^{-2(r+\epsilon_2)}e^{i\theta}\bigl(1-q^{2(r+\epsilon_2-\epsilon_1)+1}
e^{i\theta}\bigr)\varsigma^{(r)}_l\bigl(q^2e^{i\theta})=\\
q^{-2(l+\epsilon_1)+2}\bigl(1-q^2e^{2i\theta}\bigr)s_{l-1}\bigl(q^{2\epsilon_1},
q^{2\sigma_1}\bigr)s_{l+r-1}\bigl(q^{2\epsilon_2},q^{2\sigma_2}\bigr)
\varsigma^{(r)}_{l-1}(e^{i\theta})\\
-\Bigl\{\bigl(q^{-2(l+\epsilon_1)+1}+q^{2(l+r+
\epsilon_2)}e^{i\theta}\bigr)\bigl(1-q^2e^{2i\theta}\bigr)-q^2e^{2i
\theta}\bigl(1-q^{2(r+\epsilon_2-\epsilon_1)-1}e^{-i\theta}\bigr)\cdot\\
\bigl(q^{2\sigma_1}+q^{-2\sigma_1}\bigr)+qe^{i\theta}\bigl(1-q^{2(r+\epsilon_2-
\epsilon_1)+1}e^{i\theta}\bigr)\bigl(q^{2\sigma_2}+q^{-2\sigma_2}\bigr)
\Bigr\}\varsigma^{(r)}_l(e^{i\theta}),
\end{multline}
provided that $q^{2(l+\epsilon_1\pm\sigma_1)-1}<1$. Here the definition of
$\varsigma^{(r)}_l$ according to equation \eqref{E:varsigma} has been taken into
account.

In view of the expression \eqref{E:rho1b} for $\rho^{(r)\pm}_l(e^{i\theta})$,
the relation \eqref{E:rhoco1} with the replacement $\theta\to-\theta$ is
satisfied as well. In the case $q^{2(l+\epsilon_2\pm\sigma_1)-1}<1$, it leads to
\begin{multline}\label{E:varsco1b}
q^{2(r+\epsilon_2+2)}e^{-3i\theta}\bigl(1-q^{2(r+\epsilon_2-\epsilon_1)+1}e^{
-i\theta}\bigr)\varsigma^{(r)}_l\bigl(q^{-2}e^{i\theta}\bigr)=\\
q^{-2(l+\epsilon_1)+2}\bigl(1-q^2e^{-2i\theta}\bigr)s_{l-1}\bigl(q^{2\epsilon_1},
q^{2\sigma_1}\bigr)s_{l+r-1}\bigl(q^{2\epsilon_2},q^{2\sigma_2}\bigr)\varsigma^{
(r)}_{l-1}\bigl(e^{i\theta}\bigr)\\
-\Bigl\{\bigl(q^{-2(l+\epsilon_1)+1}+q^{2(l+r+\epsilon_2)}e^{-i\theta}\bigr)
\bigl(1-q^2e^{-2i\theta}\bigr)
-q^2e^{-2i\theta}\bigl(1-q^{2(r+\epsilon_2-\epsilon_1)-1}e^{i\theta}
\bigr)\cdot\\
\bigl(q^{2\sigma_1}+q^{-2\sigma_1}\bigr)+qe^{-i\theta}\bigl(1-q^{2(r+
\epsilon_2-\epsilon_1)+1}e^{-i\theta}\bigr)
\bigl(q^{2\sigma_2}+q^{-2\sigma_2}\bigr)\Bigr\}\varsigma^{(r)}_l\bigl(e^{i\theta}
\bigr).
\end{multline} 
Alternatively, the expression
\eqref{E:varsig2} for $\varsigma^{(r)}_l(e^{i\theta})$ can be employed. 
Then the relations \eqref{E:3phi2cont2} and
\eqref{E:3phi2cont3} with $A\to q^{-2(r+\epsilon_2-\epsilon_1)+1}e^{i\theta}$,
$B\to q^{2(\sigma_2-\sigma_1)+1}e^{i\theta}$, $C\to q^{-2(\sigma_1+\sigma_2)
+1}e^{i\theta}$, $D\to-q^{-2(l+r+\epsilon_2+\sigma_1)+2}e^{i\theta}$, $E\to 
q^2e^{2i\theta}$ give rise to the equations \eqref{E:varsco1a} and \eqref{E:varsco1b} in the case $q^{-2(l+\epsilon_1-\sigma_1)+1}<1$.

The contiguous relation \eqref{E:rholll} combines $\rho^{(r)\pm
}_{l-1}(e^{i\theta})$, $\rho^{(r)\pm}_{l+1}(e^{i\theta})$ and $\rho^{(r)\pm}_l
(e^{i\theta})$. A linear combination of both sign options following equation \eqref{E:varsigma} yields the corresponding relation for 
$\varsigma^{(r)}_l(e^{i\theta})$ given by
\begin{multline}\label{E:varsco2}
q^{-2(l+\epsilon_1)+2}s_{l-1}\bigl(q^{2\epsilon_1},q^{2\sigma_1}\bigr)s_{l+r-1}
\bigl(q^{2\epsilon_2},q^{2\sigma_2}\bigr)\varsigma^{(r)}_{l-1}\bigl(e^{i\theta}
\bigr)\\
\shoveright{
+q^{-2(l+\epsilon_1+1)}s_l\bigl(q^{2\epsilon_1},q^{2\sigma_1}\bigr)s_{l+r}\bigl(
q^{2\epsilon_2},q^{2\sigma_2}\bigr)\varsigma^{(r)}_{l+1}\bigl(e^{i\theta}\bigr)}\\
-\Bigl\{q^{-2(l+\epsilon_1)-1}(1+q^2)+q^{2(l+r+\epsilon_2)}\bigl(e^{i\theta}+e^{
-i\theta}\bigr)+q^{2\sigma_1}+q^{-2\sigma_1}+q^{2(r+\epsilon_2-\epsilon_1)}\cdot\\
\bigl(q^{2\sigma_2}+q^{-2\sigma_2}\bigr)\Bigr\}\varsigma^{(r)}_l\bigl(e^{i\theta}
\bigr)=0.
\end{multline}

According to the definition of the coproduct, the modules $W^{(\epsilon_1,
q^{2\sigma_1})}(\tilde z^{\frac{1}{2}})$ and $W^{(\epsilon_2,q^{2\sigma_2})*}(
\tilde z^{\,-\frac{1}{2}})$
by \eqref{E:co}, \eqref{E:Wmod} and \eqref{E:Wdual}, the action of $\Delta_{
\tilde z^{\frac{1}{2}}}(e_1e_0+y_{\pm}e_0e_1)$ on the finite sum
\begin{equation*}
\sum_{l=-M}^N\varsigma^{(r)}_l(e^{i\theta})w^{(\epsilon_1,q^{2\sigma_1})}_l
\otimes w^{(\epsilon_2,q^{2\sigma_2})*}_{l+r}
\end{equation*}
is given by
\begin{multline}\label{E:e1e0act1}
(1-q^2)^2\Delta_{\tilde z^{\frac{1}{2}}}
\bigl(e_1e_0+y_{\pm}e_0e_1\bigr)\sum_{l=-M}^N\varsigma^{(r)}_l(e^{i\theta})
w^{(\epsilon_1,q^{2\sigma_1})}_l\otimes w^{(\epsilon_2,q^{2\sigma_2})*}_{l+r}=\\
\sum_{l=-M}^N\Bigl\{(1+y_{\pm}q^2)\tilde z^{\frac{1}{2}}q^{-2(l+\epsilon_1)+2}
s_{l-1}
\bigl(q^{2\epsilon_1},q^{2\sigma_1}\bigr)s_{l+r-1}\bigl(q^{2\epsilon_2},q^{2
\sigma_2}\bigr)\varsigma^{(r)}_{l-1}(e^{i\theta})\\
+(q^2+y_{\pm})\tilde z^{\,-\frac{1}{2}}q^{-2(l+r+\epsilon_2)}s_l\bigl(q^{2
\epsilon_1},q^{2\sigma_1}\bigr)s_{l+r}\bigl(q^{2\epsilon_2},q^{2\sigma_2}\bigr)
\varsigma^{(r)}_{l+1}(e^{i\theta})\\
\shoveleft{
-\Bigl[\tilde z^{\frac{1}{2}}q^{-2(l+\epsilon_1)+1}\Bigl(s^2_l\bigl(q^{2
\epsilon_1},q^{2\sigma_1}\bigr)+y_{\pm}q^2s^2_{l-1}\bigl(q^{2\epsilon_1},q^{2\sigma_1}
\bigr)\Bigr)}\\
+\tilde z^{\,-\frac{1}{2}}q^{-2(l+r+\epsilon_2)+1}\Bigl(q^2s^2_{l+r-1}\bigl(
q^{2\epsilon_2},q^{2\sigma_2}\bigr)+y_{\pm}s^2_{l+r}\bigl(q^{2\epsilon_2},q^{2\sigma_2}
\bigr)\Bigr)\Bigr]\varsigma^{(r)}_l(e^{i\theta})\Bigr\}\\
\shoveright{
\cdot w^{(\epsilon_1,q^{2\sigma_1})}_l\otimes w^{(\epsilon_2,q^{2\sigma_2})*}_{
l+r}}\\
-(1+y_{\pm}q^2)\tilde z^{\frac{1}{2}}q^{2(M-\epsilon_1+1)}s_{-M-1}\bigl(q^{2
\epsilon_1},q^{2\sigma_1}\bigr)s_{-M-1+r}\bigl(q^{2\epsilon_2},q^{2\sigma_2}
\bigr)\varsigma^{(r)}_{-M-1}(e^{i\theta})\cdot\\
\shoveright{w^{(\epsilon_1,q^{2\sigma_1})}_{-M}
\otimes w^{(\epsilon_2,q^{2\sigma_2})*}_{-M+r}}\\
+(q^2+y_{\pm})\tilde z^{\,-\frac{1}{2}}q^{2(M-r-\epsilon_2+1)}s_{-M-1}\bigl(q^{
2\epsilon_1},q^{2\sigma_1}\bigr)s_{-M-1+r}\bigl(q^{2\epsilon_2},q^{2\sigma_2}
\bigr)\varsigma^{(r)}_{-M}(e^{i\theta})\cdot\\
\shoveright{w^{(\epsilon_1,q^{2\sigma_1})}_{-M-1}
\otimes w^{(\epsilon_2,q^{2\sigma_2})*}_{-M-1+r}}\\
+(1+y_{\pm}q^2)\tilde z^{\frac{1}{2}}q^{-2(N+\epsilon_1)}s_N\bigl(q^{2
\epsilon_1},q^{2
\sigma_1}\bigr)s_{N+r}\bigl(q^{2\epsilon_2},q^{2\sigma_2}\bigr)\varsigma^{(r)}_N
(e^{i\theta})w^{(\epsilon_1,q^{2\sigma_1})}_{N+1}\otimes w^{(\epsilon_2,q^{2
\sigma_2})*}_{N+r+1}\\
-(q^2+y_{\pm})\tilde z^{\,-\frac{1}{2}}q^{-2(N+r+\epsilon_2)}s_N\bigl(q^{2
\epsilon_1},q^{2\sigma_1}\bigr)s_{N+r}\bigl(q^{2\epsilon_2},q^{2\sigma_2}\bigr)
\varsigma^{(r)}_{N+1}(e^{i\theta})\cdot\\
w^{(\epsilon_1,q^{2\sigma_1})}_N\otimes w^{(\epsilon_2,q^{2\sigma_2}
)*}_{N+r}.
\end{multline}
By means of the relations \eqref{E:varsco1a}, \eqref{E:varsco1b} and \eqref{E:varsco2}, the sum on the rhs is expressed by
\begin{multline}\label{E:e1e0act2}
\sum_{l=-M}^N\biggl\{\Bigl[(1+{y}_{\pm}q^2)\tilde z^{\frac{1}{2}}-(q^2+y_{\pm})\tilde z^{\,
-\frac{1}{2}}q^{-2(r+\epsilon_2-\epsilon_1)+2}\Bigr]e^{\pm3i
\theta}\cdot\\
\shoveright{\lambda_{\pm}
\bigl(1-q^{2(r+\epsilon_2-\epsilon_1)+1}e^{\pm i\theta}\bigr)\varsigma^{(r)
}_l\bigl(q^{\pm2}e^{i\theta}\bigr)}\\
-\tilde z^{\frac{1}{2}}\bigl(1-z^{-1}qe^{\mp i\theta}\bigr)
\Bigl[q^2\bigl(1-q^{2(r+\epsilon_2-\epsilon_1)-3}e^{\pm i\theta}\bigr)+y_{\pm}\bigl(1-
q^{2(r+\epsilon_2-\epsilon_1)+1}e^{\pm i\theta}\bigr)\Bigr]\cdot\\
\shoveright{q^{2(l+\epsilon_1)+1}\varsigma^{(r)}_l(e^{i\theta})}\\
-(1+y_{\pm})q^2\Bigl[\tilde z^{\frac{1}{2}}\bigl(q^{2\sigma_1}+q^{-2\sigma_1}\bigr)
+\tilde z^{\,-\frac{1}{2}}\bigl(q^{2\sigma_2}+q^{-2\sigma_2}\bigr)\Bigr]
\varsigma^{(r)}_l(e^{i\theta})\\
-q^2e^{\pm2i\theta}\Bigl[(1+y_{\pm}q^2)\tilde z^{\frac{1}{2}}\bigl(q^{2\sigma_1}+
q^{-2\sigma_1}\bigr)+(q^2+y_{\pm})\tilde z^{\,-\frac{1}{2}}
q^{-2(r+\epsilon_2-\epsilon_1)+1}e^{\mp i\theta}
\bigl(q^{2\sigma_2}+q^{-2\sigma_2}\bigr)\Bigr]\cdot\\
\shoveright{
\frac{1-q^{2(r+\epsilon_2-\epsilon_1)-1}e^{\mp i\theta}}{1-q^2e^{\pm2i\theta}}
\varsigma^{(r)}_l(e^{i\theta})}\\
+qe^{\pm i\theta}\Bigl[(1+y_{\pm}q^2)\tilde z^{\frac{1}{2}}\bigl(q^{2\sigma_2}
+q^{-2\sigma_2}\bigr)+(q^2+y_{\pm})\tilde z^{\,-\frac{1}{2}}q^{-2(r+
\epsilon_2-\epsilon_1)
+1}e^{\mp i\theta}\bigl(q^{2\sigma_1}+q^{-2\sigma_1}\bigr)\Bigr]\cdot\\
\frac{1-q^{2(r+\epsilon_2-\epsilon_1)+1}e^{\pm i\theta}}{1-q^2e^{\pm2i\theta}}
\varsigma^{(r)}_l(e^{i\theta})\biggr\}w^{(\epsilon_1,q^{2\sigma_1})}_l\otimes 
w^{(\epsilon_2,q^{2\sigma_2})*}_{l+r},
\end{multline}
where $\lambda_-=q^{2(r+\epsilon_2+2)}$ and
\begin{multline*}
\lambda_+=q^{-2(r+\epsilon_2)}e^{-2i\theta}\cdot\\
\bigl(1-q^{2(\sigma_1+\sigma_2)+1}e^{i\theta}\bigr)\bigl(1-q^{2(\sigma_1-\sigma_2)
+1}e^{i\theta}\bigr)\bigl(1-q^{2(\sigma_2-\sigma_1)+1}e^{i\theta}\bigr)\bigl(1-
q^{-2(\sigma_1+\sigma_2)+1}e^{i\theta}\bigr).
\end{multline*}
Setting 
\begin{equation}\label{E:xych}
y_{\pm}=-\frac{q^2\bigl(1-q^{2(r+\epsilon_2-\epsilon_1)-3}e^{\pm i\theta}\bigr)}{1-q^{2(r+\epsilon_2-\epsilon_1)+1}e^{\pm i\theta}},
\end{equation}
the second contribution to \eqref{E:e1e0act2} vanishes. With the choice
\eqref{E:xych}, multiplication by $a_r(e^{i\theta})$ and
application of $R(\tilde z^{\frac{1}{2}},z^{\frac{
1}{2}})$ on both sides of equation \eqref{E:e1e0act1} making use of
\eqref{E:e1e0act2} yields
\begin{multline}\label{E:Re1e0}
\frac{1-q^2}{1+q^2}R(\tilde z^{\frac{1}{2}},z^{\frac{1}{2}})\Delta_{
\tilde z^{\frac{1}{2}}}\bigl(e_1e_0+y_{\pm}e_0e_1)a_r(e^{i\theta})\sum_{l=-M}^N
\varsigma^{(r)}_l(e^{i\theta})w^{(\epsilon_1,q^{2\sigma_1})}_l\otimes 
w^{(\epsilon_2,q^{2\sigma_2})*}_{l+r}\\
=\sum_{k=-\infty}^{\infty}w^{(\epsilon_3,q^{2\sigma_3})}_k\otimes w^{(\epsilon_4,
q^{2\sigma_4})*}_{k+r}\cdot\\
\shoveleft{\biggl\{\lambda_{\pm}
\tilde z^{\frac{1}{2}}\bigl(1-z^{-1}qe^{\pm i\theta}\bigr)
e^{\pm3i\theta}\frac{a_r(e^{i\theta})}{a_r(q^{\pm2}e^{i\theta})}
\boldsymbol{\tau}^{(r,k;N,M)}
\bigl(q^{\pm2}e^{i\theta}\bigr)}\\
-\frac{q^2\bigl(1+e^{\pm2i\theta}\bigr)}{(1+q^2)\bigl(1-q^2e^{\pm2i\theta}\bigr)}
\Bigl[\tilde z^{\frac{1}{2}}\bigl(q^{2\sigma_1}+q^{-2\sigma_1}\bigr)+
\tilde z^{\,-\frac{1}{2}}\bigl(q^{2\sigma_2}+q^{-2\sigma_2}\bigr)\Bigr]
\boldsymbol{\tau}^{(r,k;N,M)}(e^{i\theta})\\
+\frac{qe^{\pm i\theta}}{1-q^2e^{\pm2i\theta}}\Bigl[\tilde z^{\frac{1}{2}}\bigl(
q^{2\sigma_2}+q^{-2\sigma_2}\bigr)+\tilde z^{\,-\frac{1}{2}}\bigl(q^{2\sigma_1}
+q^{-2\sigma_1}\bigr)\Bigr]\boldsymbol{\tau}^{(r,k;N,M)}(e^{i\theta})\\
+a_r(e^{i\theta})\bigl(1-q^{2(r+\epsilon_2-\epsilon_1)+1}e^{\pm i\theta}\bigr)^{-1}
q^{-2(N+\epsilon_1)}s_N\bigl(q^{2\epsilon_1},q^{2\sigma_1}
\bigr)s_{N+r}\bigl(q^{2\epsilon_2},q^{2\sigma_2}\bigr)\cdot\\
\shoveright{
\Bigl[\tilde z^{\frac{1}{2}}\mathsf R^{k,k+r^*}_{N+1,N+r+1^*}\varsigma^{(r)}_N
(e^{i\theta})-\tilde z^{\,-\frac{1}{2}}q^{-1}e^{\pm i\theta}
\mathsf R_{N,N+r^*}^{k,k+r^*}\varsigma^{(r)}_{N+1}(e^{i\theta})\Bigr]}\\
-a_r(e^{i\theta})\bigl(1-q^{2(r+\epsilon_2-\epsilon_1)+1}e^{\pm i\theta}\bigr)^{-1}
q^{2(M-\epsilon_1+1)}s_{-M-1}\bigl(q^{2\epsilon_1},q^{2\sigma_1}\bigr)s_{-M-1+r}
\bigl(q^{2\epsilon_2},q^{2\sigma_2}\bigr)\\
\cdot\Bigl[\tilde z^{\frac{1}{2}}\mathsf R^{k,k+r^*}_{-M,-M+r^*}\varsigma^{
(r)}_{-M-1}(e^{i\theta})-\tilde z^{\,-\frac{1}{2}}q^{-1}e^{\pm i\theta}\mathsf R_{
-M-1,-M-1+r}^{k,k+r^*}\varsigma^{(r)}_{-M}(e^{i\theta})\Bigr]\biggr\}.
\end{multline} 
Referring to the intertwining property \eqref{E:int}
and the definition of $\boldsymbol{\tau}^{(r,k;N,M)}(e^{i\theta})$ by
\eqref{E:taufin}, the lhs of equation \eqref{E:Re1e0} is rewritten by
\begin{multline}\label{E:e1e0R}
\frac{1-q^2}{1+q^2}\Delta_{z^{-\frac{1}{2}}}\bigl(e_1e_0+y_{\pm}e_0e_1\bigr)\sum_{
k=-\infty}^{\infty}\boldsymbol{\tau}^{(r,k;N,M)}(e^{i\theta})w^{(\epsilon_4,
q^{2\sigma_4})}_k\otimes w^{(\epsilon_3,q^{2\sigma_3})*}_{k+r}=\\
\bigl(1-q^{2(r+\epsilon_2-\epsilon_1)+1}e^{\pm i\theta}\bigr)^{-1}\cdot\\
\sum_{k=-\infty}^{\infty}\biggl\{z^{-\frac{1}{2}}q^{-2(k+\epsilon_4)+2}s_{k-1}
\bigl(q^{2\epsilon_4},q^{2\sigma_4}\bigr)s_{k+r-1}\bigl(q^{2\epsilon_3},q^{2
\sigma_3}\bigr)\boldsymbol{\tau}^{(r,k-1;N,M)}(e^{i\theta})\\
+z^{\frac{1}{2}}q^{-2(k+\epsilon_4)-1}e^{\pm i\theta}s_k\bigl(q^{2\epsilon_4},q^{2
\sigma_4}\bigr)s_{k+r}\bigl(q^{2\epsilon_3},q^{2\sigma_3}\bigr)
\boldsymbol{\tau}^{(r,k+1;N,M)}(e^{i\theta})\\
-z^{\frac{1}{2}}\Bigl[q^{-2(k+\epsilon_4)}e^{\pm i\theta}+\tfrac{q^2}{1+q^2}\bigl(1+q^{2(
r+\epsilon_2-\epsilon_1)-1}e^{\pm i\theta}\bigr)\bigl(q^{2\sigma_3}+
q^{-2\sigma_3}\bigr)+q^{2(k+r+\epsilon_3)+1}\Bigr]\cdot\\
\shoveright{\boldsymbol{\tau}^{(r,k;N,M)}(e^{i\theta})}\\
-z^{-\frac{1}{2}}\Bigl[q^{-2(k+\epsilon_4)+1}+\tfrac{q^2}{1+q^2}\bigl(1+q^{
2(r+\epsilon_2-\epsilon_1)-1}e^{\pm i\theta}\bigr)\bigl(q^{2\sigma_4}+q^{-2\sigma_4}
\bigr)+q^{2(k+r+\epsilon_3)}e^{\pm i\theta}\Bigr]\cdot\\
\boldsymbol{\tau}^{(r,k;N,M)}(e^{i\theta})\biggr\}w^{(\epsilon_4,q^{2\sigma_4})*}_k
\otimes w^{(\epsilon_3,q^{2\sigma_3})*}_{k+r}.
\end{multline}
Equating the coefficients of $w^{(\epsilon_4,q^{2\sigma_4})}_k\otimes w^{(
\epsilon_3,q^{2\sigma_3})*}_{k+r}$ on the rhs of \eqref{E:Re1e0} and \eqref{E:e1e0R}, the limit $N,M\to\infty$ gives rise to an inhomogeneous relation satisfied by the sums $\boldsymbol{\tau}^{(r,k)}(e^{i\theta})$ and $\boldsymbol{\tau}^{(r,k)}(q^{\pm2}e^{i\theta})$.
According to the results \eqref{E:Rllim-} and \eqref{E:rholim-comb},
the last two lines of the expression \eqref{E:Re1e0}
vanish in the limit $M\to\infty$ provided that $\vert zq^{\pm1}e^{i\theta}
\vert<1$. In the limit $N\to\infty$, the two preceeding lines 
tend towards a finite limit expressed by
\begin{multline*}
\tilde z^{\frac{1}{2}}\bigl(1-z^{-1}q^{-1}e^{\pm i\theta}\bigr)
\bigl(1-q^{2(r+\epsilon_2-\epsilon_1)+1}e^{\pm i\theta}\bigr)^{-1}
q^{-2\epsilon_1+1}\cdot\\
\lim_{N\to\infty}\bigl(q^{-N}\mathsf R^{k,k+r^*}_{N,N+r^*}
\bigr)\lim_{N'\to\infty}\bigl(q^{-N'}a_r(e^{i\theta})\varsigma^{(r)}_{N'}(e^{i
\theta})\bigr)
\end{multline*}
with the two limits in the second line specified by
\eqref{E:Rllim+} and \eqref{E:varsigma} supplemented by \eqref{E:rholim+}.
The first limit depends on the parameters $\epsilon_1$, $\epsilon_2$
and $q^{2\sigma_1}$, $q^{2\sigma_2}$
only through the difference $\epsilon_2-\epsilon_1$ and through the factors
$q^{2(\pm\sigma_1+\sigma_2)}=q^{2(\pm\sigma_3+\sigma_4)}$ and $q^{2(\pm\sigma_1
-\sigma_2)}=q^{2(\pm\sigma_3-\sigma_4)}$. Due to the particular choices of
$a_r(e^{i\theta})$, the second limit factorizes into a part with the same
properties and a part invariant under the exchange $\epsilon_1\leftrightarrow
\epsilon'_1$, $\epsilon_2\leftrightarrow\epsilon'_2$, $q^{2\sigma_1}
\leftrightarrow\alpha\alpha'q^{2\sigma_1}$, $q^{2\sigma_2}\leftrightarrow\alpha 
\alpha'q^{2\sigma_2}$.
Therefore, multiplying the 
inhomogeneous relation by $\alpha q^{2\epsilon_1}$ and subtracting the 
same relation for the parameters $\epsilon'_1,\epsilon'_2$, $q^{2
\sigma'_1}=\alpha\alpha'q^{2\sigma_2}$, $q^{2\sigma'_2}=\alpha\alpha'q^{2
\sigma_2}$ with $\epsilon'_2-\epsilon'_1=\epsilon_2-\epsilon_1$ yields
the homogeneous relation for $\boldsymbol{\tau}^{
(r,k)}(e^{i\theta})-\alpha\alpha'q^{2(\epsilon'_1-\epsilon_1)}\boldsymbol{\tau}'^{
(r,k)}(e^{i\theta})={\Xi}^{(r,k)}(e^{i\theta})$ and ${\Xi}^{(r,k)}(q^{\pm2}e^{i\theta})$ given by \eqref{E:Xitheta}.

\subsection{A further relation}\label{S:Xiz}
In the following
the $z$-dependence of $R^{k,k+r^*}_{l,l+r^*}$ and $r^{k,k+r^*}_{l,l+r^*}$ is
indicated writing $R^{k,k+r^*}_{l,l+r^*}(z)$ and $r^{k,k+r^*}_{l,l+r^*}(z)$.
Moreover, the last sums in \eqref{E:hatnot1} and \eqref{E:hatnot2} are denoted as $\widehat{\tau}^{(r,k)}(z,e^{i\theta})$ and 
$\widehat{\tau}'^{(r,k)}(z,e^{i\theta})$.
Referring to the expression \eqref{E:RW1} for $R^{k,k+r^*}_{l,l+r^*}(z)$ in the case $\vert q^{-2(k+r+\epsilon_3-\sigma_3)+1}\vert<1$,
the contiguous relation [\cite{isra}:2.2] with $a\to -\alpha q^{2(l+
\epsilon_1-\sigma_3+2\sigma_4)+1}$, $b\to zq^{2(\sigma_4-\sigma_3+1)}$, $c\to z^{-1}q^{2(\sigma_4-\sigma_3)}$, $d\to-\alpha q^{2(l+\epsilon_1-\sigma_3)+1}$,
$e\to-\alpha q^{2(l+r+\epsilon_2+\sigma_4)+1}$, $f\to-q^{2(k+\epsilon_4+\sigma_4
)+1}$ combined with [\cite{isra}:2.3] with $a\to-\alpha q^{2(l+\epsilon_1-\sigma_3+2
\sigma_4)+1}$, $b\to-q^{2(k+\epsilon_4+\sigma_4)+3}$, $c\to z^{-1}q^{2(\sigma_4
-\sigma_3)}$, $d\to-\alpha q^{2(l+\epsilon_1-\sigma_3)+1}$, $e\to-\alpha q^{2(
l+r+\epsilon_2+\sigma_4)+1}$, $f\to zq^{2(\sigma_4-\sigma_3+1)}$ yields
\begin{multline}\label{E:Rzz}
\frac{\bigl(1-zq^{-2(r+\epsilon_2-\epsilon_1)}\bigr)\bigl(1-z^{-1}q^{-2(
r+\epsilon_2-\epsilon_1)}\bigr)\bigl(1-zq^{2(\sigma_3-\sigma_4)}\bigr)
\bigl(1-zq^{-2(\sigma_3+\sigma_4)}\bigr)}{1-zq^{-2(\epsilon_2-\epsilon_1)}}\cdot\\
q^{2\sigma_4-2k}\,R^{k,k+r^*}_{l,l+r^*}(zq^{-2})=\\
\nu_k\bigl(1-\alpha zq^{2(l-k+\epsilon_1-\epsilon_4)}\bigr)R^{k,k+r^*}_{l,l+r^*}(z)
+z^2\lambda_kq^{2(k+\epsilon_4+1)}\bigl(1-\alpha z^{-1}q^{2(l-k+
\epsilon_1-\epsilon_4-1)}\bigr)\cdot\\
R^{k+1,k+r+1^*}_{l,l+r^*}(z)
\end{multline}
with $\lambda_k=(z-z^{-1})s_{-k-r-1}(q^{-2\epsilon_3},q^{2\sigma_3})s_{-k-1}(q^{-2
\epsilon_4},q^{2\sigma_4})$ and
\begin{multline*}
\nu_k=-z^{-1}q^{-2\sigma_3}\bigl(1-zq^{
-2(r+\epsilon_2-\epsilon_1)}\bigr)\bigl(1-zq^{2(\sigma_3+\sigma_4)}\bigr)
\bigl(1-zq^{2(\sigma_3-\sigma_4)}\bigr)\\
-q^{-2\sigma_3}(z-z^{-1})\bigl(1+q^{-2(k+r+\epsilon_3-\sigma_3)-1}\bigr)
\bigl(1+zq^{2(k+\epsilon_4+\sigma_3)+1}\bigr).
\end{multline*}
For $\vert q^{2(k+r+\epsilon_3+\sigma_3)+1}\vert<1$, equation \eqref{E:Rzz}
is shown applying the relation [\cite{isra}:2.3] followed by [\cite{isra}:2.2] 
to the $_8W_7$-series in \eqref{E:hatR} and \eqref{E:breveR}.

According to the definition by \eqref{E:hatRel}, the rhs of \eqref{E:Rzz} with 
each $R^{m,m+r^*}_{l,l+r^*}(\tilde z)$ replaced by 
$r^{m,m+r^*}_{l,l+r^*}(\tilde z)$ equals
\begin{multline*}
\alpha\alpha'z^{-1}q^{-2(k+\epsilon_3-\epsilon_2-\epsilon'_2)+1}\frac{\bigl(
1-zq^{-2(r+\epsilon_2-\epsilon_1)}\bigr)\bigl(1-z^{-1}q^{-2(r+\epsilon_2-\epsilon_1
)}\bigr)}{1-zq^{2(\epsilon_1-\epsilon_2)}}\cdot\\
\bigl(1-zq^{2(\sigma_3+\sigma_4)}\bigr)\bigl(1-zq^{2(\sigma_3-\sigma_4)}\bigr)
\bigl(1-zq^{2(\sigma_4-\sigma_3)}\bigr)\bigl(1-zq^{-2(\sigma_3+\sigma_4)}\bigr)
r^{k,k+r^*}_{l,l+r^*}(zq^{-2}).
\end{multline*}
Dividing the relations \eqref{E:varsco1a} and \eqref{E:varsco1b} by
$1-q^2e^{2i\theta}$ and $1-q^2e^{-2i\theta}$, respectively, the difference
reads
\begin{multline}\label{E:varsigtheta}
\lambda_+e^{3i\theta}\frac{1-q^{2(r+\epsilon_2-\epsilon_1)+1}e^{i\theta}}{\bigl(1-q^2
e^{2i\theta}\bigr)\bigl(e^{i\theta}-e^{-i\theta}\bigr)}\varsigma^{(r)}_l(q^2e^{i
\theta})\\
-\lambda_-e^{-3i\theta}\frac{1-q^{2(r+\epsilon_2-\epsilon_1)+1}e^{-i\theta}}{
\bigl(1-q^2
e^{-2i\theta}\bigr)\bigl(e^{i\theta}-e^{-i\theta}\bigr)}\varsigma^{(r)}_l(q^{-2}
e^{i\theta})+q^{2(l+r+\epsilon_2)}\varsigma^{(r)}_l(e^{i\theta})=\\
\frac{\alpha q^2}{\bigl(1-q^2e^{2i\theta}\bigr)\bigl(1-q^2e^{-2i\theta}\bigr)}
\Bigl\{\Bigl[e^{i\theta}+e^{-i\theta}-q^{2(r+\epsilon_2-\epsilon_1)}(q+q^{-1})
\Bigr]\bigl(q^{2\sigma_3}+q^{-2\sigma_3}\bigr)\\
-\Bigl[q+q^{-1}-q^{2(r+\epsilon_2-\epsilon_1)}(e^{i\theta}+e^{-i\theta})\Bigr]
\bigl(q^{2\sigma_4}+q^{-2\sigma_4}\bigr)\Bigr\}\varsigma^{(r)}_l(e^{i\theta})\\
\end{multline} 
with $\varsigma^{(r)}(e^{i\theta})$ defined by equation \eqref{E:varsigma}.
Multiplying the contiguous relation for $r^{m,m+r^*}_{l,l+r^*}
(z)$ by $a_r(e^{i\theta})\varsigma^{(r)}_l(e^{i\theta})$ and making use of \eqref{E:varsigtheta}
to eliminate the factor $q^{2l}$ leads to
\begin{multline}\label{E:Xiz}
\alpha\alpha'z^{-1}q^{-2(k+\epsilon_3-\epsilon_2-\epsilon'_2)+1}
\frac{\bigl(1-zq^{-2(r+\epsilon_2-\epsilon_1)}\bigr)\bigl(1-z^{-1}q^{-2(r+\epsilon_2
-\epsilon_1)}\bigr)}{1-zq^{-2(\epsilon_2-\epsilon_1)}}\cdot\\
\bigl(1-zq^{2(\sigma_3+\sigma_4)}\bigr)\bigl(1-zq^{2(\sigma_3-\sigma_4)}\bigr)
\bigl(1-zq^{2(\sigma_4-\sigma_3)}\bigr)\bigl(1-zq^{-2(\sigma_3+\sigma_4)}\bigr)
\widehat{\tau}^{(r,k)}(zq^{-2},e^{i\theta})\\
\shoveleft{
=\nu_k\,\widehat{\tau}^{(r,k)}(z,e^{i\theta})+z^2\lambda_kq^{2(k+
\epsilon_4+1)}\,\widehat{\tau}^{(r,k+1)}(z,e^{i\theta})}\\
\shoveleft{+\alpha z q^{-2(r+\epsilon_2-\epsilon_1)}
\biggl\{\lambda_+e^{3i\theta}\frac{1-q^{2(r+
\epsilon_2-\epsilon_1)+1}e^{i\theta}}{\bigl(1-q^2e^{2i\theta}\bigr)\bigl(e^{i\theta}
-e^{-i\theta}\bigr)}\frac{a_r(e^{i\theta})}{a_r(q^2e^{i\theta})}\cdot}\\
\shoveright{\Bigl[\lambda_k\widehat{\tau}^{(r,k+1)}(z,q^2e^{2i\theta})
+\nu_kq^{-2(k+\epsilon_4)}\widehat{\tau}^{(r,k)}
(z,q^2e^{i\theta})\Bigr]}\\
\shoveleft{-\lambda_-e^{-3i\theta}\frac{1-q^{2(r+\epsilon_2-\epsilon_1)+1}e^{
-i\theta}}{\bigl(1-q^2e^{-2i\theta}\bigr)\bigl(e^{i\theta}-e^{-i\theta}\bigr)}
\frac{a_r(e^{i\theta})}{a_r(q^{-2}e^{i\theta})}\cdot}\\
\shoveright{
\Bigl[\lambda_k\widehat{\tau}^{(r,k+1)}(z,q^{-2}e^{i\theta})+\nu_kq^{-2(k+\epsilon_4
)}\widehat{\tau}^{(r,k)}(z,q^{-2}e^{i\theta})\Bigr]}\\
-\frac{q^2}{\bigl(1-q^2e^{2i\theta}\bigr)\bigl(1-q^2e^{-2i\theta}\bigr)}\biggl(
\Bigl[e^{i\theta}+e^{-i\theta}-q^{2(r+\epsilon_2-\epsilon_1)}(q+q^{-1})\Bigr]
\bigl(q^{2\sigma_3}+q^{-2\sigma_3}\bigr)\\
\shoveright{
-\Bigl[q+q^{-1}-q^{2(r+\epsilon_2-\epsilon_1)}(e^{i\theta}+e^{-i\theta})\Bigr]
\bigl(q^{2\sigma_4}+q^{-2\sigma_4}\bigr)\biggr)\cdot}\\
\bigl[\lambda_k\widehat{\tau}^{(r,k+1)}(z,e^{i\theta})+\nu_kq^{-2(k+\epsilon_4)}
\widehat{\tau}^{(r,k)}(z,e^{i\theta})\Bigr]\biggr\}.
\end{multline}
Due to the specifications by \eqref{E:wprop1}, \eqref{E:ardef3}, \eqref{E:ardef4} and \eqref{E:Xitheta}, the products $\alpha\lambda_+a_r(e^{i\theta})/a_r(q^2e^{i
\theta})$ and $\alpha\lambda_-a_r(e^{i\theta})/a_r(q^{-2}e^{i\theta})$
with $e^{i\theta}=\alpha\alpha'q^{2(s+\epsilon'_1+\epsilon_2)+1}$ or with
$e^{i\theta}=q^{2(s'+\epsilon_2-\epsilon_1)-1}$
coincide for the two sets \eqref{E:twosets} underlying ${\Xi}^{(r,k)}$.
Therefore the sums $\widehat{\tau}^{(r,k)}(\tilde z,e^{i\tilde{\theta}})$
in \eqref{E:Xiz} can be substituted by $\widehat{\tau}^{(r,k)}(\tilde z,e^{i\tilde{\theta}})-\alpha\alpha'q^{2(\epsilon'_2-\epsilon_2)}\widehat{
\tau}'^{(r,k)}(\tilde z,e^{i\tilde{\theta}})$. 

\vskip 0.5cm

\emph{Acknowledgment.} The author would like to thank the referee for 
drawing her attention to reference \cite{isma}.

\end{document}